\PassOptionsToPackage{table}{xcolor}

\documentclass{SciPostTwoColumn}
\usepackage[T1]{fontenc}      
\usepackage[utf8]{inputenc}   
\usepackage{lmodern}
\usepackage{babel}            
\usepackage{hyphenat}         

\usepackage{titlesec}

\titleformat{\section}
  {\normalfont\Large\bfseries\raggedright}
  {\thesection}
  {1em}
  {}

\titleformat{\subsection}
  {\normalfont\large\bfseries\raggedright}
  {\thesubsection}
  {0.5em}
  {}

\titleformat{\subsubsection}
  {\normalfont\normalsize\bfseries\raggedright}
  {\thesubsubsection}
  {0.5em}
  {}

  \titleformat{\paragraph}[runin]
  {\normalfont\normalsize\bfseries\raggedright}
  {\theparagraph}
  {1em}
  {}

  \titlespacing*{\paragraph}
  {0pt}
  {0.8em}
  {0.5em}

\usepackage{amsmath}          
\usepackage{amsfonts}         
\usepackage{amssymb}          
\usepackage{bm}               
\usepackage{braket}           
\usepackage{siunitx}          
\usepackage[version=4]{mhchem}           
\usepackage{mathrsfs}

\usepackage{graphicx}         
\usepackage[table]{xcolor}    
\usepackage{tikz}             
\usetikzlibrary{shapes.geometric, arrows.meta, positioning, shadows, chains, decorations.pathmorphing}
\usepackage{booktabs}         
\usepackage{tabularx}         
\usepackage{multirow}         
\usepackage{makecell}         
\usepackage{array}
\usepackage{ragged2e}         
\usepackage{placeins}
\usepackage{verbatim}         
\hypersetup{
    colorlinks=true,
    citecolor=blue,
    linkcolor=blue,
    urlcolor=blue
}

\usepackage{cleveref}

\newcommand{\I}{\mathrm{i}}

\usepackage{orcidlink}

\usepackage[most]{tcolorbox}
\usepackage{xcolor}

\usepackage[most]{tcolorbox}
\tcbset{
    beamerexample/.style={
        enhanced,
        breakable,
        colback=blue!4,
        colframe=blue!75!black,
        fonttitle=\bfseries,
        coltitle=white,
        top=1.4mm,
        bottom=1.4mm,
        left=1.6mm,
        right=1.6mm,
        before skip=5pt,
        after skip=5pt,
        sharp corners,
        boxrule=0.7pt
    },
    plainlanguage/.style={
        beamerexample,
        colback=black!2,
        colframe=black!65
    },
    discoveryguide/.style={
        beamerexample,
        colback=green!3,
        colframe=green!45!black
    }
}
\newtcolorbox{example}[1][]{beamerexample,title={#1}}

\begin{document}
\makeatletter
\twocolumn[
\begin{@twocolumnfalse}

\begin{center}
{\Large\textbf{Engineering Quantum Interactions: From Effective Models to Validated Physical Predictions}}
\end{center}

\begin{center}
Sinara S. Dourado\textsuperscript{1},
Ciro Micheletti Diniz\textsuperscript{1}~\orcidlink{0000-0002-7602-0468},
Gabriel P. L. M. Fernandes\textsuperscript{1},\\
Rog\'erio J. de Assis\textsuperscript{1},
G. D. de Moraes Neto\textsuperscript{2,$\star$}~\orcidlink{0000-0003-4273-8380},
and Celso J. Villas-Boas\textsuperscript{1}
\end{center}

\begin{center}
\textsuperscript{1}Departamento de F\'isica, Universidade Federal de S\~ao Carlos,
13565-905 S\~ao Carlos, S\~ao Paulo, Brazil\\
\textsuperscript{2}Department of Fundamental Sciences, Hainan Bielefeld
University of Applied Sciences, Danzhou, Hainan 578101, China\par
\vspace{2pt}

\qquad
\textsuperscript{$\star$}\small\sf gentil.neto@hibiuh.edu.cn 
\end{center}

\section*{Abstract}
Effective models make complex quantum dynamics tractable by restricting the
description to the states, processes, and timescales relevant to a specified
physical question. A reliable effective theory, however, requires more than a
truncated Hamiltonian or reduced generator. The retained manifold, initial
state, observables, micromotion, dissipative channels, and eliminated degrees
of freedom must be treated consistently if the resulting model is to predict
measurable quantities. We develop a unified framework for constructing and
validating such predictions using projection methods, Schrieffer--Wolff
transformations, Magnus expansions, adiabatic elimination, small rotations,
time-coarse-graining methods, and Floquet theory. The quantum Rabi model and
its dispersive limits provide recurring benchmarks for identifying control
parameters, virtual processes, spectral corrections, observable dressing,
and breakdown. Higher-order resonances and applications to collective
light--matter coupling, Raman processes, spin-chain buses, lossy mediators,
and driven lattices then show how effective interactions can be interpreted
and designed. Throughout, a physical prediction is organized as three
order-consistent steps -- encoding into the retained description, evolution
under the effective dynamics, and reconstruction of the physical observable.
This structure allows methods to be compared at the level of predictions and
exposes failures caused by leakage, small denominators, missing micromotion,
inconsistent observables, memory effects, or incorrectly transformed
dissipation. The presentation consequently develops three interconnected skills: deriving effective interactions, interpreting the mechanisms that generate them, and using those mechanisms to discover new processes.
Companion notebooks provide worked QuTiP implementations, convergence tests,
and numerical comparisons between microscopic and effective dynamics.
\end{@twocolumnfalse}
]
\makeatother

\vspace{10pt}

\noindent\rule{\linewidth}{1pt}

\tableofcontents

\noindent\rule{\linewidth}{1pt}

\vspace{10pt}

\section{Introduction}
\label{sec:introduction}
An effective Hamiltonian can reproduce the correct energy shift and still
predict the wrong laboratory trajectory. The discrepancy may arise because a
transformed Hamiltonian is combined with an untransformed initial state, the
measured observable is represented in the wrong frame, an eliminated variable
is not reconstructed, or experimentally resolved micromotion has been
discarded. The question addressed throughout this article is therefore not
only how to derive an effective generator, but what additional information is
required for that generator to define a physical prediction.

An effective description must therefore accomplish two complementary tasks: (i) retain the states, frequencies, and processes relevant to the regime of interest, and (ii) connect this reduced description to state preparation and measurement.
This connection determines how states and observables are represented, how eliminated variables are reconstructed, and how changes of frame and dissipative channels are handled. Without it, the spectrum may be correct to the stated order while transition probabilities, phases, or measured signals are not.

Effective descriptions have a long history in nuclear, atomic, molecular,
condensed-matter, and open-system physics
\cite{VanVleck1929,Bloch_Horowitz1958,Feshbach1958,Lwdin1962,Klein1974,Brandow1979,Shavitt1980,Bravyi2011}.
For a Hamiltonian $\hat H=\hat H_0+\lambda\hat V$, ordinary perturbation theory
expands eigenvalues and eigenvectors around the solvable problem $\hat H_0$
\cite{Sakurai,Griffiths_mec_quant}. Degeneracies, near resonances, separated
timescales, periodic driving, dissipation, and dense spectra require more
structured reductions. Projection methods represent the influence of an
eliminated sector through resolvents or memory kernels
\cite{Bloch_Horowitz1958,Feshbach1958,Lwdin1962,GonzalezBallestero2024Projector};
Schrieffer--Wolff transformations perturbatively decouple retained and
eliminated blocks \cite{Schrieffer1966,Bravyi2011}; Magnus and Floquet
constructions organize time-dependent evolution
\cite{Magnus1954,Blanes2009,Bukov2015,Eckardt2017}; and adiabatic elimination
follows the response of fast variables to slow dynamics
\cite{Lugiato1984,Kessler2012DissipativeSW,Reiter2012Effective}. Small
rotations and time-coarse-graining methods exploit closed operator algebras and
harmonic structure in quantum-optical settings \cite{Klimov2000,James2007}.
Modern symbolic constructions extend the range of transformations that can be
generated and checked systematically \cite{reascos2024universal,Huang2026}.

These methods grew out of different physical problems and are often presented
with different notation, objectives, and standards of approximation. Focused
pedagogical accounts introduce projector methods, master equations,
parametrically engineered interactions, and the practical use and limitations
of Lindblad descriptions
\cite{GonzalezBallestero2024Projector,Campaioli2024MasterEquations,Metelmann2023Parametric,Stefanini2026IsLindblad}.
When moving between these subjects, one soon encounters a basic difficulty:
the methods do not always return the same kind of object. An energy-dependent
projected operator, a block-diagonal Hamiltonian in a dressed frame, a
time-averaged generator, a Floquet Hamiltonian, and a reduced Liouvillian may
agree on one quantity through a given order while keeping the remaining
information in different places. Throughout the article, we return to the same
questions: What is retained and what is eliminated? Which virtual transitions
or temporal harmonics generate the effective term? Which states and
observables must be transformed? Where do dressing, micromotion, memory, and
eliminated amplitudes appear? Finally, what comparison with the microscopic
model is needed for the prediction at hand?

To keep those questions explicit, we organize a prediction as
\begin{equation}
    \hat{\rho}_{\mathrm{rec}}(t)
    =
    \left[
        \mathcal{R}_t
        \circ
        \Phi_{\mathrm{eff}}(t,0)
        \circ
        \mathcal{E}_0
    \right]
    [\hat{\rho}(0)].
    \label{eq:introduction_complete_prediction}
\end{equation}
Here $\mathcal E_0$ encodes the physical initial condition in the retained
description, $\Phi_{\mathrm{eff}}(t,0)$ generates the reduced evolution, and
$\mathcal R_t$ returns the result to the laboratory description. One may
instead transform the measured observable, but applying both descriptions in
the same expectation value would count the endpoint correction twice.
Depending on the method, $\mathcal E_0$ and $\mathcal R_t$ may be projections,
changes of frame, perturbative coordinate maps, or reconstruction rules; they
need not be physical quantum channels on their own. Similarly, a GKLS (Gorini–Kossakowski–Sudarshan–Lindblad) master equation form
guarantees complete positivity for the reduced semigroup under its stated
assumptions~\cite{Gorini1976,Lindblad1976}, but agreement with the microscopic dynamics must still be tested.

The discussion is organized around three connected abilities:
\emph{derivation}, \emph{interpretation}, and \emph{discovery}. Derivation
means selecting the retained description and calculating the effective
generator and endpoint corrections to a declared order. Interpretation means
tracing an effective term to virtual transitions, energy denominators,
temporal harmonics, symmetry constraints, or dissipative response, and then
relating it to an observable. Discovery means using those structures to
identify candidate interactions and resonance conditions before testing them
against exact or converged dynamics. In practice, this means listing the
allowed paths, using symmetry to discard forbidden ones, identifying resonant
combinations, accounting for diagonal shifts and competing channels, and then
testing the resulting process.

The quantum Rabi model serves as the recurring benchmark because the same
microscopic Hamiltonian supports several controlled descriptions that retain
different states and processes. These include near-resonant Jaynes--Cummings
exchange, dispersive shifts and mediated interactions, displaced-oscillator
descriptions, multiphoton resonances, and the critical scaling associated with
a large frequency ratio. The dispersive Jaynes--Cummings limit~\cite{Meystre1992,Brune1990QND,Blais2004} then provides a
common second-order calculation whose exact excitation blocks permit direct
cross-method comparisons. It also makes an important point about validity: in
the $n$-excitation sector, the relevant ratio is
$g\sqrt{n+1}/|\Delta|$, not merely $g/|\Delta|$. This means that the coupling strength \(g\), the photon number \(n\), and the detuning \(\Delta\) between the mode and qubit frequencies all play a relevant role. More generally, the local
control parameter is set by matrix elements relative to the spectral or
dynamical gaps of the occupied and retained sectors. Collective enhancement,
spectral crowding, critical gaps, locality, and Liouvillian gaps modify that
assessment and will reappear in the examples that follow.

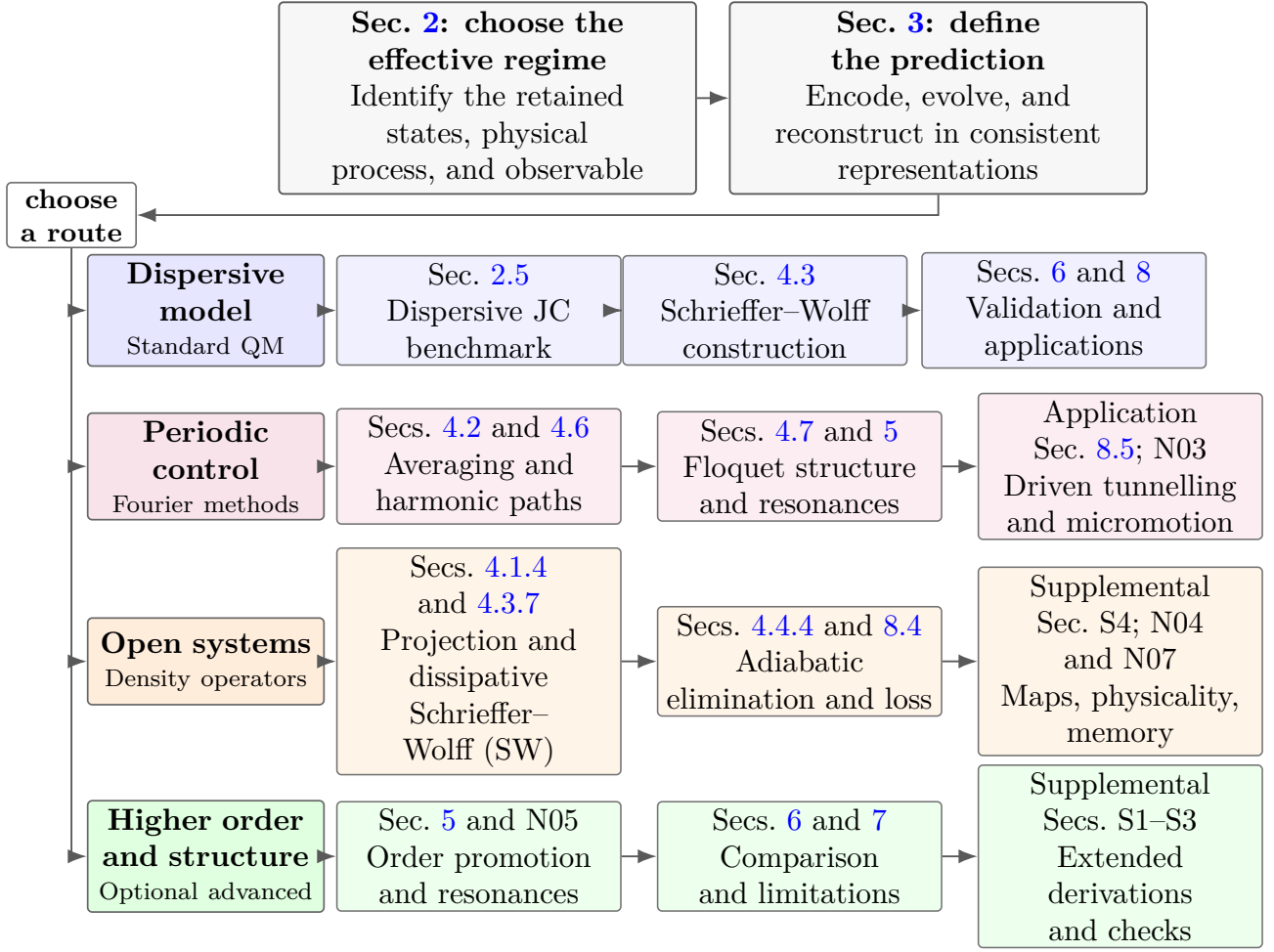
\begin{figure*}[t]
\centering
\resizebox{\linewidth}{!}{%
\begin{tikzpicture}[
    font=\small,
    >={Latex[length=2.2mm]},
    common/.style={
        draw=black!65,
        fill=black!3,
        rounded corners=1.5pt,
        line width=0.6pt,
        minimum height=13mm,
        text width=4.6cm,
        align=center,
        inner sep=3pt
    },
    route head/.style={
        draw=black!60,
        rounded corners=1.5pt,
        line width=0.55pt,
        minimum height=10mm,
        text width=2.55cm,
        align=center,
        inner sep=2.5pt
    },
    route/.style={
        draw=black!55,
        rounded corners=1.5pt,
        line width=0.5pt,
        minimum height=10mm,
        text width=3.10cm,
        align=center,
        inner sep=2.5pt
    },
    choice/.style={
        draw=black!60,
        fill=white,
        rounded corners=1.5pt,
        line width=0.5pt,
        text width=1.35cm,
        align=center,
        inner sep=2pt,
        font=\footnotesize\bfseries
    },
    flow/.style={->,line width=0.6pt,draw=black!65},
    trunk/.style={line width=0.6pt,draw=black!65}
]

\node[common] (qrm) at (3.25,3.70)
{\textbf{Sec.~\ref{sec:rabi_reductions}: choose the effective regime}\\
Identify the retained states, physical process, and observable};

\node[common] (prediction) at (8.45,3.70)
{\textbf{Sec.~\ref{sec:common_framework}: define the prediction}\\
Encode, evolve, and reconstruct in consistent representations};

\draw[flow] (qrm) -- (prediction);
\node[choice] (hub) at (-1.55,2.35) {choose\\a route};
\draw[flow] (prediction.south) |- (hub.east);

\node[route head,fill=blue!10] (dh) at (0,1.25)
{\textbf{Dispersive model}\\[-1pt]
{\scriptsize Standard QM}};
\node[route,fill=blue!6] (d1) at (3.15,1.25)
{Sec.~\ref{subsec:dispersive_benchmark}\\
Dispersive JC benchmark};
\node[route,fill=blue!6] (d2) at (6.45,1.25)
{Sec.~\ref{subsec:schrieffer_wolff}\\
Schrieffer--Wolff construction};
\node[route,fill=blue!6] (d3) at (9.90,1.25)
{Secs.~\ref{sec:methods_comparison} and \ref{sec:applications}\\
Validation and applications};

\node[route head,fill=purple!11] (ph) at (0,-0.55)
{\textbf{Periodic control}\\[-1pt]
{\scriptsize Fourier methods}};
\node[route,fill=purple!7] (p1) at (3.15,-0.55)
{Secs.~\ref{subsec:magnus_expansion} and
\ref{subsec:james_jerke_method}\\
Averaging and harmonic paths};
\node[route,fill=purple!7] (p2) at (6.85,-0.55)
{Secs.~\ref{subsec:floquet_method} and
\ref{sec:higher_order_expansions}\\
Floquet structure and resonances};
\node[route,fill=purple!7] (p3) at (10.55,-0.55)
{Application Sec.~\ref{subsec:application_floquet_tunnelling}; N03\\
Driven tunnelling and micromotion};

\node[route head,fill=orange!14] (oh) at (0,-2.80)
{\textbf{Open systems}\\[-1pt]
{\scriptsize Density operators}};
\node[route,fill=orange!9] (o1) at (3.15,-2.80)
{Secs.~\ref{subsec:projection_open_systems} and
\ref{subsec:sw_open_systems}\\
Projection and dissipative Schrieffer--Wolff (SW)};
\node[route,fill=orange!9] (o2) at (6.85,-2.80)
{Secs.~\ref{subsec:ae_open_systems} and
\ref{subsec:application_lossy_mediator}\\
Adiabatic elimination and loss};
\node[route,fill=orange!9] (o3) at (10.55,-2.80)
{Supplemental Sec.~S4; N04 and N07\\
Maps, physicality, memory};

\node[route head,fill=green!12] (ah) at (0,-5.05)
{\textbf{Higher order and structure}\\[-1pt]
{\scriptsize Optional advanced}};
\node[route,fill=green!7] (a1) at (3.15,-5.05)
{Sec.~\ref{sec:higher_order_expansions} and N05\\
Order promotion and resonances};
\node[route,fill=green!7] (a2) at (6.85,-5.05)
{Secs.~\ref{sec:methods_comparison} and
\ref{sec:advanced_effective_dynamics}\\
Comparison and limitations};
\node[route,fill=green!7] (a3) at (10.55,-5.05)
{Supplemental Secs.~S1--S3\\
Extended derivations and checks};

\draw[trunk] (hub.south) -- (-1.55,-5.05);
\foreach \head/\first/\second/\third in
{dh/d1/d2/d3,ph/p1/p2/p3,oh/o1/o2/o3,ah/a1/a2/a3}{
    \draw[flow] (-1.55,0 |- \head.west) -- (\head.west);
    \draw[flow] (\head) -- (\first);
    \draw[flow] (\first) -- (\second);
    \draw[flow] (\second) -- (\third);
}

\end{tikzpicture}%
}
\caption{\textbf{Suggested routes through the article.} Each route begins with the
regime selection developed in Sec.~\ref{sec:rabi_reductions} and the
encoding--evolution--reconstruction framework of
Sec.~\ref{sec:common_framework}. The first three rows provide direct paths to
a calculation and an application; the final row leads to higher-order and
structural topics. Arrows show the suggested reading order. Labels N03--N07
refer to the companion notebooks described below and in Supplemental Sec.~S5.}
\label{fig:introduction_reading_routes}
\end{figure*}

Open systems add a further complication. Eliminating a lossy degree of freedom
generally produces coherent corrections and dissipative channels at the same
time. Jump operators must therefore be transformed along with the Hamiltonian,
and reconstructing an eliminated mode may reveal information that is absent
from a Markovian reduced generator. A Hamiltonian-only treatment can miss this
physics even when it gives the correct dispersive shift. It is also useful to
keep three tests separate: positivity of a generator, complete positivity of
a finite-time map, and agreement with the microscopic evolution.

These questions now arise in experimentally accessible settings. Circuit-QED
experiments have realized or resolved processes that
convert one resonator photon into two artificial-atom excitations
\cite{Tomonaga2025TwoArtificialAtoms}, hybridize one- and two-photon states
through a detuned artificial atom \cite{Wang2025OneTwoPhoton}, and generate
three-photon down-conversion and tripartite non-Gaussian correlations
\cite{Chang2020ThreePhotonSPDC,JarvisFrain2026Tripartite}. Nonlinear reservoir
engineering similarly targets multicomponent cat manifolds and higher-order
bosonic dissipation \cite{Rojkov2026NLRE,Vanselow2026Quartets}. Although the
microscopic systems differ, the same questions return: Which states should be
retained? Which paths generate the desired interaction? How do diagonal
corrections shift the resonance? Which competing channels remain, and which
observable provides the clearest test?

The applications give these questions a concrete setting. A multiqubit cavity
shows how counter-rotating paths alter the
available operator sectors; three-level systems show how topology and signed
detunings select exchange or pair processes; a spin-chain bus exposes the role
of spectral isolation and retained-mode choice; a lossy mediator requires
coherent and dissipative effects to be derived together; and a driven
two-site lattice separates stroboscopic dynamics from continuous-time
micromotion. Together they show how the same habits of derivation,
interpretation, and validation carry across rather different physical
platforms.

The appropriate validation depends on the claim. Agreement of selected
eigenvalues supports a spectral claim, but not necessarily a trajectory.
Agreement for one initial state does not establish a channel approximation. A
continuous-time Floquet prediction requires micromotion, whereas a
stroboscopic prediction may not. A long-time prediction must state the window
over which phase errors, leakage, and memory remain controlled. We therefore
use exact diagonalization, direct time evolution, convergence tests,
block-decoupling or equation-of-motion residuals, and finite-map diagnostics
according to the quantity under examination. In each case, we want to know
what the reduced model predicts, what gives rise to its operator structure,
and where its predictions begin to fail.

\subsection{Audience, prerequisites, and routes through the article}
\label{subsec:introduction_reader_guide}

This article is intended for graduate students and researchers who want to
construct and test effective descriptions of quantum dynamics. We assume
familiarity with
\begin{itemize}
    \item the Schr\"odinger equation, unitary time evolution, and Dirac
    notation;
    \item operators, commutators, eigenvalue problems, and changes of basis;
    \item density operators, partial traces, and expectation values; and
    \item nondegenerate Rayleigh--Schr\"odinger perturbation theory through
    second order.
\end{itemize}
No prior knowledge of the effective-model methods developed here is needed.
For the periodic-control route, familiarity with interaction pictures and
Fourier series is useful. The open-system route introduces GKLS generators,
quantum jumps, and finite-time dynamical maps as they arise.

Two sections provide a common starting point. Section~\ref{sec:rabi_reductions}
uses the quantum Rabi model to show that the appropriate effective theory
depends on the parameter range, retained states, observable, and timescale.
It covers near-resonant exchange, dispersive dynamics, displaced-oscillator
descriptions, multiphoton resonances, and the critical
large-frequency-ratio limit. Section~\ref{sec:common_framework} then separates
state preparation, reduced evolution, and reconstruction. From there, the
article may be read linearly or along one of the four routes in
Fig.~\ref{fig:introduction_reading_routes}. Readers looking for a particular
construction can follow a single row; those interested in comparing methods
can continue through Secs.~\ref{sec:methods}--\ref{sec:methods_comparison}.

Section~\ref{sec:advanced_effective_dynamics} examines questions that become
important beyond finite-dimensional, weakly coupled settings, including
operator domains, many-body locality, Liouvillian gaps, memory kernels, and
generator logarithms. Readers focused on practical constructions can continue
directly to the applications and return to this section when one of these
issues arises.

The companion notebooks offer a more gradual computational route through
selected problems. N01 treats beyond-RWA mediated interactions, N02 a spin-chain
quantum bus, N03 Floquet-controlled tunnelling, N04 elimination of a lossy
mediator, N05 a higher-order three-photon resonance, N06 three-level
quantum-mixer configurations, N07 open-system conventions and finite-map
tests, and N08 the quantum Rabi regime map and excitation diagnostics. N00
provides the contents and links for the complete set. Each notebook states its
assumptions, derives the model used in the calculation, compares microscopic
and effective dynamics, and examines convergence and breakdown.

Appendix~\ref{app:order_consistent_reconstruction} derives the order-by-order
relation among state preparation, effective evolution, observable
transformation, and final reconstruction. The Supplemental Material contains
exact frame transformations and extended algebra in Secs.~S1--S3,
Liouville-space and dynamical-map conventions in Sec.~S4, the notebook guide
in Sec.~S5, and a selective glossary in Sec.~S6.

\FloatBarrier


\section{Rabi-Model Reductions and Effective Regimes}
\label{sec:rabi_reductions}
\label{GRWA}
\label{sec:rwa_grwa}

The quantum Rabi model~\cite{Rabi1936} is an ideal place to learn how an effective theory is
chosen. Its Hamiltonian is simple enough to analyze in detail, yet it supports
several distinct reductions whose usefulness depends on the frequencies,
coupling strength, populated states, observable, and timescale. The Rabi and
Jaynes--Cummings (JC) models~\cite{Rabi1936,JaynesCummings1963} will therefore
reappear throughout the article.

\begin{figure*}[t]
    \centering
    \includegraphics[
        width=\textwidth
    ]{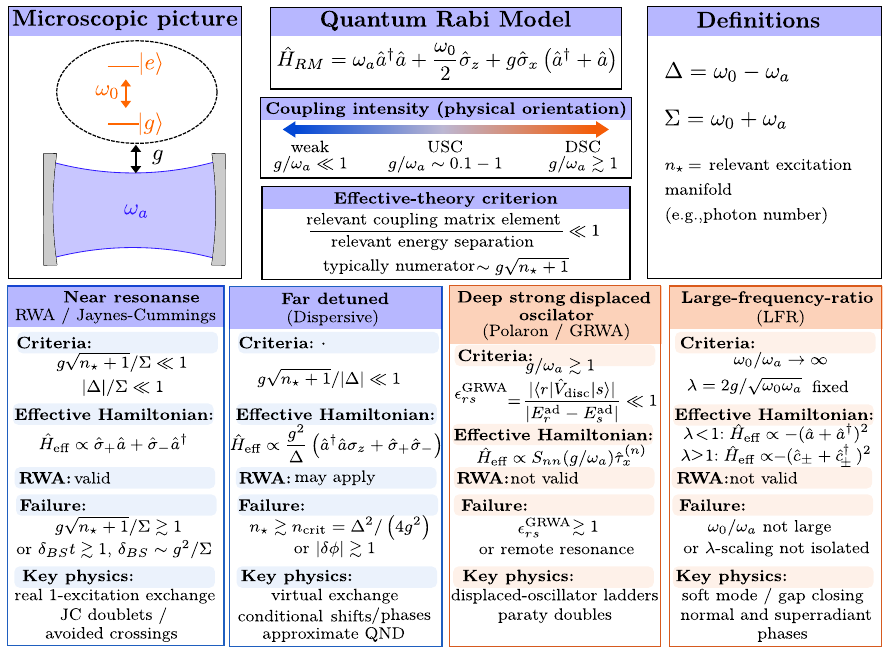}
    \caption{
    \textbf{One microscopic model, several effective theories.}
    The quantum Rabi Hamiltonian admits distinct controlled descriptions
    because different parameter regimes select different natural bases and
    retained processes. Near resonance, the Jaynes--Cummings reduction
    retains real one-excitation exchange. Far from resonance, exchange is
    virtual and produces conditional shifts, phases, mediated interactions,
    and dressed dissipation. At intermediate and deep-strong coupling, a
    displaced-oscillator basis exposes parity doublets and distinguishes bare,
    displaced, and detector-defined excitations. In the singular
    large-frequency-ratio limit at fixed
    \(\lambda=2g/\sqrt{\omega_0\omega_a}\), quadratic normal- and
    superradiant-phase descriptions exhibit soft-mode criticality. Each branch
    is tied to a different scale separation, retained structure, observable,
    and first failure mode.
    }
    \label{fig:qrm-effective-regime-map}
\end{figure*}

Figure~\ref{fig:qrm-effective-regime-map} gives the organizing picture. Near
resonance and at weak coupling, the rotating-wave approximation retains real
one-excitation exchange and leads to the JC model. Far from resonance, that
exchange becomes virtual and produces conditional shifts, phases, mediated
interactions, and induced dissipation. At stronger coupling, the bare product
basis gives way to qubit-conditioned displaced oscillators. A multiphoton
resonance can in turn isolate a small manifold of nearly degenerate
same-parity states, while a singular large-frequency-ratio scaling produces a
soft oscillator mode and a quantum critical point. These descriptions are
adapted to different physical questions; they are not successive corrections
to one universal approximation.

The dispersive JC Hamiltonian will serve as the common second-order benchmark
for the methods developed later. Here we first derive it from the exact JC
blocks and examine the conditional shifts, quantum-nondemolition (QND) structure, phases, mediated
couplings, and dressed decay that make it useful. This also prepares the
applications in Sec.~\ref{sec:applications}, where virtual mediation is tested
in multiqubit and lossy settings. Changes of basis will already show why
states, observables, and environmental couplings must be transformed together,
a point developed systematically in Sec.~\ref{sec:common_framework}.

We use the detuning $\Delta\equiv\omega_0-\omega_a$ and the sum frequency
$\Sigma\equiv\omega_0+\omega_a$, where $\omega_0$ is the qubit transition
frequency and $\omega_a$ is the oscillator frequency.
The resulting reorganizations of the same microscopic model are summarized in
Fig.~\ref{fig:qrm-effective-regime-map}.

The four branches represent broad scale separations. The multiphoton resonance
discussed later is instead a local spectral reorganization in which a small
same-parity manifold becomes nearly degenerate and must be retained explicitly.

\subsection{The quantum Rabi model}
\label{subsec:quantum_rabi_model}

The quantum Rabi model describes a two-level system coupled to a single
bosonic mode,
\begin{equation}
    \hat H_{\mathrm{RM}}
    =
    \omega_a \hat a^{\dagger}\hat a
    +\frac{\omega_0}{2}\hat\sigma_z
    +g\hat\sigma_x
    \left(
        \hat a^{\dagger}+\hat a
    \right),
    \label{rabi}
\end{equation}
where $\hat a$ and $\hat a^\dagger$ annihilates and creates oscillator
excitations, $\hat\sigma_{x,z}$ act on the two-level system, and $g$ is
the coupling strength. We set $\hbar=1$ throughout this manuscript.

Equation~\eqref{rabi} is already a reduced light--matter model. Its use
presupposes a mode truncation and a reduction of the matter sector to two
levels, whose accuracy is logically distinct from the validity of the
rotating-wave approximation. In ultrastrong-coupling settings,
electromagnetic-gauge transformations and finite-level truncations
generally do not commute. A few-level model must therefore be derived and
validated consistently in the chosen gauge, retaining the corresponding
self-energy terms and transforming observables accordingly
\cite{DeBernardis24}. We return to this issue in
Sec.~\ref{sec:advanced_effective_dynamics}.

With
$\hat\sigma_\pm=(\hat\sigma_x\pm\I\hat\sigma_y)/2$, the total bare-excitation
operator is
$\hat N=\hat a^\dagger\hat a+\hat\sigma_+\hat\sigma_-$. The Rabi Hamiltonian
does not conserve $\hat N$, but it does conserve excitation parity,
$\hat\Pi=e^{\I\pi\hat N}$, so that
$[\hat H_{\mathrm{RM}},\hat\Pi]=0$. The Hilbert space consequently separates
into two invariant parity sectors. Braak used this discrete structure to obtain an analytic
characterization of the spectrum for arbitrary coupling and
detuning~\cite{braak2011integrability}. Exact spectral solvability,
however, does not diminish the practical value of effective models, which
can expose the relevant transitions, conserved quantities, and control
parameters more transparently.

\subsection{The rotating-wave approximation}
\label{subsec:rwa}

Using $\hat\sigma_x=\hat\sigma_++\hat\sigma_-$, the interaction in
Eq.~\eqref{rabi} separates into co-rotating and counter-rotating
contributions,
\begin{equation*}
\hat V_{\mathrm{co}}
    =
    g\left(
        \hat\sigma_+\hat a
        +
        \hat\sigma_-\hat a^\dagger
    \right), 
    \hat V_{\mathrm{cr}}
    =
    g\left(
        \hat\sigma_+\hat a^\dagger
        +
        \hat\sigma_-\hat a
    \right).
\end{equation*}
Their physical distinction becomes transparent in the interaction picture
generated by
$\hat H_0=\omega_a\hat a^\dagger\hat a+(\omega_0/2)\hat\sigma_z$, using the representation-change convention stated in
Sec.~\ref{sec:common_framework}. The interaction Hamiltonian is
\begin{align}
    \hat H_{\mathrm{RM}}^{(I)}(t)
    &=  g\left(
        \hat\sigma_-\hat a^\dagger e^{-\I\Delta t} + \hat\sigma_+\hat a e^{\I\Delta t}\right) \nonumber\\
    &\quad+g\left( \hat\sigma_+\hat a^\dagger e^{\I\Sigma t}+\hat\sigma_-\hat a e^{-\I\Sigma t}\right).
    \label{eq:rabi_interaction_picture}
\end{align}
The first line exchanges one excitation between the two subsystems and
oscillates at the detuning frequency $|\Delta|$. The second line creates
or annihilates one qubit excitation and one oscillator excitation
simultaneously and oscillates at the sum frequency $\Sigma$.

Suppose that the initial state and subsequent dynamics have negligible
oscillator population above a cutoff $n_\star$. A useful state-dependent
measure of counter-rotating mixing is then
$\epsilon_{\mathrm{cr}}(n_\star)=g\sqrt{n_\star+1}/\Sigma$.
Suppressing the counter-rotating processes requires
\begin{equation}
    \epsilon_{\mathrm{cr}}(n_\star)\ll1.
    \label{eq:rwa_counter_condition}
\end{equation}
For the standard near-resonant use of the JC model, one additionally
requires
\begin{equation}
    \frac{|\Delta|}{\Sigma}\ll1,
    \label{eq:rwa_near_resonance_condition}
\end{equation}
so that the co-rotating terms vary slowly compared with the discarded
terms~\cite{cohen1998atom,scully_zubairy}. Together, these inequalities define
the usual near-resonant JC regime. Far from resonance, both harmonic families
may be fast, so the decision to retain one and eliminate the other depends on
the spectral or dynamical quantity of interest. The conditions are more
informative than $g/\omega_a$ alone: the matrix elements grow with excitation
number, and a small instantaneous correction can accumulate over a long
evolution.

Neglecting the second term of Eq.~\eqref{eq:rabi_interaction_picture} and returning
to the Schr\"odinger picture gives the JC Hamiltonian
\begin{equation}
    \hat H_{\mathrm{JC}}
    =
    \omega_a\hat a^\dagger\hat a
    +
    \frac{\omega_0}{2}\hat\sigma_z
    +
    g\left(
        \hat\sigma_-\hat a^\dagger
        +
        \hat\sigma_+\hat a
    \right).
    \label{eq:jc_hamiltonian}
\end{equation}
Unlike the Rabi Hamiltonian, $\hat H_{\mathrm{JC}}$ conserves $\hat N$
and decomposes into independent two-dimensional excitation manifolds.
This block structure makes the model exactly solvable and physically
transparent; for a historical and conceptual review, see
Ref.~\cite{larson2021jaynes}.

The statement that rapidly oscillating terms ``average to zero'' must be
interpreted perturbatively. When a coarse-graining window exists with
\begin{equation*}
    \Sigma^{-1}
    \ll
    \tau
    \ll
    \left[
        \max
        \left\{
            |\Delta|,
            g\sqrt{n_\star+1}
        \right\}
    \right]^{-1},
\end{equation*}
the first-order average of the counter-rotating terms is suppressed.
Such a window exists only when the counter-rotating components oscillate
much faster than the retained dynamics,
\(\Sigma \gg \max \left\{ |\Delta|, g\sqrt{n_\star+1} \right\}.\)
Their virtual action nevertheless produces finite higher-order
corrections, including frequency renormalization and the Bloch--Siegert
shift~\cite{FornDaz2010}. The RWA is therefore a leading-order time-averaging, often called
secular, approximation,\footnote{The term \emph{secular} is inherited
from perturbation theory, where it denotes contributions that do not
average into bounded rapid oscillations and can accumulate over long
times~\cite{BenderOrszag1999}. In the present Hamiltonian setting, a
secular approximation retains stationary or slowly varying near-resonant
terms and neglects terms whose rapid oscillations suppress their
accumulated contribution over the timescale of interest. In open-system
master-equation derivations, the same nomenclature is used more
specifically for neglecting terms oscillating at differences between
distinct Bohr frequencies; see Ref.~\cite{breuer2002theory}.} not a
statement that the discarded interaction has no physical consequences.

\subsection{Validity, state dependence, and breakdown of the RWA}
\label{subsec:rwa_validity}

A meaningful assessment of the RWA must specify the initial state, the
target observable, the occupied excitation manifold, and the comparison
time. A short-time calculation makes this state dependence explicit. Let
the Rabi and JC evolutions begin from the same laboratory-frame pure state
$\ket{\psi_0}$, with
$\hat H_{\mathrm{RM}}=\hat H_{\mathrm{JC}}+\hat V_{\mathrm{cr}}$.
Because both Hamiltonians in this comparison are expressed in the same
Schr\"odinger-picture operator basis, no additional reconstruction map is
required.

The relative propagator has the short-time expansion
\begin{align}
    e^{\I\hat H_{\mathrm{JC}}t}
    e^{-\I\hat H_{\mathrm{RM}}t}
    &=
    \hat I \nonumber 
    -
    \I t\hat V_{\mathrm{cr}} \nonumber \\
    &+\frac{t^2}{2}
    \left(
        [
            \hat H_{\mathrm{JC}},
            \hat V_{\mathrm{cr}}
        ]
        -
        \hat V_{\mathrm{cr}}^2
    \right) \nonumber \\
    &\quad+
    \mathcal O(t^3).
    \label{eq:rwa_relative_propagator}
\end{align}
The expectation value of the commutator in
Eq.~\eqref{eq:rwa_relative_propagator} is purely imaginary for Hermitian
$\hat H_{\mathrm{JC}}$ and $\hat V_{\mathrm{cr}}$. It can affect the
phase of the overlap amplitude at this order, but not its squared
modulus. The fidelity is therefore
\begin{align}
    F(t) & \equiv\left|\langle\psi_{\mathrm{JC}}(t)| \psi_{\mathrm{RM}}(t)\rangle\right|^2 \nonumber \\
    &=1-t^2\operatorname{Var}_{\psi_0}
    \left(
        \hat V_{\mathrm{cr}}
    \right)
    +
    \mathcal O(t^3), 
\end{align}
where
$\operatorname{Var}_{\psi_0}(\hat V_{\mathrm{cr}})
=\langle\hat V_{\mathrm{cr}}^2\rangle_0
-\langle\hat V_{\mathrm{cr}}\rangle_0^2$.
For the state $\ket{\psi_0}=\ket e\otimes\ket\alpha$, one finds
$\langle\hat V_{\mathrm{cr}}\rangle_0=0$ and
$\langle\hat V_{\mathrm{cr}}^2\rangle_0=g^2|\alpha|^2$, and hence
\begin{equation*}
    F(t)
    =
    1
    -
    g^2|\alpha|^2t^2
    +
    \mathcal O(t^3).
\end{equation*}
The coefficient depends on both the qubit and field states. For example, the
quadratic coefficient vanishes for $\ket{e,0}$, even though the Rabi and JC
dynamics need not remain identical at later times. A fidelity calculated
for one initial state therefore cannot define a global RWA validity
regime.

The labels weak, ultrastrong, and deep-strong coupling provide useful physical
orientation, but they are not error bounds for the RWA. Ultrastrong
coupling is commonly associated with an interaction that is a
non-negligible fraction of the bare frequencies, often of order
$g/\omega\sim0.1$, whereas deep-strong coupling refers to interaction
strengths comparable to or larger than the bare
frequencies~\cite{FornDiaz2019,DeBernardis24}. Away from resonance, both
$g/\omega_a$ and $g/\omega_0$, together with the detuning, should be
specified. In engineered quantum simulators, an effective Rabi-model
coupling ratio should also be distinguished from the microscopic
electromagnetic coupling of the underlying hardware. Quantitative validation
must still specify the state, observable, and comparison time.

When Eqs.~\eqref{eq:rwa_counter_condition}
and~\eqref{eq:rwa_near_resonance_condition} cease to describe the
relevant dynamics, counter-rotating processes can produce observable
spectral shifts, modify selection rules, and dress the
eigenstates~\cite{LeBoit2020,FornDiaz2010,DeBernardis24}. In the
ultrastrong-coupling regime, the Rabi ground state can contain
qubit--field entanglement and nonzero occupation of the bare oscillator
mode. Such bare-mode excitations should not automatically be identified
with directly detectable output photons: their operational meaning
depends on the system--detector coupling, the gauge-consistent observable,
and the measurement protocol. This distinction illustrates why
observables must be reconstructed consistently after a change of
representation, as discussed in Sec.~\ref{sec:common_framework}.

Higher-order time averaging and block diagonalization recover systematic
corrections to the RWA. Their small parameters, retained sectors, and
reconstruction maps are developed in Sec.~\ref{sec:methods} and compared in
Sec.~\ref{sec:higher_order_expansions}.
\begin{figure*}[t]
    \centering
    \includegraphics[
        width=0.95\textwidth
    ]{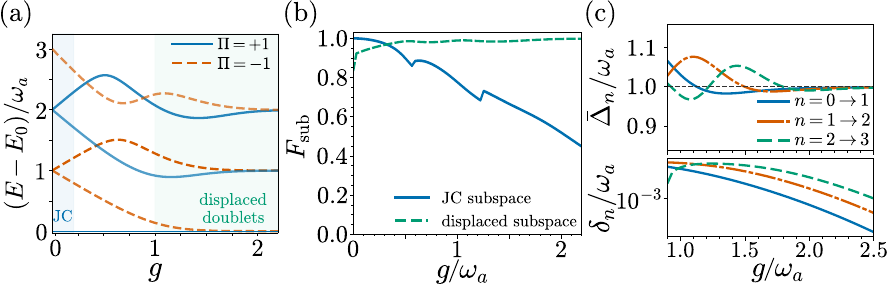}
    \caption{
    \textbf{Changing spectral organization of the quantum Rabi model.}
    \textbf{(a)}
    Lowest exact excitation energies of the resonant quantum Rabi model,
    $\omega_0=\omega_a$, as a function of the normalized coupling
    $g/\omega_a$.
    Solid and dashed curves denote the two exact parity sectors,
    $\Pi=+1$ and $\Pi=-1$, respectively.
    At weak coupling, the spectrum retains the organization inherited from
    Jaynes--Cummings excitation manifolds.
    At deep-strong coupling, it reorganizes into nearly degenerate
    opposite-parity pairs associated with qubit-conditioned displaced
    oscillators.
    \textbf{(b)}
    Normalized low-energy subspace fidelity
    \(
        F_{\mathrm{sub}}
        =
        d^{-1}
        \operatorname{Tr}
        (P_{\mathrm{exact}}P_{\mathrm{ref}})
    \)
    between the exact six-dimensional low-energy subspace and,
    respectively, the corresponding Jaynes--Cummings and displaced
    adiabatic reference subspaces.
    The comparison is invariant under rotations among degenerate or
    nearly degenerate eigenvectors.
    \textbf{(c)}
    Upper panel: spacings $\bar{\Delta}_n =  \bar E_{n+1} -  \bar E_n$ between the centers
    \(
        \bar E_i=(E_{i,+}+E_{i,-})/2
    \)
    of the first displaced-manifold doublets.
    Their convergence to $\omega_a$ demonstrates the recovery of a
    harmonic displaced ladder.
    Lower panel: opposite-parity splittings
    \(
        \delta_n=|E_{n,+}-E_{n,-}|
    \),
    which decrease as the overlap between the two oppositely displaced
    oscillator branches is suppressed.
    Exact states are assigned one-to-one to displaced targets using parity
    and maximal state overlap; assignments with squared overlap below
    $0.50$ are excluded.
    The oscillator Hilbert space is truncated at $N=70$, for which the
    displayed low-energy quantities are converged. 
    }
    \label{fig:qrm-spectral-reorganization}
\end{figure*}

\subsection{The generalized rotating-wave approximation}
\label{subsec:grwa}

The generalized rotating-wave approximation (GRWA) first changes the basis
and only then applies a rotating-wave-type truncation~\cite{Irish2007}. The
zeroth-order states already contain the qubit-dependent displacement of the
oscillator, and the retained couplings connect nearby dressed manifolds. The
displacement itself is exact; the approximation enters when remote matrix
elements are discarded in the displaced basis. This reorganization can remain
useful beyond the weak-coupling domain of the ordinary RWA, with an accuracy
set by the detuning, populated manifold, and nearby resonances.

Define
\(\eta = \tfrac{g}{\omega_a}\)
and introduce the spin-dependent displacement
\begin{equation*}
    \hat U_{\mathrm p}
    =
    \exp
    \left[
        -\eta\hat\sigma_x
        \left(
            \hat a^\dagger-\hat a
        \right)
    \right].
\end{equation*}
Using the representation convention of
Sec.~\ref{sec:common_framework}, the transformed Hamiltonian is
\begin{align}
  \widetilde{\hat H}_{\mathrm{RM}}
    &=
    \hat U_{\mathrm p}^\dagger
    \hat H_{\mathrm{RM}}
    \hat U_{\mathrm p} \nonumber\\ 
    &=
    \omega_a\hat a^\dagger\hat a
    -
    \frac{g^2}{\omega_a}
    \nonumber\\
    &\quad
    +
    \frac{\omega_0}{2}
    \hat\sigma_z
    \exp
    \left[
        -2\eta\hat\sigma_x
        \left(
            \hat a^\dagger-\hat a
        \right)
    \right].
    \label{eq:grwa_transformed_hamiltonian}   
\end{align}
The compact last factor is Hermitian even though its Hermiticity is not
visually obvious. Equivalently,
\begin{align}
  \hat U_{\mathrm p}^\dagger\hat\sigma_z\hat U_{\mathrm p}
&=
\hat\sigma_z
\cosh\!\left[2\eta(\hat a^\dagger-\hat a)\right] \nonumber\\
&\quad-
\I\hat\sigma_y
\sinh\!\left[2\eta(\hat a^\dagger-\hat a)\right].  
\end{align}
The linear interaction has disappeared from the oscillator sector, but
its effect is retained nonperturbatively in the transformed qubit
operator and in the overlaps between oppositely displaced oscillator
states.

Let $\ket{\pm_x}$ denote the eigenstates of $\hat\sigma_x$. The
corresponding displaced basis in the original Hilbert space is
\begin{equation}
    \ket{\pm_x,n_\pm}
    =
    \ket{\pm_x}
    \otimes
    \hat D(\mp\eta)
    \ket n,
    \label{eq:grwa_displaced_basis}
\end{equation}
where
$\hat D(\alpha)=\exp(\alpha\hat a^\dagger-\alpha^*\hat a)$.
For $\omega_0=0$, the states in
Eq.~\eqref{eq:grwa_displaced_basis} form degenerate pairs with energies
\(E_n^{(0)} = n\omega_a - \tfrac{g^2}{\omega_a}.\)
The qubit term couples the two displaced ladders through the overlaps
$S_{mn}(\eta)\equiv\braket{m_-|n_+}$.
With the phase convention of Ref.~\cite{Irish2007}, for $m\le n$,
\begin{equation*}
\begin{aligned}
    S_{mn}(\eta)
    ={}&
    e^{-2\eta^2}
    (2\eta)^{n-m}
    \sqrt{\frac{m!}{n!}}
    L_m^{(n-m)}
    \left(
        4\eta^2
    \right),
\end{aligned}
\end{equation*}
where $L_m^{(k)}$ is an associated Laguerre polynomial. In particular,
\begin{equation}
\begin{aligned}
    S_{nn}(\eta)
    &=
    e^{-2\eta^2}
    L_n
    \left(
        4\eta^2
    \right),
    \\
    S_{n,n+1}(\eta)
    &=
    e^{-2\eta^2}
    \frac{2\eta}{\sqrt{n+1}}
    L_n^{(1)}
    \left(
        4\eta^2
    \right).
    \label{eq:grwa_relevant_overlaps}
\end{aligned}
\end{equation}

Diagonalizing the qubit term within each degenerate displaced pair gives
the adiabatic states
\begin{equation*}
    \ket{\Psi_{\pm,n}}
    =
    \frac{1}{\sqrt2}
    \left(
        \ket{+_x,n_+}
        \pm
        \ket{-_x,n_-}
    \right),
\end{equation*}
with energies
\begin{equation}
    E_{\pm,n}^{\mathrm{ad}}
    =
    n\omega_a
    -
    \frac{g^2}{\omega_a}
    \pm
    \frac{\omega_0}{2}
    S_{nn}(\eta).
    \label{eq:grwa_adiabatic_energies}
\end{equation}
The GRWA retains the near-resonant couplings between adjacent adiabatic
manifolds. For $n\ge0$, the retained block in the basis
$\{\ket{\Psi_{+,n}},\ket{\Psi_{-,n+1}}\}$ is
\begin{equation*}
    \hat H_n^{\mathrm{GRWA}}
    =
    \begin{pmatrix}
        E_{+,n}^{\mathrm{ad}}
        &
        \dfrac{\omega_0}{2}
        S_{n,n+1}(\eta)
        \\
        \dfrac{\omega_0}{2}
        S_{n,n+1}(\eta)
        &
        E_{-,n+1}^{\mathrm{ad}}
    \end{pmatrix}.
\end{equation*}
Its eigenvalues are
\begin{align}
    E_{n,\pm}^{\mathrm{GRWA}}
    &=\frac{
        E_{+,n}^{\mathrm{ad}}
        +
        E_{-,n+1}^{\mathrm{ad}}
    }{2} \nonumber\\
    &\quad{}\pm\frac{1}{2}
    \left[
    \begin{aligned}
        &\left(
            E_{+,n}^{\mathrm{ad}}
            -E_{-,n+1}^{\mathrm{ad}}
        \right)^2\\
        &\quad{}+\omega_0^2
        \left|S_{n,n+1}(\eta)\right|^2
    \end{aligned}
    \right]^{1/2}.
\end{align}
The isolated lowest state has energy
$E_{-,0}^{\mathrm{ad}}$ within this construction.

The GRWA is controlled by the separation between retained and discarded
couplings in the adiabatic basis, rather than by the bare ratio
$g/\omega_a$ alone. A useful local diagnostic is
\begin{equation}
    \epsilon_{rs}^{\mathrm{GRWA}}
    =
    \frac{
        \left|
            \langle r|
            \hat V_{\mathrm{disc}}
            |s\rangle
        \right|
    }{
        \left|
            E_r^{\mathrm{ad}}
            -
            E_s^{\mathrm{ad}}
        \right|
    }
    \ll1,
    \label{eq:grwa_control_condition}
\end{equation}
for every discarded pair of adiabatic states
$\ket r$ and $\ket s$ that is appreciably connected to the populated
manifold. Here \(\hat V_{\mathrm{disc}}\) is the part of the Rabi
Hamiltonian omitted by the selected GRWA block. This form makes the logic
explicit: the numerator is a discarded matrix element, while the
denominator is the energy separation that suppresses it. Near a remote
avoided crossing or a multiphoton resonance, a denominator can become
small and the retained block must be enlarged.

Finally, Eq.~\eqref{eq:grwa_transformed_hamiltonian} is written in a displaced frame. The exact
transformation leaves the spectrum invariant, but laboratory observables
generally acquire dressing corrections. For example,
\begin{equation*}
    \widetilde{\hat a}
    =
    \hat U_{\mathrm p}^\dagger
    \hat a
    \hat U_{\mathrm p}
    =
    \hat a-\eta\hat\sigma_x,
\end{equation*}
and hence
\begin{equation*}
    \widetilde{
        \hat a^\dagger\hat a
    }
    =
    \left(
        \hat a^\dagger-\eta\hat\sigma_x
    \right)
    \left(
        \hat a-\eta\hat\sigma_x
    \right).
\end{equation*}
For a transformed-frame state
$\widetilde{\hat\rho}$, the laboratory oscillator occupation is therefore
\begin{equation*}
\begin{aligned}
    \left\langle
        \hat a^\dagger\hat a
    \right\rangle_{\mathrm{lab}}
    =
    \operatorname{Tr}
    \left[
        \left(
            \hat a^\dagger-\eta\hat\sigma_x
        \right)
        \left(
            \hat a-\eta\hat\sigma_x
        \right)
        \widetilde{\hat\rho}
    \right].
\end{aligned}
\end{equation*}
Thus, unlike a simple phase rotation, the polaron transformation changes
even the operator representation of the oscillator occupation.
Predictions for amplitudes, occupations, and correlations must therefore
use the transformed observables or be reconstructed in the laboratory
frame, as developed systematically in
Sec.~\ref{sec:common_framework}.

The displaced-basis strategy has been extended to biased and multi-qubit
Rabi models and combined with symmetric and squeezing
transformations~\cite{PhysRevA.91.013814,Zhang2013,Albert2011,Wu2022,Zhang2016,PhysRevA.96.063821}. These extensions correspond to different
choices of zeroth-order dressing and retained blocks; their accuracy must
be assessed in the parameter and state manifold of interest.

\subsubsection{Deep-strong-coupling limit and displaced-oscillator physics}
\label{subsubsec:grwa_dsc_limit}

The displaced basis used by the GRWA is not only a calculational device.
It identifies the natural effective degrees of freedom when the coupling
becomes comparable to or larger than the oscillator frequency. To make
this structure explicit, split the Rabi Hamiltonian as
\begin{align}
 &\hat H_{\mathrm{RM}}
    =
    \hat H_{\mathrm{DSC}}^{(0)}
    +
    \frac{\omega_0}{2}\hat\sigma_z, \nonumber \\
    \qquad
    &\hat H_{\mathrm{DSC}}^{(0)}
    =
    \omega_a\hat a^\dagger\hat a
    +
    g\hat\sigma_x
    \left(
        \hat a+\hat a^\dagger
    \right).   
\end{align}
For a fixed eigenvalue $s=\pm1$ of \(\hat\sigma_x\), define the displaced
oscillator $\hat b_s=\hat a+s g/\omega_a$.
Then
\begin{equation*}
    \hat H_{\mathrm{DSC}}^{(0)}
    =
    \omega_a\hat b_s^\dagger\hat b_s
    -
    \frac{g^2}{\omega_a}
\end{equation*}
within that spin sector. The dominant interaction therefore does not
produce JC-like doublets. It produces two oscillator ladders displaced in
opposite directions in phase space.

The qubit term couples these displaced ladders. If couplings between
different oscillator indices are neglected at leading adiabatic order,
the block spanned by
\(\{\ket{+_x,n_+},\ket{-_x,n_-}\}\) is
\begin{equation}
    \hat H_{\mathrm{ad}}^{(n)}
    =
    \left(
        n\omega_a
        -
        \frac{g^2}{\omega_a}
    \right)
    \hat I_n
    +
    \frac{\omega_0}{2}
    S_{nn}\!\left(\frac{g}{\omega_a}\right)
    \hat\tau_x^{(n)},
    \label{eq:dsc_adiabatic_block}
\end{equation}
where \(\hat\tau_x^{(n)}\) exchanges the two displaced states and
\(S_{nn}\) is the overlap in
Eq.~\eqref{eq:grwa_relevant_overlaps}. Its eigenvalues are precisely the
adiabatic energies in Eq.~\eqref{eq:grwa_adiabatic_energies}. For low
oscillator indices and large \(g/\omega_a\), the overlap envelope is
exponentially suppressed, producing near-degenerate parity doublets.

This representation also gives a simple dynamical picture. In the
limit \(\omega_0\to0\), \(\hat\sigma_x\) is conserved and the oscillator
follows one of two conditional phase-space trajectories. A spin
superposition therefore becomes entangled with two separating oscillator
wave packets; their overlap controls the apparent collapse of spin
coherence, while the periodic return of the oscillator produces
revivals. A finite \(\omega_0\) couples the two trajectories through the
displaced-state overlaps and can either restore dynamics or leave it
strongly suppressed. Spectroscopic access to coupling ratios beyond the
ultrastrong regime has been demonstrated in superconducting
qubit--oscillator circuits, while digital and analog quantum simulators
have observed deep-strong-coupling dynamics, large bosonic excitation,
conditional nonclassical states, and dynamical freezing
\cite{Yoshihara2017DSC,Langford2017DSC,Koch2023DSC}.

The ratio $g/\omega_a$ locates the deep-strong-coupling regime, while the
quality of a particular adiabatic block is set by the displaced-state matrix
elements that were discarded, the populated oscillator manifold, and nearby
avoided crossings. These ingredients enter directly through the local
diagnostic in Eq.~\eqref{eq:grwa_control_condition}.

The RWA and GRWA address different approximation problems. The RWA
discards counter-rotating processes in the bare interaction picture,
whereas the GRWA changes the zeroth-order basis before selecting a
near-resonant block. The dispersive regime considered next is different
again: it begins from the JC Hamiltonian and eliminates an exchange
process that is retained by the RWA but rendered off resonant by a large
detuning.
The changing natural organization of the same microscopic Hamiltonian is
summarized in Fig.~\ref{fig:qrm-spectral-reorganization}: bare excitation
manifolds provide the economical description near the Jaynes--Cummings
regime, whereas qubit-conditioned displaced oscillators organize the
deep-strong-coupling spectrum.

\begin{figure*}[!t]
    \centering
    \includegraphics[
        width=0.89\textwidth
    ]{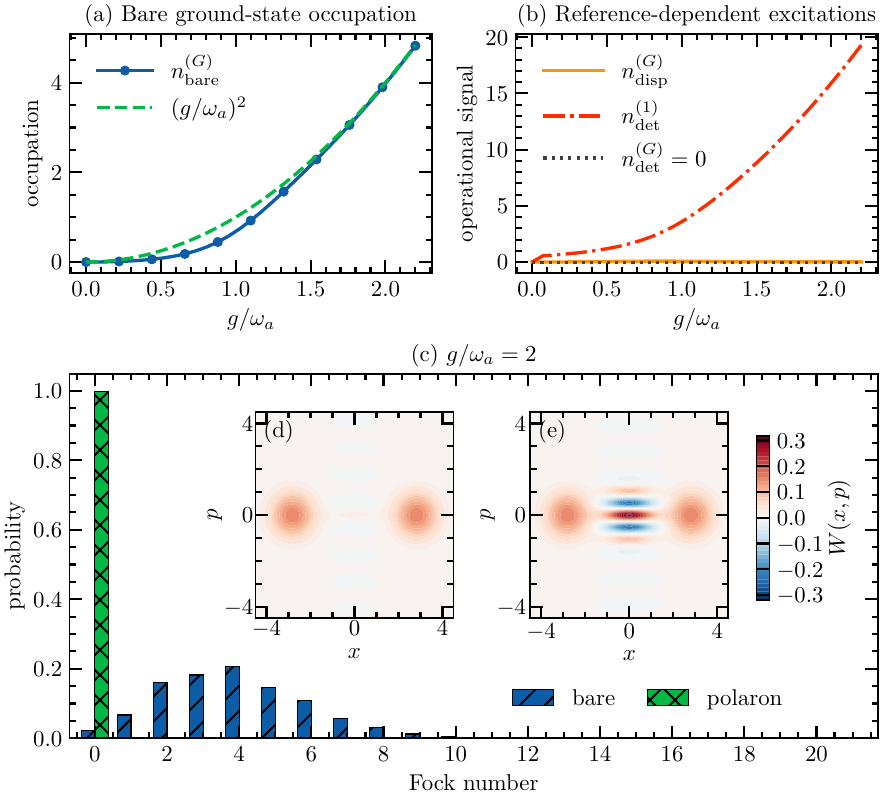}
    \caption{
    \textbf{Bare, displaced, and detector-defined excitations in the
    deep-strong-coupling regime.}
    \textbf{(a)} Bare ground-state occupation \( n_{\mathrm{bare}}^{(G)} = \langle G|\hat a^\dagger \hat a|G\rangle \) approaching \( (g/\omega_a)^2 \) at strong coupling.
 \textbf{(b)} Displaced-frame occupation, \( n_{\mathrm{disp}}^{(G)} = \langle G|\hat b^\dagger \hat b|G\rangle \), with \( b= \hat a+(g/\omega_a)\hat \sigma_x \), and detector-defined excitation signal \( n_{\mathrm{det}}^{(1)} = \langle 1|\hat X^- \hat X^+|1\rangle \). Here \( \hat X= \hat a+\hat a^\dagger \), and its positive-frequency component in the exact interacting eigenbasis is \( \hat X^+ = \sum_{E_k>E_j} \langle j|\hat X|k\rangle |j\rangle\langle k|\). Consequently, \( n_{\mathrm{det}}^{(G)}=0 \) although \( n_{\mathrm{bare}}^{(G)}>0 \). \textbf{(c)} Oscillator Fock distributions at $g/\omega_a=2$ in the bare and polaron frames, showing localization near the polaron vacuum. Inserts: \textbf{(d)} Reduced oscillator Wigner function obtained by tracing out the qubit. The two positive lobes reflect the oppositely displaced oscillator branches correlated with the qubit. \textbf{(e)} Wigner function conditioned on $|g\rangle$, where interference fringes reveal coherence between the displaced branches. Panels (d) and (e) share the same phase-space axes and color normalization. These results demonstrate that bare occupation, excitations above a transformed reference state, and detector counts are distinct operational quantities.
    }
    \label{fig:qrm-virtual-photons}
\end{figure*}

The displaced basis also separates three operationally different notions
of excitation. A bare occupation \(\langle a^\dagger a\rangle\) measures
quanta relative to the uncoupled oscillator vacuum; an occupation built
from \(b=a+(g/\omega_a)\sigma_x\) measures fluctuations above the
qubit-conditioned displaced wells; and a detector responds to the
positive-frequency part of the physical coupling operator in the exact
interacting eigenbasis~\cite{Ridolfo2012PhotonBlockade}.
Figure~\ref{fig:qrm-virtual-photons} compares these
quantities and shows why a nonzero bare ground-state population is a
signature of dressing rather than a flux of stationary output photons.

\subsection{The dispersive Jaynes--Cummings benchmark}
\label{subsec:dispersive_benchmark}

Consider the JC Hamiltonian in Eq.~\eqref{eq:jc_hamiltonian} in the
far-detuned regime. Conservation of the excitation number decomposes the
Hilbert space into the manifolds
\begin{equation*}
    \mathcal H_n
    =
    \operatorname{span}
    \left\{
        \ket{e,n},
        \ket{g,n+1}
    \right\},
    \qquad
    n=0,1,\ldots .
\end{equation*}
In this basis,
\begin{equation*}
    \begin{aligned}
    \hat H_{\mathrm{JC}}^{(n)}
    &=
    \left(
        n+\frac12
    \right)
    \omega_a\hat I_2\\
    &\quad{}+
    \frac12
    \begin{pmatrix}
        \Delta & 2g\sqrt{n+1}
        \\
        2g\sqrt{n+1} & -\Delta
    \end{pmatrix},
    \end{aligned}
\end{equation*}
and the exact eigenvalues are
\begin{equation*}
    E_{n,\pm}
    =
    \left(
        n+\frac12
    \right)
    \omega_a
    \pm
    \frac12
    \sqrt{
        \Delta^2
        +
        4g^2(n+1)
    }.
\end{equation*}

Suppose that the initial state and the subsequent dynamics have
negligible support above an oscillator occupation \(n_\star\). The
dispersive expansion is controlled locally by
\begin{equation*}
\begin{aligned}
    \eta_n
    &=
    \frac{
        g\sqrt{n+1}
    }{
        |\Delta|
    },
    \\
    \eta_\star
    &=
    \max_{0\leq n\leq n_\star}
    \eta_n
    =
    \frac{
        g\sqrt{n_\star+1}
    }{
        |\Delta|
    },
    \\
    4\eta_\star^2
    &\ll1.
\end{aligned}
\end{equation*}
Expanding the exact branches adiabatically connected to the bare states
gives
\begin{subequations}
\begin{align}
    E_{e,n}
    &=
    n\omega_a
    +
    \frac{\omega_0}{2}
    \nonumber \\
    &\quad{}+
    \frac{
        g^2(n+1)
    }{
        \Delta
    }
    +
    \mathcal O
    \left(
        |\Delta|\eta_n^4
    \right),
    \label{eq:dispersive_energy_en}
    \\
    E_{g,n+1}
    &=
    (n+1)\omega_a
    -
    \frac{\omega_0}{2} \nonumber \\
    &\quad{}-
    \frac{
        g^2(n+1)
    }{
        \Delta
    }
    +
    \mathcal O
    \left(
        |\Delta|\eta_n^4
    \right).
    \label{eq:dispersive_energy_gn}
\end{align}
\end{subequations}
The operator that reproduces these energies through second order is
\begin{equation}
\begin{aligned}
    \hat H_{\mathrm{disp}}
    &=
    \omega_a
    \hat a^\dagger\hat a
    +
    \frac{\omega_0}{2}
    \hat\sigma_z
    \\
    &\quad{}+
    \chi
    \left(
        \hat a^\dagger\hat a\,\hat\sigma_z
        +
        \hat\sigma_+\hat\sigma_-
    \right),  \\
    \chi
    &=
    \frac{g^2}{\Delta}.
    \label{eq:section2_dispersive_hamiltonian}
\end{aligned}
\end{equation}
Equivalently, up to an additive identity term,
\begin{equation*}
    \hat H_{\mathrm{disp}}
    \simeq
    \left(
        \omega_a+\chi\hat\sigma_z
    \right)
    \hat a^\dagger\hat a
    +
    \frac{
        \omega_0+\chi
    }{2}
    \hat\sigma_z.
\end{equation*}
The sign of \(\chi\) follows the sign of the detuning. The same formulas
therefore describe whether the dressed cavity and qubit frequencies are
shifted upward or downward.

Matching the exact JC spectrum gives a benchmark that does not depend on a
particular construction. Projection, Schrieffer--Wolff transformations, small
rotations, Magnus expansions, time coarse graining, and adiabatic elimination
will later recover the same second-order spectral normal form while describing
the virtual motion and reconstruction in different ways.

\subsubsection{Dressed states and the meaning of virtual exchange}
\label{subsubsec:dispersive_dressed_states}

The diagonal form of
Eq.~\eqref{eq:section2_dispersive_hamiltonian} does not mean that the
physical subsystems are completely unmixed. To leading order, the exact
eigenstates contain virtual components,
\begin{subequations}
\begin{align}
    \ket{\widetilde{e,n}}
    &=
    \ket{e,n} \nonumber \\
    &\quad{}+
    \frac{
        g\sqrt{n+1}
    }{
        \Delta
    }
    \ket{g,n+1}
    +
    \mathcal O(\eta_n^2),
    \label{eq:dispersive_dressed_en}
    \\
    \ket{\widetilde{g,n+1}}
    &=
    \ket{g,n+1} \nonumber \\
    &\quad{}-
    \frac{
        g\sqrt{n+1}
    }{
        \Delta
    }
    \ket{e,n}
    +
    \mathcal O(\eta_n^2).
    \label{eq:dispersive_dressed_gn}
\end{align}
\end{subequations}
These amplitudes are virtual in the sense that the large energy mismatch
suppresses resonant population exchange, but they remain physically
important. They generate the dispersive shifts, modify measured observables,
and allow an environmental channel acting on one subsystem to influence the
other. The effective Hamiltonian is diagonal, but the corresponding physical
states and observables remain dressed.

\subsection{Physical content and uses of the dispersive interaction}
\label{subsec:dispersive_physics}

The dispersive Hamiltonian turns microscopic exchange into conditional
frequencies and phases. This simple structure underlies nondemolition
measurement, photon-number resolution, mediated interactions, oscillator
control, and several experimentally implemented gates. We now follow these
consequences from the effective operator to the observables it controls. The
multiqubit and lossy-mediator applications in
Secs.~\ref{subsec:application_beyond_rwa_multi_qubit} and
\ref{subsec:application_lossy_mediator} will return to the same virtual
hybridization with the full reconstruction and dynamical tests included.

\subsubsection{Conditional cavity and qubit frequencies}
\label{subsubsec:dispersive_conditional_frequencies}

For a fixed qubit eigenstate, the cavity transition frequency is
\begin{equation}
    \omega_a^{(e/g)}
    =
    \omega_a
    \pm
    \chi.
    \label{eq:dispersive_conditional_cavity_frequency}
\end{equation}
The cavity response therefore carries information about the qubit
state. Conversely, the qubit transition frequency in the \(n\)-photon
sector is
\(\omega_q(n) = \omega_0 + \chi + 2\chi n.\)
The term \(\chi\) is the vacuum dispersive shift, whereas \(2\chi n\) is
the photon-number-dependent ac-Stark shift. When the separation \(2|\chi|\)
between adjacent photon-number peaks exceeds the relevant cavity and
qubit linewidths, the qubit spectrum resolves the photon-number
distribution of the oscillator.

The ac-Stark shift and its accompanying dephasing were measured in early
circuit-QED experiments~\cite{Schuster2005}, and the strong-dispersive
regime was subsequently used to resolve individual cavity photon-number
states~\cite{Schuster2007PhotonNumber}. In atomic cavity QED, dispersive
light shifts and phase accumulation provided direct probes of vacuum and
subphoton fields and enabled nondestructive photon detection
\cite{Brune1990QND,Nogues1999SinglePhoton,Brune1994LightShifts}.

\subsubsection{Approximate quantum-nondemolition structure}
\label{subsubsec:dispersive_qnd}

At this effective order,
\begin{equation}
    [
        \hat H_{\mathrm{disp}},
        \hat\sigma_z
    ]
    =
    0,
    \qquad
    [
        \hat H_{\mathrm{disp}},
        \hat a^\dagger\hat a
    ]
    =
    0.
    \label{eq:dispersive_qnd_commutators}
\end{equation}
The same Hamiltonian can therefore support two complementary QND measurements.

A weak cavity probe measures the qubit through the conditional cavity
frequency in
Eq.~\eqref{eq:dispersive_conditional_cavity_frequency}. In a frame
rotating at the probe frequency, the conditional coherent amplitudes obey
\begin{equation*}
    \dot\alpha_{\pm}
    =
    -
    \left[
        \frac{\kappa}{2}
        +
        \I
        \left(
            \Delta_d
            \pm
            \chi
        \right)
    \right]
    \alpha_{\pm}
    -
    \I\varepsilon_d,
\end{equation*}
where \(\Delta_d=\omega_a-\omega_d\), \(\varepsilon_d\) is the probe
amplitude, and the sign labels the qubit state. The stationary pointer
amplitudes are
\begin{equation*}
    \alpha_\pm^{\mathrm{ss}}
    =
    \frac{-\I\varepsilon_d}
    {\kappa/2+\I(\Delta_d\pm\chi)}.
\end{equation*}
Their separation determines the available readout signal. Once the
conditional coherent states have reached steady state, an ideal
single-mode cavity with vacuum input produces the measurement-induced
dephasing rate~\cite{Gambetta2006Measurement}
\begin{equation*}
    \Gamma_{\phi,\mathrm m}
    =
    \frac{\kappa}{2}
    \left|
        \alpha_+^{\mathrm{ss}}
        -
        \alpha_-^{\mathrm{ss}}
    \right|^2.
\end{equation*}
Detection efficiency and the measured output quadrature determine how
much of this acquired information becomes a useful measurement record.

Dispersive readout has been used experimentally to approach high-visibility
control of superconducting qubits and to demonstrate QND measurement in
superconducting devices
\cite{Wallraff2005DispersiveReadout,Lupascu2007QND}. In the reciprocal
use of the interaction, a far-detuned atom accumulates a
photon-number-dependent phase while leaving the cavity occupation
approximately unchanged. Repeated atomic probes enabled nondestructive
single-photon detection, observation of photon birth and death, and
progressive QND photon counting
\cite{Nogues1999SinglePhoton,Gleyzes2007QuantumJumps,Guerlin2007PhotonCounting}.

The commutators in
Eq.~\eqref{eq:dispersive_qnd_commutators} describe an approximate QND
structure, not an exact property of the microscopic model. The full JC
Hamiltonian does not commute with \(\hat\sigma_z\), and the Rabi
Hamiltonian does not conserve photon number. Residual exchange,
counter-rotating terms, multilevel transitions, measurement-induced
mixing, and environmental dressing therefore limit the nondemolition
character of real measurements.

\subsubsection{Conditional phases, parity mapping, and oscillator control}
\label{subsubsec:dispersive_conditional_phase}

In an interaction picture with respect to the shifted free frequencies,
the essential dispersive evolution is
\begin{equation*}
    \hat U_{\chi}(t)
    =
    \exp
    \left[
        -\I
        \chi t
        \hat a^\dagger\hat a
        \hat\sigma_z
    \right].
\end{equation*}
The phase acquired by one subsystem depends on the quantum state of the
other. This single unitary provides a common physical explanation for
several protocols.

If the qubit is prepared in a superposition, the two qubit components
rotate the oscillator in opposite directions in phase space. Conversely,
a Fock component \(\ket n\) produces a qubit relative phase
\(2\chi nt\). At
\(t_{\Pi} = \tfrac{\pi}{2|\chi|},\)
the qubit phase distinguishes even and odd photon-number sectors, which
allows oscillator parity to be mapped onto an ancilla qubit. This
mechanism has been used to track individual photon jumps by repeated QND
parity measurements and became a central ingredient in bosonic quantum
error correction
\cite{Sun2014ParityTracking,Ofek2016BosonicQEC}.

Photon-number-resolved dispersive shifts also allow selective phase
operations on individual Fock components. Experiments have used an
off-resonantly coupled qubit to implement selective
number-dependent arbitrary-phase gates and to synthesize nonclassical
oscillator states~\cite{Heeres2015SNAP}. These protocols illustrate a
general lesson: the effective term
\(\chi\hat n\hat\sigma_z\) is not tied to one specific application.
Readout, parity mapping, conditional displacements, cat-state control,
and bosonic-code operations are different ways of converting the same
conditional phase into a measurable or computational resource.
\begin{table*}[t]
    \centering
    \caption{\justifying
    Physical information exposed by the dispersive interaction. The
    first-limitation column identifies the leading structure that must be
    restored when the ideal two-level dispersive model ceases to be sufficient.
    }
    \label{tab:section2_dispersive_physics}
    \small
    \renewcommand{\arraystretch}{1.16}
    \setlength{\tabcolsep}{3pt}
    \begin{tabular}{@{}cccc@{}}
        \toprule
        \parbox[t]{0.17\textwidth}{\RaggedRight\textbf{Effective structure}}
        &
        \parbox[t]{0.22\textwidth}{\RaggedRight\textbf{Physical meaning}}
        &
        \parbox[t]{0.30\textwidth}{\RaggedRight\textbf{Representative experimental use}}
        &
        \parbox[t]{0.25\textwidth}{\RaggedRight\textbf{First limitation}}
        \\
        \midrule

        \parbox[t]{0.17\textwidth}{\RaggedRight\(\chi\hat n\hat\sigma_z\)}
        &
        \parbox[t]{0.22\textwidth}{\RaggedRight Qubit-state-dependent cavity frequency and photon-number-dependent qubit frequency.}
        &
        \parbox[t]{0.30\textwidth}{\RaggedRight Dispersive readout, ac-Stark spectroscopy, and photon-number resolution \cite{Schuster2005,Wallraff2005DispersiveReadout,Schuster2007PhotonNumber}.}
        &
        \parbox[t]{0.25\textwidth}{\RaggedRight Linewidths, photon-number fluctuations, multilevel corrections, and \(n/n_{\mathrm{crit}}\).}
        \\

        \parbox[t]{0.17\textwidth}{\RaggedRight\([\hat H_{\mathrm{disp}},\hat n]=[\hat H_{\mathrm{disp}},\hat\sigma_z]=0\)}
        &
        \parbox[t]{0.22\textwidth}{\RaggedRight Approximate nondemolition measurement of photon number or qubit energy.}
        &
        \parbox[t]{0.30\textwidth}{\RaggedRight Atomic QND photon counting and repeated photon-jump monitoring \cite{Brune1990QND,Nogues1999SinglePhoton,Gleyzes2007QuantumJumps,Guerlin2007PhotonCounting}; superconducting-qubit QND readout \cite{Lupascu2007QND}.}
        &
        \parbox[t]{0.25\textwidth}{\RaggedRight Residual exchange, measurement-induced transitions, counter-rotating terms, and open-system dressing.}
        \\

        \parbox[t]{0.17\textwidth}{\RaggedRight\(e^{-\I\chi t\hat n\hat\sigma_z}\)}
        &
        \parbox[t]{0.22\textwidth}{\RaggedRight Conditional phase between oscillator and qubit sectors.}
        &
        \parbox[t]{0.30\textwidth}{\RaggedRight Photon-parity mapping, bosonic error-syndrome tracking, and number-selective phase control \cite{Sun2014ParityTracking,Heeres2015SNAP,Ofek2016BosonicQEC}.}
        &
        \parbox[t]{0.25\textwidth}{\RaggedRight Ancilla decoherence, self-Kerr evolution, calibration errors, and residual cavity population.}
        \\

        \parbox[t]{0.17\textwidth}{\RaggedRight\(J_{12}(\hat\sigma_+^{(1)}\hat\sigma_-^{(2)}+\mathrm{H.c.})\)}
        &
        \parbox[t]{0.22\textwidth}{\RaggedRight Coherent interaction mediated by virtual cavity occupation.}
        &
        \parbox[t]{0.30\textwidth}{\RaggedRight Cavity-bus state transfer and driven resonator-induced phase gates \cite{Majer2007CavityBus,Paik2016RIPGate}.}
        &
        \parbox[t]{0.25\textwidth}{\RaggedRight Spectator modes, residual photons, cavity loss, and competing multilevel transitions.}
        \\

        \parbox[t]{0.17\textwidth}{\RaggedRight\(\widetilde a=a+(g/\Delta)\sigma_-+\cdots\)}
        &
        \parbox[t]{0.22\textwidth}{\RaggedRight The virtual dressing responsible for the coherent shift also dresses environmental channels.}
        &
        \parbox[t]{0.30\textwidth}{\RaggedRight Observation and engineering of Purcell-limited qubit relaxation \cite{Houck2008Purcell}.}
        &
        \parbox[t]{0.25\textwidth}{\RaggedRight Bath spectral structure, cavity filtering, and consistency of transformed jump operators.}
        \\

        \bottomrule
    \end{tabular}
\end{table*}
\subsubsection{Virtual buses and entangling interactions}
\label{subsubsec:dispersive_virtual_bus}

The same virtual-exchange physics can mediate interactions between
several systems coupled to a common off-resonant mode. For two qubits,
eliminating the cavity to second order generates the exchange term
\begin{equation*}
    \hat H_{12}^{\mathrm{ex}}
    =
    J_{12}
    \left(
        \hat\sigma_+^{(1)}
        \hat\sigma_-^{(2)}
        +
        \hat\sigma_-^{(1)}
        \hat\sigma_+^{(2)}
    \right),
\end{equation*}
with
\begin{equation*}
    J_{12}
    =
    \frac{
        g_1g_2
    }{2}
    \left(
        \frac{1}{\Delta_1}
        +
        \frac{1}{\Delta_2}
    \right)
\end{equation*}
in the simplest two-level model. The cavity remains only virtually
populated while carrying phase and excitation information between the
qubits. The operator is generated whenever the dispersive elimination is
controlled, but an appreciable coherent swap additionally requires the
two dressed qubit transitions to be near resonance, or to be brought into
resonance by an external control. This mechanism was used experimentally
to couple spatially separated superconducting qubits through a microwave
cavity bus \cite{Majer2007CavityBus}.

A driven common resonator can also trace qubit-state-dependent paths in
phase space and return close to its initial state while leaving behind a
conditional phase. Resonator-induced phase gates exploit this mechanism
to implement multiqubit entangling operations
\cite{Paik2016RIPGate}. These examples show how an effective Hamiltonian
identifies a process that is difficult to infer directly from the full
light--matter dynamics: the mediator can be nearly unoccupied and still
produce coherent transfer or entanglement.

The effective operator exposes the mechanism, but it does not by itself
determine entangling performance. Gate time, residual cavity population,
leakage, spectator modes, collective enhancement, and accumulated phase
errors depend on the chosen operating manifold. In
Sec.~\ref{subsec:application_beyond_rwa_multi_qubit}, we therefore derive the
interaction first for two qubits, compare the RWA and full-Rabi
reductions, reconstruct the laboratory observables, and use the resulting
dynamics to quantify entanglement generation. The same construction is
then generalized to nonidentical and many-qubit systems, for which the
relevant control parameters include collective matrix elements and the
proximity of additional resonances rather than a single ratio
\(g/\Delta\).

\subsubsection{Virtual hybridization, Purcell decay, and measurement backaction}
\label{subsubsec:dispersive_purcell}

The useful coherent shift and the leading induced dissipation have the
same microscopic origin. Under the dispersive transformation, the cavity
annihilation operator acquires a qubit component,
\begin{equation*}
    \widetilde{\hat a}
    =
    \hat a
    +
    \frac{g}{\Delta}
    \hat\sigma_-
    +
    \mathcal O
    \left(
        \frac{g^2}{\Delta^2}
    \right),
\end{equation*}
up to the phase and sign convention of the transformation. If the cavity has linewidth \(\kappa\) and is coupled to an
approximately Markovian vacuum bath, its fast susceptibility gives the
complex qubit self-energy
\begin{equation*}
    \delta\omega_q
    -
    \frac{\I}{2}\Gamma_{\mathrm P}
    =
    \frac{g^2}{\Delta+\I\kappa/2}.
\end{equation*}
Consequently,
\begin{subequations}
\begin{align}
    \delta\omega_q
    &=
    \frac{g^2\Delta}
    {\Delta^2+(\kappa/2)^2},
    \\
    \Gamma_{\mathrm P}
    &=
    \frac{g^2\kappa}
    {\Delta^2+(\kappa/2)^2}
    \xrightarrow{|\Delta|\gg\kappa}
    \kappa\frac{g^2}{\Delta^2}.
    \label{eq:section2_purcell_rate}
\end{align}
\end{subequations}
The Lorentzian form is still an ideal single-mode result; bath spectral
structure, Purcell filtering, and multilevel matrix elements modify it.
The frequency dependence of this decay and its control through circuit
design have been observed experimentally in circuit QED
\cite{Houck2008Purcell}.

The physical message is broader than the estimate in
Eq.~\eqref{eq:section2_purcell_rate}. The same virtual admixture that
creates the useful dispersive shift also opens a channel through which
cavity loss can relax the qubit. Likewise, increasing the probe field
improves the separation of the readout pointer states but also increases
measurement-induced dephasing and can activate transitions omitted by the
ideal dispersive model. Effective coherent and dissipative descriptions
must therefore be derived together.

Section~\ref{subsec:application_lossy_mediator} follows this connection in a
damped bosonic mediator. The real and imaginary parts of the same fast
susceptibility generate the coherent interaction and the effective jump
operator, and the resulting prediction is tested against the full master
equation, including the output field, virtual mediator occupation, cutoff
convergence, detuning, and linewidth.

\subsubsection{Critical occupation and breakdown}
\label{subsubsec:dispersive_breakdown}

The exact JC splitting shows directly that the dispersive series is an
occupation-dependent expansion. A characteristic scale is
\(n_{\mathrm{crit}} = \tfrac{ \Delta^2 }{ 4g^2 },\)
for which
\begin{equation*}
    \frac{
        4g^2(n+1)
    }{
        \Delta^2
    }
    =
    \frac{
        n+1
    }{
        n_{\mathrm{crit}}
    }.
\end{equation*}
The two-level dispersive expansion requires the populated photon sectors
to remain well below this scale. The numerical value of a practical
threshold depends on the desired observable and accuracy; the condition
\(n\ll n_{\mathrm{crit}}\) should therefore be interpreted as a scaling
criterion rather than a universal sharp boundary.

This limitation has an immediate physical interpretation. Increasing the
cavity occupation can accelerate readout and enlarge a conditional phase,
but it also strengthens qubit--cavity mixing and eventually invalidates
the diagonal effective description. Weakly anharmonic multilevel systems
introduce additional denominators and critical occupations, together
with self- and cross-Kerr terms
\cite{Koch2007Transmon,Blais2021}. The two-level JC result remains the
clean benchmark, but experimental use requires the populated manifold and
the multilevel structure to be checked explicitly.

\subsection{The critical large-frequency-ratio limit}
\label{subsec:qrm_critical_limit}

A second large-frequency separation of the Rabi model leads to physics
that is qualitatively different from the ordinary dispersive JC limit.
Let
\begin{equation}
    \lambda
    =
    \frac{2g}{\sqrt{\omega_0\omega_a}},
    \qquad
    \frac{\omega_0}{\omega_a}
    \longrightarrow
    \infty,
    \label{eq:qrm_critical_scaled_coupling}
\end{equation}
while \(\lambda\) is held fixed. Here \(\lambda\) denotes the
physical scaled Rabi coupling, not a formal perturbative bookkeeping
parameter. In this scaling,
\(g/|\Delta|\sim\sqrt{\omega_a/\omega_0}\to0\), but
\(g^2/\omega_0=\lambda^2\omega_a/4\) remains finite. The limit is
therefore not a trivial decoupling limit. Both co- and counter-rotating
virtual processes contribute to the same oscillator quadrature and must
be retained.

For \(\lambda<1\), the qubit remains predominantly in its lower
state. Eliminating its virtual excitations gives the normal-phase
Hamiltonian
\begin{equation*}
\begin{aligned}
    \hat H_{\mathrm{np}}
    ={}&
    \omega_a\hat a^\dagger\hat a
    -
    \frac{\lambda^2\omega_a}{4}
    \left(
        \hat a+\hat a^\dagger
    \right)^2
    -
    \frac{\omega_0}{2},
\end{aligned}
\end{equation*}
up to corrections that vanish as
\(\omega_a/\omega_0\to0\). A squeezing transformation diagonalizes this
quadratic Hamiltonian, whose excitation gap is
\(\varepsilon_{\mathrm{np}} = \omega_a \sqrt{1-\lambda^2}.\)
The softening of this gap at \(\lambda=1\) signals that the expansion
about the undisplaced oscillator vacuum has lost stability. With
\(\hat X=(\hat a+\hat a^\dagger)/\sqrt2\) and
\(\hat P=\I(\hat a^\dagger-\hat a)/\sqrt2\), the normal-phase ground state
has
\begin{equation*}
    \langle \hat X^2\rangle_{
    \mathrm{np}}
    =
    \frac{1}{2\sqrt{1-\lambda^2}},
    \qquad
    \langle \hat P^2\rangle_{
    \mathrm{np}}
    =
    \frac{\sqrt{1-\lambda^2}}{2}.
\end{equation*}
The soft quadrature therefore diverges while the conjugate quadrature is
squeezed. This is a qualitatively different use of large detuning from the
ordinary dispersive JC expansion: the virtual processes leave a finite
quadratic interaction and drive a critical mode rather than a small
frequency correction.

For \(\lambda>1\), the appropriate low-energy expansion first
displaces the oscillator and rotates the qubit. The two parity-related
semiclassical displacements have magnitude
\begin{equation*}
    \alpha_{\lambda}
    =
    \sqrt{
        \frac{\omega_0}{4\omega_a}
        \left(
            \lambda^2
            -
            \lambda^{-2}
        \right)
    }.
\end{equation*}
Writing the oscillator fluctuation operator in the two sectors as
\(\hat c_{\pm}=\hat a\mp\alpha_{\lambda}\), the leading
superradiant-phase Hamiltonian is
\begin{align*}
    \hat H_{\mathrm{sp}}^{(\pm)}
    &=
    \omega_a
    \hat c_{\pm}^\dagger
    \hat c_{\pm}
    -
    \frac{\omega_a}{4\lambda^4}
    \left(
        \hat c_{\pm}
        +
        \hat c_{\pm}^\dagger
    \right)^2 \\
    &\quad{}-
    \frac{\omega_0}{4}
    \left(
        \lambda^2
        +
        \lambda^{-2}
    \right),
\end{align*}
with excitation gap
\(\varepsilon_{\mathrm{sp}} = \omega_a \sqrt{1-\lambda^{-4}}.\)
The two signs correspond to symmetry-related minima. At a finite
frequency ratio, the exact Hamiltonian retains parity and its eigenstates
are parity-resolved superpositions of the two displaced sectors. A true
nonanalytic quantum phase transition and spontaneous symmetry breaking
emerge only in the singular limit in
Eq.~\eqref{eq:qrm_critical_scaled_coupling}; large but finite ratios show
a crossover and finite-frequency critical scaling.

This limit was used to establish the normal-to-superradiant quantum phase
transition of the single-qubit Rabi model theoretically
\cite{Hwang2015QRMQPT}. Its order parameters and gap closing were later
probed in a single trapped-ion quantum simulator
\cite{Cai2021QRMQPT}. The interpretation applies directly to an
engineered realization of the QRM. For a microscopic electromagnetic
system, the few-level truncation, gauge choice, and self-energy terms must
be treated consistently before transferring conclusions about an
equilibrium superradiant transition to the underlying hardware.
\begin{table*}[t]
    \centering
    \caption{\justifying
    Principal effective descriptions of the quantum Rabi model discussed
    in this section. The conditions shown are organizing scalings rather
    than universal numerical error thresholds; the populated manifold,
    target observable, retained sector, and comparison time must also be
    specified.
    }
    \label{tab:qrm_effective_regimes}
    \small
    \renewcommand{\arraystretch}{1.18}
    \setlength{\tabcolsep}{3pt}
    \begin{tabular}{@{}cccc@{}}
        \toprule
        \parbox[t]{0.17\textwidth}{\RaggedRight\textbf{Regime and retained structure}}
        &
        \parbox[t]{0.21\textwidth}{\RaggedRight\textbf{Organizing condition}}
        &
        \parbox[t]{0.24\textwidth}{\RaggedRight\textbf{Effective description}}
        &
        \parbox[t]{0.32\textwidth}{\RaggedRight\textbf{Physical content}}
        \\
        \midrule

        \parbox[t]{0.17\textwidth}{\RaggedRight
        Near-resonant weak coupling; bare excitation manifolds}
        &
        \parbox[t]{0.21\textwidth}{\RaggedRight
        \(g\sqrt{n_\star+1}/\Sigma\ll1\) and
        \(|\Delta|/\Sigma\ll1\)}
        &
        \parbox[t]{0.24\textwidth}{\RaggedRight
        Jaynes--Cummings Hamiltonian}
        &
        \parbox[t]{0.32\textwidth}{\RaggedRight
        Real one-excitation exchange, avoided crossings, and JC
        doublets.}
        \\

        \parbox[t]{0.17\textwidth}{\RaggedRight
        Far-detuned JC regime; bare states with virtual admixtures}
        &
        \parbox[t]{0.21\textwidth}{\RaggedRight
        \(g\sqrt{n_\star+1}/|\Delta|\ll1\)}
        &
        \parbox[t]{0.24\textwidth}{\RaggedRight
        Dispersive Hamiltonian
        \(\chi\hat n\hat\sigma_z\) with reconstruction}
        &
        \parbox[t]{0.32\textwidth}{\RaggedRight
        Conditional shifts and phases, approximate QND measurement,
        mediated interactions, and dressed dissipation.}
        \\

        \parbox[t]{0.17\textwidth}{\RaggedRight
        Selected multiphoton resonance; nearly degenerate same-parity
        states}
        &
        \parbox[t]{0.21\textwidth}{\RaggedRight
        \(\omega_0-3\omega_a=\mathcal O(g^2/\omega_0)\), with
\(|\delta_3|\lesssim|G_3|
=\mathcal O(g^3/\omega_0^2)\)}
        &
        \parbox[t]{0.24\textwidth}{\RaggedRight
        Resonant few-state Hamiltonian with a second-order level shift
        and a third-order coupling
        \(G_3\propto g^3\)}
        &
        \parbox[t]{0.32\textwidth}{\RaggedRight
        Coherent conversion
        \(\ket{e,0}\leftrightarrow\ket{g,3}\), parity selection,
        counter-rotating virtual paths, a shifted avoided crossing, and
        static or Floquet engineering of multiphoton interactions.}
        \\

        \parbox[t]{0.17\textwidth}{\RaggedRight
        Deep-strong coupling; displaced oscillator ladders}
        &
        \parbox[t]{0.21\textwidth}{\RaggedRight
        \(g/\omega_a\gtrsim1\), together with small discarded
        displaced-basis matrix elements}
        &
        \parbox[t]{0.24\textwidth}{\RaggedRight
        Adiabatic displaced blocks and GRWA}
        &
        \parbox[t]{0.32\textwidth}{\RaggedRight
        Conditional phase-space motion, parity doublets, strong
        light--matter entanglement, collapse, revival, and freezing.}
        \\

        \parbox[t]{0.17\textwidth}{\RaggedRight
        Critical large-frequency-ratio scaling; soft oscillator
        quadrature}
        &
        \parbox[t]{0.21\textwidth}{\RaggedRight
        \(\omega_0/\omega_a\to\infty\) at fixed
        \(\lambda=2g/\sqrt{\omega_0\omega_a}\)}
        &
        \parbox[t]{0.24\textwidth}{\RaggedRight
        Normal- and superradiant-phase squeezed-oscillator
        Hamiltonians}
        &
        \parbox[t]{0.32\textwidth}{\RaggedRight
        Gap closing, diverging quadrature fluctuations, displaced
        minima, and quantum critical scaling.}
        \\

        \bottomrule
    \end{tabular}
\end{table*}

\subsection{Higher-order resonant processes in the quantum Rabi model}
\label{subsec:qrm_higher_order_resonances}

The quantum Rabi model also supports resonant interactions that first appear
at higher perturbative order. Near
\(\omega_0\simeq3\omega_a\), the same-parity states \(\ket{e,0}\) and
\(\ket{g,3}\) are connected by three successive insertions of the microscopic
interaction. We refer to each insertion as a \emph{vertex} when discussing a
virtual path. One of the three vertices is counter rotating, so the ordinary
RWA removes the path that generates the target interaction. This is not a
small correction to an existing coupling: the effective coupling itself first
appears at higher order.

For selected initial and final states, the leading matrix element is a
coherent sum over virtual paths,
\begin{equation*}
G_{fi}^{(r)}
=
\sum_{\mathcal P:i\rightarrow f}
A_{\mathcal P}^{(r)}.
\end{equation*}
The number of paths alone is not predictive: their matrix-element signs,
phases, intermediate denominators, and symmetry relations determine
whether they add constructively, interfere destructively, or vanish.
This path structure is exposed differently by resolvent, Schrieffer--Wolff,
time-averaging, and Floquet constructions, and it is one reason that the
diagonal resonance shift must be derived together with the off-diagonal
coupling \cite{Wang2017CounterRotatingTwoQubits,Huang2026}.

The three-photon resonance will serve as the common higher-order benchmark.
Section~\ref{sec:higher_order_expansions} derives its shifted resonance,
cubic avoided-crossing coupling, and reconstruction, while
Sec.~\ref{sec:methods_comparison} compares the information exposed by the
different constructions. Full cross-method calculations are given in Notebook
N05, \path{N05_Three-Photon-Rabi-Resonance}, and Supplemental Sec.~S3.

Higher-order effective interactions are now experimentally resolvable.
Single-photon-driven high-order sidebands were observed in an
ultrastrongly coupled circuit-QED device
\cite{Chen2017HighOrderSidebands}. More recently, spectroscopy of two
flux qubits coupled to one resonator revealed an avoided crossing
dominated by \(\ket{gg,1}\) and \(\ket{ee,0}\), providing evidence for
the one-photon/two-atom channel predicted in
Ref.~\cite{Garziano2016OnePhotonTwoAtoms}, although not yet its
time-domain Rabi oscillation \cite{Tomonaga2025TwoArtificialAtoms}. A
complementary multimode experiment resolved strong coupling between
\(\ket{2,0,g}\) and \(\ket{0,1,g}\) through a detuned flux qubit
\cite{Wang2025OneTwoPhoton}. Pumped superconducting cavities have likewise
realized three-photon down-conversion and genuine tripartite non-Gaussian
output entanglement
\cite{Chang2020ThreePhotonSPDC,JarvisFrain2026Tripartite}. These devices
differ microscopically, but their analysis poses the same effective-theory
questions: which resonant manifold is retained, which paths contribute,
which diagonal shifts move the resonance, and which observables and noise
channels must be reconstructed.

\subsection{One microscopic model, several effective theories}
\label{subsec:qrm_effective_theory_map}

Table~\ref{tab:qrm_effective_regimes} brings the regimes together. The
microscopic Hamiltonian is the same in every row, but the scale separation,
retained states, natural basis, observable, and comparison time are not.
Changing any of these choices can change both the perturbative parameter and
the operator structure of the effective model.

The examples also locate the information that lies outside the reduced
generator. The RWA requires a return from the interaction picture, the GRWA
uses dressed states and observables in a displaced basis, and the dispersive
Hamiltonian hides virtual admixtures that remain visible in populations,
transition operators, and dissipative channels. The next section organizes
these operations as encoding, effective evolution, and reconstruction.

\section{From an Effective Hamiltonian to a Physical Prediction}
\label{sec:common_framework}

Section~\ref{sec:rwa_grwa} compared several effective descriptions of the
same microscopic Hamiltonian. Choosing one of them is only the beginning
of a prediction. We must also represent the prepared state in the chosen
description and translate the result back into the quantity measured in
the laboratory.

The Rabi--Jaynes--Cummings examples make the distinction concrete. An
interaction picture or displaced basis is an exact change of
representation; discarding counter-rotating harmonics or retaining only
selected dressed blocks is an approximation. After the reduced dynamics
has been solved, the state and observables must be returned to the physical
frame and, when necessary, to the full Hilbert space. These operations are
often hidden inside the notation \(\hat H_{\mathrm{eff}}\), although each
can introduce a different error.

We now separate the calculation into three steps: encoding the physical
preparation, evolving it with the effective model, and reconstructing the
physical prediction. The same organization applies to perturbative
dressing, projected or eliminated variables, Floquet micromotion, and
open-system channels. Method-specific derivations follow in
Sec.~\ref{sec:methods}; here the emphasis is on the information needed to
use their results consistently.

\subsection{The three-step idea: encode, evolve, reconstruct}
\label{subsec:framework_three_step}
We write the complete effective prediction as
\begin{equation}
    \hat\rho_{\mathrm{rec}}(t)
    =
    \left[
        \mathcal R_t
        \circ
        \Phi_{\mathrm{eff}}(t,0)
        \circ
        \mathcal E_0
    \right]
    [\hat\rho(0)].
    \label{eq:effective_prediction_map}
\end{equation}
The maps act from right to left:
\((\mathcal A\circ\mathcal B)[X]=\mathcal A[\mathcal B[X]]\). The
encoding map \(\mathcal E_0\) represents the physical preparation in the
chosen effective description. The propagator
\(\Phi_{\mathrm{eff}}(t,0)\) supplies the reduced evolution and may be
generated by an effective Hamiltonian, by a Liouvillian
\(\mathcal L(t)\) through \(\dot\rho=\mathcal L(t)[\rho]\), or by a
directly constructed finite-time map. The reconstruction map
\(\mathcal R_t\) returns the result to the physical description. It can
contain an inverse frame transformation, perturbative dressing, embedding
of a retained state, reconstruction of eliminated variables, transformed
measurement operators, or Floquet micromotion.

The maps \(\mathcal E_0\) and \(\mathcal R_t\) are bookkeeping maps, not
necessarily laboratory operations or completely positive trace-preserving
channels. A projection can make the encoded state \emph{subnormalized}:
its trace is smaller than unity because part of the physical state lies
outside the retained space. Dividing by that trace defines a normalized
conditional state, but it represents a physical postselection only when
such a measurement is actually performed. Likewise, reconstruction need
not be a literal inverse. An eliminated amplitude can be restored
perturbatively even though the reduced model never evolves that amplitude
as an independent variable.

When the claim concerns one observable rather than the full state and
\(\mathcal R_t\) is linear, the same reconstruction can be performed on
the operator side. The Hilbert--Schmidt adjoint of \(\mathcal R_t\) is
defined by
\begin{equation*}
    \operatorname{Tr}
    \left[
        \hat O\,\mathcal R_t[\hat X]
    \right]
    =
    \operatorname{Tr}
    \left[
        \mathcal R_t^{\dagger}[\hat O] \, \hat X
    \right].
\end{equation*}
Then
\begin{equation*}
\begin{aligned}
    \langle\hat O\rangle_{\mathrm{rec}}(t)
    ={}&
    \operatorname{Tr}
    \Bigl[
        \hat O_{\mathrm{eff}}(t)
        \Phi_{\mathrm{eff}}(t,0)
        \mathcal E_0[\hat\rho(0)]
    \Bigr],
    \\
    \hat O_{\mathrm{eff}}(t)
    ={}&
    \mathcal R_t^{\dagger}[\hat O].
\end{aligned}
\end{equation*}
State and observable reconstruction are two ways of making the same
physical comparison; the shorter route depends on the quantity of
interest. Their order-by-order equivalence is derived in
Appendix~\ref{app:order_consistent_reconstruction}.

\begin{tcolorbox}[
    beamerexample,
    breakable=false,
    title={Operational principle}
]
\textbf{An effective generator supplies the middle step of a prediction.}
The calculation must also answer three questions:
\begin{enumerate}
    \item How is the physical preparation represented in the effective
    frame, basis, retained sector, or set of slow variables?
    \item How does that encoded state evolve under the effective
    Hamiltonian, Liouvillian, or finite-time map?
    \item How are the laboratory frame, eliminated amplitudes, dressed
    observables, dissipative channels, and fast motion reconstructed?
\end{enumerate}
If the encoding is omitted, the reduced dynamics may begin from the wrong
state. If reconstruction is omitted, the spectrum may be accurate while a
laboratory phase, virtual population, transition matrix element,
dissipative rate, or micromotion is not.
\end{tcolorbox}

\subsubsection{First example: the RWA as a complete prediction}
\label{subsubsec:framework_rwa_map}

\begin{figure*}[!ht]
    \centering
    \includegraphics[
        width=1.0\textwidth
    ]{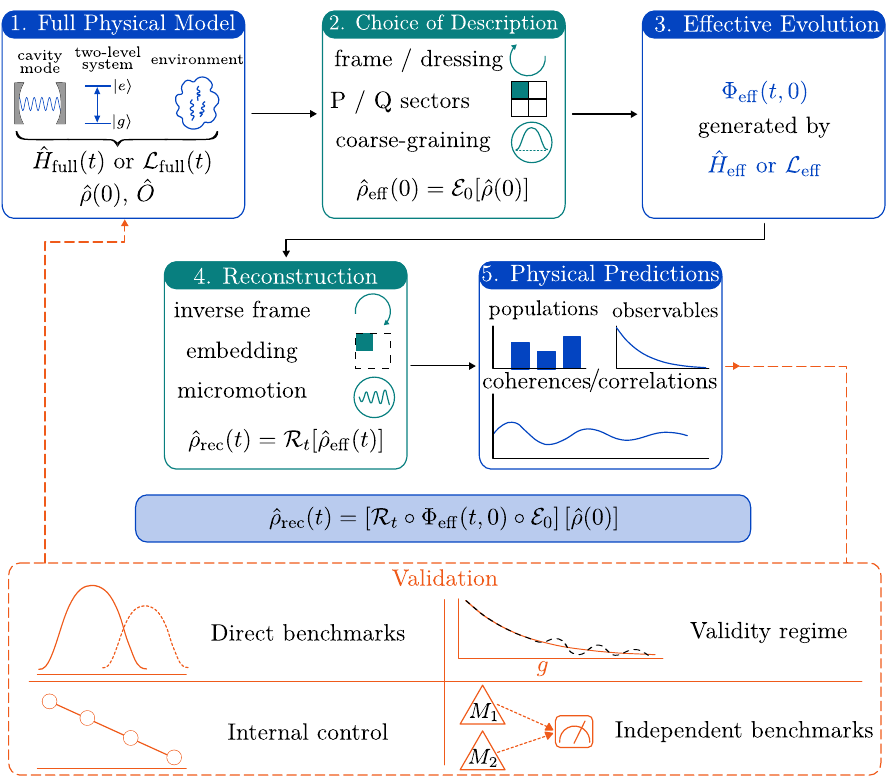}
    \captionof{figure}{\justifying
    \textbf{A complete effective prediction.} The preparation is encoded in the
    selected description, evolved with the effective map, and reconstructed
    in the physical representation. The lower loop compares that prediction
    with direct benchmarks, controlled limits, internal diagnostics, and
    independent evidence.
    }
    \label{fig:effective_prediction_pipeline}
\end{figure*}

The interaction-picture RWA is the simplest example. With
\(\hat U_0(t)=e^{-\I\hat H_0t}\) and
\(\hat U_0(0)=\hat I\), the initial state needs no transformation. The
effective propagator \(\hat U_{I,\mathrm{RWA}}(t,0)\) is obtained after
discarding the counter-rotating harmonics, and \(\hat U_0(t)\) restores
the laboratory frame. The complete prediction is
\begin{align*}
    \hat\rho_{\mathrm{rec}}^{\mathrm{RWA}}(t)
    =
    \hat U_0(t)
    \hat U_{I,\mathrm{RWA}}(t,0)
    \hat\rho(0) \\
    \times
    \hat U_{I,\mathrm{RWA}}^\dagger(t,0)
    \hat U_0^\dagger(t).
\end{align*}
The interaction-picture transformation and its inverse are exact; the
approximation is the removal of the fast harmonics. The same RWA can be
written directly in the Schr\"odinger picture with
\(\hat H_{\mathrm{JC}}\), where both endpoint maps are the identity. The
intermediate Hamiltonians look different, but their predictions agree
after they are expressed in the same frame. Figure~\ref{fig:effective_prediction_pipeline} summarizes this construction.

Here the \emph{full model} is the Hamiltonian, Liouvillian, or dynamical
map used as the reference for the approximation; it need not be a
fundamental microscopic theory.

\subsection{Exact changes of representation}
\label{subsec:exact_change_of_representation}

For a possibly time-dependent unitary transformation \(\hat U(t)\), we
use the convention
\begin{equation}
    \widetilde{\hat\rho}(t)
    =
    \hat U^\dagger(t)
    \hat\rho(t)
    \hat U(t).
    \label{eq:framework_transformed_state}
\end{equation}
The transformed Hamiltonian is
\begin{equation*}
    \widetilde{\hat H}(t)
    =
    \hat U^\dagger(t)
    \hat H(t)
    \hat U(t)
    -
    \I
    \hat U^\dagger(t)
    \dot{\hat U}(t).
\end{equation*}
No approximation has yet been made. It enters when transformed terms are
truncated, blocks are decoupled perturbatively, harmonics are filtered, or
degrees of freedom are removed. For the frame change itself, the endpoint
maps are
\begin{subequations}
\label{eq:framework_exact_unitary_maps}
\begin{align}
    \mathcal E_0^{(U)}[\hat\rho]
    & =
    \hat U^\dagger(0)\hat\rho\hat U(0),
    \\
    \mathcal R_t^{(U)}[\hat X]
    & =
    \hat U(t)\hat X\hat U^\dagger(t).
\end{align}
\end{subequations}
The common simplification \(\mathcal E_0^{(U)}=\mathcal I\) is valid only
when \(\hat U(0)=\hat I\), or when the physical preparation has already
been specified in the transformed frame. Detailed derivations of the
transformation formula, the interaction picture, and representative
rotating frames are collected in Supplemental Material, Sec.~S1.

\subsubsection{Sequential frames and the order of reconstruction}
\label{subsubsec:framework_sequential_frames}

Hamiltonian-engineering protocols often use several transformations in
sequence. If
\(\hat U_{\mathrm{tot}}(t)=\hat U_1(t)\hat U_2(t)\cdots\hat U_m(t)\),
then
\(\widetilde{\hat\rho}=\hat U_{\mathrm{tot}}^\dagger\hat\rho
\hat U_{\mathrm{tot}}\). Reconstruction applies the inverse
transformations in reverse operational order,
\begin{align*}
    \hat\rho(t)
    =
    &\hat U_1(t)\hat U_2(t)\cdots\hat U_m(t)
    \widetilde{\hat\rho}(t)
    \\
    &\times
    \hat U_m^\dagger(t)\cdots\hat U_2^\dagger(t)
    \hat U_1^\dagger(t).
\end{align*}
Because the transformations generally do not commute, their order is part
of the physical specification. A final Hamiltonian written in a doubly or
multiply dressed frame is incomplete unless this ordered reconstruction is
also recorded.

A laboratory observable must be represented in the same frame,
\(\widetilde{\hat O}(t)=\hat U^\dagger(t)\hat O\hat U(t)\), so that
\begin{equation*}
    \langle\hat O\rangle_{\mathrm{lab}}(t)
     =
    \operatorname{Tr}
    [
        \hat O\hat\rho(t)
    ]
    =
    \operatorname{Tr}
    [
        \widetilde{\hat O}(t)
        \widetilde{\hat\rho}(t)
    ].
\end{equation*}
Using the untransformed operator \(\hat O\) with the transformed state
\(\widetilde\rho\) generally describes a different measurement.

A simple phase rotation illustrates the observable transformation. For
\(\hat U_\phi=e^{-\I\phi\hat n}\), with
\(\hat n=\hat a^\dagger\hat a\), the number operator is unchanged because
it commutes with \(\hat U_\phi\). The field operator instead becomes
\(\hat U_\phi^\dagger\hat a\hat U_\phi=e^{-\I\phi}\hat a\), so the phase
factor must be retained when a laboratory amplitude is evaluated in the
rotated frame. Nothing physical has changed; the state and the operator
have simply been written in new coordinates.

\subsubsection{Example: the displaced GRWA description}
\label{subsubsec:framework_grwa_map}

The generalized rotating-wave approximation (GRWA) from
Sec.~\ref{subsec:grwa} gives a less trivial example. Let
\(\hat U_{\mathrm p}\) be the spin-dependent displacement and let
\(\hat P_{\mathrm G}\) select the dressed blocks retained by the GRWA.
A schematic encoding is
\begin{equation*}
    \mathcal E_0^{\mathrm G}[\hat\rho]
    =
    \hat P_{\mathrm G}
    \hat U_{\mathrm p}^\dagger
    \hat\rho
    \hat U_{\mathrm p}
    \hat P_{\mathrm G}.
\end{equation*}
The effective map evolves this state with the retained GRWA blocks,
\begin{equation*}
    \Phi_{\mathrm G}(t,0)[\hat X]
    =
    e^{-\I\hat H_{\mathrm{GRWA}}t}
    \hat X
    e^{\I\hat H_{\mathrm{GRWA}}t}.
\end{equation*}
If \(\mathcal I_{\mathrm G}\) denotes embedding of the retained blocks
into the displaced Hilbert space, the reconstructed state is
\begin{equation*}
    \mathcal R_t^{\mathrm G}[\hat X]
    \simeq
    \hat U_{\mathrm p}
    \mathcal I_{\mathrm G}[\hat X]
    \hat U_{\mathrm p}^\dagger.
\end{equation*}
The displacement is exact, whereas the block selection is approximate.
This distinction is visible in laboratory observables. With the
convention of Sec.~\ref{subsec:grwa},
\begin{align*}
    \hat U_{\mathrm p}^\dagger\hat a\hat U_{\mathrm p}
    &=\hat a-\eta\hat\sigma_x,
    \\
    \hat U_{\mathrm p}^\dagger\hat a^\dagger\hat a\hat U_{\mathrm p}
    &=(\hat a^\dagger-\eta\hat\sigma_x)
      (\hat a-\eta\hat\sigma_x).
\end{align*}
Evaluating the bare operator \(\hat a^\dagger\hat a\) in the GRWA frame
would omit the displacement contribution even if the GRWA energies were
accurate.

\subsection{Dressing, projection, and eliminated variables}
\label{subsec:projection_reconstruction_framework}

A projector \(\hat P\) selects the retained space, and
\(\hat Q=\hat I-\hat P\) selects its complement. If a unitary
transformation approximately decouples them, the Hamiltonian and a
physical observable become
\begin{align*}
    \hat H_{\mathrm{eff}}
    &=\hat P\widetilde{\hat H}\hat P,
    \\
    \hat O_{\mathrm{eff}}^{(U)}
    &=\hat P\hat U^\dagger\hat O\hat U\hat P.
\end{align*}
The expression above is exact only when the transformation and block
decoupling are exact on the relevant sector. In perturbation theory, the
Hamiltonian, state, and observable transformations must be truncated
consistently.

A projection method can instead store the eliminated component in a wave
operator \(\hat\Omega\), which maps retained vectors to their full-space
counterparts.\footnote{The normalization and Hermitization of wave
operators depend on the projection convention; see
Sec.~\ref{subsec:projection_methods}.} In that language, physical
matrix elements depend schematically on
\begin{equation*}
    \hat P
    \hat\Omega^\dagger
    \hat U^\dagger
    \hat O
    \hat U
    \hat\Omega
    \hat P,
\end{equation*}
rather than on \(\hat P\hat O\hat P\) alone.

\subsubsection{Example: dispersive JC dressing through Schrieffer-Wolff (SW)}
\label{subsubsec:framework_sw_jc_map}

For the dispersive Jaynes--Cummings model, define
\begin{equation}
    \epsilon
    =
    \frac{g}{\Delta},
    \qquad
    \hat S_1
    =
    \epsilon
    \left(
        \hat a^\dagger\hat\sigma_-
        -
        \hat a\hat\sigma_+
    \right),
    \label{eq:framework_sw_generator_preview}
\end{equation}
with \(\hat U_{\mathrm{SW}}=e^{\hat S_1}\). We use the transformed Hamiltonian
\(\widetilde{\hat H}=e^{-\hat S_1}\hat H e^{\hat S_1}\). The equally valid
replacement \(\hat S_1\mapsto-\hat S_1\) reverses every dressing formula;
what matters is that the Hamiltonian, state, observable, and jump-operator
conventions are changed together. The detailed derivation is given in
Sec.~\ref{subsec:sw_dispersive_jc}. At the level needed here, the three
maps are
\begin{subequations}
\label{eq:framework_sw_three_maps}
\begin{align}
    \mathcal E_0^{\mathrm{SW}}[\hat\rho]
    & =
    e^{-\hat S_1}
    \hat\rho
    e^{\hat S_1},
    \\
    \Phi_{\mathrm{disp}}(t,0)[\hat X]
    & =
    e^{-\I\hat H_{\mathrm{disp}}t}
    \hat X
    e^{\I\hat H_{\mathrm{disp}}t},
    \\
    \mathcal R_t^{\mathrm{SW}}[\hat X]
    & =
    e^{\hat S_1}
    \hat X
    e^{-\hat S_1}.
\end{align}
\end{subequations}
Projection onto a smaller dispersive sector can be inserted after the
encoding and before the reconstruction when required by the claim.

The role of reconstruction is easiest to see for an effective-frame state
\(\ket{e,0}\). Its physical dressed counterpart is
\begin{equation*}
\begin{aligned}
    e^{\hat S_1}\ket{e,0}
    & =
    \ket{e,0}
    +
    \epsilon\ket{g,1}
    +
    \mathcal O(\epsilon^2).
\end{aligned}
\end{equation*}
The reconstructed laboratory state therefore contains a virtual one-photon
component of order \(\epsilon\), with photon population of order
\(\epsilon^2\). Conversely, a physically prepared bare state
\(\ket{e,0}\) is encoded as
\begin{equation*}
    e^{-\hat S_1}\ket{e,0}
    =
    \ket{e,0}
    -
    \epsilon\ket{g,1}
    +
    \mathcal O(\epsilon^2).
\end{equation*}
Using \(\ket{e,0}\) for both the physical preparation and the effective
initial state therefore changes the protocol at first order. Accurate
dressed energies alone do not guarantee an accurate trajectory or
eliminated-sector population.

\subsubsection{Example: reconstructing an eliminated Raman level}
\label{subsubsec:framework_raman_map}

Consider a three-level system with long-lived states \(\ket{1}\) and
\(\ket{2}\), an auxiliary state \(\ket{e}\), and the rotating-frame
Hamiltonian
\begin{equation*}
    \hat H_{\Lambda}=
    \Delta\ket{e}\!\bra{e}+
    \frac{1}{2}
    \left(
        \Omega_1\ket{e}\!\bra{1}
        +
        \Omega_2\ket{e}\!\bra{2}
        +
        \mathrm{H.c.}
    \right).
\end{equation*}
For \(|\Omega_j/\Delta|\ll1\), the effective Hamiltonian acting on
\(\{\ket{1},\ket{2}\}\) is
\begin{align*}
    \hat H_{\mathrm{Raman}}
    =
    &-\frac{|\Omega_1|^2}{4\Delta}
    \ket{1}\!\bra{1}
    -\frac{|\Omega_2|^2}{4\Delta}
    \ket{2}\!\bra{2}
    \\
    &-\frac{1}{4\Delta}
    \left(
        \Omega_1^*\Omega_2
        \ket{1}\!\bra{2}
        +
        \mathrm{H.c.}
    \right).
\end{align*}
Writing the slow state as
\begin{equation*}
    \ket{\psi_s(t)}
    =
    c_1(t)\ket{1}+c_2(t)\ket{2},
\end{equation*}
the leading reconstruction of the eliminated amplitude is
\begin{equation}
    c_e(t)
    \simeq
    -
    \frac{
        \Omega_1c_1(t)+\Omega_2c_2(t)
    }{
        2\Delta
    }.
    \label{eq:framework_raman_reconstructed_amplitude}
\end{equation}
Thus,
\begin{equation*}
    \ket{\psi_{\mathrm{rec}}(t)}
    \simeq
    c_1(t)\ket{1}
    +
    c_2(t)\ket{2}
    +
    c_e(t)\ket{e}.
\end{equation*}
The eliminated state is not retained as an independent dynamical variable,
but it is not exactly empty. If the physical preparation has
\(c_e(0)=0\) while Eq.~\eqref{eq:framework_raman_reconstructed_amplitude}
predicts a nonzero slow-manifold value, a short fast transient appears.
This short, preparation-dependent boundary layer is called
\emph{initial slip}. The Raman calculation is developed further in the
adiabatic-elimination discussion and in Sec.~\ref{sec:applications}.

\subsubsection{Leakage and error inside the retained space}
\label{subsubsec:framework_leakage}

If the full transformed state develops support in \(\hat Q\), define the
leakage
\begin{equation*}
    L(t)
    =
    1
    -
    \operatorname{Tr}
    \left[
        \hat P
        \widetilde{\hat\rho}_{\mathrm{full}}(t)
        \hat P
    \right].
\end{equation*}
Leakage measures failure of the retained space to remain closed. It is
not the same quantity as the error of the dynamics conditioned on
remaining inside that space. When \(L(t)<1\), the normalized conditional
retained state is
\begin{equation}
    \widetilde{\hat\rho}_{P}^{\,\mathrm{full}}(t)
    =
    \frac{
        \hat P
        \widetilde{\hat\rho}_{\mathrm{full}}(t)
        \hat P
    }{
        1-L(t)
    }.
    \label{eq:framework_conditional_full_state}
\end{equation}
A useful validation reports both the leakage and the error between
Eq.~\eqref{eq:framework_conditional_full_state} and the effective state.
The conditional dynamics may remain accurate even after too much
probability has left the retained space.

Different block-diagonalization conventions may also produce effective
Hamiltonians related by a unitary transformation acting entirely inside
\(\hat P\). This \emph{block-gauge freedom} means that raw term-by-term
agreement is not required; invariant spectra and consistently reconstructed
observables are more meaningful comparison targets
\cite{Bravyi2011}.\footnote{Here \emph{block gauge} means a change of
basis internal to the retained space. It is distinct from electromagnetic
gauge freedom and from Floquet gauge.}

\subsection{Periodically driven systems: the missing micromotion}
\label{subsec:driven_framework}

For a periodic Hamiltonian,
\begin{equation*}
    \hat H(t+T)=\hat H(t),
\end{equation*}
Floquet theory gives an exact factorization
\begin{equation}
    \hat U_{\mathrm{full}}(t,0)
    =
    \hat U_{\mathrm{mic}}(t)
    e^{-\I\hat H_Ft}
    \hat U_{\mathrm{mic}}^\dagger(0),
    \label{eq:framework_floquet_factorization}
\end{equation}
where \(\hat H_F\) is a Floquet Hamiltonian and
\(\hat U_{\mathrm{mic}}(t+T)=\hat U_{\mathrm{mic}}(t)\) is the periodic
micromotion operator
\cite{Floquet1883,GoldmanDalibard2014,Eckardt2015,Bukov2015,Eckardt2017}. 
Equation~\eqref{eq:framework_floquet_factorization} is already in the form
of Eq.~\eqref{eq:effective_prediction_map}:
\begin{subequations}
\label{eq:framework_floquet_three_maps}
\begin{align}
    \mathcal E_0^{F}[\hat\rho]
    & =
    \hat U_{\mathrm{mic}}^\dagger(0)
    \hat\rho
    \hat U_{\mathrm{mic}}(0),
    \\
    \Phi_F(t,0)[\hat X]
    & =
    e^{-\I\hat H_Ft}
    \hat X
    e^{\I\hat H_Ft},
    \\
    \mathcal R_t^{F}[\hat X]
    & =
    \hat U_{\mathrm{mic}}(t)
    \hat X
    \hat U_{\mathrm{mic}}^\dagger(t).
\end{align}
\end{subequations}
Periodicity implies
\(\hat U_{\mathrm{mic}}(nT)=\hat U_{\mathrm{mic}}(0)\). In a
stroboscopic gauge with \(\hat U_{\mathrm{mic}}(0)=\hat I\), the endpoint
micromotion disappears and \(\hat H_F\) governs the stroboscopic
trajectory. At intermediate times, omitting
\(\mathcal R_t^{F}\) discards real laboratory motion. High driving
frequency is not required for exact Floquet factorization; it is an
assumption used by particular perturbative constructions of \(\hat H_F\)
and \(\hat U_{\mathrm{mic}}(t)\). The quasienergy branch and Floquet gauge
are not unique, but reconstructed observables are unchanged when
\(\hat H_F\) and the micromotion are transformed consistently.

\subsection{Open systems: transform the dissipative channels as well}
\label{subsec:open_framework}

For an open system, the time-local generator and the finite-time map it
produces play different roles. Accessible discussions of the
approximations behind Markovian master equations and of the dynamical-map
viewpoint are given in
Refs.~\cite{Stefanini2026IsLindblad,Szankowski2023OpenSystems}. We use
\begin{equation}
    \mathcal D[\hat L]\hat\rho
    =
    \hat L\hat\rho\hat L^\dagger
    -
    \frac{1}{2}
    \left\{
        \hat L^\dagger\hat L,
        \hat\rho
    \right\},
    \label{eq:global_dissipator_convention}
\end{equation}
and write a time-local Gorini--Kossakowski--Lindblad--Sudarshan (GKLS)
master equation as
\begin{equation*}
    \dot{\hat\rho}
    =
    -\I
    [
        \hat H(t),
        \hat\rho
    ]
    +
    \sum_k
    \gamma_k
    \mathcal D[\hat L_k(t)]
    \hat\rho.
\end{equation*}
This is the standard GKLS structure
\cite{Gorini1976,Lindblad1976,breuer2002theory,Manzano2020}; a pedagogical
route from master-equation derivations to numerical implementation is
given in Ref.~\cite{Campaioli2024MasterEquations}. A jump operator
represents a channel in the generator. Reading individual jumps as
measurement records requires an additional choice of stochastic
unravelling. Complete positivity, meanwhile, is a property of the
finite-time map and does not follow from trace preservation alone.

Under the exact unitary transformation of
Eq.~\eqref{eq:framework_transformed_state}, the jump operators transform
with the Hamiltonian,
\begin{equation*}
    \widetilde{\hat L}_k(t)
    =
    \hat U^\dagger(t)
    \hat L_k(t)
    \hat U(t),
\end{equation*}
and the transformed equation is
\begin{equation*}
\begin{aligned}
    \dot{\widetilde{\hat\rho}}
    ={}
    -\I
    [
        \widetilde{\hat H}(t),
        \widetilde{\hat\rho}
    ]+
    \sum_k
    \gamma_k
    \mathcal D[\widetilde{\hat L}_k(t)]
    \widetilde{\hat\rho}.
\end{aligned}
\end{equation*}
Transforming the Hamiltonian while leaving the jump operators bare
therefore describes a different open system.

\subsubsection{Example: cavity loss after a dispersive SW transformation}
\label{subsubsec:framework_dressed_jump_example}

Consider the lossy Jaynes--Cummings equation
\begin{equation*}
    \dot{\hat\rho}
    =
    -\I
    [
        \hat H_{\mathrm{JC}},
        \hat\rho
    ]
    +
    \kappa
    \mathcal D[\hat a]
    \hat\rho.
\end{equation*}
Using the SW generator in
Eq.~\eqref{eq:framework_sw_generator_preview}, the transformed cavity
operator is
\begin{equation*}
    \widetilde{\hat a}
    =
    e^{-\hat S_1}
    \hat a
    e^{\hat S_1}
    =
    \hat a
    +
    \epsilon\hat\sigma_-
    +
    \mathcal O(\epsilon^2).
\end{equation*}
The complete effective generator therefore contains both the dispersive
Hamiltonian and the dressed loss channel,
\begin{equation*}
    \dot{\widetilde{\hat\rho}}
    =
    -\I
    [
        \hat H_{\mathrm{disp}},
        \widetilde{\hat\rho}
    ]
    +
    \kappa
    \mathcal D
    [
        \hat a+\epsilon\hat\sigma_-
    ]
    \widetilde{\hat\rho}
    +
    \ldots
\end{equation*}
If the cavity is subsequently restricted to its vacuum sector, the
\(\epsilon\hat\sigma_-\) component produces the familiar induced qubit
relaxation, with rate scaling as \(\kappa g^2/\Delta^2\) when
\(|\Delta|\gg\kappa\). A
Hamiltonian-only dispersive model reproduces the frequency shift but
misses this inherited dissipative channel. This is the open-system
counterpart of the same consistency requirement: information carried by
the eliminated cavity appears in the effective jump operator rather than
only in the effective Hamiltonian.

An exact unitary change of frame preserves the GKLS form. A perturbative
projection or elimination requires a joint derivation of the coherent and
dissipative terms, and a finite-order generator is not automatically of
GKLS form. Physicality tests are developed in
Sec.~\ref{sec:validation_protocol}.

Table~\ref{tab:framework_examples} shows where the principal constructions
store information that is absent from the displayed effective generator.

\begin{table*}[t]
    \centering
    \caption{Concrete realizations of the encode--evolve--reconstruct
    framework and the endpoint information that accompanies each effective
    evolution.}
    \label{tab:framework_examples}
    \small
    \renewcommand{\arraystretch}{1.16}
    \setlength{\tabcolsep}{5pt}
    \begin{tabular}{@{}
        >{\RaggedRight\arraybackslash}p{0.18\textwidth}
        >{\RaggedRight\arraybackslash}p{0.27\textwidth}
        >{\RaggedRight\arraybackslash}p{0.47\textwidth}
        @{}}
        \toprule
        Method
        & Effective evolution
        & Encoding, reconstruction, and typical omission
        \\
        \midrule
        Interaction-picture RWA
        & Retained co-rotating dynamics
        & Represent the preparation in the interaction picture and return
          the result to the laboratory frame; otherwise laboratory phases
          and observables can be incorrect.
        \\
        GRWA
        & Evolution within retained displaced blocks
        & Apply the spin-dependent displacement, select and later embed the
          dressed blocks, and undo the displacement; otherwise photon number,
          quadratures, and transition matrix elements can be incorrect.
        \\
        Dispersive SW
        & Block-diagonal dispersive Hamiltonian
        & Dress the initial state and inversely dress states or observables;
          otherwise virtual populations and control-operator corrections are
          lost.
        \\
        Adiabatic elimination
        & Slow Hamiltonian, Liouvillian, or finite-time map
        & Select the slow variables, prepare the slow manifold when needed,
          and reconstruct fast variables and initial slip; otherwise an
          eliminated level may be incorrectly set to zero.
        \\
        Floquet theory
        & Quasienergy evolution under \(\hat H_F\)
        & Apply the initial micromotion transformation and restore the
          time-periodic micromotion; otherwise the stroboscopic result may be
          correct while the continuous trajectory is not.
        \\
        Open-system reduction
        & Effective Liouvillian or reduced map, including induced channels
        & Transform or project the preparation and restore dressed observables
          and eliminated-sector quantities; otherwise induced jumps can be
          missed and unconditional dynamics can become nonphysical.
        \\
        \bottomrule
    \end{tabular}
\end{table*}

\subsection{Comparing physical predictions}
\label{subsec:fair_comparison}

When a reference calculation is available, denote it by
\begin{equation*}
    \hat\rho_{\mathrm{full}}(t)
    =
    \Phi_{\mathrm{full}}(t,0)[\hat\rho(0)].
\end{equation*}
A direct comparison with \(\hat\rho_{\mathrm{rec}}(t)\) requires the same
physical initial state, frame, Hilbert space, normalization, and observable
convention. Otherwise an error curve mixes the accuracy of the effective
generator with a change of preparation or measurement. Five distinctions
are especially useful:
\begin{enumerate}
    \item \textbf{Frame and gauge.} Compare reconstructed states or
    invariant quantities, not raw Hamiltonians written in different
    frames or block gauges.

    \item \textbf{Retained-space error and leakage.} Report loss of
    weight from the retained space separately from the conditional error
    inside that space.

    \item \textbf{Stroboscopic and continuous-time claims.} A Floquet
    generator can be accurate at \(t=nT\) while an unreconstructed
    trajectory is inaccurate between periods.

    \item \textbf{Closed- and open-system claims.} A spectral match does
    not validate dissipative rates, steady states, or complete positivity.

    \item \textbf{State, observable, and time domain.} Accuracy for one
    initial state or one observable does not imply a uniform approximation
    over every populated manifold and protocol duration.
\end{enumerate}
Direct benchmarks, residuals, convergence tests, and validity maps are
developed in Sec.~\ref{sec:validation_protocol}. A norm-based separation
of encoding, evolution, and reconstruction errors is given in
Appendix~\ref{app:order_consistent_reconstruction}.

For normalized density operators, a useful global metric is the trace
distance
\begin{equation*}
    D(t)
    =
    \frac{1}{2}
    \left\|
        \hat\rho_{\mathrm{full}}(t)
        -
        \hat\rho_{\mathrm{rec}}(t)
    \right\|_1,
\end{equation*}
where \(\|X\|_1=\operatorname{Tr}\sqrt{X^\dagger X}\) is the trace
norm. For density operators, the trace distance is the maximum difference
in probability assigned by the two states to the same binary measurement
\cite{NielsenChuang2010}. For a laboratory observable \(\hat O\),
define
\begin{equation*}
\begin{aligned}
    \epsilon_O(t)
    =
    \Big|
        \operatorname{Tr}
        [
            \hat O\hat\rho_{\mathrm{full}}(t)
        ]
        -
        \operatorname{Tr}
        [
            \hat O\hat\rho_{\mathrm{rec}}(t)
        ]
    \Big|.
\end{aligned}
\end{equation*}
The metric should be selected from the physical claim. A state-transfer
protocol requires phase- and coherence-sensitive channel information; a
spectroscopic model requires transition frequencies and matrix elements;
a projected theory requires leakage; and an open-system model requires
physicality and steady-state checks in addition to state trajectories.

A usable effective model should therefore record the reference problem and
physical claim, the retained description, the local control parameter, the
encoding and reconstruction maps, the first omitted structure, and the
evidence supporting its validity region. Together, these elements specify
what the derivation predicts.

Section~\ref{sec:methods} now applies this framework to projection,
Magnus and time-coarse-grained expansions, Schrieffer--Wolff and
small-rotation transformations, adiabatic elimination, and Floquet theory.
The methods produce different intermediate objects and often store the
same eliminated physics in different places. The framework developed here
will let us compare their reconstructed predictions on equal footing.
\section{Methods for Constructing Effective Models}
\label{sec:methods}

Section~\ref{sec:common_framework} separated three parts of an effective
description: constructing the reduced generator, representing states and
observables, and testing the resulting physical predictions. We now turn to
the first of these tasks, keeping the other two in view throughout.

The methods approach reduction from different directions. Projection methods
eliminate a complementary subspace through a resolvent; unitary
transformations decouple blocks perturbatively; time-averaging methods remove
selected frequency components; adiabatic elimination exploits a hierarchy of
dynamical timescales; and Floquet theory separates stroboscopic evolution from
micromotion. Their intermediate operators need not be identical. An
energy-dependent projected Hamiltonian, a block-diagonal Hamiltonian in a
dressed frame, and a reduced dynamical generator can describe the same
selected spectrum or observable while organizing the calculation differently.

We return several times to the dispersive Jaynes--Cummings model and the
lossy-cavity problem. Comparing methods on common examples makes it easier to
see where each construction stores the eliminated physics, how that physics
re-enters measurable quantities, and which approximation fails first.

\subsection{Projection and Model-Space Reduction}
\label{subsec:projection_methods}

\begin{figure*}[t]
    \centering
    \includegraphics[
        width=1.0\linewidth
    ]{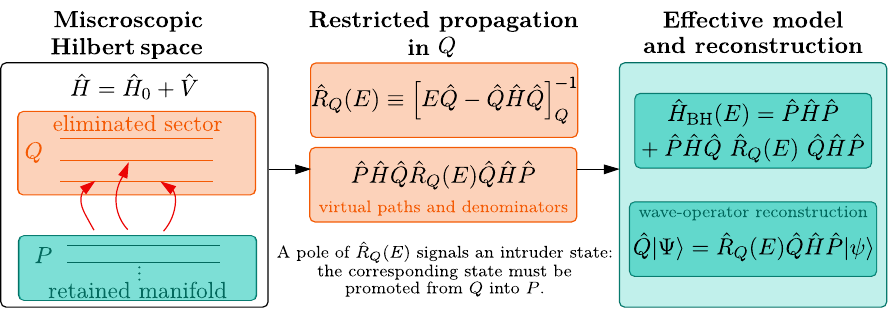}
    \caption{\justifying
    \textbf{Projection-method construction.} The Hilbert space is divided into a
    retained model space $\hat P$ and its complement
    $\hat Q=\hat I-\hat P$. Excursions into $\hat Q$ enter the
    Bloch--Horowitz operator through the restricted resolvent
    $(E-\hat Q\hat H\hat Q)^{-1}$. The wave operator then reconstructs
    the eliminated component of the state. The construction becomes ill
    conditioned when a complementary-space eigenvalue approaches the target
    energy.
    }
    \label{fig:projection_method}
\end{figure*}

Suppose that a physical question singles out a manifold of states, but the
Hamiltonian can still carry amplitude outside it. Projection theory describes
the selected manifold while retaining the feedback from these excursions. A
restricted resolvent supplies their energy denominators, and a wave operator
reconstructs the component removed from the model space. This viewpoint is
particularly useful when the intermediate states and virtual paths are part of
the physical explanation.
Figure~\ref{fig:projection_method} summarizes the construction. One first
chooses the retained space \(P\) and its complement \(Q\), then uses the
restricted resolvent to propagate an excursion through \(Q\). Its return to
\(P\) generates the effective interaction, while the wave operator restores
the eliminated part of the physical state.

Write the Hamiltonian as $\hat H=\hat H_0+\lambda\hat V$, where
$\hat H_0$ is solvable, $\hat V$ contains the remaining coupling, and
$\lambda$ keeps track of perturbative order
\cite{Sakurai,Griffiths_mec_quant}. Nondegenerate
Rayleigh--Schr\"odinger perturbation theory is well suited to an isolated
eigenstate of $\hat H_0$. A projection is more natural when the relevant
physics occupies a degenerate or nearly degenerate set of states, or whenever
a physically chosen model space must be retained while the other states enter
virtually
\cite{Feshbach1958,Kohn1959,Feshbach1962,Lwdin1962,FESHBACH1964,Shapiro1972,Lwdin1980,Klein1974,Bloch1958,Takayanagi2016,Takayanagi2016_2,Ren2019}.

\subsubsection{Exact projected eigenvalue problem}
\label{subsec:projection_exact}

Let $\hat P$ project onto a $d$-dimensional model space and let
$\hat Q=\hat I-\hat P$ project onto its complement. For a perturbative
expansion about $\hat H_0$, it is convenient to choose
\begin{equation*}
\begin{aligned}
    \hat P
    &=
    \sum_{i=1}^{d}\ket{i}\bra{i},
    &
    \hat H_0\ket{i}
    &=
    \epsilon_i\ket{i},
    \\
    \hat Q
    &=
    \sum_{\alpha=d+1}^{M}\ket{\alpha}\bra{\alpha},
    &
    \hat H_0\ket{\alpha}
    &=
    \epsilon_\alpha\ket{\alpha}.
\end{aligned}
\end{equation*}
Then $[\hat H_0,\hat P]=[\hat H_0,\hat Q]=0$, while
$\hat P^2=\hat P$, $\hat Q^2=\hat Q$, $\hat P\hat Q=0$, and
$\hat P+\hat Q=\hat I$. The exact Feshbach construction itself does not
require $\hat P$ to be assembled from eigenstates of $\hat H_0$; that choice
only prepares the later perturbative expansion.

For an exact eigenstate $\hat H\ket{\Psi_m}=E_m\ket{\Psi_m}$, define its
retained and complementary components by
\begin{equation*}
    \ket{\Psi_m}
    =
    \underbrace{\hat P\ket{\Psi_m}}_{\ket{\psi_m}}
    +
    \underbrace{\hat Q\ket{\Psi_m}}_{\ket{\phi_m}}.
\end{equation*}
Projecting the eigenvalue equation gives
\begin{subequations}
\begin{align}
    \left(\hat P\hat H\hat P-E_m\hat P\right)\ket{\psi_m}
    +\hat P\hat H\hat Q\ket{\phi_m}
    &=0,
    \\
    \hat Q\hat H\hat P\ket{\psi_m}
    +\left(\hat Q\hat H\hat Q-E_m\hat Q\right)\ket{\phi_m}
    &=0.
\end{align}
\end{subequations}

\paragraph{The resolvent and its physical meaning.}

The resolvent is the operator version of dividing by $z-a$ when solving
$(z-a)x=b$. For an operator $\hat A$, it is
$\hat R_A(z)=(z\hat I-\hat A)^{-1}$ whenever the inverse exists. If
$\hat A$ is finite-dimensional and Hermitian, with eigenvalues $a_\nu$ and
eigenvectors $\ket{\nu}$, then
\begin{equation}
    \hat R_A(z)
    =
    \sum_\nu
    \frac{\ket{\nu}\bra{\nu}}{z-a_\nu}.
    \label{eq:projection_resolvent_spectral_form}
\end{equation}
The denominators show which modes respond strongly. As $z$ approaches an
eigenvalue, the corresponding response grows and the resolvent eventually
develops a pole.

Here the operator to invert is the Hamiltonian restricted to the eliminated
space. We therefore define
\begin{align*}
    \hat R_Q(E)
    &\equiv
    \left[E\hat Q-\hat Q\hat H\hat Q\right]^{-1}_{Q} \\
    &=
    \sum_{\alpha\in Q}
    \frac{\ket{\alpha_Q}\bra{\alpha_Q}}
    {E-\varepsilon^{(Q)}_\alpha},
\end{align*}
where
$\hat Q\hat H\hat Q\ket{\alpha_Q}
=\varepsilon^{(Q)}_\alpha\ket{\alpha_Q}$ and the inverse is taken on
$\operatorname{Ran}\hat Q$. The coupling $\hat Q\hat H\hat P$ creates
an amplitude in the complementary space, $\hat R_Q(E)$ propagates it with
the appropriate energy denominators, and $\hat P\hat H\hat Q$ can return it
to the retained space.

For example, if $P$ and $Q$ each contain one state, $\ket p$ and $\ket q$,
with $\bra q\hat H\ket q=\varepsilon_q$ and
$\bra q\hat H\ket p=v$, then
\begin{equation*}
    \braket{q|\Psi}
    =
    \frac{v}{E-\varepsilon_q}\braket{p|\Psi},
\end{equation*}
 feedback in $P
    \propto
    \dfrac{|v|^2}{E-\varepsilon_q}$.
This familiar second-order denominator also gives an immediate warning: if
$E\simeq\varepsilon_q$, the state $\ket q$ should normally be included in
the retained space rather than treated as a distant virtual state.

Provided $E_m$ is not in the spectrum of $\hat Q\hat H\hat Q$, the second
projected equation gives
\begin{align}
    \ket{\phi_m}
    &=
    \hat R_Q(E_m)\hat Q\hat H\hat P\ket{\psi_m} \nonumber
    \\
    &=
    \left(E_m-\hat Q\hat H\hat Q\right)^{-1}_{Q}
    \hat Q\hat H\hat P\ket{\psi_m}.
    \label{eq:projection_complement_state}
\end{align}
This is a well-defined inverse on the eliminated subspace, not the singular
inverse of $E_m-\hat H$ on the full Hilbert space
\cite{Kato1980,Economou2006}.

Substitution into the first projected equation yields
\begin{equation}
    \hat H_{\mathrm{BH}}(E_m)\ket{\psi_m}
    =
    E_m\ket{\psi_m},
    \label{eq:projection_bh_eigenproblem}
\end{equation}
with the Bloch--Horowitz operator
\begin{align}
    \hat H_{\mathrm{BH}}(E)
    &=
    \hat P\hat H\hat P \nonumber\\
    &\quad{}+\hat P\hat H\hat Q
    \left(E-\hat Q\hat H\hat Q\right)^{-1}_{Q}
    \hat Q\hat H\hat P.
    \label{eq:projection_bloch_horowitz}
\end{align}
This step is exact. The price of eliminating $Q$ is the dependence on the
unknown energy: the physical roots satisfy
$\det_P[E\hat P-\hat H_{\mathrm{BH}}(E)]=0$. Thus one solves a nonlinear
eigenvalue problem rather than diagonalizing one fixed matrix.

The same construction provides the missing part of the eigenstate. Define the
energy-dependent wave operator
\begin{equation}
    \hat\Omega(E)
    =
    \hat P
    +
    \hat Q
    \left(E-\hat Q\hat H\hat Q\right)^{-1}_{Q}
    \hat Q\hat H\hat P.
    \label{eq:projection_wave_operator}
\end{equation}
Then
\begin{equation}
    \ket{\Psi_m}
    =
    \hat\Omega(E_m)\ket{\psi_m}.
    \label{eq:projection_state_reconstruction}
\end{equation}
If $\ket{\psi_m}$ is normalized only inside the model space, an observable
must be evaluated as
\begin{equation*}
    \langle\hat O\rangle_m
    =
    \frac{
        \bra{\psi_m}\hat\Omega^\dagger(E_m)
        \hat O\hat\Omega(E_m)\ket{\psi_m}
    }{
        \bra{\psi_m}\hat\Omega^\dagger(E_m)
        \hat\Omega(E_m)\ket{\psi_m}
    }.
\end{equation*}
Using only $\hat P\hat O\hat P$ would omit the dressing carried by the
eliminated sector.

\paragraph{Intruder states and the choice of model space.}

The problem is not confined to exact degeneracy. When an eigenvalue of
$\hat Q\hat H\hat Q$ approaches the target energy, the restricted resolvent
becomes large and the projected problem becomes ill conditioned. This is the
intruder-state problem. A useful diagnostic is the retained weight
$w_m=\bra{\Psi_m}\hat P\ket{\Psi_m}$. A small $w_m$ means that the chosen
model space does not represent the physical state well, even if its energy can
still be reproduced algebraically. The usual remedy is to enlarge or redefine
$P$, not simply to increase the perturbative order.

\subsubsection{Energy-independent perturbative Hamiltonians}
\label{subsec:projection_energy_independent}

Calculations often call for a single energy-independent operator acting in
$P$. Such an operator is not unique. Bloch, L\"owdin, canonical, and related
constructions can differ in normalization, Hermiticity, and block gauge while
reproducing the same selected eigenvalues to a stated order
\cite{Bloch1958,Lwdin1962,Bravyi2011,Takayanagi2016,Takayanagi2016_2,Ren2019}.

A convenient Hermitian second-order form for a nearly degenerate model space
is
\begin{align}
    \bra{i}\hat H_{\mathrm{eff}}\ket{j}
    &=
    \epsilon_i\delta_{ij}
    +\lambda V_{ij} \nonumber \\
    &\quad{}+
    \frac{\lambda^2}{2}
    \sum_{\alpha\in Q}
    \frac{V_{i\alpha}V_{\alpha j}}
         {\epsilon_i-\epsilon_\alpha}
    \nonumber\\
    &\quad{}+
    \frac{\lambda^2}{2}
    \sum_{\alpha\in Q}
    \frac{V_{i\alpha}V_{\alpha j}}
         {\epsilon_j-\epsilon_\alpha}
    \nonumber\\
    &\quad{}+\mathcal O(\lambda^3),
    \label{eq:projection_lowdin_second_order}
\end{align}
where $V_{\mu\nu}=\bra{\mu}\hat V\ket{\nu}$. The symmetric
denominators make the result Hermitian when $\hat V$ is Hermitian.

If all retained states have the same unperturbed energy $\epsilon_P$, this
reduces to
\begin{align*}
    \hat H_{\mathrm{eff}}
    &=\epsilon_P\hat P
    +\lambda\hat P\hat V\hat P \\
    &\quad{}+
    \lambda^2\hat P\hat V\hat Q
    \left(\epsilon_P-\hat Q\hat H_0\hat Q\right)^{-1}_{Q}
    \nonumber\\
    &\qquad{}\times
    \hat Q\hat V\hat P
    +\mathcal O(\lambda^3).
\end{align*}
Higher-order expressions can be generated recursively, but their detailed
form depends on the wave-operator and Hermitization convention. Before two
higher-order Hamiltonians are compared term by term, their block gauges and
reconstruction maps must therefore be aligned.

The local control parameter compares a coupling matrix element with the gap
across the $P$--$Q$ boundary,
\begin{equation*}
    \epsilon_{i\alpha}
    =
    \lambda
    \frac{|V_{i\alpha}|}{|\epsilon_i-\epsilon_\alpha|},
    \qquad
    i\in P,\quad \alpha\in Q.
\end{equation*}
The relevant values must be small for the populated manifold. A quantity such
as the smallest eigenvalue of $\hat H_0$ cannot serve as the denominator,
because the zero of energy is arbitrary.

\subsubsection{Example: the dispersive Jaynes--Cummings model}
\label{subsec:projection_dispersive_jc}

The closed Jaynes--Cummings Hamiltonian in
Eq.~\eqref{eq:jc_hamiltonian} provides a compact example. Let
$\Delta=\omega_0-\omega_a$ and move to a frame rotating at the cavity
frequency with
$\hat U_a(t)=\exp[-\mathrm{i}\omega_a t
(\hat a^\dagger\hat a+\hat\sigma_z/2)]$. Using the convention of
Sec.~\ref{subsec:exact_change_of_representation} gives
\begin{equation*}
    \widetilde{\hat H}_{\mathrm{JC}}
    =
    \frac{\Delta}{2}\hat\sigma_z
    +
    g\left(\hat a\hat\sigma_+
    +\hat a^\dagger\hat\sigma_-\right).
\end{equation*}
We take $\hat H_0=\Delta\hat\sigma_z/2$ and the remaining interaction as
$\hat V$. The unperturbed Hamiltonian is degenerate in photon number, so the
ground and excited atomic sectors form natural model spaces.

For
$\hat P_g=\sum_{n\geq0}\ket{g,n}\bra{g,n}$, the interaction obeys
$\hat P_g\hat V\hat P_g=0$ and
$\hat V\ket{g,n}=g\sqrt n\ket{e,n-1}$. For
$\hat P_e=\sum_{n\geq0}\ket{e,n}\bra{e,n}$, it gives
$\hat V\ket{e,n}=g\sqrt{n+1}\ket{g,n+1}$. Equation~\eqref{eq:projection_lowdin_second_order}
therefore yields
\begin{equation*}
    \delta E_{g,n}^{(2)}
    =-\frac{g^2n}{\Delta},
    \qquad
    \delta E_{e,n}^{(2)}
    =\frac{g^2(n+1)}{\Delta}.
\end{equation*}
Combining the two sectors gives the rotating-frame dispersive Hamiltonian
\begin{align}
    \widetilde{\hat H}_{\mathrm{disp}}
    &=
    \frac{\Delta}{2}\hat\sigma_z
    +
    \frac{g^2}{\Delta} \left( \hat a^\dagger\hat a\hat\sigma_z +\hat\sigma_+\hat\sigma_- \right)
    \nonumber\\
    & \quad{}
    +
    \mathcal O\!\left(\frac{g^4}{\Delta^3}\right).
    \label{eq:projection_rotating_dispersive}
\end{align}
Returning to the Schr\"odinger picture yields
\begin{align}
    \hat H_{\mathrm{disp}}
    &=
    \omega_a\hat a^\dagger\hat a
    +\frac{\omega_0}{2}\hat\sigma_z \nonumber \\
    &\quad{}+
    \frac{g^2}{\Delta}
    \left(
        \hat a^\dagger\hat a\hat\sigma_z
        +\hat\sigma_+\hat\sigma_-
    \right) \nonumber \\
    &\quad{}+
    \mathcal O\!\left(\frac{g^4}{\Delta^3}\right).
    \label{eq:projection_jc_dispersive_schrodinger}
\end{align}

The remainder in Eqs.~\eqref{eq:projection_rotating_dispersive}
and~\eqref{eq:projection_jc_dispersive_schrodinger} refers to the
block-diagonal energies in a fixed populated manifold. Each elementary
interaction flips the atomic branch, so an odd number of interactions cannot
return to the same model space. The third-order diagonal correction vanishes,
and the first omitted energy term scales as
$g^4(n_\star+1)^2/|\Delta|^3$. The dressed eigenstate nevertheless has a
first-order component outside $P$.

Indeed, expanding Eq.~\eqref{eq:projection_wave_operator} gives
\begin{equation*}
\begin{aligned}
    \hat\Omega_g
    &=
    \hat P_g-\frac{1}{\Delta}\hat Q_g\hat V\hat P_g
    +\mathcal O\!\left(\frac{g^2}{\Delta^2}\right),
    \\
    \hat\Omega_e
    &=
    \hat P_e+\frac{1}{\Delta}\hat Q_e\hat V\hat P_e
    +\mathcal O\!\left(\frac{g^2}{\Delta^2}\right),
\end{aligned}
\end{equation*}
and hence
\begin{equation*}
\begin{aligned}
    \ket{g,n}_{\mathrm{phys}}
    &=
    \ket{g,n}
    -\frac{g\sqrt n}{\Delta}\ket{e,n-1} \\
    &\quad{}+\mathcal O\!\left(\frac{g^2n}{\Delta^2}\right),
    \\
    \ket{e,n}_{\mathrm{phys}}
    &=
    \ket{e,n}
    +\frac{g\sqrt{n+1}}{\Delta}\ket{g,n+1} \\
    &\quad{}+\mathcal O\!\left(\frac{g^2(n+1)}{\Delta^2}\right).
\end{aligned}
\end{equation*}
These virtual components are absent from the model-space Hamiltonian but enter
bare-state populations and other laboratory observables.

If the appreciably occupied photon manifold extends to $n_\star$, the useful
condition is
\begin{equation*}
    \frac{g\sqrt{n_\star+1}}{|\Delta|}\ll1.
\end{equation*}
For a coherent state, the mean occupation does not by itself bound the Fock
tail; $n_\star$ should be chosen so that the probability above it lies below
the intended error tolerance.

\subsubsection{Projection superoperators for open systems}
\label{subsec:projection_open_systems}

For an open system, the corresponding object is a projection superoperator
acting on density operators
\cite{FinkelsteinShapiro2020,Saideh2020Projection}. Accessible introductions
to projection methods and perturbative reduced dynamics are given in
Refs.~\cite{GonzalezBallestero2024Projector,Szankowski2023OpenSystems}.
Consider a bipartite system $A\otimes B$ with
$\dot{\hat\rho}_{AB}=\mathcal L\hat\rho_{AB}$ and
$\mathcal L=\mathcal L_A+\mathcal L_B+\mathcal L_{AB}$. Suppose that
$B$ relaxes rapidly to a reference state
$\hat\rho_B^{\mathrm{ss}}$ satisfying
$\mathcal L_B\hat\rho_B^{\mathrm{ss}}=0$. A standard projector is
\begin{equation*}
    \mathcal P\hat X
    =
    \operatorname{Tr}_B(\hat X)\otimes\hat\rho_B^{\mathrm{ss}},
    \qquad
    \mathcal Q=\mathcal I-\mathcal P.
\end{equation*}
Normalization of $\hat\rho_B^{\mathrm{ss}}$ ensures
$\mathcal P^2=\mathcal P$. This projector need not be orthogonal in the
Hilbert--Schmidt inner product: its job is to retain
$\hat\rho_A=\operatorname{Tr}_B\hat\rho_{AB}$ while placing the fast
subsystem in its reference state.

The Liouvillian resolvent
$\mathcal G(z)=(z-\mathcal L)^{-1}$ propagates the Laplace-transformed
state,
\begin{equation*}
\begin{aligned}
    \widehat{\hat\rho}(z)
    &=
    \int_0^\infty e^{-zt}\hat\rho(t)\,\mathrm dt
    =
    \mathcal G(z)\hat\rho(0),
    \\
    \hat\rho(t)
    &=
    \frac{1}{2\pi\mathrm i}
    \int_{c-\mathrm i\infty}^{c+\mathrm i\infty}
    e^{zt}\mathcal G(z)\hat\rho(0)\,\mathrm dz,
\end{aligned}
\end{equation*}
where $c$ lies to the right of the spectrum of $\mathcal L$. Its projected
part receives contributions from both components of the initial state,
\begin{equation}
    \mathcal P\widehat{\hat\rho}(z)
    =
    \mathcal P\mathcal G(z)\mathcal P\,
    \mathcal P\hat\rho(0)+
    \mathcal P\mathcal G(z)\mathcal Q\,
    \mathcal Q\hat\rho(0).
    \label{eq:projection_resolvent_initial_components}
\end{equation}
The second term is an initial transient when the fast subsystem begins away
from its reference manifold. We first set it aside by taking
$\mathcal Q\hat\rho(0)=0$.

The Schur complement of $z-\mathcal L$ gives
\begin{equation}
\begin{aligned}
    \mathcal P\mathcal G(z)\mathcal P
    &=
    \left[z-\mathcal L_{\mathrm{eff}}(z)\right]^{-1}\mathcal P,
    \\
    \mathcal L_{\mathrm{eff}}(z)
    ={}&
    \mathcal P\mathcal L\mathcal P
    +
    \mathcal P\mathcal L\mathcal Q
    \left(z-\mathcal Q\mathcal L\mathcal Q\right)^{-1}
    \mathcal Q\mathcal L\mathcal P.
    \label{eq:projection_liouvillian_schur_complement}
\end{aligned}
\end{equation}
This is the Liouville-space counterpart of the Bloch--Horowitz operator. It
is an exact representation of the projected resolvent, but its dependence on
$z$ means that the time-domain evolution generally contains memory and an
initial transient rather than one time-independent Markovian generator.

When the slow pole near $z=0$ is isolated from the fast poles, expand
$\mathcal L_{\mathrm{eff}}(z)=\mathcal L_0+z\mathcal L_1+
\mathcal O(z^2)$. If $\mathcal Q\mathcal L\mathcal Q$ is invertible on
the fast subspace,
\begin{equation*}
\begin{aligned}
    \mathcal L_0
    ={}&
    \mathcal P\mathcal L\mathcal P
    -
    \mathcal P\mathcal L\mathcal Q
    (\mathcal Q\mathcal L\mathcal Q)^{-1}
    \mathcal Q\mathcal L\mathcal P,
    \\
    \mathcal L_1
    ={}&
    -
    \mathcal P\mathcal L\mathcal Q
    (\mathcal Q\mathcal L\mathcal Q)^{-2}
    \mathcal Q\mathcal L\mathcal P.
\end{aligned}
\end{equation*}
With
$\mathcal S=(\mathcal I-\mathcal L_1)^{-1}$ and
$\mathcal L_{\mathrm{red}}=\mathcal S\mathcal L_0$, the long-time
solution becomes
\begin{equation*}
    \mathcal P\hat\rho(t)
    \simeq
    e^{\mathcal L_{\mathrm{red}}t}
    \mathcal S\mathcal P\hat\rho(0).
\end{equation*}
The map $\mathcal S$ is the initial-slip correction. If coupling between the
two sectors is of order $\epsilon$, then typically
$\mathcal L_1=\mathcal O(\epsilon^2)$ and
$\mathcal S=\mathcal I+\mathcal O(\epsilon^2)$. The correction may be
irrelevant for a stationary state, but it is of the same order as the leading
finite-time memory correction.

An isolated stationary state belongs to the kernel of $\mathcal L_0$ under
the same spectral assumptions, so its projected form can be correct even when
the finite-time evolution is only approximate. Neither $\mathcal L_0$ nor
$\mathcal L_{\mathrm{red}}$ is automatically in GKLS form; trace
preservation, Hermiticity preservation, and complete positivity must follow
from the reduction or be checked separately.

\subsubsection{Example: eliminating a lossy cavity mode}
\label{subsec:projection_lossy_cavity}

Consider a two-level system coupled to a cavity mode that decays into a
zero-temperature reservoir at rate $\kappa$. In the frame rotating at the
cavity frequency,
\begin{align*}
    \dot{\hat\rho}_{AB}
   & =
    -\mathrm i
    \left[
        \frac{\Delta}{2}\hat\sigma_z
        +
        g(\hat a\hat\sigma_+
        +\hat a^\dagger\hat\sigma_-),
        \hat\rho_{AB}
    \right] \\
    &\quad{}+
    \kappa\mathcal D[\hat a]\hat\rho_{AB},
\end{align*}
where the dissipator is defined in
Eq.~\eqref{eq:global_dissipator_convention}. We identify
\begin{subequations}
\begin{align}
    \mathcal L_A\hat\rho
    &=
    -\mathrm i[\Delta\hat\sigma_z/2,\hat\rho],
    \\
    \mathcal L_B\hat\rho
    &=
    \kappa\mathcal D[\hat a]\hat\rho,
    \\
    \mathcal L_{AB}\hat\rho
    &=
    -\mathrm i g
    [\hat a\hat\sigma_++\hat a^\dagger\hat\sigma_-,\hat\rho].
\end{align}
\end{subequations}
The stationary cavity state is the vacuum, so
$\mathcal P\hat X=\operatorname{Tr}_B(\hat X)\otimes\ket0\bra0$.

It is convenient to remove the retained atomic motion generated by
$\mathcal L_A$. In this interaction picture the atomic state changes little
during the cavity correlation time, while the one-photon coherences carry the
relative phase $e^{\pm\mathrm i\Delta\tau}$. In resolvent language, this
amounts to centering the residual Laplace variable at $z=0$ after evaluating
the fast response at the relevant atomic Bohr frequency.

For $\hat\rho_P=\hat\rho_A\otimes\ket0\bra0$, the projected first-order
dynamics is simply
$\mathcal P\mathcal L\mathcal P\hat\rho_P
=-\mathrm i[\Delta\hat\sigma_z/2,\hat\rho_A]\otimes\ket0\bra0$.
The interaction sends this state into one-photon coherences,
\begin{equation}
\begin{aligned}
\mathcal{Q}\mathcal{L}_{AB}\mathcal{P}\hat{\rho}_{P}
&=
-ig\Big(
    \hat{\sigma}_{-}\hat{\rho}_{A}\otimes
    \lvert 1\rangle\langle 0\rvert
    \\
&\qquad
    -
    \hat{\rho}_{A}\hat{\sigma}_{+}\otimes
    \lvert 0\rangle\langle 1\rvert
\Big).
\end{aligned}
\label{eq:projection_cavity_Q_source}
\end{equation}
During the cavity correlation time these coherences decay with
$e^{-(\kappa/2-\mathrm i\Delta)\tau}$. Their susceptibility is
\begin{equation*}
    \chi_c(\Delta)
    =
    \int_0^\infty
    e^{-(\kappa/2-\mathrm i\Delta)\tau}\,\mathrm d\tau
    =
    \frac{\kappa/2+\mathrm i\Delta}
    {\Delta^2+\kappa^2/4}.
\end{equation*}
Its real and imaginary parts produce dissipation and a frequency shift,
respectively.

To second order in $g$, the projected Liouvillian is
\begin{align*}
    \mathcal{L}_{\mathrm{eff}}(z)
&\simeq
\mathcal{P}\mathcal{L}_{A}\mathcal{P}
\\
&\quad+
\mathcal{P}\mathcal{L}_{AB}\mathcal{Q}
\Big[
z-\mathcal{Q}(\mathcal{L}_{A}+\mathcal{L}_{B})\mathcal{Q}
\Big]^{-1}
\\
&\qquad\times
\mathcal{Q}\mathcal{L}_{AB}\mathcal{P}.
\end{align*}
After the atomic interaction-picture shift, the remaining $z$ dependence can
be neglected when
$|z|\ll\sqrt{\Delta^2+\kappa^2/4}$ and the induced atomic rates are slow on
the same scale. Evaluating the resolvent at $z=0$ then turns the memory kernel
into a time-local generator.

Applying the interaction a second time and tracing out the cavity gives
\begin{equation*}
\begin{aligned}
    \dot{\hat\rho}_A
    ={}&
    -\mathrm i
    \left[
        \frac{\Delta}{2}\hat\sigma_z
        +\delta\omega\hat\sigma_+\hat\sigma_-,
        \hat\rho_A
    \right]
    +
    \Gamma\mathcal D[\hat\sigma_-]\hat\rho_A,
\end{aligned}
\end{equation*}
where
\begin{subequations}
\begin{align}
    \delta\omega
    &=
    \frac{g^2\Delta}{\Delta^2+\kappa^2/4},
    \\
    \Gamma
    &=
    \frac{g^2\kappa}{\Delta^2+\kappa^2/4}.
    \label{eq:purcell_and_shift_projector}
\end{align}
\end{subequations}
The two coefficients are the imaginary and real responses of the same cavity
susceptibility. Thus the same eliminated mode produces both the coherent
frequency shift and Purcell relaxation.

The local control parameter is
$\epsilon_{\mathrm{cav}}=g/\sqrt{\Delta^2+\kappa^2/4}$. It must be small,
and the induced atomic rates must remain slow compared with the eliminated
cavity scale. In the dispersive limit $|\Delta|\gg g,\kappa$,
\begin{equation*}
    \delta\omega\simeq\frac{g^2}{\Delta},
    \qquad
    \Gamma\simeq\frac{g^2\kappa}{\Delta^2};
\end{equation*}
at resonance in the bad-cavity limit $\kappa\gg g$, one instead finds
$\Gamma\simeq4g^2/\kappa$ and $\delta\omega=0$.

The reduced atomic state does not contain the small cavity coherences that
were eliminated. To leading order they are reconstructed as
\begin{equation*}
\begin{aligned}
    \mathcal Q\hat\rho_{AB}
    &\simeq -
    \left[
        \mathcal Q(\mathcal L_A+\mathcal L_B)\mathcal Q
    \right]^{-1} \\
    &\quad{}\times \mathcal Q\mathcal L_{AB}\mathcal P
    [\hat\rho_A\otimes\ket0\bra0].
\end{aligned}
\end{equation*}
This is the open-system counterpart of
Eq.~\eqref{eq:projection_state_reconstruction}. If the cavity does not begin
in the vacuum, the term proportional to $\mathcal Q\hat\rho_{AB}(0)$ in
Eq.~\eqref{eq:projection_resolvent_initial_components} adds a transient that
cannot be reproduced by replacing the initial cavity state with its stationary
one. For the vacuum example the reduced generator is manifestly of GKLS form;
this is a property of the susceptibility and coupling used here, not a general
guarantee of finite-order projection.

\subsection{The Magnus Expansion}
\label{subsec:magnus_expansion}

When the unwanted structure is rapid in time rather than remote in energy, it
is often better to work directly with the propagator. The Magnus expansion
collects the operator combinations that accumulate during the evolution and
places them in a single exponential. Its immediate result is a finite-time
propagator. A static Hamiltonian appears only after one specifies a temporal
resolution---for example, secular, stroboscopic, or coarse-grained.

\begin{figure*}[t]
    \centering
    \includegraphics[
        width=1.0\linewidth
    ]{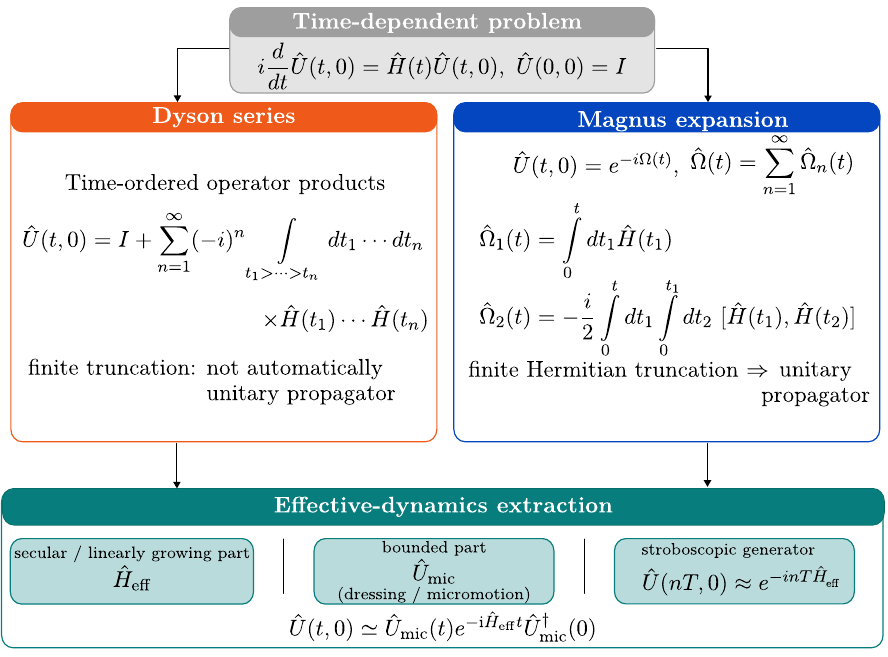}
    \caption{\justifying
    \textbf{Dyson and Magnus representations of time-dependent evolution.} The Dyson
    series expresses the propagator as time-ordered operator products. The
    Magnus expansion reorganizes the same evolution into one exponential, so a
    finite truncation remains unitary for a Hermitian Hamiltonian. A
    time-independent effective Hamiltonian requires a further secular,
    stroboscopic, or coarse-graining step. Bounded contributions describe
    dressing or micromotion and remain part of a continuous-time prediction.
    }
    \label{fig:magnus_dyson_comparison}
\end{figure*}

Figure~\ref{fig:magnus_dyson_comparison} contrasts this construction with the
Dyson series and shows the additional separation needed to obtain a slow
effective Hamiltonian. The bounded part of the evolution is not discarded: it
becomes the dressing or micromotion required for continuous-time predictions.

Magnus introduced this exponential representation in 1954
\cite{Magnus1954}. It is now widely used in magnetic resonance, quantum
control, periodically driven systems, and coarse-grained treatments of rapidly
oscillating interactions \cite{Blanes2009,Blanes2010,macri2023coarse}. For a
closed system, exponentiating a truncated Hermitian Magnus exponent gives an
exactly unitary approximation. This structural advantage does not guarantee
accuracy: convergence still depends on the chosen frame, the occupied
manifold, the coupling strength, and the time over which errors accumulate.

\subsubsection{Exact exponential representation}
\label{subsec:magnus_general}

In a chosen frame, the propagator obeys
\begin{equation*}
    \mathrm i\frac{\mathrm d}{\mathrm dt}\hat U(t,0)
    =
    \hat H(t)\hat U(t,0),
    \qquad
    \hat U(0,0)=\hat I.
\end{equation*}
The Dyson solution is
\begin{equation}
\begin{aligned}
\hat U(t,0)
&=
\hat I
\\
&+
\sum_{n=1}^{\infty}
(-i)^n
\int_0^t \mathrm{d}t_1
\int_0^{t_1} \mathrm{d}t_2
\cdots
\int_0^{t_{n-1}} \mathrm{d}t_n
\\
&\times
\hat H(t_1)
\hat H(t_2)
\cdots
\hat H(t_n).
\end{aligned}
\label{eq:magnus_dyson_series}
\end{equation}
The nested limits impose time ordering, but a finite polynomial truncation of
Eq.~\eqref{eq:magnus_dyson_series} is generally not unitary.

The Magnus ansatz instead writes
\begin{equation*}
    \hat U(t,0)
    =
    \exp[-\mathrm i\hat\Omega(t)],
    \qquad
    \hat\Omega(t)
    =
    \sum_{n=1}^{\infty}\hat\Omega_n(t).
\end{equation*}
With this convention, every $\hat\Omega_n$ is Hermitian when $\hat H(t)$ is
Hermitian. The first three terms are
\begin{subequations}
\begin{align}
    \hat\Omega_1(t)
    &=
    \int_0^t\mathrm dt_1\,\hat H(t_1),
    \label{eq:magnus_first_term}
    \\
    \hat\Omega_2(t)
    &=
    -\frac{\mathrm i}{2}
    \int_0^t\mathrm dt_1
    \int_0^{t_1}\mathrm dt_2\,
    [\hat H(t_1),\hat H(t_2)],
    \label{eq:magnus_second_term}
    \\
    \hat\Omega_3(t)
    &=
    -\frac{1}{6}
    \int_0^t\mathrm dt_1
    \int_0^{t_1}\mathrm dt_2
    \int_0^{t_2}\mathrm dt_3
    \nonumber\\
    &\quad\times
    \Big(
        [\hat H(t_1),[\hat H(t_2),\hat H(t_3)]]
        \\
        &\quad+
        [\hat H(t_3),[\hat H(t_2),\hat H(t_1)]]
    \Big).
    \label{eq:magnus_third_term}
\end{align}
\label{eq:magnus_first_three_terms}
\end{subequations}
Thus
$\hat U^{[N]}(t,0)=
\exp[-\mathrm i\sum_{n=1}^N\hat\Omega_n(t)]$
is unitary at every truncation order.

For bounded Hamiltonians, a standard sufficient convergence condition is
\begin{equation}
    \int_0^t\|\hat H(s)\|\,\mathrm ds<\pi,
    \label{eq:magnus_pi_condition}
\end{equation}
although it is not necessary \cite{Blanes2009,Blanes2010}. For an unbounded
bosonic Hamiltonian, the global operator norm makes
Eq.~\eqref{eq:magnus_pi_condition} inapplicable. One must instead work in a
finite-energy sector, a converged numerical truncation, or an asymptotic
expansion controlled on the populated manifold. Failure of the sufficient
bound does not by itself invalidate a finite-order approximation, but in an
asymptotic regime there may be an optimal truncation order beyond which the
error grows.

Dividing the truncated exponent by the final time defines
$\hat H_{\mathrm M}^{[N]}(t)=t^{-1}\sum_{n=1}^N\hat\Omega_n(t)$. This operator
reproduces the approximate propagator at that time; it is not automatically a
time-independent slow Hamiltonian. The latter is obtained by separating
linearly accumulating terms from bounded ones. Schematically,
\begin{equation}
    \hat U(t,0)
    \simeq
    \hat U_{\mathrm{mic}}(t)
    e^{-\mathrm i\hat H_{\mathrm{eff}}t}
    \hat U_{\mathrm{mic}}^\dagger(0),
    \label{eq:magnus_effective_factorization}
\end{equation}
where $\hat U_{\mathrm{mic}}(t)$ carries the bounded dressing or micromotion.
This factorization requires an order-consistent reorganization of the
exponent, not simply division by $t$. Omitting
$\hat U_{\mathrm{mic}}(t)$ restricts the prediction to stroboscopic or
coarse-grained observables.

\subsubsection{Example: the dispersive Jaynes--Cummings model}
\label{subsec:magnus_dispersive_jc}

The Jaynes--Cummings model is a useful benchmark because each excitation
manifold can also be solved exactly. In the interaction picture generated by
$\hat H_0=\omega_a\hat a^\dagger\hat a+\omega_0\hat\sigma_z/2$, and with
$\Delta=\omega_0-\omega_a$,
\begin{equation}
    \hat H_I(t)
    =
    g\left(
        \hat\sigma_+\hat a\,e^{\mathrm i\Delta t}
        +
        \hat\sigma_-\hat a^\dagger e^{-\mathrm i\Delta t}
    \right).
    \label{eq:jc_interaction_picture}
\end{equation}
Set $\hat A=\hat\sigma_+\hat a$ and
$\hat A^\dagger=\hat\sigma_-\hat a^\dagger$. The first Magnus term is
\begin{equation}
    \hat\Omega_1(t)
    =
    \frac{g}{\mathrm i\Delta}
    \left[
        \hat A(e^{\mathrm i\Delta t}-1)
        -
        \hat A^\dagger(e^{-\mathrm i\Delta t}-1)
    \right].
    \label{eq:magnus_jc_omega1}
\end{equation}
It is Hermitian and bounded. In a populated sector $n\leq n_\star$, its
matrix elements scale as $g\sqrt{n_\star+1}/|\Delta|$. This is the leading
virtual excursion: it affects continuous-time amplitudes and bare-state
observables even though it does not produce a secular energy shift.

At second order,
\begin{equation}
    [\hat H_I(t_1),\hat H_I(t_2)]
    =
    2\mathrm i g^2
    \sin[\Delta(t_1-t_2)]\,\hat K,
    \label{eq:magnus_jc_commutator}
\end{equation}
where
\begin{equation}
    \hat K
    =
    [\hat A,\hat A^\dagger]
    =
    \hat a^\dagger\hat a\hat\sigma_z
    +
    \hat\sigma_+\hat\sigma_-.
    \label{eq:magnus_jc_K}
\end{equation}
Substitution into Eq.~\eqref{eq:magnus_second_term} gives
\begin{equation}
    \hat\Omega_2(t)
    =
    g^2\left(
        \frac{t}{\Delta}
        -
        \frac{\sin(\Delta t)}{\Delta^2}
    \right)\hat K.
    \label{eq:magnus_jc_omega2_full}
\end{equation}
The first term grows linearly and yields the slow interaction-picture
Hamiltonian $\hat H_{\mathrm{disp},I}=g^2\hat K/\Delta$. The second remains
bounded,
\begin{equation}
    \hat\Omega_{2,\mathrm{osc}}(t)
    =
    -\frac{g^2\sin(\Delta t)}{\Delta^2}\hat K,
    \label{eq:magnus_jc_omega2_oscillatory}
\end{equation}
and belongs to the micromotion. Restoring the free evolution gives
Eq.~\eqref{eq:projection_jc_dispersive_schrodinger}. Projection and Magnus
therefore agree on the slow spectrum in a common frame, while storing the
eliminated motion differently: the former uses a wave operator, whereas the
latter retains it in $\hat\Omega_1$ and the bounded part of
$\hat\Omega_2$.

The local control parameter is
$\epsilon_{\mathrm{JC}}(n_\star)=
g\sqrt{n_\star+1}/|\Delta|\ll1$. Resolving a slow generator also requires
$t\gg|\Delta|^{-1}$, but $t$ must remain short enough that omitted
higher-order phase errors do not accumulate. Thus the useful long-time regime
is a time window, not the limit $t\rightarrow\infty$ at fixed order.

At $t_m=2\pi m/|\Delta|$, with $m\in\mathbb N$, both
$e^{\mathrm i\Delta t_m}-1$ and $\sin(\Delta t_m)$ vanish. The leading
interaction-picture dressing is then undone, making these times natural
stroboscopic benchmarks for the dispersive Hamiltonian. Away from them, one
must retain the micromotion or state explicitly that the comparison has been
coarse grained.

\subsubsection{Magnus expansions in Liouville space}
\label{subsec:magnus_liouville}

The same construction applies to a time-local master equation,
\begin{equation}
    \frac{\mathrm d}{\mathrm dt}\hat\rho(t)
    =
    \mathcal L(t)\hat\rho(t).
    \label{eq:magnus_liouville_equation}
\end{equation}
Vectorization is reviewed in Supplemental Material, Sec.~S4, but is not needed
here. If the dynamical map satisfies
$\dot{\mathcal P}(t,0)=\mathcal L(t)\mathcal P(t,0)$, then
\begin{equation*}
    \mathcal P(t,0)
    =
    \exp[\boldsymbol{\Lambda}(t)],
    \qquad
    \boldsymbol{\Lambda}(t)
    =
    \sum_{n=1}^{\infty}\boldsymbol{\Lambda}_n(t).
\end{equation*}
Because the evolution equation contains $\mathcal L(t)$ rather than
$-\mathrm i\hat H(t)$, the coefficients carry different factors of
$\mathrm i$:
\begin{subequations}
\begin{align}
    \boldsymbol{\Lambda}_1(t)
    &=
    \int_0^t\mathrm dt_1\,\mathcal L(t_1),
    \label{eq:magnus_liouville_lambda1}
    \\
    \boldsymbol{\Lambda}_2(t)
    &=
    \frac{1}{2}
    \int_0^t\mathrm dt_1
    \int_0^{t_1}\mathrm dt_2\,
    [\mathcal L(t_1),\mathcal L(t_2)],
    \label{eq:magnus_liouville_lambda2}
    \\
    \boldsymbol{\Lambda}_3(t)
    &=
    \frac{1}{6}
    \int_0^t\mathrm dt_1
    \int_0^{t_1}\mathrm dt_2
    \int_0^{t_2}\mathrm dt_3 \\
   &\times
    \Big(
        [\mathcal L(t_1),[\mathcal L(t_2),\mathcal L(t_3)]]
        \\
        &\qquad+
        [\mathcal L(t_3),[\mathcal L(t_2),\mathcal L(t_1)]]
    \Big).
    \label{eq:magnus_liouville_lambda3}
\end{align}
\end{subequations}

Trace and Hermiticity preservation pass to the nested commutators. Complete
positivity does not: GKLS generators are not closed under superoperator
commutators, so neither
$\boldsymbol{\Lambda}^{[N]}(t)/t$ nor
$\exp[\boldsymbol{\Lambda}^{[N]}(t)]$ is automatically a physical Markovian
generator or channel \cite{Schnell2020}. Complete positivity must follow from
a suitable reorganization or be checked independently. This is an important
difference from closed dynamics, where Hermiticity of the exponent guarantees
unitarity.

\subsubsection{Example: a lossy dispersive cavity}
\label{subsec:magnus_lossy_cavity}

Now include cavity decay at rate $\kappa$. In the interaction picture of
$\hat H_0$,
\begin{equation*}
\begin{aligned}
    \dot{\hat\rho}
    &=
    \mathcal L_I(t)\hat\rho,
    &
    \mathcal L_I(t)
    &=
    \mathcal L_\kappa+\mathcal G(t),
    \\
    \mathcal L_\kappa\hat\rho
    &=
    \kappa\mathcal D[\hat a]\hat\rho,
    &
    \mathcal G(t)\hat\rho
    &=
    -\mathrm i[\hat H_I(t),\hat\rho].
\end{aligned}
\end{equation*}
The free phase of $\hat a$ cancels inside $\mathcal D[\hat a]$, so the
dissipator remains time independent. The first Magnus term contains the
secular damping $t\mathcal L_\kappa$ and the bounded coherent dressing
$\int_0^t\mathcal G(t_1)\,\mathrm dt_1$.

The coherent--coherent part of the second term is
\begin{equation*}
    \boldsymbol{\Lambda}_2^{(gg)}(t)\hat\rho
    =
    -\mathrm i g^2
    \left(
        \frac{t}{\Delta}
        -
        \frac{\sin(\Delta t)}{\Delta^2}
    \right)
    [\hat K,\hat\rho],
\end{equation*}
whose secular part is the dispersive Hamiltonian already obtained above. The
mixed commutators
$[\mathcal L_\kappa,\mathcal G(t)]$ are generally nonzero. They say that the
near-identity transformation which removes the off-resonant coherent coupling
must also act on the loss channel.

This content is most transparent after transforming the Hamiltonian and jump
operator together. Let
\begin{equation*}
    \lambda=\frac{g}{\Delta},
    \qquad
    \hat S_1
    =
    \lambda(\hat a^\dagger\hat\sigma_-
    -\hat a\hat\sigma_+),
    \qquad
    \hat U_{\mathrm d}=e^{\hat S_1}.
\end{equation*}
This is the Schrieffer--Wolff convention used in
Sec.~\ref{subsec:schrieffer_wolff}: with
$\widetilde H=e^{-\hat S_1}He^{\hat S_1}$,
$[\hat H_0,\hat S_1]=-\hat V$ cancels the off-resonant interaction at first
order. Using
$\widetilde{\hat\rho}=\hat U_{\mathrm d}^\dagger\hat\rho\hat U_{\mathrm d}$,
one obtains
\begin{equation*}
    \widetilde{\hat H}_{\mathrm{disp},I}
    =
    \frac{g^2}{\Delta}\hat K
    +
    \mathcal O\!\left(\frac{g^3}{\Delta^2}\right),
\end{equation*}
while the cavity jump becomes
\begin{equation}
    \widetilde{\hat a}
    =
    \hat U_{\mathrm d}^\dagger\hat a\hat U_{\mathrm d}=
    \hat a
    +\lambda\hat\sigma_-
    +\frac{\lambda^2}{2}\hat a\hat\sigma_z
    +\mathcal O(\lambda^3).
    \label{eq:magnus_open_dressed_jump}
\end{equation}
The term $\lambda\hat\sigma_-$ generates the mixed dissipative dressing at
order $\kappa g/\Delta$ and, at second order, the atomic relaxation channel.
The dressed-frame equation is therefore
\begin{equation}
\begin{aligned}
    \frac{\mathrm d}{\mathrm dt}\widetilde{\hat\rho}
    ={}&
    -\mathrm i[
        \widetilde{\hat H}_{\mathrm{disp},I},
        \widetilde{\hat\rho}
    ]
    +
    \kappa\mathcal D[\widetilde{\hat a}]
    \widetilde{\hat\rho},
    \label{eq:Leff_open_magnus_final}
\end{aligned}
\end{equation}
understood to the retained order in $\lambda$.

Keeping the dissipator in this compact form makes the GKLS structure
manifest, but also partially resums higher-order terms. A strict expansion
contains the original cavity loss, cross terms of order
$\kappa g/\Delta$, second-order corrections to cavity loss, and
$\kappa g^2\mathcal D[\hat\sigma_-]/\Delta^2$. The last contribution gives
the large-detuning Purcell rate
$\Gamma_{\mathrm P}\simeq\kappa g^2/\Delta^2$. This is the
$|\Delta|\gg\kappa$ limit of
Eq.~\eqref{eq:purcell_and_shift_projector}; unlike the projection result, the
present expansion does not retain the full Lorentzian denominator
$\Delta^2+\kappa^2/4$.

Equation~\eqref{eq:Leff_open_magnus_final} still evolves the full
atom--cavity state in a dressed frame; it has not eliminated the cavity. An
atomic-only master equation requires a further projection or adiabatic
elimination. Returning to the laboratory frame must likewise transform the
Hamiltonian, state, observables, and jump operator together. Restoring only the
Hamiltonian would miss the dissipative dressing in
Eq.~\eqref{eq:magnus_open_dressed_jump}.

\subsection{The Schrieffer--Wolff Transformation}
\label{subsec:schrieffer_wolff}

The Schrieffer-Wolff (SW) transformation removes weak coupling between chosen
sectors without discarding its physical influence. The sectors decouple in a
dressed frame, where energies, interactions, states, observables, and
environmental couplings have all been transformed. This makes SW especially
useful when one needs both a control-ready Hamiltonian and a map back to
laboratory quantities.

\begin{figure*}[t]
    \centering
    \includegraphics[
        width=\linewidth
    ]{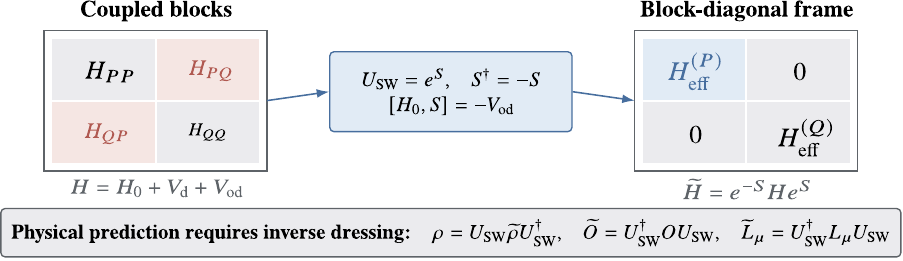}
    \caption{\justifying
    \textbf{Schrieffer--Wolff block diagonalization.} The Hilbert space is divided into
    a retained sector $\hat P$ and its complement
    $\hat Q=\hat I-\hat P$. A near-identity unitary
    $\hat U_{\mathrm{SW}}=e^{\hat S}$, generated by an anti-Hermitian,
    block-off-diagonal $\hat S$, removes the intersector coupling
    perturbatively. The $\hat P$ block of the transformed Hamiltonian defines
    the effective dynamics; physical states, observables, and dissipative
    channels require the inverse dressing.
    }
    \label{fig:schrieffer_wolff_schematic}
\end{figure*}

Figure~\ref{fig:schrieffer_wolff_schematic} shows the construction: a
near-identity unitary suppresses the off-diagonal blocks order by order, and
the retained block becomes the effective Hamiltonian. The same unitary supplies
the state and observable reconstruction.

Schrieffer and Wolff originally used this transformation to reduce the
Anderson impurity model to the Kondo model when virtual charge fluctuations
are energetically costly \cite{Schrieffer1966}. The same construction now
appears in condensed-matter physics, quantum optics, superconducting circuits,
and quantum information
\cite{Bravyi2011,Benito2019,Jager2022BosonicModes}. It is closely related to
L\"owdin partitioning and quasi-degenerate perturbation theory
\cite{Lowdin1951,Winkler2003}. Unlike a secular truncation, SW is a
perturbative change of basis. Its control parameter is therefore a
matrix-element-to-gap ratio evaluated on the occupied manifold, not simply the
oscillation frequency of a term in one interaction picture.

\subsubsection{Two-block formulation and canonical gauge}
\label{subsec:sw_general_formalism}

Write $\hat H=\hat H_0+\lambda\hat V$, where $\lambda$ records perturbative
order. Let $\hat P+\hat Q=\hat I$, $\hat P\hat Q=0$, and assume that
$\hat H_0$ is block diagonal. Separate the perturbation into
\begin{subequations}
\begin{align}
    \hat V_{\mathrm d}
    &=
    \hat P\hat V\hat P+\hat Q\hat V\hat Q,
    \label{eq:sw_Vd}
    \\
    \hat V_{\mathrm{od}}
    &=
    \hat P\hat V\hat Q+\hat Q\hat V\hat P.
    \label{eq:sw_Vod}
\end{align}
\end{subequations}
Following Sec.~\ref{sec:common_framework}, we use
$\hat U_{\mathrm{SW}}=e^{\hat S}$ and
\[
    \widetilde{\hat H}
    =
    \hat U_{\mathrm{SW}}^\dagger\hat H\hat U_{\mathrm{SW}}
    =
    e^{-\hat S}\hat H e^{\hat S},
    \qquad
    \hat S^\dagger=-\hat S.
\]
Expand
$\hat S=\lambda\hat S_1+\lambda^2\hat S_2+\cdots$. Adding a
block-diagonal anti-Hermitian term to $\hat S_n$ would only rotate states
within $P$ or $Q$. The canonical gauge removes this freedom by requiring
$\hat P\hat S_n\hat P=\hat Q\hat S_n\hat Q=0$ at every order
\cite{Bravyi2011}.

The Baker--Campbell--Hausdorff expansion gives
\begin{align}
    \widetilde{\hat H}
    &=
    \hat H_0
    +
    \lambda\bigl(
        \hat V+[\hat H_0,\hat S_1]
    \bigr) \nonumber\\
    &\quad+
    \lambda^2\left(
        [\hat H_0,\hat S_2]
        +
        [\hat V,\hat S_1]
        +
        \frac12[[\hat H_0,\hat S_1],\hat S_1]
    \right) \nonumber \\
    &\quad+
    \mathcal O(\lambda^3).
    \label{eq:sw_bch_second_order}
\end{align}
The first off-diagonal block vanishes when
\begin{equation}
    [\hat H_0,\hat S_1]
    =
    -\hat V_{\mathrm{od}}.
    \label{eq:sw_first_order_generator_equation}
\end{equation}
At second order,
$[\hat V_{\mathrm{od}},\hat S_1]$ is block diagonal, whereas
$[\hat V_{\mathrm d},\hat S_1]$ is off diagonal. Choosing
$[\hat H_0,\hat S_2]=-[\hat V_{\mathrm d},\hat S_1]$ removes the latter.
The retained block is therefore
\begin{equation}
    \hat H_{\mathrm{eff}}^{(2)}
    =
    \hat P\hat H_0\hat P
    +
    \lambda\hat P\hat V\hat P
    +
    \frac{\lambda^2}{2}
    \hat P[\hat V_{\mathrm{od}},\hat S_1]\hat P
    +
    \mathcal O(\lambda^3).
    \label{eq:sw_heff2}
\end{equation}
The corresponding full transformed Hamiltonian contains both approximately
decoupled blocks; Eq.~\eqref{eq:sw_heff2} retains only the one selected by
$P$.

\subsubsection{Energy denominators and the relation to L\"owdin partitioning}
\label{subsec:sw_matrix_elements}

Let $\hat H_0\ket i=\epsilon_i\ket i$ for $\ket i\in P$ and
$\hat H_0\ket\alpha=\epsilon_\alpha\ket\alpha$ for
$\ket\alpha\in Q$. Provided the two sectors are nondegenerate with one
another, Eq.~\eqref{eq:sw_first_order_generator_equation} gives
\begin{subequations}
\begin{align}
    \langle i|\hat S_1|\alpha\rangle
    &=
    \frac{\langle i|\hat V|\alpha\rangle}
    {\epsilon_\alpha-\epsilon_i},
    \label{eq:sw_S1_matrix_PQ}
    \\
    \langle\alpha|\hat S_1|i\rangle
    &=
    -\langle i|\hat S_1|\alpha\rangle^*.
    \label{eq:sw_S1_matrix_QP}
\end{align}
\end{subequations}
Inserting these matrix elements into Eq.~\eqref{eq:sw_heff2} reproduces the
Hermitian second-order L\"owdin expression in
Eq.~\eqref{eq:projection_lowdin_second_order}. Projection theory inserts a
resolvent between visits to $Q$; SW instead rotates away the coupling. Their
model-space Hamiltonians agree to this order when the partition, energy
convention, and block gauge agree.

The local perturbative ratio is
\[
    \epsilon_{i\alpha}^{\mathrm{SW}}
    =
    \lambda
    \frac{|V_{i\alpha}|}{|\epsilon_i-\epsilon_\alpha|}.
\]
It must be small for the states and virtual paths that are appreciably
occupied. A global norm may be unnecessarily restrictive in a many-body
problem and undefined for an unbounded bosonic Hamiltonian; the practical
condition then belongs to a finite-energy or occupation manifold. A small
denominator signals that the corresponding state should normally be moved
into an enlarged retained space.

\subsubsection{State and observable reconstruction}
\label{subsec:sw_reconstruction}

With the present convention, the encoded state, its reduced evolution, and
the reconstructed physical state are
\begin{equation*}
\begin{aligned}
    \widetilde{\hat\rho}_P(0)
    &=
    \hat P e^{-\hat S}\hat\rho(0)e^{\hat S}\hat P,
    \\
    \widetilde{\hat\rho}_P(t)
    &=
    e^{-\mathrm i\hat H_{\mathrm{eff}}t}
    \widetilde{\hat\rho}_P(0)
    e^{\mathrm i\hat H_{\mathrm{eff}}t},
    \\
    \hat\rho_{\mathrm{rec}}(t)
    &=
    e^{\hat S}
    [\widetilde{\hat\rho}_P(t)\oplus\hat 0_Q]
    e^{-\hat S}.
\end{aligned}
\end{equation*}
If the encoded state is subnormalized, its trace is the initial model-space
weight; normalization should be imposed only for a stated conditional
prediction.

A physical observable is represented by
\[
    \hat O_{\mathrm{eff}}
    =
    \hat P e^{-\hat S}\hat O e^{\hat S}\hat P,
\]
with
\[
    e^{-\hat S}\hat Oe^{\hat S}
    =
    \hat O+[\hat O,\hat S]
    +\frac12[[\hat O,\hat S],\hat S]+\cdots.
\]
Thus a dressed Hamiltonian combined with a bare observable generally mixes
orders and misses the leading correction. At finite order, the generator,
Hamiltonian, states, observables, and residual block coupling must all be
truncated and tested consistently.

\subsubsection{An operator-basis construction of the generator}
\label{subsec:sw_operator_basis_generator}

For hybrid finite-dimensional--bosonic systems, occupation-dependent
denominators can make Eq.~\eqref{eq:sw_first_order_generator_equation}
awkward to solve by inspection. An operator-basis construction treats a broad
class of such problems directly \cite{reascos2024universal}. Let
\begin{align}
    &\hat H_0
    =
    \sum_{\mu=0}^{d_f-1}
    f_\mu(\hat N)\hat\sigma_{\mu\mu},
    \label{eq:sw_operator_basis_H0}
\end{align}
$\hat N=\hat a^\dagger\hat a$, $\hat\sigma_{\mu\nu}=\ket\mu\bra\nu$.
At order $j$, write the known Hermitian right-hand side of the generator
equation as
\begin{equation}
    \hat Z^{(j)}
    =
    \sum_{\mu,\nu}\sum_{\ell\geq0}
    z_{\mu\nu}^{(\ell,j)}(\hat N)
    \hat a^\ell\hat\sigma_{\mu\nu}
    +
    \mathrm{H.c.},
    \label{eq:sw_operator_basis_Z}
\end{equation}
so that $[\hat H_0,\hat S_j]=\hat Z^{(j)}$; at first order,
$\hat Z^{(1)}=-\hat V_{\mathrm{od}}$. A compatible anti-Hermitian generator is
\begin{equation}
    \hat S_j
    =
    \sum_{\mu,\nu}\sum_{\ell\geq0}
    s_{\mu\nu}^{(\ell,j)}(\hat N)
    \hat a^\ell\hat\sigma_{\mu\nu}
    -
    \mathrm{H.c.},
    \label{eq:ordered_SW_generator}
\end{equation}
where
\begin{align}
    &s_{\mu\nu}^{(\ell,j)}(\hat N)
    =
    -\frac{
        z_{\mu\nu}^{(\ell,j)}(\hat N)
    }{
        \omega_{\mu\nu}^{(\ell)}(\hat N)
    }, \nonumber\\
    &\omega_{\mu\nu}^{(\ell)}(\hat N)
    =
    f_\nu(\hat N+\ell)-f_\mu(\hat N).
    \label{eq:universal_generator}
\end{align}
The denominator is the occupation-dependent mismatch of the process
$\hat a^\ell\hat\sigma_{\mu\nu}$. It must remain nonzero and large compared
with the corresponding matrix elements on the populated manifold.

Ordering matters because
\begin{align}
   &\hat a^\ell f(\hat N)
    =
    f(\hat N+\ell)\hat a^\ell, \nonumber \\
    &f(\hat N)\hat a^\ell
    =
    \hat a^\ell f(\hat N-\ell).
    \label{eq:sw_number_shift_identity} 
\end{align}
The explicit Hermitian-conjugate term in
Eq.~\eqref{eq:ordered_SW_generator} preserves anti-Hermiticity in this
ordering. Multiple modes can be included by replacing $N$ with an occupation
multi-index, while Hamiltonians outside this class may be better handled with
an eigenoperator or Sylvester-equation formulation.

\subsubsection{Time-dependent Schrieffer--Wolff transformations}
\label{subsec:sw_time_dependent}

For $\hat U_{\mathrm{SW}}(t)=e^{\hat S(t)}$, the exact transformed Hamiltonian
contains a derivative term,
\begin{equation*}
    \widetilde{\hat H}(t)
    =
    e^{-\hat S(t)}\hat H(t)e^{\hat S(t)}
    -
    \mathrm i e^{-\hat S(t)}
    \frac{\mathrm d}{\mathrm dt}e^{\hat S(t)}.
\end{equation*}
The first-order cancellation equation is therefore
\[
    \hat V_{\mathrm{od}}(t)
    +
    [\hat H_0,\hat S_1(t)]
    -
    \mathrm i\dot{\hat S}_1(t)
    =
    0.
\]
Neglecting $\dot{\hat S}_1$ is justified only if it is of higher
perturbative order than the retained generator.

For an operator-basis coefficient
$z_k(\hat N,t)=\sum_n z_k^{(n)}(\hat N)e^{\mathrm in\Omega_k t}$, where
$k=(\mu,\nu,\ell)$, a periodic solution is
\begin{equation}
    s_k(\hat N,t)
    =
    -
    \sum_{n=-\infty}^{\infty}
    \frac{
        z_k^{(n)}(\hat N)e^{\mathrm in\Omega_k t}
    }{
        \omega_k(\hat N)-n\Omega_k
    }.
    \label{eq:sw_time_dependent_generator}
\end{equation}
The denominator is the quasienergy mismatch of the virtual process and is
understood spectrally on the occupied number states. The local condition is
\[
    \frac{|z_k^{(n)}(r)|}
    {|\omega_k(r)-n\Omega_k|}
    \ll1
\]
for every appreciably occupied $r$. Near
$\omega_k(r)=n\Omega_k$, the corresponding harmonic is resonant and should
normally be retained rather than eliminated. Continuous-time observables also
require the explicitly time-dependent reconstruction $e^{\hat S(t)}$.

\subsubsection{Example: the dispersive Jaynes--Cummings model}
\label{subsec:sw_dispersive_jc}

For the Jaynes--Cummings model,
\begin{align}
     &\hat H_0
    =
    \omega_a\hat a^\dagger\hat a
    +
    \frac{\omega_0}{2}\hat\sigma_z, \nonumber\\
    \qquad
    &\hat V
    =
    g(\hat a\hat\sigma_+
    +\hat a^\dagger\hat\sigma_-),
\end{align}
with $\Delta=\omega_0-\omega_a$. Since
$[\hat H_0,\hat a\hat\sigma_+]=
\Delta\hat a\hat\sigma_+$, the first generator is
\[
    \hat S_1
    =
    \frac{g}{\Delta}
    (\hat a^\dagger\hat\sigma_-
    -\hat a\hat\sigma_+),
    \qquad
    [\hat H_0,\hat S_1]=-\hat V.
\]
Because $\hat V$ is entirely off diagonal, Eq.~\eqref{eq:sw_heff2} reduces to
$\widetilde{\hat H}^{(2)}=\hat H_0+[\hat V,\hat S_1]/2$ and gives the
dispersive Hamiltonian in
Eq.~\eqref{eq:projection_jc_dispersive_schrodinger}. The residual off-diagonal
operator begins at order $g^3/\Delta^2$, while the first omitted JC energy
correction is of order $g^4/\Delta^3$ because the exact energies contain only
even powers.

For occupation up to $n_\star$, the condition is
$g\sqrt{n_\star+1}/|\Delta|\ll1$. The physical states reconstructed from
$e^{\hat S_1}$ are
\begin{subequations}
\begin{align}
    \ket{g,n}_{\mathrm{phys}}
    &=
    \ket{g,n}
    -
    \frac{g\sqrt n}{\Delta}\ket{e,n-1} \nonumber
    \\
    &\quad{}+
    \mathcal O\!\left(\frac{g^2n}{\Delta^2}\right),
    \label{eq:sw_jc_dressed_ground}
    \\
    \ket{e,n}_{\mathrm{phys}}
    &=
    \ket{e,n}
    +
    \frac{g\sqrt{n+1}}{\Delta}\ket{g,n+1} \nonumber\\
    &\quad+
    \mathcal O\!\left(\frac{g^2(n+1)}{\Delta^2}\right).
    \label{eq:sw_jc_dressed_excited}
\end{align}
\end{subequations}
They coincide with the projection wave-operator result. The transformed
annihilation operator is the one already displayed in
Eq.~\eqref{eq:magnus_open_dressed_jump}; field observables and cavity-loss
channels therefore cannot be evaluated by transforming the Hamiltonian alone.

\subsubsection{Generalized SW transformations for open systems}
\label{subsec:sw_open_systems}

Block diagonalization can also be performed directly in Liouville space
\cite{Kessler2012DissipativeSW}. Let
$\mathcal L=\mathcal L_0+\lambda\mathcal V$, and let $\mathcal P$ project
onto the slow stationary manifold of $\mathcal L_0$, so that
$\mathcal P\mathcal L_0=\mathcal L_0\mathcal P=0$ and
$\mathcal Q=\mathcal I-\mathcal P$. Assume that
$\mathcal L_Q=\mathcal Q\mathcal L_0\mathcal Q$ is invertible on the fast
sector and spectrally separated from zero.

The decoupling is now a generally nonunitary similarity transformation,
\begin{equation}
    \widetilde{\mathcal L}
    =
    e^{-\boldsymbol{\mathcal S}}
    \mathcal L
    e^{\boldsymbol{\mathcal S}},
    \label{eq:sw_liouville_similarity}
\end{equation}
with
$\mathcal P\boldsymbol{\mathcal S}\mathcal P=
\mathcal Q\boldsymbol{\mathcal S}\mathcal Q=0$. At first order,
$[\mathcal L_0,\boldsymbol{\mathcal S}_1]=-\mathcal V_{\mathrm O}$, where
$\mathcal V_{\mathrm O}=
\mathcal P\mathcal V\mathcal Q+
\mathcal Q\mathcal V\mathcal P$. A solution is
\[
    \boldsymbol{\mathcal S}_1
    =
    \mathcal P\mathcal V\mathcal Q\mathcal L_Q^{-1}\mathcal Q
    -
    \mathcal Q\mathcal L_Q^{-1}\mathcal Q\mathcal V\mathcal P.
\]
The slow generator is
\begin{equation}
    \mathcal L_{\mathrm{eff}}
    =
    \lambda\mathcal P\mathcal V\mathcal P-
    \lambda^2
    \mathcal P\mathcal V\mathcal Q
    (\mathcal Q\mathcal L_0\mathcal Q)^{-1}
    \mathcal Q\mathcal V\mathcal P
    +
    \mathcal O(\lambda^3).
    \label{eq:sw_liouville_effective_second_order}
\end{equation}
This is the zero-frequency second-order Schur complement found by projection,
now obtained through a similarity transformation.

Effective coordinates are reconstructed according to
\begin{equation}
    \ket{\!\ket{\rho_{\mathrm{rec}}}}
    \simeq
    e^{\boldsymbol{\mathcal S}}
    \mathcal P
    \ket{\!\ket{\widetilde\rho_{\mathrm{eff}}}}.
    \label{eq:sw_liouville_reconstruction}
\end{equation}
Because $e^{\boldsymbol{\mathcal S}}$ need not arise from a Hilbert-space
unitary or a completely positive map, the intermediate effective coordinate
need not itself be a normalized positive density operator. The reconstructed
operator and any reduced dynamical map must be tested for trace, Hermiticity,
positivity, and, when required, complete positivity. Likewise,
Eq.~\eqref{eq:sw_liouville_effective_second_order} is not automatically a
GKLS generator.

\subsubsection{Example: eliminating a lossy cavity}
\label{subsec:sw_lossy_cavity}

In a frame rotating at the atomic frequency, the lossy Jaynes--Cummings model
is
\begin{align}
    \dot{\hat\rho}
    ={}&
    -\mathrm i[
        -\Delta\hat a^\dagger\hat a
        +
        g(\hat a\hat\sigma_+
        +\hat a^\dagger\hat\sigma_-),
        \hat\rho
    ]
    +
    \kappa\mathcal D[\hat a]\hat\rho.
    \label{eq:lossy_jc_common_benchmark}
\end{align}
Separate
\begin{subequations}
\begin{align}
    \mathcal L_0\hat\rho
    &=
    -\mathrm i[-\Delta\hat a^\dagger\hat a,\hat\rho]
    +
    \kappa\mathcal D[\hat a]\hat\rho,
    \label{eq:sw_open_jc_L0}
    \\
    \mathcal V\hat\rho
    &=
    -\mathrm i g[
        \hat a\hat\sigma_+
        +\hat a^\dagger\hat\sigma_-,
        \hat\rho
    ].
    \label{eq:sw_open_jc_V}
\end{align}
\end{subequations}
The slow projector is
$\mathcal P\hat\rho=
\operatorname{Tr}_c(\hat\rho)\otimes\ket0\bra0$, for which
$\mathcal P\mathcal V\mathcal P=0$. The fast one-photon coherences have
rates $-\kappa/2\pm\mathrm i\Delta$ and susceptibility
$\chi_c(\Delta)=(\kappa/2-\mathrm i\Delta)^{-1}$.

Equation~\eqref{eq:sw_liouville_effective_second_order} then gives
\begin{align}
    \dot{\hat\rho}_{\mathrm{atom}}
    &=
    -\mathrm i[
        \dfrac{\omega_0}{2}\hat\sigma_z
        +
        \delta\omega\hat\sigma_+\hat\sigma_-,
        \hat\rho_{\mathrm{atom}}
    ] \nonumber\\
    &\quad{}+\Gamma\mathcal D[\hat\sigma_-]\hat\rho_{\mathrm{atom}},
    \label{eq:SW_effective_liouville}
\end{align}
where $\delta\omega$ and $\Gamma$ are given in
Eq.~\eqref{eq:purcell_and_shift_projector}. This agreement with projection is
expected because both calculations use the same slow projector and the same
inverse fast Liouvillian.

The local open-system parameter is
$g/\sqrt{\Delta^2+\kappa^2/4}$, or
$g\sqrt{n_\star+1}/\sqrt{\Delta^2+\kappa^2/4}$ when higher cavity
occupations matter. The leading eliminated component is
\begin{equation}
    \mathcal Q\hat\rho_{AB}
    \simeq
    -
    (\mathcal Q\mathcal L_0\mathcal Q)^{-1}
    \mathcal Q\mathcal V\mathcal P
    [\hat\rho_A\otimes\ket0\bra0].
    \label{eq:sw_open_cavity_reconstruction}
\end{equation}
The atomic generator in Eq.~\eqref{eq:SW_effective_liouville} is explicitly
GKLS, but this is a feature of the present model. Its completely positive
atomic map does not remove the need to validate the perturbative
reconstruction of cavity observables and correlations. Related
operator-valued displacement methods can produce manifestly Lindblad
system-only equations for rapidly relaxing bosonic mediators, subject to their
own weak-displacement and timescale-separation conditions
\cite{Jager2022BosonicModes}.

\subsubsection{Symbolic and numerical block diagonalization}
\label{subsec:sw_numerical_tools}

Nested commutators proliferate rapidly beyond low order. The open-source
package \texttt{Pymablock} implements perturbative quasi-degenerate block
diagonalization for symbolic expressions and dense or sparse numerical
matrices \cite{ArayaDay2025}. In the same two-block, block-off-diagonal gauge,
its construction is equivalent to canonical SW; multiblock and selective
diagonalization need not coincide with a sequence of two-block
transformations. Documented workflows and version-specific capabilities are
available in the
\href{https://pymablock.readthedocs.io/}{package documentation}.

Automation checks algebra, not the physical partition. A calculation should
still verify the $P$--$Q$ split and block gauge, cancellation of the
off-diagonal block to the requested order, Hermiticity, scaling of the first
omitted term, convergence with any bosonic cutoff, consistent reconstruction,
and sensitivity to small denominators.

\subsection{Adiabatic Elimination}
\label{subsec:adiabatic_elimination}

Adiabatic elimination (AE) is useful when part of a system adjusts much
faster than the motion one wishes to describe. The fast variables are not
assumed to vanish. Instead, after a short transient or rapid oscillatory
adjustment, they remain close to values fixed by the instantaneous slow
state. Inserting this relation into the slow equations gives a reduced
model and, equally importantly, a formula for reconstructing the eliminated
variables.

\begin{figure*}[t]
    \centering
    \includegraphics[width=1.0\linewidth]
    {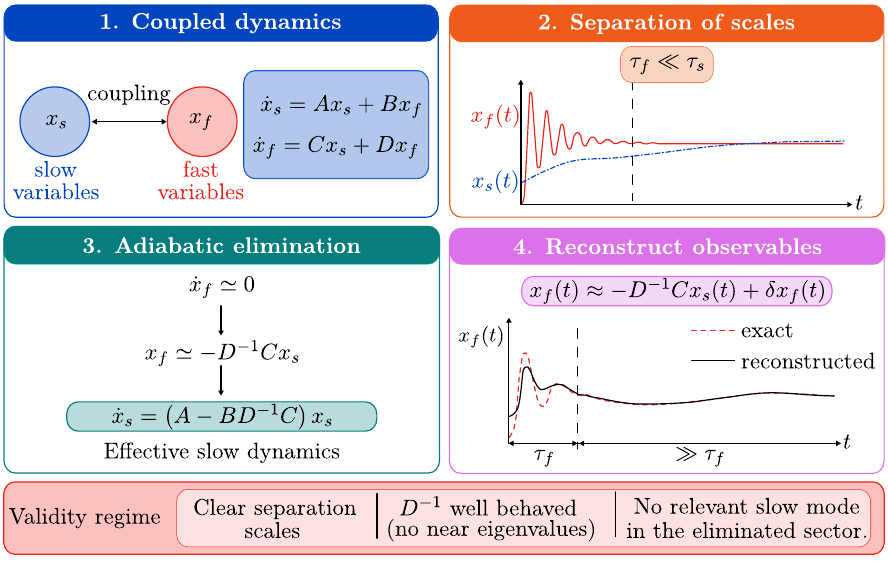}
    \caption{\justifying
    \textbf{Operational workflow for adiabatic elimination.} The variables are first
    separated into a retained sector and a rapidly relaxing or strongly
    detuned sector. For the linear system shown, the leading fast response is
    $x_{\mathrm f}^{\mathrm{ss}}[x_{\mathrm s}]
    =-D^{-1}C x_{\mathrm s}$, and substitution gives the Schur-complement
    generator $\dot{x}_{\mathrm s}=(A-BD^{-1}C)x_{\mathrm s}$. The same
    response reconstructs observables of the eliminated sector. The
    approximation fails when the inverse fast block becomes large, a fast
    pole approaches the slow spectrum, or the initial fast transient is
    important on the observation timescale.}
    \label{fig:adiabatic_elimination_workflow}
\end{figure*}

Here \emph{adiabatic} refers to this separation of dynamical timescales. It is
unrelated to thermodynamic adiabaticity and should not be confused with the
quantum adiabatic theorem. Figure~\ref{fig:adiabatic_elimination_workflow}
shows the calculation in its simplest linear form. The nonlinear and quantum
constructions below follow the same steps: choose the slow variables, solve
for the fast response, substitute it into the slow equation, and retain the
response for later reconstruction.

AE has long been used in laser and maser theory and now appears throughout
quantum optics, cavity QED, reservoir engineering, driven multilevel
systems, and quantum control
\cite{Graham1968,Lugiato1984,Klimov2002master,FinkelsteinShapiro2020}.
Its familiar instruction---set the fast derivative to zero---is only the
leading step. The result also depends on the chosen frame, the stability or
detuning of the fast sector, the initial condition, and the reconstruction
order \cite{Brion2007,Mirrahimi2009,Reiter2012Effective,Azouit2017}.

\subsubsection{Slow manifolds and timescale separation}
\label{subsec:ae_slow_manifold}

A generic slow--fast system can be written as
\begin{subequations}
\begin{align}
    \dot{x}
    &=F(x,y,t;\epsilon),
    \label{eq:ae_generic_slow}
    \\
    \epsilon\dot{y}
    &=G(x,y,t;\epsilon),
    \label{eq:ae_generic_fast}
\end{align}
\end{subequations}
where $x$ and $y$ denote the slow and fast variables and
$\epsilon=\tau_{\mathrm f}/\tau_{\mathrm s}\ll1$. At leading order, the
relation $G(x,h_0(x,t),t;0)=0$ defines a slow manifold $y=h_0(x,t)$, and
the reduced equation is $\dot{x}=F(x,h_0(x,t),t;0)$.

The derivative has not disappeared; it determines the corrections to this
leading manifold. If
$y=h(x,t;\epsilon)=h_0+\epsilon h_1+\cdots$, invariance requires
\begin{equation*}
    \epsilon\left(
        \partial_t h
        +D_xh\,F(x,h,t;\epsilon)
    \right)
    =G(x,h,t;\epsilon).
\end{equation*}
Expanding this equation in $\epsilon$ gives $h_1,h_2,\ldots$ and makes
precise which derivative terms were neglected by the stationary
approximation.

For a dissipative fast sector, the manifold must attract nearby motion.
The relevant eigenvalues of
$J_{\mathrm f}=\partial_yG|_{y=h_0}$ must have negative real parts well
separated from the slow rates. An eigenvalue approaching the imaginary
axis identifies critical slowing down, an instability, or another variable
that belongs in the slow description. A closed Hamiltonian system is
different: a strongly detuned component generally oscillates rather than
relaxes. Any initial component away from the dressed slow manifold then
survives as bounded fast motion and must be restored when continuous-time
observables are calculated.

The slow--fast split is therefore a choice made in a specified frame, not
an immutable division into bare subsystems. Before using AE, one should
identify the retained variables, the fast spectral scale, the largest
coupling in the populated manifold, the initial layer or fast oscillation,
and the time interval of interest. Changing frame can change which
variables appear stationary and can invalidate a derivative-zero argument
made in the original coordinates \cite{Rodrigues2006,Brion2007}.

\subsubsection{Hamiltonian block elimination}
\label{subsec:ae_hamiltonian_blocks}

Let $\hat P$ select the retained amplitudes and
$\hat Q=\hat I-\hat P$. In a suitable rotating or interaction frame, the
two blocks of the Schr\"odinger equation are
\begin{subequations}
\begin{align}
    \mathrm{i}\frac{\mathrm d}{\mathrm{d}t}\ket{\psi_P}
    &=
    \hat H_{PP}\ket{\psi_P}
    +\hat H_{PQ}\ket{\psi_Q},
    \label{eq:ae_block_P}
    \\
    \mathrm{i}\frac{\mathrm d}{\mathrm{d}t}\ket{\psi_Q}
    &=
    \hat H_{QP}\ket{\psi_P}
    +\hat H_{QQ}\ket{\psi_Q},
    \label{eq:ae_block_Q}
\end{align}
\end{subequations}
where $\ket{\psi_A}=\hat A\ket\psi$ and
$\hat H_{AB}=\hat A\hat H\hat B$. Suppose that $\hat H_{QQ}$ is
invertible in the populated eliminated manifold and that its frequencies
are large compared with both $\hat H_{QP}$ and the retained evolution.
The leading solution of Eq.~\eqref{eq:ae_block_Q} is
\begin{equation}
    \ket{\psi_Q}
    \simeq
    -\hat H_{QQ}^{-1}\hat H_{QP}\ket{\psi_P}.
    \label{eq:ae_closed_reconstruction}
\end{equation}
Substitution in Eq.~\eqref{eq:ae_block_P} gives
\begin{equation}
    \hat H_{\mathrm{AE}}
    =
    \hat H_{PP}
    -\hat H_{PQ}\hat H_{QQ}^{-1}\hat H_{QP}.
    \label{eq:ae_closed_effective_hamiltonian}
\end{equation}

Equation~\eqref{eq:ae_closed_effective_hamiltonian} is the zero-frequency
Schur complement in the selected frame. Retaining the frequency dependence
of the eliminated resolvent leads to the Bloch--Horowitz construction,
whereas a Hermitian perturbative block diagonalization leads to the SW
form. The three results agree only after the same partition, frame, and
order have been chosen.

The companion relation~\eqref{eq:ae_closed_reconstruction} is part of the
prediction. It defines the approximate embedding
$\hat\Omega_{\mathrm{AE}}
=\hat P-\hat Q\hat H_{QQ}^{-1}\hat H_{QP}$ and shows why a retained
amplitude should not automatically be normalized as if the eliminated
component were absent. If the initial state does not lie on this manifold,
the mismatch produces a decaying boundary layer in a relaxing system or
persistent fast motion in a closed one. A small derivative is thus a
property of the chosen slow solution, not of every allowed initial state.

\subsubsection{Example: the dispersive Jaynes--Cummings model}
\label{subsec:ae_dispersive_jc}

The dispersive Jaynes--Cummings model makes this distinction concrete. In
each excitation manifold the coupling is $g\sqrt{n+1}$ and the branch
separation is $\Delta=\omega_0-\omega_a$, so the local control parameter is
$\epsilon_{\mathrm{AE}}(n_\star)
=g\sqrt{n_\star+1}/|\Delta|\ll1$. Because the system is closed, the
off-resonant branch is fast but not attracting.

For the branch connected to $\ket{g,n}$, remove the bare phase and write
the two amplitude equations as
\begin{subequations}
\begin{align}
    \mathrm{i}\dot c_{g,n}
    &=g\sqrt n\,c_{e,n-1},
    \label{eq:ae_jc_ground_branch_slow}
    \\
    \mathrm{i}\dot c_{e,n-1}
    &=\Delta c_{e,n-1}+g\sqrt n\,c_{g,n}.
    \label{eq:ae_jc_ground_branch_fast}
\end{align}
\end{subequations}
The stationary fast response is
$c_{e,n-1}\simeq-(g\sqrt n/\Delta)c_{g,n}$, which shifts the retained
energy by $-g^2n/\Delta$. The branch connected to $\ket{e,n}$ similarly
obeys
\begin{subequations}
\begin{align}
    \mathrm{i}\dot c_{e,n}
    &=g\sqrt{n+1}\,c_{g,n+1},
    \label{eq:ae_jc_excited_branch_slow}
    \\
    \mathrm{i}\dot c_{g,n+1}
    &=-\Delta c_{g,n+1}+g\sqrt{n+1}\,c_{e,n},
    \label{eq:ae_jc_excited_branch_fast}
\end{align}
\end{subequations}
and gives $c_{g,n+1}\simeq(g\sqrt{n+1}/\Delta)c_{e,n}$ and the shift
$g^2(n+1)/\Delta$.

Combining the two branches recovers the dispersive Hamiltonian in
Eq.~\eqref{eq:projection_jc_dispersive_schrodinger}. The calculation also
shows why the reconstruction and the energy expansion have different
leading orders. The eliminated amplitude is first order in
$g\sqrt{n+1}/\Delta$, while the first correction to the energy is second
order and the next JC energy correction is fourth order. In particular,
the excited-state vacuum shift cannot be recovered from population
equations alone; phase or coherence information is needed. The slaved
amplitudes reproduce the dressed states in
Eqs.~\eqref{eq:sw_jc_dressed_ground} and
\eqref{eq:sw_jc_dressed_excited}.
The three-photon Rabi resonance of Sec.~\ref{subsec:higher_order_three_photon} is less forgiving: the two virtual intermediate amplitudes must be eliminated successively to generate the third-order endpoint coupling, while the second-order endpoint shifts must be retained first because they move the resonance by more than the cubic gap. It is therefore a useful example of why AE must sometimes be iterated order by order rather than applied as a single stationary substitution.
\subsubsection{Open-system adiabatic elimination}
\label{subsec:ae_open_systems}

For open dynamics, the slow and fast variables are operator-space
coordinates. Write $\mathcal L=\mathcal L_0+\lambda\mathcal V$, let
$\mathcal P$ project onto the slow manifold, and set
$\mathcal Q=\mathcal I-\mathcal P$. We take
$\mathcal P\mathcal L_0=\mathcal L_0\mathcal P=0$ and assume that
$\mathcal L_Q=\mathcal Q\mathcal L_0\mathcal Q$ is invertible on the
fast sector, with its nonzero spectrum separated from the slow rates.
For $\hat\rho_{\mathrm s}=\mathcal P\hat\rho$ and
$\hat\rho_{\mathrm f}=\mathcal Q\hat\rho$, the exact block equations are
\begin{subequations}
\begin{align}
    \dot{\hat\rho}_{\mathrm s}
    ={}&
    \lambda\mathcal P\mathcal V\mathcal P\hat\rho_{\mathrm s}
    +\lambda\mathcal P\mathcal V\mathcal Q\hat\rho_{\mathrm f},
    \label{eq:ae_open_slow_equation}
    \\
    \dot{\hat\rho}_{\mathrm f}
    ={}&
    \mathcal L_Q\hat\rho_{\mathrm f}
    +\lambda\mathcal Q\mathcal V\mathcal P\hat\rho_{\mathrm s}
    +\lambda\mathcal Q\mathcal V\mathcal Q\hat\rho_{\mathrm f}.
    \label{eq:ae_open_fast_equation}
\end{align}
\end{subequations}
The leading fast response is
\begin{equation}
    \hat\rho_{\mathrm f}
    \simeq
    -\lambda\mathcal L_Q^{-1}
    \mathcal Q\mathcal V\mathcal P\hat\rho_{\mathrm s},
    \label{eq:ae_open_fast_reconstruction}
\end{equation}
and the reduced generator becomes
\begin{equation}
    \mathcal L_{\mathrm{AE}}
    =
    \lambda\mathcal P\mathcal V\mathcal P
    -
    \lambda^2\mathcal P\mathcal V\mathcal Q
    (\mathcal Q\mathcal L_0\mathcal Q)^{-1}
    \mathcal Q\mathcal V\mathcal P
    +\mathcal O(\lambda^3).
    \label{eq:ae_open_effective_liouvillian}
\end{equation}

This is the same second-order slow generator obtained from the projected
zero-frequency resolvent and Liouvillian SW transformation. AE emphasizes
the dynamical content of the inverse: after the fast transient, the
eliminated operator follows the slow state according to
Eq.~\eqref{eq:ae_open_fast_reconstruction}. If the initial state is off
that manifold, the homogeneous term
$e^{\mathcal L_Qt}\mathcal Q\hat\rho(0)$ produces a boundary layer and can
leave an initial-slip correction in the slow coordinate.

Neither Eq.~\eqref{eq:ae_open_effective_liouvillian} nor a stationary
substitution guarantees GKLS form. Trace and positivity may be lost if the
slow coordinate and its embedding are truncated inconsistently.
Structure-preserving reductions construct the reduced generator together
with the embedding under conditions that make their physical meaning clear
\cite{Mirrahimi2009,Reiter2012Effective,Azouit2017}.

\subsubsection{Effective-operator formalism}
\label{subsec:ae_effective_operator_formalism}

When a weak drive couples a slow ground manifold to a rapidly decaying
excited manifold, the effective-operator formalism gives a compact
structure-preserving reduction \cite{Reiter2012Effective}. Decompose
\begin{align*}
    &\hat H
    =\hat H_{\mathrm g}+\hat H_{\mathrm e}
    +\hat V_++\hat V_-, \\
    &\hat V_+=\hat P_{\mathrm e}\hat H\hat P_{\mathrm g},\\
    & \hat V_-=\hat V_+^\dagger.
\end{align*}
Using physical jump operators $\hat J_k=\sqrt{\gamma_k}\hat L_k$ and the
dissipator convention of Eq.~\eqref{eq:global_dissipator_convention},
define the no-jump Hamiltonian in the fast manifold,
\begin{equation*}
    \hat H_{\mathrm{NH}}
    =\hat H_{\mathrm e}
    -\frac{\mathrm i}{2}\sum_k\hat J_k^\dagger\hat J_k.
\end{equation*}
If $\hat H_{\mathrm{NH}}$ is invertible on the relevant fast states and
the excitation is weak, then
\begin{equation}
\begin{aligned}
    \hat H_{\mathrm{eff}}
    ={}&
    \hat H_{\mathrm g}
    -\frac12\hat V_-
    \left[
        \hat H_{\mathrm{NH}}^{-1}
        +(\hat H_{\mathrm{NH}}^{-1})^\dagger
    \right]
    \hat V_+,
    \label{eq:ae_reiter_effective_hamiltonian}
\end{aligned}
\end{equation}
and
\begin{equation}
    \hat J_{k,\mathrm{eff}}
    =\hat J_k\hat H_{\mathrm{NH}}^{-1}\hat V_+.
    \label{eq:ae_reiter_effective_jump}
\end{equation}
The slow master equation built from $\hat H_{\mathrm{eff}}$ and these
jumps is manifestly of GKLS form. The Hermitian part of the fast
susceptibility generates a shift or mediated interaction; its decaying
part generates the effective jumps.

The same inverse reconstructs the leading no-jump fast amplitude,
\begin{equation}
    \ket{\psi_{\mathrm e}}
    \simeq
    -\hat H_{\mathrm{NH}}^{-1}\hat V_+
    \ket{\psi_{\mathrm g}}.
    \label{eq:ae_reiter_fast_amplitude}
\end{equation}
This trajectory-level relation explains the virtual excursion that either
returns coherently or ends in a jump. It is not, by itself, an
unconditional density-operator reconstruction: fast populations,
slow--fast coherences, and the initial boundary layer require the
corresponding operator-space embedding.

\subsubsection{Example: eliminating a lossy cavity}
\label{subsec:ae_lossy_cavity}

Consider the shared lossy Jaynes--Cummings benchmark of
Eq.~\eqref{eq:lossy_jc_common_benchmark}. In the frame rotating jointly at
the atomic frequency, the cavity vacuum is the slow manifold and the
one-photon sector is the leading fast manifold. The relevant operators are
\begin{align*}
   & \hat J=\sqrt{\kappa}\,\hat a,
    \qquad \hat H_{\mathrm{NH}}
    =-(\Delta+\mathrm i\kappa/2),\\
   &\hat V_+=g\hat a^\dagger\hat\sigma_-,\qquad
    \hat V_-=g\hat a\hat\sigma_+.
\end{align*}
Equations~\eqref{eq:ae_reiter_effective_hamiltonian} and
\eqref{eq:ae_reiter_effective_jump} give
\begin{equation*}
    \delta\omega
    =\frac{g^2\Delta}{\Delta^2+\kappa^2/4},
    \qquad
    \Gamma
    =\frac{g^2\kappa}{\Delta^2+\kappa^2/4},
\end{equation*}
and the Schr\"odinger-picture atomic equation
\begin{equation}
\begin{aligned}
    \dot{\hat\rho}_A
    ={}&
    -\mathrm i\left[
        \frac{\omega_0}{2}\hat\sigma_z
        +\delta\omega\hat\sigma_+\hat\sigma_-,
        \hat\rho_A
    \right]
    +\Gamma\mathcal D[\hat\sigma_-]\hat\rho_A.
    \label{eq:ae_lossy_cavity_atom_master}
\end{aligned}
\end{equation}

The two coefficients are the reactive and dissipative parts of the same
cavity susceptibility. At large detuning they reduce to
$\delta\omega\simeq g^2/\Delta$ and
$\Gamma\simeq g^2\kappa/\Delta^2$; at resonance,
$\delta\omega=0$ and $\Gamma=4g^2/\kappa$. The local elimination
parameter is
$g\sqrt{n_\star+1}/\sqrt{\Delta^2+\kappa^2/4}\ll1$, and the induced
atomic rate and shift must also remain slow relative to the complex cavity
scale.

The reduced equation is explicitly GKLS, but it does not say that the
cavity is empty in the physical state. Cavity observables and
atom--cavity correlations follow from
Eq.~\eqref{eq:ae_open_fast_reconstruction}, or conditionally from
Eq.~\eqref{eq:ae_reiter_fast_amplitude}. Projection, Liouvillian SW, and
AE give the same shift and Purcell rate here because they evaluate the same
second-order susceptibility. AE contributes the useful dynamical picture:
after an initial layer, the cavity follows the atomic motion rather than
being identically absent.

\subsubsection{Initial conditions, normalization, and physicality}
\label{subsec:ae_initial_conditions}

It is helpful to distinguish the coordinate evolved by the reduced
equation, the projection of the physical state, and the reconstructed
state. They need not have the same trace or positivity properties at a
finite order. If population initially occupies the eliminated manifold,
$\mathcal P\hat\rho$ is subnormalized; renormalizing it gives a conditional
state. Conversely, a trace-preserving slow equation does not reconstruct
the weight and coherences stored in the fast sector.

Frequency-domain and singular-perturbation constructions make this
distinction explicit by deriving the slow dynamics and its embedding
together \cite{FinkelsteinShapiro2020,Azouit2017}. In practice, one should
check the trace, Hermiticity, and positivity of the reconstructed state;
the complete positivity of a reduced map when such a claim is made; the
residual fast population; sensitivity to the initial-slip correction; and
convergence as the slow manifold is refined. A reduced coordinate may
predict selected slow observables accurately without being the literal
laboratory density operator. Its operational meaning then comes from the
reconstruction map.
\subsection{Method of Small Rotations}
\label{subsec:method_small_rotations}

\begin{figure*}[t]
    \centering
    \includegraphics[width=1.0\linewidth]{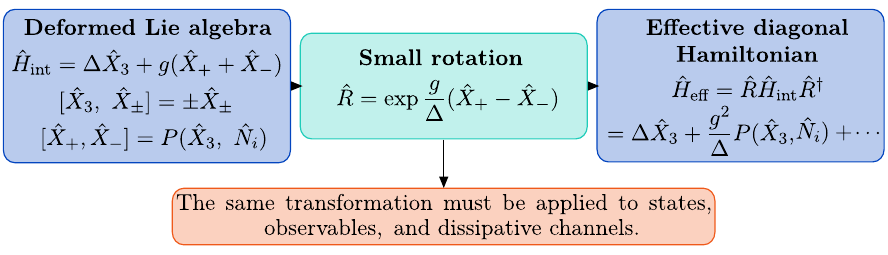}
    \caption{\justifying
    Algebraic workflow of the method of small rotations. Interaction
    operators that close a polynomially deformed $\mathfrak{su}(2)$ algebra
    are removed by a near-identity nonlinear rotation. Their commutators
    generate an effective Hamiltonian that is diagonal in the joint
    eigenbasis of $\hat X_3$ and the integrals of motion $\{\hat N_i\}$.
    Successive, smaller rotations remove higher-order residual terms. The
    same transformation must be applied to states, observables, and
    dissipative channels before laboratory predictions are reconstructed.}
    \label{fig:msr_schematic}
\end{figure*}

The method of small rotations (MSR) is most useful when the interaction
operators close under commutation. Instead of beginning with an explicit
matrix partition, it asks which operators appear when a weak off-diagonal
term is removed by a near-identity nonlinear rotation. If the algebra closes
on a small polynomial set, each induced term can be traced back to a short
sequence of commutators. Figure~\ref{fig:msr_schematic} summarizes this
algebraic route from the microscopic interaction to the diagonal effective
Hamiltonian and its laboratory reconstruction.

MSR was developed for quantum-optical models whose ladder operators form
polynomial deformations of Lie algebras
\cite{Klimov2000,Klimov2002,Karassiov1992,Karassiov1994,
KarassiovandKlimov1994,Delgado2000}. It has been used for multiphoton and
multilevel interactions, nonlinear optics, dispersive light--matter
coupling, and effective spin models
\cite{Klimov2003,Zueco2009,MedinaGonzlez2020,Villanueva2020}. A broader
tutorial introduction to Lie-algebraic organization of quantum dynamics is
given in Ref.~\cite{Qvarfort2025LieAlgebra}.

At leading order, MSR is an algebra-adapted Schrieffer--Wolff (SW)
transformation. Its practical advantage is not a different perturbative
principle, but the ability to read the generator and the induced operators
directly from a closed nonlinear algebra. This is especially convenient
when bosonic, spin, or multilevel operators look complicated in a matrix
basis but have only a few independent commutators.

\subsubsection{Polynomially deformed algebra and the first small rotation}
\label{subsec:msr_general_formalism}

Consider
\begin{equation}
    \hat H
    =
    \hat H_{\mathrm c}(\hat N_1,\hat N_2,\ldots)
    +\Delta\hat X_3
    +g(\hat X_++\hat X_-),
    \label{eq:msr_general_hamiltonian}
\end{equation}
where $\hat X_-=\hat X_+^\dagger$ and
\begin{subequations}
\begin{align}
    [\hat X_3,\hat X_\pm]
    &=\pm\hat X_\pm,
    \label{eq:msr_X3_commutator}
    \\
    [\hat X_+,\hat X_-]
    &=P(\hat X_3,\hat N_1,\hat N_2,\ldots).
    \label{eq:msr_polynomial_commutator}
\end{align}
\label{eq:msr_deformed_algebra}
\end{subequations}
The integrals of motion $\hat N_i$ commute with $\hat X_3$ and
$\hat X_\pm$, so $\hat H_{\mathrm c}$ is unchanged by the rotation. For
real $g$, the sum $\hat X_++\hat X_-$ makes the interaction Hermitian;
the anti-Hermitian difference belongs in the generator.

Set $\epsilon=g/\Delta$, assume $|\epsilon|\ll1$, and define
$\hat Y=\hat X_+-\hat X_-$. We use
\begin{align}
    &\hat U_{\mathrm{MSR}}
    =e^{\hat S_{\mathrm{MSR}}},
    \qquad
    \hat S_{\mathrm{MSR}}=-\epsilon\hat Y, \nonumber
    \\
    &\widetilde{\hat H}
    =\hat U_{\mathrm{MSR}}^\dagger
     \hat H\hat U_{\mathrm{MSR}}.
    \label{eq:msr_unitary_convention}
\end{align}
Thus $\widetilde{\hat H}=e^{\epsilon\hat Y}\hat H e^{-\epsilon\hat Y}$.
Some MSR papers denote $\hat U_{\mathrm{MSR}}^\dagger$ by $\hat R$;
this accounts for an apparent reversal of signs when conventions are
compared.

The two commutators needed initially are
\begin{subequations}
\begin{align}
    [\hat Y,\hat X_3]
    &=-(\hat X_++\hat X_-),
    \label{eq:msr_Y_X3}
    \\
    [\hat Y,\hat X_++\hat X_-]
    &=2P(\hat X_3,\{\hat N_i\}).
    \label{eq:msr_Y_ladder_sum}
\end{align}
\end{subequations}
The first cancels the term linear in $g$; the second supplies the leading
diagonal correction. Expanding with the BCH formula gives
\begin{equation}
\begin{aligned}
    \widetilde{\hat H}^{(2)}
    ={}&
    \hat H_{\mathrm c}(\{\hat N_i\})
    +\Delta\hat X_3 \\
    &+\frac{g^2}{\Delta}P(\hat X_3,\{\hat N_i\})
    +\mathcal O\!\left(\frac{g^3}{\Delta^2}\right).
    \label{eq:msr_effective_hamiltonian_second_order}
\end{aligned}
\end{equation}
Because $P$ is a function of $\hat X_3$ and the conserved quantities,
this Hamiltonian is diagonal in their joint eigenbasis through the stated
order.

The first residual term is worth retaining because it identifies the next
operator that the rotation generates:
\begin{equation}
\begin{aligned}
    \widetilde{\hat H}
    ={}&
    \hat H_{\mathrm c}
    +\Delta\hat X_3
    +\frac{g^2}{\Delta}P 
    \\
    &+\frac{2g^3}{3\Delta^2}[\hat Y,P]
    +\mathcal O\!\left(\frac{g^4}{\Delta^3}\right).
    \label{eq:msr_third_order_residual}
\end{aligned}
\end{equation}
The coefficient $2/3$ combines the second BCH contribution
$g^3[\hat Y,P]/\Delta^2$ with the third contribution
$-g^3[\hat Y,P]/(3\Delta^2)$. Since $[\hat Y,P]$ is generally off
diagonal, a smaller second rotation is required to continue the
diagonalization.

\subsubsection{Control parameter and algebraic scope}
\label{subsec:msr_control}

The ratio $|g/\Delta|$ alone is not enough when ladder-operator matrix
elements grow with excitation number. In a populated invariant sector
$\mathcal M$, the relevant local parameter is
\begin{equation*}
    \eta_{\mathrm{MSR}}(\mathcal M)
    =
    \max_{\substack{\ket\mu,\ket\nu\in\mathcal M\\
    \langle\nu|\hat X_\pm|\mu\rangle\ne0}}
    \frac{|g|\,
    |\langle\nu|\hat X_++\hat X_-|\mu\rangle|}{|\Delta|},
\end{equation*}
and one requires $\eta_{\mathrm{MSR}}(\mathcal M)\ll1$. Higher nested
commutators must remain controlled in the same sector.

For ordinary $\mathfrak{su}(2)$, a finite rotation can diagonalize a
two-level Hamiltonian exactly. For a polynomial deformation, $P$ is
nonlinear and later commutators usually generate further structures, so the
rotation is perturbative. MSR is efficient when a few ladder operators and
integrals of motion organize the relevant Hilbert space and their
commutators remain simpler than an explicit matrix calculation. If the
algebra proliferates rapidly, a general SW, projection, or numerical block
diagonalization is usually clearer.

\subsubsection{Relation to the Schrieffer--Wolff transformation}
\label{subsec:msr_relation_sw}

Writing
$\hat H_{\mathrm d}=\hat H_{\mathrm c}+\Delta\hat X_3$ and
$\hat V=g(\hat X_++\hat X_-)$, the MSR generator satisfies
\begin{equation*}
    [\hat H_{\mathrm d},\hat S_{\mathrm{MSR}}]=-\hat V.
\end{equation*}
This is the first-order canonical SW equation. Moreover,
\begin{equation*}
    \frac12[\hat V,\hat S_{\mathrm{MSR}}]
    =\frac{g^2}{\Delta}P(\hat X_3,\{\hat N_i\}),
\end{equation*}
which reproduces Eq.~\eqref{eq:msr_effective_hamiltonian_second_order}.
Thus MSR and canonical SW give the same second-order physics here. MSR
simply performs the calculation in a basis adapted to the nonlinear
algebra. At higher orders, block-internal gauges and different sequences of
rotations can redistribute terms, so comparisons must align the convention,
order, and reconstruction map.

\subsubsection{State and observable reconstruction}
\label{subsec:msr_reconstruction}

The diagonal Hamiltonian is only the dynamical core of the prediction. With
the convention in Eq.~\eqref{eq:msr_unitary_convention},
\begin{subequations}
\begin{align}
    \widetilde{\hat\rho}
    &=\hat U_{\mathrm{MSR}}^\dagger
      \hat\rho\hat U_{\mathrm{MSR}},
    \label{eq:msr_transformed_state}
    \\
    \widetilde{\hat O}
    &=\hat U_{\mathrm{MSR}}^\dagger
      \hat O\hat U_{\mathrm{MSR}}.
    \label{eq:msr_transformed_observable}
\end{align}
\end{subequations}
The transformed state evolves under $\widetilde{\hat H}^{(2)}$, and the
physical state is reconstructed with
$\hat\rho_{\mathrm{rec}}
=\hat U_{\mathrm{MSR}}\widetilde{\hat\rho}_{\mathrm{eff}}
 \hat U_{\mathrm{MSR}}^\dagger$. Equivalently,
$\langle\hat O\rangle
=\operatorname{Tr}(\widetilde{\hat O}
\widetilde{\hat\rho}_{\mathrm{eff}})$.

The BCH series begins
$\widetilde{\hat O}=\hat O+[\hat O,\hat S_{\mathrm{MSR}}]
+\tfrac12[[\hat O,\hat S_{\mathrm{MSR}}],\hat S_{\mathrm{MSR}}]+\cdots$.
An untransformed laboratory observable can therefore miss a first-order
dressing correction even when the diagonal energy is correct through
second order.

\subsubsection{Example: the dispersive Jaynes--Cummings model}
\label{subsec:msr_dispersive_jc}

For the recurring Jaynes--Cummings benchmark, it is convenient to separate
\begin{subequations}
\begin{align}
    \hat H_{\mathrm c}
    &=\omega_a\left(
        \hat a^\dagger\hat a+\frac{\hat\sigma_z}{2}
      \right),
    \label{eq:msr_jc_commuting_part}
    \\
    \hat H_{\mathrm{int}}
    &=\frac{\Delta}{2}\hat\sigma_z
      +g(\hat a\hat\sigma_+
      +\hat a^\dagger\hat\sigma_-),
    \label{eq:msr_jc_interaction_part}
\end{align}
\end{subequations}
where $\Delta=\omega_0-\omega_a$. Choose
$\hat X_3=\hat\sigma_z/2$,
$\hat X_+=\hat a\hat\sigma_+$, and
$\hat X_-=\hat a^\dagger\hat\sigma_-$. The conserved quantity
$\hat N_1=\hat a^\dagger\hat a+\hat\sigma_z/2$ makes
$\hat H_{\mathrm c}=\omega_a\hat N_1$ inert under the rotation. The
deformed commutator is
\begin{equation*}
    [\hat X_+,\hat X_-]
    =\hat a^\dagger\hat a\hat\sigma_z
     +\hat\sigma_+\hat\sigma_-.
\end{equation*}

The first rotation is generated by
\begin{equation}
    \hat S_{\mathrm{MSR}}
    =\frac{g}{\Delta}
    (\hat a^\dagger\hat\sigma_-
     -\hat a\hat\sigma_+),
    \label{eq:msr_jc_generator}
\end{equation}
and gives
\begin{equation}
\begin{aligned}
    \widetilde{\hat H}_{\mathrm{disp}}
    ={}&
    \omega_a\hat a^\dagger\hat a
    +\frac{\omega_0}{2}\hat\sigma_z
    \\
    &+\frac{g^2}{\Delta}
      (\hat a^\dagger\hat a\hat\sigma_z
       +\hat\sigma_+\hat\sigma_-)\\
    &+\mathcal O\!\left(\frac{g^3}{\Delta^2}\right).
    \label{eq:msr_jc_effective_hamiltonian}
\end{aligned}
\end{equation}
This is the same dispersive Hamiltonian obtained by projection, Magnus,
SW, and AE after their frames and conventions are aligned. MSR makes its
operator origin particularly immediate: it is the polynomial
$[\hat X_+,\hat X_-]$ generated by the first rotation.

For occupations $n\le n_\star$, the local parameter is
$g\sqrt{n_\star+1}/|\Delta|\ll1$. The physical states follow from
$\ket\psi_{\mathrm{phys}}
=\hat U_{\mathrm{MSR}}\ket\psi_{\mathrm{eff}}$ and reproduce the
first-order dressed states in Eqs.~\eqref{eq:sw_jc_dressed_ground} and
\eqref{eq:sw_jc_dressed_excited}.

\subsubsection{Extension to open quantum systems}
\label{subsec:msr_open_systems}

For a time-independent rotation, every jump operator must be transformed
with the state and Hamiltonian. Starting from the dissipator convention in
Eq.~\eqref{eq:global_dissipator_convention}, the exact transformed equation
is
\begin{equation}
\begin{aligned}
    \dot{\widetilde{\hat\rho}}
    ={}&
    -\mathrm i[\widetilde{\hat H},\widetilde{\hat\rho}]
    +\sum_m\gamma_m
    \mathcal D[\widetilde{\hat C}_m]\widetilde{\hat\rho},
    \label{eq:msr_transformed_master}
\end{aligned}
\end{equation}
where
\begin{subequations}
\begin{align}
    \widetilde{\hat H}
    &=\hat U_{\mathrm{MSR}}^\dagger
      \hat H\hat U_{\mathrm{MSR}},
    \\
    \widetilde{\hat C}_m
    &=\hat U_{\mathrm{MSR}}^\dagger
      \hat C_m\hat U_{\mathrm{MSR}}.
    \label{eq:msr_transformed_jumps}
\end{align}
\end{subequations}
No eigenoperator condition on $\hat C_m$ is needed for this exact unitary
change of representation. Keeping the transformed jumps in compact form
preserves GKLS structure. Expanding their dissipators term by term and
truncating does not automatically preserve complete positivity; keeping a
truncated jump inside $\mathcal D[\,cdot\,]$ is manifestly GKLS but partly
resums higher orders.

\subsubsection{Example: a lossy dispersive Jaynes--Cummings model}
\label{subsec:msr_open_jc}

For cavity loss, the generator in Eq.~\eqref{eq:msr_jc_generator}
transforms the annihilation operator as
\begin{align}
    \widetilde{\hat a}
    &=e^{-\hat S_{\mathrm{MSR}}}\hat a e^{\hat S_{\mathrm{MSR}}}
    \nonumber\\
    &=\hat a+\frac{g}{\Delta}\hat\sigma_-
      +\frac{g^2}{2\Delta^2}\hat a\hat\sigma_z
      +\mathcal O\!\left(\frac{g^3}{\Delta^3}\right).
    \label{eq:msr_transformed_a}
\end{align}
The full transformed atom--cavity equation is therefore
\begin{equation}
\begin{aligned}
    \dot{\widetilde{\hat\rho}}
    ={}&
    -\mathrm i[\widetilde{\hat H}_{\mathrm{disp}},
                \widetilde{\hat\rho}]
    \\&+\kappa\mathcal D\!\left[
        \hat a+\frac{g}{\Delta}\hat\sigma_-
        +\frac{g^2}{2\Delta^2}\hat a\hat\sigma_z
      \right]\widetilde{\hat\rho}
    \\
    &+\mathcal O\!\left(\kappa\frac{g^3}{\Delta^3}\right).
    \label{eq:msr_open_jc_transformed}
\end{aligned}
\end{equation}

To see the perturbative content, define
\begin{equation*}
\begin{aligned}
    \mathcal C[\hat A,\hat B]\hat\rho
    ={}&
    \hat A\hat\rho\hat B^\dagger
    +\hat B\hat\rho\hat A^\dagger
    \\
    &-\frac12\{\hat A^\dagger\hat B
                   +\hat B^\dagger\hat A,\hat\rho\}.
\end{aligned}
\end{equation*}
With $\lambda=g/\Delta$,
\begin{equation*}
\begin{aligned}
    \mathcal D[\widetilde{\hat a}]
    ={}&
    \mathcal D[\hat a]
    +\lambda\mathcal C[\hat a,\hat\sigma_-]
    \\
    &+\lambda^2\left[
        \mathcal D[\hat\sigma_-]
        +\frac12\mathcal C[\hat a,\hat a\hat\sigma_z]
      \right]
    +\mathcal O(\lambda^3).
\end{aligned}
\end{equation*}
The linear term describes dissipative dressing and atom--cavity
correlations. At second order, $\mathcal D[\hat\sigma_-]$ is the
large-detuning Purcell channel, while the remaining cross term makes the
cavity loss state dependent. Discarding the cross terms is an additional
projection, secular, or coarse-graining step; it is not part of the small
rotation itself.

If the transformed cavity remains near vacuum and that additional
projection is justified, the reduced atomic equation is
\begin{equation}
\begin{aligned}
    \dot{\hat\rho}_A
    ={}&
    -\mathrm i\left[
        \frac{\omega_0}{2}\hat\sigma_z,
        \hat\rho_A
      \right]
    \\
    &-\mathrm i\left[
        \frac{g^2}{\Delta}\hat\sigma_+\hat\sigma_-,
        \hat\rho_A
      \right]
    \\
    &+\Gamma_{\mathrm P}^{\mathrm{MSR}}
      \mathcal D[\hat\sigma_-]\hat\rho_A,
    \label{eq:msr_atom_only_master}
\end{aligned}
\end{equation}
with $\Gamma_{\mathrm P}^{\mathrm{MSR}}=\kappa g^2/\Delta^2$. This is
the $|\Delta|\gg\kappa$ limit of
$g^2\kappa/(\Delta^2+\kappa^2/4)$. The Hamiltonian rotation alone does
not retain the full complex cavity susceptibility and does not eliminate
the cavity. When $\kappa/|\Delta|$ is appreciable, projection or open-system
AE is the appropriate route to the Lorentzian denominator.

Laboratory observables still require the transformed operator. For example,
$\langle\hat a\rangle_{\mathrm{phys}}
=\operatorname{Tr}(\widetilde{\hat a}\widetilde{\hat\rho})$, and the same
$\widetilde{\hat a}$ determines the physical output field and loss channel.

\subsubsection{Higher orders and practical validation}
\label{subsec:msr_higher_orders}

Higher-order MSR applies successively smaller rotations to remove the
off-diagonal residual left by the preceding step
\cite{Klimov2000,Klimov2002}. Equation~\eqref{eq:msr_third_order_residual}
identifies the first such residual as $[\hat Y,P]$. This recursive picture is
valuable only while the new commutators remain manageable and their matrix
elements remain small in the populated sector.

For a finite invariant sector, one may monitor
\begin{equation*}
    r_{\mathrm{od}}^{[N]}
    =\left\|
      \widetilde{\hat H}^{[N]}
      -\mathcal D_{X_3}[\widetilde{\hat H}^{[N]}]
     \right\|,
\end{equation*}
where $\mathcal D_{X_3}$ deletes matrix elements between different
$\hat X_3$ eigenspaces at fixed $\{\hat N_i\}$. The norm and sector must be
specified. A controlled calculation should show the expected decrease with
rotation order and should also reproduce converged spectra, dynamics, and
reconstructed observables; excitation-cutoff convergence is a separate
test.

The three-photon Rabi resonance in
Sec.~\ref{subsec:higher_order_three_photon} is a useful higher-order
example. Successive commutators reveal how the cubic endpoint operator is
generated, but its observation also requires the second-order diagonal
shifts that locate the dressed resonance. The detailed MSR algebra and its
comparison with the other constructions are given in
Sec.~\ref{sec:higher_order_expansions} and Supplemental Material,
Secs.~S2--S3.

\subsection{The James--Jerke Time-Coarse-Graining Method}
\label{subsec:james_jerke_method}

\begin{figure*}[t]
    \centering
    \includegraphics[width=1.0\linewidth]{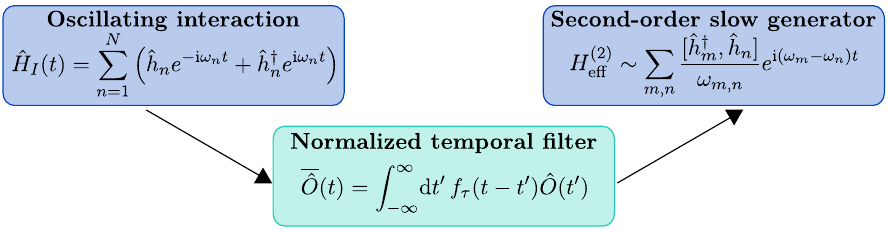}
    \caption{\justifying
    \textbf{Temporal filtering in the James--Jerke construction.} A normalized
    filter $f_\tau$ suppresses fast carriers while retaining selected slow
    components, including near-resonant beat frequencies. Products of the
    discarded carriers generate the second-order coherent Hamiltonian. The
    filtered state need not evolve unitarily, however, and continuous-time
    predictions require the corresponding micromotion or reconstruction
    data.}
    \label{fig:james_jerke_filter}
\end{figure*}

The James--Jerke method begins with the temporal resolution of the
description. Fast carrier oscillations may be invisible on the observation
timescale even though their second-order products generate measurable Stark
shifts, Raman couplings, or conditional interactions. The method applies a
specified temporal filter and asks which carriers and beat notes survive.
Figure~\ref{fig:james_jerke_filter} shows this separation and also the
information that must be restored for a continuous-time prediction.

The construction was introduced for weak, rapidly oscillating interactions
in quantum-optical and trapped-ion problems
\cite{James2000,James2007,Mlmer1999,Srensen1999,Srensen2000}. It is often
described as the approximate method for oscillating Hamiltonians (AMOH); we
use that abbreviation in the higher-order discussion, while "James--Jerke
method" better emphasizes the temporal coarse graining used here.

Time averaging removes temporal resolution, not a tensor factor of the
Hilbert space. A filtered trajectory can therefore be nonunitary even when
the microscopic system is closed, but that information loss is not, by
itself, coupling to a physical reservoir. This distinction becomes
important when the method is extended to density matrices and genuinely
open systems.

\subsubsection{Temporal coarse graining}
\label{subsec:james_temporal_filter}

For an operator $\hat O(t)$, define
\begin{equation}
\begin{aligned}
    \overline{\hat O}(t)
    {}&=\int_{-\infty}^{\infty}\!\mathrm dt'\,
      f_\tau(t-t')\hat O(t'),
    \\
    &\int_{-\infty}^{\infty}\!\mathrm ds\,f_\tau(s)=1.
\end{aligned}
    \label{eq:james_filter_definition}
\end{equation}
Its transfer function
$F_\tau(\nu)=\int\mathrm ds\,f_\tau(s)e^{\mathrm i\nu s}$ gives
$\overline{e^{-\mathrm i\nu t}}
=F_\tau(\nu)e^{-\mathrm i\nu t}$. A useful low-pass window satisfies
\begin{equation*}
    \tau_{\mathrm f}\ll\tau\ll\tau_{\mathrm s},
\end{equation*}
where $\tau_{\mathrm f}$ is the period of the discarded carriers and
$\tau_{\mathrm s}$ is the shortest retained timescale. Without such a
window, the slow--fast distinction is not controlled.

A causal real-time filter has support only for $t'\le t$. Symmetric
Gaussian filters are convenient for retrospective smoothing but should not
be interpreted as causal predictors. Nor can a perfectly sharp cutoff be
simultaneously smooth, positive, and causal. In practice, the passband is
placed between the slow beats and fast carriers; the filter shape then
changes quantitative boundaries but not the commutator structure of the
second-order Hamiltonian \cite{Milonni1995,James2007}.

Filtering and differentiation commute when the convolution and its
derivative converge and the boundary term vanishes. A finite switch-on
requires an explicit envelope or a filter restricted to the available time
interval; otherwise the boundary contribution belongs to the initial
coarse-graining layer.

\subsubsection{Second-order coarse-grained generator}
\label{subsec:james_general_derivation}

Introduce a bookkeeping parameter through
$\hat H_I\mapsto\lambda\hat H_I$ and let
$\mathrm i\partial_t\hat U=\lambda\hat H_I\hat U$ with
$\hat U(t_0,t_0)=\hat I$. If the filtered propagator is invertible on the
relevant state manifold, its formal time-local generator is
\begin{equation}
    \hat H_{\mathrm{cg}}'(t)
    =\lambda\overline{\hat H_I(t)\hat U(t,t_0)}
      \bigl(\overline{\hat U}(t,t_0)\bigr)^{-1}.
    \label{eq:james_raw_generator}
\end{equation}
If $\overline{\hat U}$ develops a kernel, this generator need not exist
uniquely and the filtered map should remain the primary object.

Write
$\hat U=\hat I+\lambda\hat U_1+\lambda^2\hat U_2+\cdots$, where
$\hat U_1(t)=-\mathrm i\int_{t_0}^{t}\hat H_I(t_1)\mathrm dt_1$ is
anti-Hermitian. The two expansions needed through second order are
\begin{equation}
    \lambda\overline{\hat H_I\hat U}
    =\lambda\overline{\hat H_I}
     +\lambda^2\overline{\hat H_I\hat U_1}
     +\mathcal O(\lambda^3),
    \label{eq:james_numerator_expansion}
\end{equation}
and
\begin{equation}
\begin{aligned}
    \bigl(\overline{\hat U}\bigr)^{-1}
    ={}&\hat I-\lambda\overline{\hat U_1}
    +\lambda^2\left[(\overline{\hat U_1})^2
                     -\overline{\hat U_2}\right]
    +\mathcal O(\lambda^3).
    \label{eq:james_filtered_inverse_expansion}
\end{aligned}
\end{equation}

The average of unitary propagators is generally not unitary. The coherent
part is therefore the Hermitian component
\begin{equation}
    \hat H_{\mathrm{eff}}
    =\frac12(\hat H_{\mathrm{cg}}'
             +\hat H_{\mathrm{cg}}'^\dagger),
    \label{eq:james_hermitian_part}
\end{equation}
which becomes
\begin{equation}
\begin{aligned}
    \hat H_{\mathrm{eff}}(t)
    ={}&\lambda\overline{\hat H_I}(t)\\
    &+\frac{\lambda^2}{2}\left(
      \overline{[\hat H_I(t),\hat U_1(t)]}
      -[\overline{\hat H_I}(t),\overline{\hat U_1}(t)]
      \right)\\
    &+\mathcal O(\lambda^3).
    \label{eq:james_general_effective_hamiltonian}
\end{aligned}
\end{equation}
This is the second-order James--Jerke Hamiltonian
\cite{James2007}. It governs the coherent part of the filtered motion; a
separate nonunitary term can remain in the density-matrix equation.

\subsubsection{Harmonic interactions and the initial-time term}
\label{subsec:james_harmonic_interactions}

Let
\begin{equation}
    \hat H_I(t)
    =\sum_{n=1}^{N}\left(
      \hat h_ne^{-\mathrm i\omega_nt}
      +\hat h_n^\dagger e^{\mathrm i\omega_nt}\right), 
    \label{eq:james_harmonic_hamiltonian}
\end{equation}
with $\omega_n>0$. A negative physical detuning is represented by exchanging
$\hat h_n$ and $\hat h_n^\dagger$. In a populated manifold $\mathcal M$,
the local condition is
\begin{equation*}
    \epsilon_{\mathrm J}(\mathcal M)
    =\max_{n,\mu,\nu\in\mathcal M}
      \frac{|\langle\mu|\hat h_n|\nu\rangle|}{\omega_n}
    \ll1,
\end{equation*}
with a finite-energy restriction for unbounded operators.

Integration gives $\hat U_1(t)=\hat V_1(t)-\hat V_1(t_0)$, where
\begin{equation*}
    \hat V_1(t)
    =\sum_n\frac{1}{\omega_n}\left(
      \hat h_ne^{-\mathrm i\omega_nt}
      -\hat h_n^\dagger e^{\mathrm i\omega_nt}\right),
\end{equation*}
and
\begin{equation}
    \dot{\hat V}_1(t)=-\mathrm i\hat H_I(t).
    \label{eq:james_V1_derivative}
\end{equation}
Even when the oscillatory part of $\hat V_1$ is filtered out,
$\overline{\hat U_1}$ still contains $-\hat V_1(t_0)$. This endpoint
term cancels between the two commutators in
Eq.~\eqref{eq:james_general_effective_hamiltonian}; no assumption
$\hat V_1(t_0)=0$ is required. It remains relevant, however, to the
encoding and reconstruction of a finite-time experiment.

The standard low-pass conditions are
\begin{subequations}
\begin{align}
    F_\tau(\omega_n)&\simeq0,
    \label{eq:james_carrier_filter}
    \\
    F_\tau(\omega_m+\omega_n)&\simeq0,
    \label{eq:james_sum_filter}
\end{align}
\end{subequations}
while selected differences $\omega_m-\omega_n$ remain in the passband.
Then
\begin{equation}
\begin{aligned}
    \hat H_{\mathrm{eff}}^{(2)}(t)
    ={}&\sum_{m,n}
    \frac{F_\tau(\omega_m-\omega_n)}{\overline\omega_{mn}}
    [\hat h_m^\dagger,\hat h_n]
    e^{\mathrm i(\omega_m-\omega_n)t},
    \label{eq:james_harmonic_effective}
\end{aligned}
\end{equation}
where
$1/\overline\omega_{mn}
=\tfrac12(1/\omega_m+1/\omega_n)$. The filter factor makes the
selection rule operational: a beat contributes only if it lies inside the
chosen passband.

If every nonzero difference frequency is removed as well, degenerate
harmonics should first be combined and the result reduces to
\begin{equation}
    \hat H_{\mathrm{eff}}^{(2)}
    =\sum_n\frac{1}{\omega_n}
      [\hat h_n^\dagger,\hat h_n].
    \label{eq:james_fully_secular_effective}
\end{equation}
This additional secular approximation is inappropriate when a slow Raman,
exchange, or parametric beat is the process of interest.

\subsubsection{Micromotion and reconstruction}
\label{subsec:james_relation_magnus}

The slow Hamiltonian does not reconstruct the motion discarded by the
filter. Define the exact near-identity transformation
$\hat W(t,t_0)=e^{-\hat V_1(t)}\hat U(t,t_0)e^{\hat V_1(t_0)}$.
Using Eq.~\eqref{eq:james_V1_derivative}, its generator can be separated as
\begin{equation*}
    \hat H_W(t)
    =\frac12[\hat H_I(t),\hat V_1(t)]
     +\hat r_{\mathrm J}^{(3)}(t),
\end{equation*}
where, in a finite-energy manifold,
$\|\hat r_{\mathrm J}^{(3)}\|
=\mathcal O(\|\hat H_I\|^3/\omega_{\mathrm f}^2)$. Filtering the
second-order commutator defines its slow part
$\hat H_{\mathrm{eff}}^{(2)}$ and a suppressed fast remainder.

The leading reconstructed propagator is
\begin{equation}
\begin{aligned}
    \hat U_{\mathrm{rec}}(t,t_0)
    ={}&e^{\hat V_1(t)}
    \\
    &\times
    \mathcal T\exp\left[-\mathrm i\int_{t_0}^{t}\!\mathrm ds\,
      \hat H_{\mathrm{eff}}^{(2)}(s)\right]
    \\
    &\times e^{-\hat V_1(t_0)}.
    \label{eq:james_reconstructed_propagator}
\end{aligned}
\end{equation}
It contains the complete first-order micromotion and the second-order slow
generator, but not every second-order fast correction. Its accuracy must
therefore be stated over a finite time window.

The endpoint factors encode the initial state and reconstruct the physical
state. Equivalently, laboratory observables are represented in the slow
frame by
$\widetilde{\hat O}(t)=e^{-\hat V_1(t)}\hat Oe^{\hat V_1(t)}$. This is the
same leading kick or micromotion structure found in Magnus and
time-dependent SW theory. A different kick gauge redistributes bounded
terms without changing consistently reconstructed predictions.

\subsubsection{Example: the dispersive Jaynes--Cummings model}
\label{subsec:james_dispersive_jc}

For the recurring JC benchmark,
\begin{equation*}
    \hat H_I(t)
    =g\left(
      \hat a\hat\sigma_+e^{\mathrm i\Delta t}
      +\hat a^\dagger\hat\sigma_-e^{-\mathrm i\Delta t}\right),
\end{equation*}
with $\Delta=\omega_0-\omega_a$. For $\Delta>0$, take
$\omega_1=\Delta$ and $\hat h_1=g\hat a^\dagger\hat\sigma_-$; for
$\Delta<0$, interchange $\hat h_1$ and $\hat h_1^\dagger$. The single
carrier gives
\begin{equation*}
    \hat H_{\mathrm{eff}}^{(I)}
    =\frac{g^2}{\Delta}\left(
      \hat a^\dagger\hat a\hat\sigma_z
      +\hat\sigma_+\hat\sigma_-\right),
\end{equation*}
and hence the dispersive Hamiltonian in
Eq.~\eqref{eq:projection_jc_dispersive_schrodinger}.

For $n\le n_\star$, one requires
$g\sqrt{n_\star+1}/|\Delta|\ll1$ and a window
\begin{equation*}
    |\Delta|^{-1}\ll\tau
    \ll\frac{|\Delta|}{g^2(n_\star+1)}.
\end{equation*}
The micromotion generator is
\begin{equation*}
    \hat V_1(t)=\frac{g}{\Delta}\left(
      \hat a^\dagger\hat\sigma_-e^{-\mathrm i\Delta t}
      -\hat a\hat\sigma_+e^{\mathrm i\Delta t}\right).
\end{equation*}
At $t_m-t_0=2\pi m/|\Delta|$, its endpoint values coincide, making
stroboscopic evolution a clean benchmark for the slow Hamiltonian. Between
those times, amplitudes and phases require the reconstruction.

\subsubsection{Coarse-grained density-matrix dynamics}
\label{subsec:james_density_matrix}

For a positive normalized kernel, the filtered state
$\overline{\hat\rho}(t)=\int\mathrm dt'\,
f_\tau(t-t')\hat\rho(t')$ is a convex mixture of density operators and is
therefore physical. Its Hamiltonian need not generate its complete motion.
Following Ref.~\cite{GamelJames2010}, write the filtered map as
$\mathcal E_t=\sum_k\lambda^k\mathcal E_{k,t}$ and its perturbative
inverse as $\mathcal F_t=\mathcal E_t^{-1}$. The first inverse coefficients
are
\begin{subequations}
\begin{align}
    \mathcal F_{0,t}&=\mathcal I,
    \\
    \mathcal F_{1,t}&=-\mathcal E_{1,t},
    \\
    \mathcal F_{2,t}&=-\mathcal E_{2,t}
      +\mathcal E_{1,t}\circ\mathcal E_{1,t}.
\end{align}
\label{eq:james_inverse_map_terms}
\end{subequations}
The time-local generator
$\mathcal L_{\mathrm{cg}}=\dot{\mathcal E}_t\circ\mathcal F_t$ has
\begin{subequations}
\begin{align}
    \mathcal L_0&=0,
    \\
    \mathcal L_1&=\mathrm i\dot{\mathcal E}_1,
    \\
    \mathcal L_2&=\mathrm i\dot{\mathcal E}_2
      -\mathrm i\dot{\mathcal E}_1\circ\mathcal E_1.
\end{align}
\label{eq:james_L_terms}
\end{subequations}
These terms generate both the coherent Hamiltonian and the information-loss
correction.

Under the ideal passband assumptions used in
Ref.~\cite{GamelJames2010}, the latter is
\begin{equation}
\begin{aligned}
    \mathcal N_{\mathrm{cg}}^{(2)}[\hat\rho]
    ={}&\sum_{m,n}d_{mn}e^{\mathrm i(\omega_m-\omega_n)t}
    \\
    \times
    &\Big(
      \{\hat h_m^\dagger\hat h_n,\hat\rho\}
      -2\hat h_n\hat\rho\hat h_m^\dagger
      \\
    &\quad+\{\hat h_n\hat h_m^\dagger,\hat\rho\}
      -2\hat h_m^\dagger\hat\rho\hat h_n
    \Big),
\end{aligned}
    \label{eq:james_coarse_graining_correction}
\end{equation}
where $d_{mn}=\tfrac12(1/\omega_n-1/\omega_m)$. It is trace and
Hermiticity preserving through the retained order. For one carrier,
$d_{nn}=0$, explaining why the dispersive JC example is Hamiltonian at this
order. With several unequal frequencies the correction need not vanish.

This term records information discarded by the temporal filter, not decay
into a physical reservoir. The exact filtered maps are CPTP when the kernel
is positive, normalized, and supported on physical evolution times. Their
time-local generator need not be GKLS unless the filtered family is CP
divisible, and a finite-order truncation need not preserve complete
positivity exactly.

\subsubsection{Genuinely open systems}
\label{subsec:james_open_systems}

For microscopic open dynamics, let the physical jump operators $\hat J_m$
include their rates and define
$\hat K=\tfrac12\sum_m\hat J_m^\dagger\hat J_m$. The nonunitary decaying
frame used in the open-system construction is
\begin{subequations}
\begin{align}
    \hat\rho_K&=e^{\hat Kt}\hat\rho e^{\hat Kt},
    \\
    \hat H_K&=e^{\hat Kt}\hat He^{-\hat Kt},
    \\
    \hat J_{m,K}&=e^{\hat Kt}\hat J_me^{-\hat Kt}.
\end{align}
\label{eq:james_decaying_frame}
\end{subequations}
Here $\hat K$ is the damping operator, not the no-jump Hamiltonian
$\hat H-\mathrm i\hat K$. A compact form results when
$[\hat K,\hat J_m]=-\eta_m\hat J_m/2$, so that
$\hat J_{m,K}=e^{-\eta_mt/2}\hat J_m$; this eigenoperator condition is
restrictive and must be checked model by model.

The open-system construction expands the nonunitary propagator in this
frame, filters the resulting map, and forms its time-local generator when
the inverse exists \cite{Lee2018}. Damping and oscillatory phases then
combine into complex fast denominators whose real and imaginary parts
produce dissipative and coherent corrections.

A positive normalized average of exact microscopic maps is CPTP provided
the filter is supported only at times after preparation. Support before
$t_0$ requires an explicit boundary prescription; a formal backward
propagator is not a physical channel. Even for a CPTP filtered family, its
time-local generator need not be GKLS without CP divisibility, and
finite-order approximations require separate physicality checks.

\subsubsection{Example: a lossy dispersive cavity}
\label{subsec:james_lossy_cavity}

For the lossy JC benchmark of Eq.~\eqref{eq:lossy_jc_common_benchmark}, the
fast cavity coherence has susceptibility
\begin{equation*}
    \chi_c(\Delta)
    =\frac{1}{\kappa/2-\mathrm i\Delta}
    =\frac{\kappa/2+\mathrm i\Delta}{\Delta^2+\kappa^2/4}.
\end{equation*}
After filtering the fast return and projecting the cavity onto its vacuum
reference state, the atomic second-order kernel is
\begin{align}
    \mathcal K_{\mathrm J}^{(2)}[\hat\rho_A]
    ={}&g^2\chi_c[\hat\sigma_-\hat\rho_A,\hat\sigma_+]
       +g^2\chi_c^*[\hat\sigma_-,\hat\rho_A\hat\sigma_+]
    \nonumber\\
    ={}&-\mathrm ig^2\operatorname{Im}\chi_c
       [\hat\sigma_+\hat\sigma_-,\hat\rho_A]
    \nonumber\\
    &+2g^2\operatorname{Re}\chi_c
       \mathcal D[\hat\sigma_-]\hat\rho_A.
    \label{eq:james_open_second_order_kernel}
\end{align}
Thus the reactive part of the cavity response gives the shift and the
absorptive part gives irreversible relaxation:
\begin{subequations}
\begin{align}
    \delta\omega
    &=g^2\operatorname{Im}\chi_c
      =\frac{g^2\Delta}{\Delta^2+\kappa^2/4},
    \label{eq:james_lossy_shift}
    \\
    \Gamma
    &=2g^2\operatorname{Re}\chi_c
      =\frac{g^2\kappa}{\Delta^2+\kappa^2/4}.
    \label{eq:james_lossy_rate}
\end{align}
\end{subequations}

Two operations are present here. Temporal coarse graining isolates the slow
response of the complex cavity pole; projection onto the near-vacuum cavity
manifold produces the atom-only state. Filtering alone does not remove the
cavity Hilbert space. The local parameter is
$g\sqrt{n_\star+1}/\sqrt{\Delta^2+\kappa^2/4}\ll1$, and the induced
shift and rate must remain slow relative to the same complex scale.

Simply adding the bare cavity dissipator to the closed-system James
Hamiltonian is insufficient when $\kappa/|\Delta|$ is appreciable: it keeps
$1/\Delta$ and misses both the Lorentzian denominator and the associated
dissipative response. Projection or open-system AE supplies the additional
Hilbert-space reduction.

\subsubsection{Higher orders and limitations}
\label{subsec:james_higher_orders}

The standard James--Jerke Hamiltonian is second order. Higher-order AMOH
formulas have been proposed for processes whose first nonzero coupling
appears later \cite{Shao2017,Rosado2023}, but their symmetrization,
finite-time terms, and Hermiticity depend on the filtering assumptions.
They should therefore be checked against a perturbative expansion of the
filtered propagator or map and against an independent construction.

The three-photon Rabi resonance of
Sec.~\ref{subsec:higher_order_three_photon} is a demanding example: three
fast interactions generate the cubic slow coupling, while second-order
diagonal terms shift the resonance by more than the cubic gap. A consistent
higher-order treatment must retain both effects and reconstruct the bounded
motion discarded by the filter. Increasing the formal order does not repair
large matrix-element-to-frequency ratios, a missing coarse-graining window,
or a carrier that belongs in the retained resonant manifold.

\subsubsection{Practical validation}
\label{subsec:james_validation}

A James--Jerke model should be tested as a filtered and reconstructed
prediction. The carrier and sum frequencies must lie outside the passband,
the desired beats inside it, and the result should be stable under reasonable
changes of filter width and shape. Stroboscopic agreement tests the slow
Hamiltonian; continuous-time agreement tests the micromotion. One should
also verify the endpoint terms, populated-manifold scaling, physicality of
truncated maps, and convergence against exact or numerically resolved
dynamics. Agreement with another second-order method is useful but not
independent evidence when both calculations use the same weak-coupling and
frequency-separation assumptions.


\subsection{Floquet Theory and Floquet Engineering}
\label{subsec:floquet_method}

Floquet theory applies when the Hamiltonian is periodic,
\begin{equation*}
    \hat H(t+T)=\hat H(t),
    \qquad
    \omega=\frac{2\pi}{T}.
\end{equation*}
It separates the evolution into a time-independent quasienergy problem and
periodic motion within each cycle \cite{Floquet1883}. Floquet engineering
uses this structure as a design tool: a periodic control can produce
effective couplings, gauge fields, band structures, and symmetries that are
absent from the undriven Hamiltonian
\cite{Bukov2015,Bukov2017Floquet,Eckardt2017,Floquet_handbook}.

\begin{figure*}[t]
    \centering
    \includegraphics[width=\linewidth]{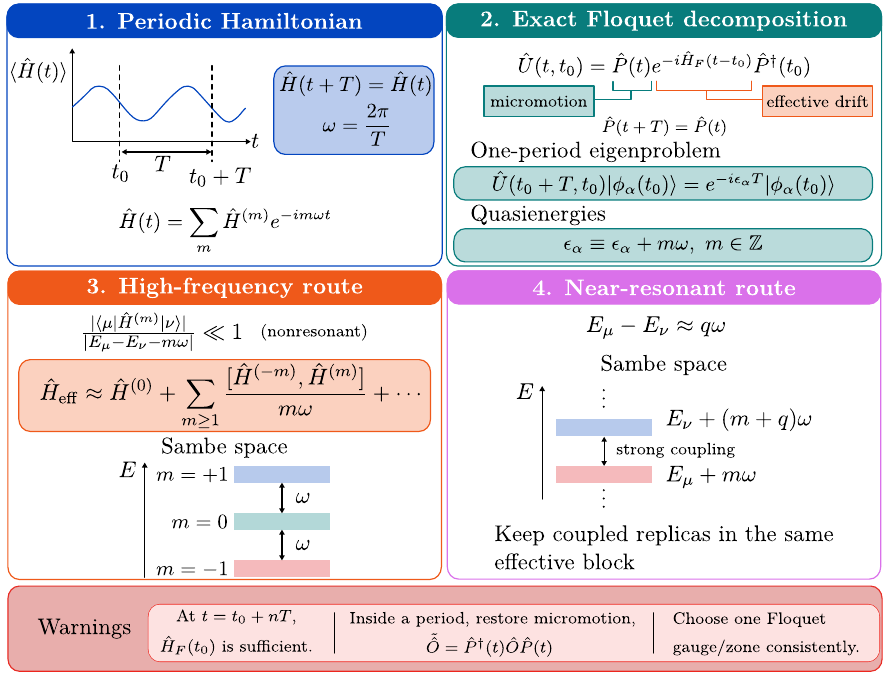}
    \caption{\justifying
    \textbf{From a periodic Hamiltonian to a physical Floquet prediction.} The
    one-period propagator $\hat U(t_0+T,t_0)$ defines quasienergies modulo
    $\omega=2\pi/T$ and, after a logarithm branch is chosen, a
    phase-referenced Floquet Hamiltonian. The exact evolution separates into
    time-independent quasienergy dynamics and periodic micromotion. At
    $t=t_0+nT$, the phase-referenced Hamiltonian gives the complete
    propagation; between those times, both endpoint micromotion factors are
    required. A high-frequency expansion approximates the nonresonant branch
    of this construction. Near a quasienergy resonance, the nearly degenerate
    Fourier replicas must instead be retained in the same effective block.}
    \label{fig:floquet_method_workflow}
\end{figure*}

This flexibility comes with two cautions. First, the effective Hamiltonian
does not by itself describe measurements made between integer numbers of
periods; the micromotion must also be restored. Second, a high-frequency
expansion is only one way of approximating Floquet dynamics. Near a
multiphoton resonance, the relevant Fourier sectors must instead be retained
and treated together. Figure~\ref{fig:floquet_method_workflow} summarizes
these two routes. Applications range from artificial gauge fields and driven
band structures to tunable qubit couplings and dynamical operating points
\cite{Li2018,Mikami2016,Li2023,Akbari2025,Deng2016,Sameti2019,
Huang2021,Gandon2022,GarcaMata2024,Nguyen2024,Chessari2025}.

\subsubsection{Exact Floquet decomposition}
\label{subsec:floquet_exact_decomposition}

Let $\hat U(t,t_0)$ be the propagator of a finite-dimensional system, or of a
converged finite-dimensional truncation. Floquet theory gives the exact
factorization
\begin{equation}
    \begin{aligned}
   &\hat U(t,t_0)
    =\hat P(t)e^{-\mathrm i\hat H_{\mathrm F}(t-t_0)}
      \hat P^\dagger(t_0),\\
    &\hat P(t+T)=\hat P(t),
    \label{eq:floquet_exact_factorization} 
\end{aligned}
\end{equation}
where $\hat P(t)$ is unitary and periodic. It describes the micromotion,
while the time-independent $\hat H_{\mathrm F}$ generates the quasienergy
dynamics.

The same evolution can be described by a stroboscopic Hamiltonian tied to a
chosen initial phase $t_0$,
\begin{equation}
\begin{aligned}
    &\hat U(t_0+T,t_0)=e^{-\mathrm iT\hat H_{\mathrm F}(t_0)},
   \\
   &\hat H_{\mathrm F}(t_0)
    =\hat P(t_0)\hat H_{\mathrm F}\hat P^\dagger(t_0).
    \label{eq:floquet_stroboscopic_hamiltonian}
\end{aligned}
\end{equation}
Consequently,
$\hat U(t_0+nT,t_0)=e^{-\mathrm inT\hat H_{\mathrm F}(t_0)}$.
This compact form is exact at stroboscopic times, but it does not determine
the motion inside a period without $\hat P(t)$.

The logarithm in Eq.~\eqref{eq:floquet_stroboscopic_hamiltonian} is
multivalued. Changing its branch shifts a quasienergy by an integer multiple
of $\omega$. A periodic unitary may also be redistributed between
$\hat P(t)$ and $\hat H_{\mathrm F}$; this is the Floquet-gauge freedom.
Neither ambiguity changes the exact propagator, but effective Hamiltonians,
states, observables, and micromotion must be expressed in the same gauge.

\subsubsection{Floquet modes and quasienergy gauge}
\label{subsec:floquet_modes}

Solutions of the Schr\"odinger equation can be chosen in the form
\begin{align*}
    &\ket{\psi_\alpha(t)}
    =e^{-\mathrm i\epsilon_\alpha t}\ket{\Phi_\alpha(t)}, \\
    &\ket{\Phi_\alpha(t+T)}=\ket{\Phi_\alpha(t)}.
\end{align*}
The periodic modes solve
\begin{equation}
    \left(\hat H(t)-\mathrm i\frac{\partial}{\partial t}\right)
    \ket{\Phi_\alpha(t)}
    =\epsilon_\alpha\ket{\Phi_\alpha(t)}.
    \label{eq:floquet_quasienergy_equation}
\end{equation}
A single Floquet-mode density operator is periodic, because its quasienergy
phase cancels. A superposition of modes is generally not periodic: relative
phases continue to accumulate from cycle to cycle.

For any integer $q$,
\begin{subequations}
\begin{align}
    \epsilon_{\alpha,q}&=\epsilon_\alpha+q\omega,
    \\
    \ket{\Phi_{\alpha,q}(t)}
    &=e^{\mathrm iq\omega t}\ket{\Phi_\alpha(t)}
\end{align}
\label{eq:floquet_quasienergy_gauge}
\end{subequations}
represent the same physical solution. Quasienergies are therefore defined
modulo $\omega$. Choosing one interval of width $\omega$---a Floquet zone---
is analogous to choosing one Brillouin zone for crystal momentum. Spectral
comparisons must first place all quasienergies in the same zone and gauge.

\subsubsection{The Sambe-space quasienergy operator}
\label{subsec:floquet_sambe_space}

The periodic eigenvalue problem becomes time independent after a Fourier
expansion. We use
\begin{subequations}
\begin{align}
    \hat H(t)&=\sum_{m\in\mathbb Z}e^{-\mathrm im\omega t}\hat H^{(m)},
    \label{eq:floquet_hamiltonian_fourier}
    \\
    \ket{\Phi_\alpha(t)}
    &=\sum_{n\in\mathbb Z}e^{-\mathrm in\omega t}
      \ket{\phi_\alpha^{(n)}}.
    \label{eq:floquet_mode_fourier}
\end{align}
\end{subequations}
Here $\hat H^{(-m)}=\hat H^{(m)\dagger}$. Substitution in
Eq.~\eqref{eq:floquet_quasienergy_equation} gives
\begin{equation*}
    \sum_n\left[\hat H^{(m-n)}-m\omega\delta_{mn}\right]
    \ket{\phi_\alpha^{(n)}}
    =\epsilon_\alpha\ket{\phi_\alpha^{(m)}}.
\end{equation*}

This matrix acts in Sambe space
$\mathscr H_{\mathrm S}=\mathscr H\otimes\mathscr T$, where
$\mathscr T$ is the space of $T$-periodic functions \cite{Sambe1973}. In
compact form, the quasienergy operator has blocks
\begin{equation*}
    \hat{\mathcal Q}_{mn}
    =\hat H^{(m-n)}-m\omega\delta_{mn}.
\end{equation*}
Each Fourier index supplies a shifted copy of the physical spectrum. We
refer to such a copy as a \emph{Floquet replica}; it is a bookkeeping sector
of the extended space, not an additional physical Hilbert space. The
Fourier component $\hat H^{(k)}$ couples replicas whose indices differ by
$k$.

The Sambe operator and the physical Floquet Hamiltonian play different
roles. $\hat{\mathcal Q}$ contains every replica and is infinite
dimensional, whereas $\hat H_{\mathrm F}$ acts on one physical copy after
the replica structure has been block diagonalized. Likewise, the $m=0$
sector alone is not a complete Floquet state. The physical mode is rebuilt
from all Fourier components through
\begin{equation*}
    \ket{\Phi_\alpha(t)}
    =\sum_m e^{-\mathrm im\omega t}\ket{\phi_\alpha^{(m)}}.
\end{equation*}

Numerically one retains $-M\leq m\leq M$. Convergence must be checked as
$M$ is increased, and a state with appreciable weight at $m=\pm M$ should
not be regarded as converged.

\subsubsection{High-frequency block diagonalization}
\label{subsec:floquet_high_frequency}

When no relevant replicas are resonant, their separation is set mainly by
integer multiples of $\omega$. They can then be block diagonalized in
powers of the Fourier amplitudes divided by the drive frequency
\cite{Bukov2015,Mikami2016,Eckardt2017}. In the phase-independent van Vleck
gauge, the micromotion generator has zero period average. Through the first
inverse-frequency correction,
\begin{equation}
    \hat H_{\mathrm{eff}}
    =\hat H^{(0)}
     +\sum_{m=1}^{\infty}
      \frac{[\hat H^{(-m)},\hat H^{(m)}]}{m\omega}
     +\mathcal O(\omega^{-2}).
    \label{eq:floquet_van_vleck_expansion}
\end{equation}
The symbol $\mathcal O(\omega^{-2})$ presumes that the relevant Fourier
matrix elements and static energy differences are held fixed. For
unbounded or many-body Hamiltonians, the statement must instead be made in
a populated finite-energy manifold or in terms of local interaction scales.

The leading anti-Hermitian micromotion generator is
\begin{align*}
    &\hat S_{\mathrm F}^{(1)}(t)
    =\sum_{m\neq0}\frac{e^{-\mathrm im\omega t}\hat H^{(m)}}{m\omega},\\ 
    &\hat P(t)=e^{\hat S_{\mathrm F}^{(1)}(t)+\mathcal O(\omega^{-2})}.
\end{align*}
It is $\hat S_{\mathrm F}$, not the unitary $\hat P(t)$, whose average is
set to zero. The reconstructed propagator is
\begin{equation*}
    \hat U_{\mathrm{rec}}(t,t_0)
    =\hat P(t)e^{-\mathrm i\hat H_{\mathrm{eff}}(t-t_0)}
      \hat P^\dagger(t_0).
\end{equation*}
Thus the initial state is encoded with $\hat P^\dagger(t_0)$ and a
laboratory observable is represented by
$\widetilde{\hat O}(t)=\hat P^\dagger(t)\hat O\hat P(t)$.

A useful first diagnostic in a populated manifold $\mathcal M$ is that
$|\langle\mu|\hat H^{(m)}|\nu\rangle|/(|m|\omega)\ll1$. The
resonance-aware condition is stronger. If
$\hat H^{(0)}\ket\mu=E_\mu\ket\mu$, eliminating a transition generated by
$\hat H^{(m)}$ requires
\begin{equation}
    \frac{|\langle\mu|\hat H^{(m)}|\nu\rangle|}
    {|E_\mu-E_\nu-m\omega|}\ll1.
    \label{eq:floquet_sambe_detuning_parameter}
\end{equation}
A small denominator identifies a resonant pair of replicas. Those states
belong in the retained block; increasing the order of a nondegenerate
high-frequency expansion will not repair the partition.

In interacting many-body systems the total bandwidth commonly grows with
system size, so it is not the useful expansion scale. Local high-frequency
expansions can nevertheless describe a long prethermal interval before
drive-induced heating becomes important, even when the infinite series is
only asymptotic.

\subsubsection{Resonant and degenerate Floquet perturbation theory}
\label{subsec:floquet_degenerate_perturbation}

Suppose $E_\alpha-E_\beta\simeq q\omega$. The Sambe states associated with
$\ket\alpha$ in one replica and $\ket\beta$ in the replica shifted by $q$
are then nearly degenerate. They must be diagonalized together, while only
well-detuned replicas are eliminated.

To make the power counting explicit, write
\begin{align*}
    &\hat H^{(0)}=\hat H_{\mathrm{ref}}+\lambda\hat V^{(0)},\\
    &\hat H^{(k)}=\lambda\hat V^{(k)}\quad(k\neq0),
\end{align*}
and decompose the Sambe operator as
$\hat{\mathcal Q}=\hat{\mathcal Q}_0+\lambda\hat{\mathcal V}$, with
\begin{align*}
    &(\hat{\mathcal Q}_0)_{mn}
    =(\hat H_{\mathrm{ref}}-m\omega)\delta_{mn}, \\
    &\hat{\mathcal V}_{mn}=\hat V^{(m-n)}.
\end{align*}
The common choice $\hat H_{\mathrm{ref}}=\hat H^{(0)}$ sets
$\hat V^{(0)}=0$; another rotating-frame partition may be better when a
strong resonant term should be solved exactly.

Let $D$ contain every nearly degenerate Sambe state and let
$\hat{\mathbb P}_D$ project onto this set. If
$\mathcal E_i$ are the eigenvalues of $\hat{\mathcal Q}_0$ and
$\mathcal V_{ij}=\langle\!\langle i|\hat{\mathcal V}|j\rangle\!\rangle$,
Hermitian quasi-degenerate perturbation theory gives, through second order,
\begin{align*}
    \langle\!\langle i|\hat{\mathcal Q}_{\mathrm{eff}}|j\rangle\!\rangle
    &=\mathcal E_i\delta_{ij}+\lambda\mathcal V_{ij} \\
    &\quad+\frac{\lambda^2}{2}\sum_{r\notin D}
      \frac{\mathcal V_{ir}\mathcal V_{rj}}
           {\mathcal E_i-\mathcal E_r}\\
    &\quad+\frac{\lambda^2}{2}\sum_{r\notin D}
      \frac{\mathcal V_{ir}\mathcal V_{rj}}
           {\mathcal E_j-\mathcal E_r}\\
    &\quad+\mathcal O(\lambda^3).
\end{align*}
The denominators are quasienergy differences in Sambe space. The
symmetrized form keeps the retained matrix Hermitian.

The reduced eigenvector must also be reconstructed. To first order, the
wave operator is
\begin{equation*}
    \hat{\mathcal W}
    =\hat{\mathbb P}_D
     +\lambda\sum_{\substack{r\notin D\\i\in D}}
      \frac{\ket{r}\!\rangle\!\rangle\mathcal V_{ri}
      \langle\!\langle i|}{\mathcal E_i-\mathcal E_r}
     +\mathcal O(\lambda^2).
\end{equation*}
For an effective vector $\ket{\!\ket{\phi_{\mathrm{eff}}}}$, the physical
periodic mode is
\begin{equation*}
    \ket{\Phi_{\mathrm{rec}}(t)}
    =\sum_m e^{-\mathrm im\omega t}
      (\hat I\otimes\bra m)\hat{\mathcal W}
      \ket{\!\ket{\phi_{\mathrm{eff}}}}.
\end{equation*}
This restores off-resonant harmonics and leakage amplitudes. A unitary
block-diagonalizing transformation provides an equivalent reconstruction in
a different block gauge; mixing its effective Hamiltonian with the wave
operator from another gauge is inconsistent.

High-order implementations of this construction can extract multiphoton
couplings, drive-induced shifts, leakage, and reconstruction operators
\cite{Huang2026,codigo_python_huang}. Their predictions still require
convergence with perturbative order and comparison with direct time-dependent
dynamics when that comparison is available.

\subsubsection{Example: the dispersive Jaynes--Cummings model}
\label{subsec:floquet_dispersive_jc}

The interaction-picture JC Hamiltonian,
\begin{equation*}
    \hat H_I(t)=g\left(
      \hat a\hat\sigma_+e^{\mathrm i\Delta t}
      +\hat a^\dagger\hat\sigma_-e^{-\mathrm i\Delta t}\right),
\end{equation*}
with $\Delta=\omega_0-\omega_a$, is periodic even though the laboratory Hamiltonian is static. For
$\Delta>0$ and the Fourier convention above,
\begin{subequations}
\begin{align}
    \hat H^{(-1)}&=g\hat a\hat\sigma_+,
    \\
    \hat H^{(1)}&=g\hat a^\dagger\hat\sigma_-,
    \\
    \hat H^{(0)}&=0.
\end{align}
\label{eq:floquet_jc_fourier_components}
\end{subequations}
For negative detuning the two nonzero components are interchanged, while
the final result retains the signed $\Delta$.

Equation~\eqref{eq:floquet_van_vleck_expansion} gives
\begin{equation*}
    \hat H_{\mathrm{eff}}^{(I)}
    =\frac{g^2}{\Delta}[\hat a\hat\sigma_+,
      \hat a^\dagger\hat\sigma_-]
    =\frac{g^2}{\Delta}\left(
      \hat a^\dagger\hat a\hat\sigma_z
      +\hat\sigma_+\hat\sigma_-\right).
\end{equation*}
Returning to the Schr\"odinger picture reproduces the dispersive
Hamiltonian in Eq.~\eqref{eq:projection_jc_dispersive_schrodinger}. The
leading micromotion is
\begin{equation}
    \hat S_{\mathrm F}^{(1)}(t)
    =\frac{g}{\Delta}\left(
      \hat a^\dagger\hat\sigma_-e^{-\mathrm i\Delta t}
      -\hat a\hat\sigma_+e^{\mathrm i\Delta t}\right).
    \label{eq:floquet_jc_micromotion}
\end{equation}
It restores the virtual exchange between $\ket{e,n}$ and
$\ket{g,n+1}$ inside each period. At times separated by
$2\pi/|\Delta|$, the endpoint micromotion agrees and the effective
Hamiltonian alone gives the stroboscopic motion. For occupations
$n\leq n_\star$, the local parameter is
$g\sqrt{n_\star+1}/|\Delta|\ll1$.

\subsubsection{Periodic open-system dynamics}
\label{subsec:floquet_open_systems}

Now let a periodic time-local generator obey
\begin{equation*}
    \dot{\hat\rho}=\mathcal L(t)\hat\rho,
    \qquad \mathcal L(t+T)=\mathcal L(t).
\end{equation*}
If $\mathcal L(t)$ is of GKLS form at every time, its propagator
$\Phi(t,t_0)$ is CPTP. The exact stroboscopic object is the one-period map
$\Phi_T(t_0)=\Phi(t_0+T,t_0)$.

In finite-dimensional Liouville space, this map is invertible as a linear
fundamental solution, so a branch of its matrix logarithm defines
\begin{equation*}
    \mathcal L_{\mathrm F}(t_0)=\frac{1}{T}\log\Phi_T(t_0),
    \qquad
    \Phi_T(t_0)=e^{T\mathcal L_{\mathrm F}(t_0)}.
\end{equation*}
The logarithm is branch dependent, and it need not be a GKLS generator.
This is the embeddability question: can the one-period CPTP map be generated
by one time-independent Markovian semigroup? Floquet theory guarantees the
map, not an affirmative answer to that additional question.

One may write the exact linear decomposition
\begin{align*}
    &\Phi(t,t_0)
    =\mathcal P(t)e^{(t-t_0)\mathcal L_{\mathrm{eff}}}
      \mathcal P^{-1}(t_0),\\
    &\mathcal P(t+T)=\mathcal P(t).
\end{align*}
For compatible logarithm branches,
$\mathcal L_{\mathrm F}(t_0)=
\mathcal P(t_0)\mathcal L_{\mathrm{eff}}\mathcal P^{-1}(t_0)$.
Unlike unitary micromotion in a closed system, the periodic similarity maps
and their inverses need not be completely positive. Only the complete exact
propagator is guaranteed to be physical. In an infinite-dimensional bosonic
space, logarithms and inverses require additional domain assumptions or a
converged numerical truncation.

With
$\mathcal L(t)=\sum_m e^{-\mathrm im\omega t}\mathcal L^{(m)}$, an
operator-space Floquet mode
$\hat\rho_\alpha(t)=e^{\mu_\alpha t}\hat R_\alpha(t)$ obeys
$(\mathcal L-\partial_t)\hat R_\alpha=\mu_\alpha\hat R_\alpha$.
The Sambe--Liouville blocks are therefore
\begin{equation}
    (\mathbb L_{\mathrm S})_{mn}
    =\mathcal L^{(m-n)}
     +\mathrm im\omega\delta_{mn}\mathcal I.
    \label{eq:floquet_sambe_liouville_matrix}
\end{equation}
The $+\mathrm im\omega$ offset follows from
$-\partial_t e^{-\mathrm im\omega t}=+\mathrm im\omega
e^{-\mathrm im\omega t}$; it differs from the Hamiltonian Sambe offset.

The direct-frame high-frequency expansion begins
\begin{equation}
    \mathcal L_{\mathrm{eff}}
    =\mathcal L^{(0)}
     +\sum_{m=1}^{\infty}
      \frac{[\mathcal L^{(m)},\mathcal L^{(-m)}]}
      {\mathrm im\omega}
     +\mathcal O(\omega^{-2}).
    \label{eq:floquet_liouvillian_high_frequency}
\end{equation}
It is not automatically of GKLS form: commutators of GKLS generators need
not be GKLS generators. A rotating-frame reorganization can produce a
physical Floquet Lindbladian in particular models, but that is a
model-dependent result rather than a consequence of periodicity alone
\cite{Schnell2020}.

\subsubsection{Example: a lossy dispersive cavity}
\label{subsec:floquet_open_jc}

For the interaction-picture lossy JC benchmark, the Fourier components are
\begin{subequations}
\begin{align}
    \mathcal L^{(0)}\hat\rho
    &=\kappa\mathcal D[\hat a]\hat\rho,
    \\
    \mathcal L^{(-1)}\hat\rho
    &=-\mathrm ig[\hat a\hat\sigma_+,\hat\rho],
    \\
    \mathcal L^{(1)}\hat\rho
    &=-\mathrm ig[\hat a^\dagger\hat\sigma_-,\hat\rho].
\end{align}
\label{eq:floquet_open_jc_fourier}
\end{subequations}
The first correction in
Eq.~\eqref{eq:floquet_liouvillian_high_frequency} is the commutator with
the dispersive interaction-picture Hamiltonian obtained above. At this
order the equation still acts on the full atom--cavity state; Floquet block
diagonalization has not eliminated the cavity.

Micromotion also dresses the loss channel. Equation
\eqref{eq:floquet_jc_micromotion} gives
\begin{equation*}
    \widetilde{\hat a}(t)
    =\hat a+\frac{g}{\Delta}e^{-\mathrm i\Delta t}\hat\sigma_-
     +\frac{g^2}{2\Delta^2}\hat a\hat\sigma_z
     +\mathcal O(g^3/\Delta^3).
\end{equation*}
The linear cross term oscillates within a period, while the quadratic
contribution contains the leading atomic loss channel. If the transformed
cavity remains near its vacuum and is subsequently projected out, the
large-detuning atom-only result is
\begin{subequations}
\begin{align}
    \delta\omega_{\mathrm F}&=\frac{g^2}{\Delta},
    \\
    \Gamma_{\mathrm F}&=\kappa\frac{g^2}{\Delta^2}.
\end{align}
\label{eq:floquet_large_detuning_shift_rate}
\end{subequations}
This second step uses a slow-manifold assumption that is not supplied by
Floquet theory itself.

Projection or open-system adiabatic elimination retains the full complex
cavity susceptibility and gives
\begin{subequations}
\begin{align}
    \delta\omega
    &=\frac{g^2\Delta}{\Delta^2+\kappa^2/4},
    \\
    \Gamma
    &=\frac{g^2\kappa}{\Delta^2+\kappa^2/4}.
\end{align}
\label{eq:floquet_exact_cavity_susceptibility_result}
\end{subequations}
The Floquet expressions in
Eq.~\eqref{eq:floquet_large_detuning_shift_rate} are the leading terms of
these formulas for $\kappa/|\Delta|\ll1$; a simple inverse-frequency
truncation does not resum the linewidth into the Lorentzian denominator.
Sambe--Liouville resolvent and continued-fraction methods can retain more
multiphoton sectors for monochromatic open systems \cite{Keliri2026}, but
the existence of such a solution still does not settle whether its
physical-space logarithm is GKLS.

\subsubsection{Practical reconstruction and validation}
\label{subsec:floquet_validation}

A Floquet approximation should be tested at three levels. First, compare
the exact or converged one-period propagator with
$e^{-\mathrm iT\hat H_{\mathrm F}(t_0)}$, after folding quasienergies into
the same Floquet zone. Second, increase the Sambe cutoff and check that the
retained modes carry negligible weight at its boundary. Third, compare both
stroboscopic observables and continuous-time observables reconstructed with
the micromotion. Near resonance, the retained Sambe block must be enlarged;
in many-body systems, the comparison must also stay inside the prethermal
time window.

A useful eigenvector diagnostic is the Sambe residual
\begin{equation*}
    r_{\mathrm S}
    =\left\|(\hat{\mathcal Q}-\epsilon_\alpha)
      \ket{\!\ket{\Phi_\alpha^{[M]}}}\right\|,
\end{equation*}
evaluated away from the artificial Fourier boundary. For open systems, the
one-period CPTP maps should be compared directly, and any claimed
time-independent generator should be checked separately for GKLS form. The
dynamical residual in Eq.~\eqref{eq:validation_dynamical_residual} then tests
the reconstructed physical prediction rather than only its quasienergies.

Floquet theory therefore adds two ideas that are easy to lose in a static
effective Hamiltonian: quasienergy is defined modulo the drive frequency,
and micromotion carries physical information between periods. The
three-photon example in Sec.~\ref{subsec:higher_order_three_photon} combines
both lessons: a resonant pair of replicas must be retained, a higher-order
coupling opens the avoided crossing, and lower-order diagonal shifts decide
where that crossing occurs.

\section{Higher-Order Effective Expansions}
\label{sec:higher_order_expansions}
\label{higher orders}

Higher order can enter an effective prediction in three different ways. It
can refine the slow spectrum, refine the transformation used to encode and
reconstruct physical quantities, or generate the first nonzero coupling
inside a resonant manifold. These improvements need not occur at the same
order. In the dispersive Jaynes--Cummings (JC) model, for example, energy
corrections contain even powers of $g$, whereas the rotation that dresses the
states contains odd powers. Near a three-photon resonance of the quantum Rabi
model, a third-order matrix element is the leading retained interaction, but
second-order energy shifts decide where that interaction becomes resonant.

Perturbative order should also be distinguished from operator degree and
excitation transfer. A cubic operator may be present in the microscopic
Hamiltonian at first order, or it may be generated by three insertions of a
linear interaction. Conversely, a second-order calculation can produce a
two-body operator. With $\hbar=1$, an $r$-step contribution to a Hamiltonian
has the typical frequency scale
$g^r/(\Omega_1\cdots\Omega_{r-1})$, while an $r$th-order state amplitude
contains one more energy denominator. An order statement is meaningful only
after the microscopic coupling, retained manifold, and denominator counting
have been specified.

For a JC manifold populated up to $n_\star$, the local expansion parameter is
\begin{equation*}
    \eta_\star=\frac{g\sqrt{n_\star+1}}{|\Delta|}\ll1.
\end{equation*}
In circuit-QED notation,
$n_{\mathrm{crit}}=\Delta^2/(4g^2)$ and
$\eta_\star^2=(n_\star+1)/(4n_{\mathrm{crit}})$ \cite{Blais2021}.
Increasing the order cannot compensate for a resonant denominator or for an
occupied manifold in which $\eta_\star$ is no longer small.

Order and observation time are likewise connected. Writing
$\epsilon=g/\Delta$, the second-order dispersive scale
$\Delta\epsilon^2$ defines
$t_{\mathrm{disp}}\sim1/(|\Delta|\epsilon^2)=|\Delta|/g^2$.
A fourth-order correction $\Delta\epsilon^4$ accumulates an
$O(\epsilon^2)$ phase over this natural slow time. A spectrally small term
can therefore be necessary for an accurate long-time prediction.

\subsection{Exact Jaynes--Cummings Benchmark}
\label{subsec:higher_order_exact_jc}

We begin with the detuning-frame Hamiltonian
\begin{equation*}
    \hat H_{\mathrm{JC}}
    =\frac{\Delta}{2}\hat\sigma_z
     +g(\hat a\hat\sigma_++\hat a^\dagger\hat\sigma_-).
\end{equation*}
Let $\hat n=\hat a^\dagger\hat a$,
$\hat P_e=\hat\sigma_+\hat\sigma_-$, and
$\hat P_g=\hat\sigma_-\hat\sigma_+$. In the invariant doublet
$\mathcal M_n=\operatorname{span}\{\ket{e,n},\ket{g,n+1}\}$,
\begin{align*}
    &\hat H_n=
    \begin{pmatrix}
        \Delta/2 & g\sqrt{n+1}\\
        g\sqrt{n+1} & -\Delta/2
    \end{pmatrix},
    \\
    &E_{\pm,n}=\pm\frac12\sqrt{\Delta^2+4g^2(n+1)}.
\end{align*}
For $\Delta>0$, expansion of the exact eigenvalues gives
\begin{align*}
    E_{\pm,n}&=\pm\Bigg[
      \frac{\Delta}{2}+\frac{g^2(n+1)}{\Delta}
      -\frac{g^4(n+1)^2}{\Delta^3}
      \\
      &\qquad+\frac{2g^6(n+1)^3}{\Delta^5}
      +\mathcal O\!\left(\frac{g^8(n+1)^4}{\Delta^7}\right)
    \Bigg].
\end{align*}
For negative detuning, the bare-state assignment of the two branches is
reversed; the signed-$\Delta$ operator expansion remains valid.

The block-diagonal Hamiltonian through fourth order is
\begin{equation}
\begin{aligned}
    \hat H_{\mathrm{bd}}^{[4]}
    ={}&\frac{\Delta}{2}\hat\sigma_z
      +\frac{g^2}{\Delta}(\hat n\hat\sigma_z+\hat P_e)\\
      &-\frac{g^4}{\Delta^3}
      \left[\hat n^2\hat\sigma_z+(2\hat n+1)\hat P_e\right].
    \label{eq:higher_order_block_diagonal_H4}
\end{aligned}
\end{equation}
It acts correctly on every JC doublet and on the uncoupled state
$\ket{g,0}$. The fourth-order term contains a qubit-dependent Kerr-like
curvature because
$\hat n^2=\hat n+(\hat a^\dagger)^2\hat a^2$. It should not be read as an
independent cavity Kerr nonlinearity: the number curvature and the atomic
projector are parts of the same dressed JC spectrum.

The exact mixing angle satisfies
\begin{align*}
    &\tan(2\theta_n)=\frac{2g\sqrt{n+1}}{\Delta},
    \\
    &\theta_n=\epsilon_n-\frac43\epsilon_n^3
      +\frac{16}{5}\epsilon_n^5+\mathcal O(\epsilon_n^7),
\end{align*}
where $\epsilon_n=g\sqrt{n+1}/\Delta$. Thus a fourth-order energy does
not imply a fourth-order state reconstruction. Starting from $\ket{e,n}$,
the exact population difference is
\begin{align*}
    \langle\hat\sigma_z(t)\rangle_{\mathrm{exact}}
    ={}&1-\frac{8g^2(n+1)}{\Delta^2+4g^2(n+1)} \\
    &\times
      \sin^2\!\left[\frac{t}{2}
      \sqrt{\Delta^2+4g^2(n+1)}\right].
\end{align*}
The phase tests the diagonal energy, while the small oscillation amplitude
tests the dressing. A diagonal generator acting on an untransformed bare
state reproduces only the first of these effects.

\subsection{Higher-Order Block Diagonalization and Reconstruction}
\label{subsec:higher_order_msr}

The exact benchmark can be recovered algebraically with successive small
rotations \cite{Klimov2000,Klimov2002,Klimov2003}. Define
\begin{equation*}
    \hat X_3=\frac{\hat\sigma_z}{2},\qquad
    \hat X_+=\hat a\hat\sigma_+,\qquad
    \hat X_-=\hat a^\dagger\hat\sigma_-,
\end{equation*}
and $\hat Y=\hat X_+-\hat X_-$, $\hat X=\hat X_++\hat X_-$. The
first anti-Hermitian generator is
\begin{equation*}
    \hat A_1=\frac{g}{\Delta}\hat Y,
    \qquad
    \widetilde{\hat H}=e^{\hat A}\hat H e^{-\hat A}.
\end{equation*}
Since
$[\hat A_1,\Delta\hat X_3]=-g\hat X$, the first BCH commutator cancels
the microscopic coupling.

Introduce the diagonal operators and the first residual structure
\begin{equation*}
\begin{aligned}
    \hat{\mathcal P}&=[\hat X_+,\hat X_-]
      =\hat n\hat\sigma_z+\hat P_e,\\
    \hat{\mathcal B}&=(\hat n+1)\hat X_+
      +\hat X_-(\hat n+1),\\
    \hat{\mathcal Q}&=\hat n^2\hat\sigma_z
      +(2\hat n+1)\hat P_e.
\end{aligned}
\end{equation*}
Their closed commutator chain is
\begin{subequations}
\begin{align}
    [\hat Y,\hat{\mathcal P}]&=-2\hat{\mathcal B},\\
    [\hat Y,\hat{\mathcal B}]&=2\hat{\mathcal Q}.
\end{align}
\label{eq:higher_order_Y_relations}
\end{subequations}
Consequently,
\begin{subequations}
\begin{align}
    [\hat A_1,g\hat X]
    &=\frac{2g^2}{\Delta}\hat{\mathcal P},
    \label{eq:higher_order_first_nested_commutator}\\
    [\hat A_1,[\hat A_1,g\hat X]]
    &=-\frac{4g^3}{\Delta^2}\hat{\mathcal B},
    \label{eq:higher_order_second_nested_commutator}\\
    [\hat A_1,[\hat A_1,[\hat A_1,g\hat X]]]
    &=-\frac{8g^4}{\Delta^3}\hat{\mathcal Q}.
    \label{eq:higher_order_third_nested_commutator}
\end{align}
\end{subequations}
The BCH coefficients $1/2$, $1/3$, and $1/8$ then give
\begin{equation}
    \widetilde{\hat H}_{\mathrm{1rot}}^{[4]}
    =\frac{\Delta}{2}\hat\sigma_z
      +\frac{g^2}{\Delta}\hat{\mathcal P}
      -\frac{4g^3}{3\Delta^2}\hat{\mathcal B}-\frac{g^4}{\Delta^3}\hat{\mathcal Q}
      +\mathcal O(g^5/\Delta^4).
    \label{eq:higher_order_msr_once_rotated}
\end{equation}

The cubic term is not a diagonal energy correction. It is the residual
intersector coupling left after the first rotation. It may be kept in this
normal form, or removed with
\begin{equation*}
    \hat A_3=-\frac{4g^3}{3\Delta^3}
      \left[(\hat n+1)\hat X_+-\hat X_-(\hat n+1)\right],
\end{equation*}
for which
$[\hat A_3,\Delta\hat\sigma_z/2]
=4g^3\hat{\mathcal B}/(3\Delta^2)$. In each JC doublet,
$\hat A_1$ and $\hat A_3$ are proportional to the same block generator,
so they commute for this benchmark. The second rotation yields
\begin{align*}
    \widetilde{\hat H}_{\mathrm{bd}}^{[4]}
    ={}&\frac{\Delta}{2}\hat\sigma_z
      +\frac{g^2}{\Delta}\hat{\mathcal P}
      -\frac{g^4}{\Delta^3}\hat{\mathcal Q}
      \\
      &+\mathcal O_{\mathrm{od}}(g^5/\Delta^4)
      +\mathcal O_{\mathrm d}(g^6/\Delta^5).
\end{align*}
The first remainder is off diagonal; the second is the next spectral
correction. Their different orders are consistent with the odd expansion of
the mixing angle and the even expansion of the eigenenergies.

With the sequential convention above, the transformed state is encoded by
$e^{\hat A_3}e^{\hat A_1}$ and reconstructed by its inverse:
\begin{equation*}
\begin{aligned}
    \widetilde{\hat\rho}^{[4]}(0)
    &=e^{\hat A_3}e^{\hat A_1}\hat\rho(0)
      e^{-\hat A_1}e^{-\hat A_3},\\
    \hat\rho_{\mathrm{rec}}^{[4]}(t)
    &=e^{-\hat A_1}e^{-\hat A_3}
      \widetilde{\hat\rho}_{\mathrm{eff}}^{[4]}(t)
      e^{\hat A_3}e^{\hat A_1}.
\end{aligned}
\end{equation*}
Observables are transformed with the same forward map. Keeping the cubic
residual or eliminating it with $\hat A_3$ defines two normal-form gauges;
they give the same physical prediction only when their endpoint maps are
used consistently.

\subsection{Higher-Order Time-Coarse-Grained Dynamics}
\label{subsec:higher_order_amoh}

At higher order, a James--Jerke or average-Hamiltonian construction depends
on the filter, retained frequency sectors, inverse filtered map, and kick
gauge \cite{James2007,Shao2017,Rosado2023}. For the single-harmonic JC
interaction, a Hermitian low-pass gauge can be aligned with the block
diagonalization above. Its fourth-order slow Hamiltonian is
\begin{equation*}
    \hat H_{\mathrm{slow},I}^{[4]}
    =\frac{g^2}{\Delta}\hat{\mathcal P}
     -\frac{g^4}{\Delta^3}\hat{\mathcal Q}
     +\mathcal O(g^6/\Delta^5).
\end{equation*}
There is no third-order diagonal term: each harmonic interaction changes
the Fourier index by $\pm1$, and an odd number of steps cannot return to the
zero-frequency sector. Returning from the interaction picture restores
$\Delta\hat\sigma_z/2$.

In a kick gauge matched to the two rotations,
\begin{equation*}
\begin{aligned}
    &\hat P_{\mathrm{AMOH}}^{[4]}(t)
    =e^{\hat K_1(t)}e^{\hat K_3(t)},\\
    &\hat K_1(t)
    =\frac{g}{\Delta}\left(
      e^{-\mathrm i\Delta t}\hat X_-
      -e^{\mathrm i\Delta t}\hat X_+\right),\\
    &\hat K_3(t)
    =\frac{4g^3}{3\Delta^3}
      e^{\mathrm i\Delta t}(\hat n+1)\hat X_+
    \\
    &\hspace{2.2cm}
      -\frac{4g^3}{3\Delta^3}
      e^{-\mathrm i\Delta t}\hat X_-(\hat n+1).
\end{aligned}
\end{equation*}
At $t=0$, $\hat K_1=-\hat A_1$ and $\hat K_3=-\hat A_3$. The reconstructed
interaction-picture propagator is
\begin{equation*}
    \hat U_{\mathrm{rec}}^{(I),[4]}(t,0)
    =\hat P_{\mathrm{AMOH}}^{[4]}(t)
      e^{-\mathrm i\hat H_{\mathrm{slow},I}^{[4]}t}
      \hat P_{\mathrm{AMOH}}^{[4]\dagger}(0).
\end{equation*}
The diagonal middle factor refines the accumulated phase; the endpoint
kicks restore the virtual population oscillation. Different kick gauges may
redistribute bounded terms, but not the complete prediction through the
controlled order.

\subsection{Order-Consistent Dynamical Accuracy}
\label{subsec:higher_order_fair_comparison}

A fair comparison uses the full prediction map
\begin{equation*}
    \hat\rho_{\mathrm{rec}}^{[N]}(t)
    =\left[\mathcal R_t^{[N]}\circ
      \Phi_{\mathrm{eff}}^{[N]}(t,0)\circ
      \mathcal E_0^{[N]}\right][\hat\rho(0)],
\end{equation*}
with encoding, evolution, reconstruction, and observables truncated in the
same convention. For pure states, useful finite-window metrics are
\begin{subequations}
\begin{align}
    &\epsilon_F^{\max}
    =\max_{0\leq t\leq t_{\max}}[1-F^{[N]}(t)],\\
    &\overline{\epsilon}_F
    =\frac{1}{t_{\max}}\int_0^{t_{\max}}\!\mathrm dt\,
      [1-F^{[N]}(t)],
\end{align}
\label{eq:higher_order_fidelity_metrics}
\end{subequations}
where
$F^{[N]}=|\langle\psi_{\mathrm{exact}}|
\psi_{\mathrm{rec}}^{[N]}\rangle|^2$. For mixed states, the trace distance
$D^{[N]}=\|\hat\rho_{\mathrm{exact}}-
\hat\rho_{\mathrm{rec}}^{[N]}\|_1/2$ bounds the error of a bounded
laboratory observable by $2\|\hat O\|D^{[N]}$.

The exact JC solution predicts
\begin{subequations}
\begin{align}
    \delta E^{[2]}&=O(|\Delta|\eta_\star^4),&
    \delta E^{[4]}&=O(|\Delta|\eta_\star^6),\\
    \delta\theta^{[1]}&=O(\eta_\star^3),&
    \delta\theta^{[3]}&=O(\eta_\star^5).
\end{align}
\label{eq:higher_order_expected_scalings}
\end{subequations}
On $t_{\mathrm{disp}}\sim1/(|\Delta|\eta_\star^2)$, the omitted
fourth-order energy gives an $O(\eta_\star^2)$ phase error, whereas the
omitted sixth-order term gives $O(\eta_\star^4)$. Spectral accuracy,
dressing accuracy, and comparison time should therefore be reported
separately. Convergence with perturbative order and convergence with the
oscillator cutoff are also independent tests.

\subsection{Three-Photon Resonance as a Leading Third-Order Interaction}
\label{subsec:higher_order_three_photon}
\label{subsec:higher_order_resonant_processes}

Higher-order terms do not always refine an existing lower-order coupling.
Near a multiphoton degeneracy, a term first generated at third or higher
order can be the leading interaction in the retained manifold
\cite{ma2015three,Garziano2015Multiphoton,Huang2026}. Consider the quantum
Rabi Hamiltonian
\begin{equation*}
    \hat H_{\mathrm R}
    =\omega_a\hat a^\dagger\hat a
     +\frac{\omega_0}{2}\hat\sigma_z
     +g(\hat a+\hat a^\dagger)(\hat\sigma_++\hat\sigma_-),
\end{equation*}
with $\Delta=\omega_0-\omega_a$ and $\Sigma=\omega_0+\omega_a$.
The same-parity states
$\ket A=\ket{e,0}$ and $\ket B=\ket{g,3}$ approach degeneracy near
$\omega_0=3\omega_a$. Their shortest connecting sequence is
\begin{equation*}
    \ket{e,0}\xrightarrow{g}\ket{g,1}
    \xrightarrow{\sqrt2g}\ket{e,2}
    \xrightarrow{\sqrt3g}\ket{g,3}.
\end{equation*}
The middle step uses $\hat a^\dagger\hat\sigma_+$ and is removed by the
ordinary rotating-wave approximation.

The retained projector is
$\hat P_3=\ket A\!\bra A+\ket B\!\bra B$. The endpoint shifts are
\begin{subequations}
\begin{align}
    \delta E_{e,0}^{(2)}&=\frac{g^2}{\Delta},\\
    \delta E_{g,3}^{(2)}&=-\frac{3g^2}{\Delta}-\frac{4g^2}{\Sigma}.
\end{align}
\label{eq:higher_order_three_photon_shifts}
\end{subequations}
The $1/\Sigma$ contribution comes from $\ket{e,4}$. Although this state is
not on the shortest transfer path, it shifts the $\ket B$ endpoint and must
be included when locating the resonance.

In the retained space,
\begin{equation}
    \hat H_{\mathrm{eff},3}
    =\bar E_3\hat I_3+\frac{\delta_3}{2}\hat\tau_z
     +G_3\hat\tau_++G_3^*\hat\tau_-,
    \label{eq:higher_order_three_photon_normal_form}
\end{equation}
where
\begin{equation}
    \delta_3=\omega_0-3\omega_a
      +4g^2\left(\frac{1}{\Delta}+\frac{1}{\Sigma}\right)
      +\mathcal O(g^4/\omega_a^3),
    \label{eq:higher_order_three_photon_detuning}
\end{equation}
and
\begin{equation}
    G_3=-\frac{\sqrt6\,g^3}{2\omega_a\Delta}
      +\mathcal O(g^5/\omega_a^4).
    \label{eq:higher_order_three_photon_coupling}
\end{equation}
Equivalently, the retained operator is
$-g^3[(\hat a^\dagger)^3\hat\sigma_-+
\hat a^3\hat\sigma_+]/(2\omega_a\Delta)$. The factor $\sqrt6$ in $G_3$
is the matrix element between $\ket0$ and $\ket3$, not part of the operator
coefficient.

Solving $\delta_3=0$ gives
\begin{equation}
    \omega_a^{\mathrm{res}}
    =\frac{\omega_0}{3}+\frac{3g^2}{\omega_0}
      +\mathcal O(g^4/\omega_0^3).
    \label{eq:higher_order_three_photon_resonance}
\end{equation}
At the leading crossing,
$|G_3|=\sqrt6g^3/(4\omega_a^2)=9\sqrt6g^3/(4\omega_0^2)$,
the avoided-crossing gap is $2|G_3|$, and the resonant population period is
$T_{\mathrm{pop}}^{(3\gamma)}=\pi/|G_3|$. Starting from $\ket A$,
\begin{equation*}
    P_{g,3}^{\mathrm{eff}}(t)
    =\frac{4|G_3|^2}{\delta_3^2+4|G_3|^2}
      \sin^2\!\left[\frac{t}{2}
      \sqrt{\delta_3^2+4|G_3|^2}\right].
\end{equation*}
Large conversion requires $|\delta_3|\lesssim|G_3|$. Since the resonance
shift is $O(g^2/\omega_0)$ but the target gap is
$O(g^3/\omega_0^2)$, imposing the bare condition leaves a detuning larger
than the interaction being sought. Lower-order diagonal shifts must be used
before the higher-order coupling is tested.

\begin{figure*}[!ht]
    \centering
\includegraphics[width=\linewidth]{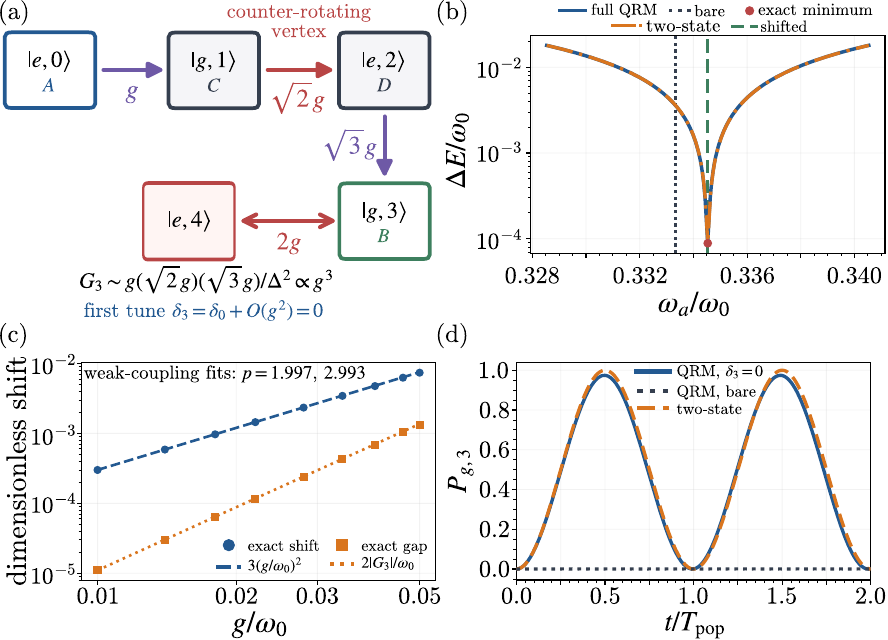}
    \caption{\justifying
    \textbf{Three-photon resonance as a resonance-promoted third-order interaction.}
    (a) The retained states $\ket{e,0}$ and $\ket{g,3}$ are connected through
    $\ket{g,1}$ and $\ket{e,2}$; the counter-rotating step
    $\hat a^\dagger\hat\sigma_+$ is essential. The additional
    $\ket{g,3}\leftrightarrow\ket{e,4}$ branch shifts the endpoint without
    belonging to the shortest transfer path. (b) Exact selected-doublet and
    two-state gaps distinguish the bare condition, the perturbatively shifted
    condition $\delta_3=0$, and the exact avoided-crossing minimum. The exact
    doublet is tracked by overlap with the retained states. (c) The resonance
    displacement is quadratic in $g$, whereas the minimum gap is cubic; fits
    to the four weakest couplings give powers $1.997$ and $2.993$.
    (d) At the bare condition the $O(g^2)$ detuning suppresses conversion. At
    the analytical shifted condition, the full Rabi dynamics and the
    two-state prediction agree over the first transfer cycles without fitting
    the exact resonance. Panels (b) and (d) use $\omega_0=1$,
    $g/\omega_0=0.02$, and $N_{\mathrm{cav}}=16$; time is measured in
    $T_{\mathrm{pop}}^{(3\gamma)}=\pi/|G_3|$. Reconstruction and preparation
    tests are given in Supplemental Material, Sec.~S3, and the calculation is
    reproduced in Notebook N05.}
    \label{fig:three_photon_mechanism_validation}
\end{figure*}

Figure~\ref{fig:three_photon_mechanism_validation} checks this ordering at
the level of pathway, spectrum, scaling, and dynamics. The shifted condition
used in panel (d) is obtained analytically from
Eq.~\eqref{eq:higher_order_three_photon_detuning}; the exact avoided-crossing
minimum is used only as an independent benchmark.

The retained amplitudes alone do not give bare laboratory populations. To
leading order, the reconstructed endpoints are
\begin{subequations}
\begin{align}
    \ket{A}_{\mathrm{rec}}
    ={}&\left(1-\frac{g^2}{2\Delta^2}\right)\ket{e,0}
      +\frac{g}{\Delta}\ket{g,1}\nonumber\\
    &+\mathcal O[(g/\omega_a)^2]_Q,\\
    \ket{B}_{\mathrm{rec}}
    ={}&\left[
      \begin{aligned}
      1&-\frac{3g^2}{2\Delta^2}\\[-2pt]
       &-\frac{2g^2}{\Sigma^2}
      \end{aligned}
      \right]\ket{g,3}\nonumber\\
    &-\frac{\sqrt3g}{\Delta}\ket{e,2}
      -\frac{2g}{\Sigma}\ket{e,4}\nonumber\\
    &+\mathcal O[(g/\omega_a)^2]_Q.
\end{align}
\label{eq:higher_order_three_photon_reconstruction}
\end{subequations}
For a retained state $c_A\ket A+c_B\ket B$, the leading eliminated-sector
population is
\begin{equation*}
    P_Q=|c_A|^2\frac{g^2}{\Delta^2}
      +|c_B|^2\left(\frac{3g^2}{\Delta^2}
      +\frac{4g^2}{\Sigma^2}\right)
      +\mathcal O[(g/\omega_a)^3].
\end{equation*}
These populations are the structured dressing of the two endpoints, not
automatically uncontrolled leakage. A bare preparation also contains a fast
component relative to this dressed slow manifold and therefore shows initial
slip.

The first omitted resonance shift is $O(g^4/\omega_a^3)$, whereas the first
omitted coupling correction is $O(g^5/\omega_a^4)$. Since the conversion
time is $O(\omega_a^2/g^3)$, the resonance error can produce visible phase
drift before the relative correction to the minimum gap. The second-order
shift is sufficient to reveal the cubic process and its first conversion
cycles; using an exactly calibrated resonance extends the time window but
should be reported as an additional calibration, not hidden inside the
perturbative prediction.

\subsection{Scope and Validation of Higher-Order Expansions}
\label{subsec:higher_order_validation}

A higher-order calculation should answer three separate questions. Does it
reproduce the spectrum and dressed states in the populated manifold? Has the
transformation removed the intended coupling with the predicted power of the
small parameter? After encoding and reconstruction, does it reproduce the
physical dynamics over the stated time window? Passing one test does not
guarantee the others.

\subsubsection{Spectral and state benchmarks}

The error powers in Eq.~\eqref{eq:higher_order_expected_scalings} should be
tested over $0\leq n\leq n_\star$, not only in the vacuum doublet. Near a
selected resonance, the intended eigenpair should be tracked by symmetry and
overlap with the retained states rather than by sorted energy index. This is
essential at avoided crossings, where energy ordering exchanges the physical
character of the two states.

\subsubsection{Did the block transformation work?}

For the nonresonant JC benchmark, define
\begin{equation*}
    \mathcal D_0[\hat O]
    =\hat P_e\hat O\hat P_e+\hat P_g\hat O\hat P_g.
\end{equation*}
This keeps terms diagonal in the atomic detuning sectors; it is not a
projection onto the excitation doublets. The unwanted coupling is measured by
\begin{equation*}
    r_{\mathrm{od}}^{[N]}
    =\left\|\widetilde{\hat H}^{[N]}
      -\mathcal D_0[\widetilde{\hat H}^{[N]}]\right\|_{\mathcal M},
\end{equation*}
with the norm restricted to the populated finite-energy manifold. After the
first rotation the cubic residual scales as
$g^3(n_\star+1)^{3/2}/\Delta^2$; after the cubic rotation the leading
off-diagonal remainder scales as
$g^5(n_\star+1)^{5/2}/\Delta^4$. A log--log fit provides a direct numerical
test of the cancellation order.

\subsubsection{Does the complete prediction satisfy the original dynamics?}

The reconstructed trajectory can be tested without diagonalizing the full
model by substituting it into the microscopic equation:
\begin{equation*}
    r_{\mathrm{dyn}}^{[N]}(t)
    =\left\|\dot{\hat\rho}_{\mathrm{rec}}^{[N]}(t)
      +\mathrm i[\hat H,\hat\rho_{\mathrm{rec}}^{[N]}(t)]\right\|_1.
\end{equation*}
When an exact reference is available, this residual should be complemented by
observable errors and maximum or time-averaged infidelity. Stroboscopic
agreement alone is insufficient for an observable measured continuously,
because kick errors can disappear at selected sampling times.

\subsubsection{A practical validation sequence}

In practice: specify $n_\star$ and the local control parameter; inspect every
denominator before expanding; benchmark both energies and dressing; measure
the remaining off-diagonal block; reconstruct the initial state and
observable; test the natural slow time and the onset of phase drift; and vary
the Hilbert-space cutoff independently of perturbative order. For a resonant
process, report the analytical resonance, any spectroscopic or numerical
calibration, and the associated gap separately.

\subsubsection{Scope of the different constructions}

Projection, Schrieffer--Wolff, small rotations, and time-coarse-grained
methods agree on the physical spectrum in their common regime once frames and
gauges are aligned. They store the same fast information in different places:
a wave operator, a unitary rotation, a residual coupling, or a kick.
Higher-order adiabatic elimination additionally requires a chosen slow
embedding and, in open systems, consistent effective jumps
\cite{Brion2007,Azouit2017}. Magnus truncations remain unitary for closed
systems, although their nested integrals grow rapidly \cite{ture2024simple}.
Periodic van Vleck, Floquet--Magnus, Brillouin--Wigner, and Sambe reductions
distribute drive-phase dependence differently between the generator and
micromotion \cite{Eckardt2015,Mikami2016,Bukov2017Floquet}.

Higher order is not guaranteed to improve a result outside the asymptotic
regime. A worse fourth-order curve may reveal a large populated-manifold
parameter, a small denominator, inconsistent reconstruction, an insufficient
cutoff, or an observation time beyond the perturbative window. The useful
question is then which assumption failed. The order of the generator alone
cannot answer it; the encoding, reconstruction, resonance condition,
populated manifold, and comparison time belong to the same prediction.

\section{From Effective Descriptions to Physical Insight, Discovery, and Validation}
\label{sec:methods_comparison_merged}
\label{sec:methods_comparison}

The preceding sections developed several ways to construct effective models.
We now turn to the questions that arise when those calculations are used:
which method makes the relevant physics easiest to see, which parts of an
answer depend on representation, how can a desired interaction be recognized
before a long calculation, and what evidence is needed before the result can
be trusted?

The comparison is made at the level of the reconstructed prediction
\begin{equation*}
    \hat\rho_{\mathrm{rec}}(t)
    =\left[\mathcal R_t\circ\Phi_{\mathrm{eff}}(t,0)
      \circ\mathcal E_0\right][\hat\rho(0)],
\end{equation*}
not at the level of an isolated Hamiltonian. Two order-$N$ descriptions are
operationally equivalent on a declared state, parameter range, and time
window when
\begin{equation}
\begin{aligned}
&
\Bigl\|
\bigl(
\mathcal{R}'_{t}\circ\Phi'_{\mathrm{eff}}(t,0)\circ\mathcal{E}'_{0}
-
\mathcal{R}_{t}\circ\Phi_{\mathrm{eff}}(t,0)\circ\mathcal{E}_{0}
\bigr)
[\hat{\rho}]
\Bigr\|
\\
&=
\mathcal{O}\!\left(
\epsilon^{N+1}
\right).
\end{aligned}
\label{eq:...}
\end{equation}
using a state or map norm, or a stated family of observables, appropriate to
the claim. Their generators may differ by a block rotation, a micromotion
gauge, or a change of retained coordinates. Conversely, identical-looking
generators can predict different laboratory outcomes when paired with
inconsistent preparations, observables, jump operators, eliminated-variable
reconstructions, or micromotion.

\begin{tcolorbox}[
    plainlanguage,
    title={Plain-language guide: what an effective method reorganizes}
]
An effective method does more than drop small terms. It chooses a description
in which the slow or relevant physics is easier to see. Projection organizes
virtual excursions outside a retained space; Schrieffer--Wolff organizes
dressed degrees of freedom; Magnus and James--Jerke organize temporal
harmonics; Floquet theory organizes quasienergy replicas; and adiabatic
elimination organizes slow and fast variables. Reconstruction translates the
answer back to the laboratory.
\end{tcolorbox}

\subsection{What each method reorganizes}
\label{subsec:comparison_framework}

Table~\ref{tab:compact_method_lenses} summarizes the questions for which each
method is especially transparent. It is a guide to choosing a starting point,
not a ranking of methods.

\begin{table*}[t]
    \centering
    \caption{\justifying
    Complementary method lenses. The last column combines the endpoint
    information needed for a laboratory prediction with the first structural
    failure to inspect.}
    \label{tab:compact_method_lenses}
    \footnotesize
    \renewcommand{\arraystretch}{1.13}
    \setlength{\tabcolsep}{3.5pt}
    \begin{tabularx}{\linewidth}{
        @{}
        >{\RaggedRight\arraybackslash}p{0.18\linewidth}
        >{\RaggedRight\arraybackslash}p{0.36\linewidth}
        >{\RaggedRight\arraybackslash}X
        @{}
    }
        \toprule
        \textbf{Method} & \textbf{What it shows most directly}
        & \textbf{Reconstruction and first failure}\\
        \midrule
        Projection/\allowbreak resolvent
        & Virtual paths, denominators, self-energies, and intruder states
        & Restore eliminated amplitudes and the model-space metric when
        required; inspect small denominators and leakage.\\
        Schrieffer--Wolff
        & Dressed states, induced operators, and residual block coupling
        & Transform states, controls, observables, and jumps; inspect the block
        gap and near-identity condition.\\
        Small rotations
        & Commutator ancestry, algebraic selection rules, and nonlinear terms
        & Undo rotations in reverse order; inspect algebraic proliferation and
        occupation-enhanced matrix elements.\\
        Magnus
        & Time ordering, coherent accumulation, and pulse symmetry
        & Restore kick terms; inspect convergence of the integrated interaction
        and long-time accumulation.\\
        James--Jerke
        & Slow beats, phase matching, bandwidth, and boundary terms
        & Restore filter and endpoint corrections; inspect whether one filter
        can separate the target beat from nearby frequencies.\\
        Floquet/van Vleck
        & Quasienergy replicas, sideband resonances, and drive phase
        & Restore micromotion in the same gauge; inspect omitted resonant
        replicas and the heating or prethermal time window.\\
        Adiabatic elimination
        & Slaving, complex susceptibility, induced loss, initial slip, and memory
        & Reconstruct fast variables and the initial layer; inspect closure of
        the fast spectral gap and overlap of timescales.\\
        \bottomrule
    \end{tabularx}
\end{table*}

Complementary methods are useful when they answer different questions.
Projection and Schrieffer--Wolff form a natural pair: one displays virtual
routes and denominators, while the other supplies a dressed representation
for states, controls, observables, and environmental couplings. Magnus or
James--Jerke exposes temporal phase matching, whereas Floquet theory tests
the same process in the quasienergy spectrum and reconstructs its periodic
motion. Adiabatic elimination is especially transparent for dynamical
following; Liouvillian projection can reveal the spectral origin of the same
coherent and dissipative terms. Repeating one truncation in different
notation is not independent triangulation.

\subsection{Cross-method agreement and representation dependence}
\label{subsec:comparison_triangulation}

Agreement can be tested at three levels. A \emph{spectral} comparison uses
energy or quasienergy splittings, poles, and spectral projectors. A
\emph{dynamical} comparison uses reconstructed states, reduced maps, and
transition amplitudes over a stated interval. An \emph{operational}
comparison uses measured observables, gate or channel actions, emitted
fields, and decay rates. A correct dressed spectrum need not give a correct
bare-state trajectory, and a correct stroboscopic map does not determine the
motion inside a period.

Even coefficients inside an effective Hamiltonian can depend on the retained
coordinates. If
$\hat\rho_{\mathrm{eff}}'=\hat W(t)\hat\rho_{\mathrm{eff}}
\hat W^\dagger(t)$, then
\begin{equation}
    \hat H_{\mathrm{eff}}'(t)
    =\hat W(t)\hat H_{\mathrm{eff}}(t)\hat W^\dagger(t)
     -\mathrm i\hat W(t)\dot{\hat W}^\dagger(t),
    \label{eq:comparison_retained_unitary}
\end{equation}
while the endpoint maps must change as
\begin{equation}
    \mathcal E_0'=\mathcal W_0\circ\mathcal E_0,
    \qquad
    \mathcal R_t'=\mathcal R_t\circ\mathcal W_t^{-1}.
    \label{eq:comparison_encoding_reconstruction_gauge}
\end{equation}
The complete prediction is unchanged even though terms have moved between
the generator and the retained representation. The same bookkeeping appears
in Floquet gauges, projected-space metrics, and invertible Liouvillian
similarities. Electromagnetic gauge is more delicate after finite-level
truncation because the truncation need not be gauge covariant; we return to
this point in Sec.~\ref{sec:advanced_effective_dynamics}.

\subsubsection{Three-photon resonance as a compact cross-method example}
\label{subsec:comparison_three_photon_case_study}

The $\ket{e,0}\leftrightarrow\ket{g,3}$ resonance illustrates why these
levels must be separated. Every consistent construction gives the two-state
normal form in Eq.~\eqref{eq:higher_order_three_photon_normal_form} after the
second-order endpoint shifts are included. The invariant tests are the
shifted resonance, the gap $2|G_3|$, the projectors onto reconstructed
endpoint states, the conversion dynamics, and population outside the
retained pair. The phase of $G_3$ and the placement of higher-order terms
inside the two-state Hamiltonian depend on convention.

Projection displays the connecting paths and endpoint self-energies;
Schrieffer--Wolff supplies a unitary dressed basis; small rotations identify
the commutator chain; time-coarse-grained methods show how three fast
interaction steps produce a slow term; adiabatic elimination must iterate the
slaving relation to third order; and Floquet--Sambe theory places the
commensurability in the quasienergy spectrum. Agreement on the cubic
coefficient is only the beginning: the methods must also agree on the shifted
resonance and reconstructed physical prediction. The detailed comparison is
given in Supplemental Material, Sec.~S3.

\subsection{From physical intuition to interaction discovery}
\label{sec:effective_dynamics_discovery}

Effective theory is often taught as a forward calculation from a microscopic
Hamiltonian. Interaction engineering also poses an inverse problem:
\begin{align*}
    &\text{observation or target}
    \longrightarrow \text{slow structure} 
    \longrightarrow \\
    &\text{operator and paths}
    \longrightarrow \text{physical prediction}.
\end{align*}
An avoided crossing, state-dependent shift, linewidth, collective response,
or desired transition constrains the effective model but does not uniquely
identify its microscopic origin. A candidate mechanism must still pass tests
of symmetry, perturbative scaling, resonance placement, path interference,
open-system completion, and reconstruction.

Experiments provide useful examples of this inverse reasoning:

\begin{itemize}
    \item A photon-number-dependent qubit shift and measurement-induced
    dephasing suggest
    $\chi\hat a^\dagger\hat a\hat\sigma_z$ together with dressed measurement
    and dephasing channels \cite{Schuster2005}. The derivation must connect
    the spectral shift to the virtual path and connect the linewidth to
    cavity fluctuations and loss.

    \item Coherent exchange between separated qubits through an almost empty
    cavity suggests
    $J(\hat\sigma_+^{(1)}\hat\sigma_-^{(2)}+\mathrm{H.c.})$
    \cite{Majer2007CavityBus}. The same calculation must identify the virtual
    cavity states and accompanying Stark and Purcell terms.

    \item Hybridization of one resonator photon with two artificial-atom
    excitations suggests
    $G(\hat a\hat\sigma_+^{(1)}\hat\sigma_+^{(2)}+\mathrm{H.c.})$
    \cite{Tomonaga2025TwoArtificialAtoms}. The observed gap constrains the
    coherent sum of counter-rotating, longitudinal, and direct spin--spin
    paths allowed by symmetry.

    \item Hybridization of $\ket{2,0,g}$ and $\ket{0,1,g}$ suggests
    $G(\hat a_1^2\hat a_2^\dagger+\mathrm{H.c.})$
    \cite{Wang2025OneTwoPhoton}. The derivation must show how a virtually
    excited artificial atom mediates the process and how Lamb shifts move the
    one-to-two-photon resonance.

    \item Photon-number-selective loss of four excitations suggests an
    effective four-photon jump produced by six-wave mixing
    \cite{Vanselow2026Quartets}. Explaining the observation requires the
    auxiliary loss, residual number-dependent Hamiltonian terms, and the
    restricted manifold over which an $\hat a^4$-type jump is accurate.
\end{itemize}

The following sequence turns such clues into a calculation.

\paragraph{1. State the physical claim.}
Name the initial and final states, observable, response pole, decay channel,
or control task. "Derive an effective Hamiltonian" is not yet a physical
claim; "explain this avoided crossing" is.

\paragraph{2. Choose a representation that exposes the slow structure.}
A static near-degeneracy suggests a retained spectral manifold; a periodic
problem suggests harmonics or Sambe replicas; a lossy auxiliary suggests a
slow--fast Liouvillian split. A many-body problem may require local clusters
or symmetry sectors rather than a global norm.

\paragraph{3. Enumerate allowed paths and exclusions.}
For a transition $i\rightarrow f$, the schematic nonresonant scaling is
\begin{equation*}
    G_{fi}^{(r)}=\sum_{\mathcal P:i\rightarrow f}A_{\mathcal P}^{(r)},
    \qquad
    A_{\mathcal P}^{(r)}\sim
    \frac{V_{fm_{r-1}}\cdots V_{m_1i}}
    {\Delta_{m_{r-1}}\cdots\Delta_{m_1}}.
\end{equation*}
Symmetry and selection rules determine which paths exist; phases and
denominators determine whether they add or cancel. This expression gives the
power counting, not a universal quasi-degenerate perturbation formula.

\paragraph{4. Infer the operator, its order, and the retained manifold.}
The endpoint changes suggest an operator, and the shortest nonzero path sets
its leading perturbative order. A state connected by a small denominator
belongs in the retained space rather than in a large coefficient. For an
appreciably occupied retained set $P_\star$, a useful local diagnostic is
\begin{equation}
    \eta_\star
    =\max_{\substack{p\in P_\star\\m\in Q}}
      \frac{|\langle m|\hat V|p\rangle|}{|E_p-E_m|}.
    \label{eq:compact_local_control_parameter}
\end{equation}
Occupation factors, collective enhancement, drive harmonics, and linewidths
must be included when they change the coupling or spectral separation. In a
dissipative problem, the relevant denominator is a complex pole or
Liouvillian separation.

\paragraph{5. Distinguish generated from resonantly active terms.}
An operator may be present but rotate too rapidly to drive slow conversion.
For
$G_{fi}\hat{\mathcal O}_{fi}e^{\mathrm i\delta_{fi}t}+\mathrm{H.c.}$,
the mismatch must include all lower-order diagonal shifts. Large coherent
conversion requires $|\delta_{fi}|\lesssim2|G_{fi}|$ and a coupling that
exceeds the relevant decoherence rates. A weaker process may still be
resolved spectroscopically if the linewidth and measurement time allow it.

\paragraph{6. Choose complementary calculations.}
Use the method whose natural organization matches the physical question.
Add a second method only if it contributes new information about paths,
dressing, temporal filtering, micromotion, or noise.

\paragraph{7. Complete the prediction.}
Transform the preparation, observables, controls, and jumps; restore
eliminated variables, micromotion, or the initial layer; and state the first
omitted structure. The proposed interaction becomes useful only when it
produces a falsifiable laboratory prediction.

The applications in Sec.~\ref{sec:applications} follow this sequence in
different settings: Raman paths through a virtual level, a spin-chain bus,
the complex susceptibility of a lossy mediator, and Floquet-controlled
tunnelling. Their microscopic details differ, but the inverse questions are
the same.

\subsection{Validation, failure, and repair}
\label{sec:validation_protocol}
\label{subsec:comparison_validation}

Validation is claim dependent, a viewpoint shared by recent pedagogical work
on open-system approximations \cite{Stefanini2026IsLindblad}. A model may
reproduce a selected spectrum but fail for a bare population, reproduce
stroboscopic dynamics but fail inside a period, or reproduce a coherent shift
while missing an induced jump. The relevant question is therefore: does the
reconstructed prediction support the stated physical claim for the declared
preparations, observables, parameters, and time window?

\begin{tcolorbox}[
    plainlanguage,
    breakable=false,
    title={Plain-language guide: validate the claim, not the method name}
]
If the claim concerns a resonance, test the dressed crossing and its gap. If
it concerns a trajectory, test reconstructed states or observables over the
stated interval. If it concerns dissipation, test the map, steady state, jump
structure, and physicality. Passing one test does not validate a different
claim.
\end{tcolorbox}

\begin{table*}[!ht]
    \centering
    \caption{\justifying
    Claim-sensitive validation. Direct evidence is compared with the most
    common structural repair when the test fails.}
    \label{tab:compact_validation_claims}
    \footnotesize
    \renewcommand{\arraystretch}{1.13}
    \setlength{\tabcolsep}{3.5pt}
    \begin{tabularx}{\linewidth}{
        @{}
        >{\RaggedRight\arraybackslash}p{0.20\linewidth}
        >{\RaggedRight\arraybackslash}p{0.39\linewidth}
        >{\RaggedRight\arraybackslash}X
        @{}
    }
        \toprule
        \textbf{Claim} & \textbf{Direct evidence} & \textbf{Typical repair}\\
        \midrule
        Spectrum or resonance
        & Dressed energies, gap, projector overlap, and pole location
        & Include diagonal shifts, enlarge the resonant manifold, or align
        effective gauges.\\
        Closed-system trajectory
        & Reconstructed state distance, observables, leakage, and phase drift
        & Transform endpoints, restore micromotion, or retain the next
        frequency correction.\\
        Control operation
        & Physical-frame gate or channel fidelity and parameter sensitivity
        & Transform control operators and redesign the pulse in the dressed
        frame.\\
        Open-system reduction
        & Finite-time map, Choi positivity, trace preservation, jumps, steady
        state, and correlations
        & Restore effective jumps, reverse processes, initial slip, or a
        memory-bearing auxiliary.\\
        Higher-order process
        & Scaling of the gap or rate, shifted resonance, interference, and
        leakage
        & Rechoose the retained space, enumerate all paths, and compare with
        converged spectra and dynamics.\\
        \bottomrule
    \end{tabularx}
\end{table*}

The metric should live at the same level as the claim. The trace distance
$D(\hat\rho,\hat\sigma)=\|\hat\rho-\hat\sigma\|_1/2$ bounds every bounded
observable through
\begin{equation*}
    |\operatorname{Tr}[\hat O(\hat\rho-\hat\sigma)]|
    \leq2\|\hat O\|_\infty D(\hat\rho,\hat\sigma),
\end{equation*}
whereas fidelity is often more intuitive for target-state preparation.
Leakage tests the adequacy of the retained manifold but not errors within it.
Near a degeneracy, compare spectral projectors as well as eigenvalues because
individual eigenvectors are basis dependent inside the nearly degenerate
space.

The reconstructed dynamical residual
\begin{equation}
    r_{\mathrm{dyn}}(t)
    =\left\|\dot{\hat\rho}_{\mathrm{rec}}(t)
      -\mathcal G_{\mathrm{full}}(t)
      [\hat\rho_{\mathrm{rec}}(t)]\right\|_1
    \label{eq:validation_dynamical_residual}
\end{equation}
tests whether the complete trajectory nearly satisfies the microscopic
equation. It should be accompanied by an integrated state or observable
error over the intended interval. For a finite-dimensional open-system map,
complete positivity is tested with its Choi matrix and trace preservation
with the corresponding partial trace; a generator that merely resembles
GKLS form is not a substitute for these map-level tests.

A reference calculation must itself be converged. Schematically,
\begin{equation*}
    \epsilon_{\mathrm{reported}}
    \lesssim\epsilon_{\mathrm{model}}
      +\epsilon_{\mathrm{reference}}+\epsilon_{\mathrm{numerics}}.
\end{equation*}
The last contribution includes Hilbert-space and Floquet cutoffs, time steps,
solver tolerances, and optimization error. The validity region should be
mapped over the parameters that control the local problem, including
occupation, collective sector, detuning, linewidth, drive frequency,
observation time, and system size. An expected
$O(\epsilon^{N+1})$ residual should display that power before numerical noise
or a nearby resonance takes over.

Failure often identifies the repair. A small denominator calls for a larger
retained manifold. Accurate stroboscopic but inaccurate continuous dynamics
calls for micromotion. Immediate trajectory error points to encoding, initial
slip, or preparation. Missing decay points to transformed jumps, reverse
processes, or memory. Growth linear in time usually indicates a missing
frequency or quasienergy correction. The repair should follow the failed
structure rather than consist automatically of one more perturbative order.

\subsection{Compact decision map}
\label{subsec:compact_decision_map}

Before accepting an effective description, one should be able to answer:
\begin{enumerate}
    \item What physical claim is being made, over what time and parameter
    domain?
    \item Which states, modes, harmonics, or slow variables must remain
    explicit?
    \item Which microscopic paths and symmetries generate or suppress the
    target operator?
    \item What local matrix-element-to-gap or timescale ratio controls the
    approximation?
    \item Which method exposes that structure, and what independent insight
    should a complementary method add?
    \item How are preparations, observables, controls, jumps, eliminated
    variables, and micromotion reconstructed?
    \item Which converged reference, metric, scaling law, failure example,
    and repair support the claim?
\end{enumerate}

These questions connect derivation, interpretation, and discovery. They do
not force every problem through every method. Their purpose is to identify
where the necessary information is stored and what evidence turns an
effective operator into a trustworthy physical prediction. The applications
that follow put this workflow to work on concrete quantum-optical and
quantum-engineering problems.

\section{Advanced Structural Issues in Effective Dynamics}
\label{sec:advanced_effective_dynamics}

\textbf{Reading guide.}
This section is intended for readers who are already comfortable with the
effective-model constructions, encoding maps, reconstruction procedures,
and validation principles introduced above. It may be skipped on a first
reading without interrupting the main development. Rather than deriving
another effective Hamiltonian, we ask what remains to be checked after one
has been obtained. The questions come from mathematical physics,
many-body theory, open-system dynamics, and non-Hermitian quantum
mechanics. Definitions of terms used repeatedly here and elsewhere are
collected in the glossary in Supplemental Material, Sec.~S6.

The derivation of an effective generator is only one part of an effective
description. Once a Hamiltonian, Liouvillian, or finite-time map has been
obtained, its physical use can depend on structures that are not visible
from the truncated algebra alone. A representation change may cease to be
equivalent after a finite-dimensional truncation; a locally accurate
Hamiltonian may fail to be bounded from below; a symmetric bosonic
polynomial may not determine a unique self-adjoint evolution; a formal
perturbation series may be useful only up to an optimal order; and a
completely positive stroboscopic map may have no time-independent
logarithm of GKLS form. Non-Hermitian operators introduce a further layer
of interpretation because the same complex spectral structure can encode
resonance escape, conditional survival, quasiparticle damping, or
Liouvillian decay.

These checks are not needed for every dispersive shift or few-level gate
calculation. They become important when a prediction is
sensitive to a representation or subsystem partition, when a bosonic
polynomial is extrapolated far beyond the fitted spectral window, when a
large Fock cutoff is used as evidence of convergence, when an expansion is
applied near a dense resonance or in an extended system, or when a map or
non-Hermitian spectrum is interpreted as a physical generator. In such
cases, the structural question is part of the approximation rather than a
mathematical afterthought.

Table~\ref{tab:advanced_structural_questions} summarizes the questions
reviewed in this section. Their common theme is that algebraic agreement
inside a retained sector does not, by itself, determine the global
spectral, dynamical, or operational meaning of the effective model.

\begin{table*}[!t]
    \footnotesize
    \centering
    \caption{\justifying
    Structural questions that arise after an effective Hamiltonian,
    Liouvillian, or dynamical map has been constructed. Each diagnostic
    complements perturbative residuals and direct dynamical benchmarks.
    }
    \label{tab:advanced_structural_questions}

    \renewcommand{\arraystretch}{1.26}
    \setlength{\tabcolsep}{4pt}

    \begin{tabularx}{\textwidth}{
        @{}
        >{\RaggedRight\arraybackslash}p{0.18\textwidth}
        >{\RaggedRight\arraybackslash}p{0.37\textwidth}
        >{\RaggedRight\arraybackslash}X
        @{}
    }
        \toprule
        \textbf{Issue}
        &
        \textbf{What must be established}
        &
        \textbf{Consequence and diagnostic}
        \\
        \midrule

        Representation and gauge
        &
        Does the representation change commute with the selected
        matter-level, mode, or model-space truncation?
        &
        Gauge- or partition-dependent spectra, observables, and bath
        couplings can appear in a reduced model. Test convergence across
        retained levels, modes, and representations, using consistently
        transformed observables.
        \\

        Spectral admissibility
        &
        Is the reduced Hamiltonian bounded from below, and does
        $e^{-\beta H_{\mathrm{eff}}}$ define a trace-class operator when
        equilibrium thermodynamics is invoked?
        &
        A locally accurate normal form may fail as a global Hamiltonian or
        Gibbs model. Test the spectral window and cutoff dependence, and
        compare with the bounded microscopic theory.
        \\

        Operator domains
        &
        Does the formal bosonic operator possess a unique self-adjoint
        realization on the stated domain?
        &
        Different domains or numerical truncations can select different
        unitary evolutions. State the domain and any deficiency-index
        information; test parity, basis sensitivity, and boundary
        population.
        \\

        Asymptoticity and resonances
        &
        Does increasing perturbative order improve the approximation over
        the populated manifold and observation time?
        &
        Small denominators, secular accumulation, or factorial growth can
        limit the useful order. Compare residuals order by order, verify
        that all resonances are retained, and study optimal truncation.
        \\

        Locality and system size
        &
        Which local, collective, or sector-dependent scale controls the
        expansion in an extended system?
        &
        Longer-range and many-body interactions emerge, while gaps and
        control parameters can change with size. Examine local norms,
        linked-cluster structure, range growth, and finite-size or
        symmetry-sector scaling.
        \\

        Maps and generator logarithms
        &
        Does the reduced map admit a regular time-local generator or a
        time-independent GKLS logarithm?
        &
        Memory, singular time-local rates, branch ambiguity, and failure
        of semigroup embeddability can occur. Test map invertibility and
        Choi positivity, search logarithm branches, and compare
        memory-kernel and time-local descriptions.
        \\

        Non-Hermitian reduction
        &
        Which object is propagated or diagonalized, and what does its
        imaginary part represent?
        &
        Resonance widths, survival probabilities, quasiparticle
        lifetimes, and Liouvillian decay modes are not interchangeable.
        Identify the normalization, recycling terms, energy dependence,
        postselection, and measured observable.
        \\
        \bottomrule
    \end{tabularx}
\end{table*}

\subsection{Representation Covariance, Gauge Freedom, and Truncation}
\label{subsec:advanced_gauge_truncation}

The word \emph{gauge} appears in several effective-model settings and
should be interpreted according to context. Electromagnetic gauge freedom
relates full light--matter Hamiltonians through unitary transformations;
block-gauge freedom refers to unitary changes of basis within an effective
model space; and Floquet gauge freedom redistributes information between a
static quasienergy generator and periodic micromotion. These freedoms all
express nonuniqueness of representation, but they act on different
objects. The corresponding definitions are summarized in Supplemental
Material, Sec.~S6.12.

For a full-space unitary representation change,
\begin{equation*}
    \hat H_{g}
    =
    \hat U_{g}
    \hat H
    \hat U_{g}^{\dagger},
    \qquad
    \hat O_{g}
    =
    \hat U_{g}
    \hat O
    \hat U_{g}^{\dagger},
\end{equation*}
physical predictions are unchanged when states, observables, interactions,
and system--environment couplings are transformed consistently. In
nonrelativistic cavity QED, the Power--Zienau--Woolley transformation
provides the standard connection between minimal-coupling and multipolar
representations. The exact equivalence belongs to the complete theory,
including the quadratic diamagnetic or self-polarization terms required
by the chosen representation
\cite{babiker1983derivation,stokes2022implications,Schafer2020SelfPolarization}.

A finite projection changes this statement. If $\hat P$ selects a fixed
matter-level, field-mode, or model-space truncation, then, in general,
\begin{equation*}
\begin{aligned}
    \hat P
    \hat U_g
    \hat H
    \hat U_g^{\dagger}
    \hat P
    \neq{}&
    \left(
        \hat P\hat U_g\hat P
    \right)
    \left(
        \hat P\hat H\hat P
    \right)
    \left(
        \hat P\hat U_g^{\dagger}\hat P
    \right).
\end{aligned}
\end{equation*}
The projected operator $\hat P\hat U_g\hat P$ is generally not unitary on
the retained space, because the full transformation mixes retained and
eliminated states. Gauge transformation and a naive fixed projection
therefore need not commute. This observation underlies the gauge
sensitivity of few-level models in ultrastrong-coupling cavity and circuit
QED
\cite{stokes2019gauge,distefano2019resolution,taylor2020resolution,StokesNazir2024GaugeTruncation}.

This noncommutation is not an impossibility theorem for reduced models.
Gauge-consistent truncation schemes can instead derive the finite theory
while enforcing the gauge principle, including simultaneous matter and
mode truncation in structured or lossy media
\cite{Gustin2023GaugeInvariantTruncation}. Their practical lesson is that
the truncation prescription, photodetection operator, and
system--environment coupling must belong to the same representation; a
comparison of reduced Hamiltonians alone is insufficient.

The practical consequences depend on what has been truncated. A dipole-like
representation can be advantageous when only a small number of material
levels is retained, while a Coulomb-like representation can converge more
rapidly with the number of field modes or in other frequency ranges. In
multimode or multiband problems, the representation that minimizes one
truncation error need not minimize another, and the best finite model can
depend on the target spectrum, observable, and dynamical timescale
\cite{Li2020GaugeTightBinding,TaylorMandalHuo2022,arwas2023metrics,DeBernardis24,Qing2025OptimalDynamicalGauge}.
The lesson is not that one gauge is universally best. Enlarge the retained
matter and field spaces, compare operational observables, and transform
every measured and dissipative operator in the same representation.

Quadratic light--matter terms deserve particular attention in this
comparison. In minimal-coupling formulations, the diamagnetic term is tied
to the linear momentum--field interaction; in multipolar formulations, the
corresponding self-polarization structure appears differently. Removing
these contributions independently of the accompanying truncation can alter
stability, bound-state structure, and response functions
\cite{Schafer2020SelfPolarization,stokes2022implications}. Statements about
photon number, entanglement between nominal subsystems, or a superradiant
transition also require the subsystem partition and the operationally
measured observable to be specified, because the decomposition of the
full light--matter system is representation dependent even though the
complete physical prediction remains gauge invariant.

The same logic extends beyond electromagnetic gauge. Two
energy-independent effective Hamiltonians related by a unitary acting
inside the retained block can have different operator coefficients while
predicting the same reconstructed observables. Likewise, different
Floquet gauges assign different bounded terms to the kick operator and to
the static generator. Comparisons across these representations are made at
the level of the complete prediction, with the encoding and reconstruction
maps transformed together, rather than by matching isolated coefficients.

\paragraph{Further reading.}
The broad conceptual framework of gauge freedom, subsystem relativity, and
observable reconstruction is reviewed in
Ref.~\cite{stokes2022implications}. Few-level and multimode truncation
problems are analyzed from complementary perspectives in
Refs.~\cite{stokes2019gauge,distefano2019resolution,StokesNazir2024GaugeTruncation,Li2020GaugeTightBinding,TaylorMandalHuo2022,arwas2023metrics,Qing2025OptimalDynamicalGauge}.
Gauge-consistent matter and mode truncation in arbitrary media is
developed in Ref.~\cite{Gustin2023GaugeInvariantTruncation}. The role of
quadratic diamagnetic
and self-polarization terms is reviewed in
Ref.~\cite{Schafer2020SelfPolarization}.

\subsection{Spectral Windows, Boundedness, and Thermodynamic Use}
\label{subsec:advanced_boundedness}

An effective Hamiltonian can reproduce selected transition frequencies
without defining a globally admissible many-level theory. Several
properties that coincide automatically for a finite Hermitian matrix
separate in infinite dimension: symmetry on a chosen domain,
self-adjointness, boundedness from below, and the trace-class character of
$e^{-\beta \hat H}$. Self-adjointness supplies a well-defined unitary
evolution; a lower bound supplies spectral stability; and
$e^{-\beta\hat H}$ being trace class is the additional requirement for a
normalizable canonical Gibbs state. None of these implications should be
reversed without further assumptions on the spectrum and its
multiplicities.

A familiar example is the negative-Kerr normal form
\begin{equation}
    \hat H_{\mathrm{Kerr}}^{(-)}
    =
    \omega
    \hat n
    -
    \chi
    \hat n^2,
    \qquad
    \hat n=\hat a^{\dagger}\hat a,
    \qquad
    \chi>0.
    \label{eq:advanced_negative_kerr}
\end{equation}
Its eigenvalues are
\(E_n = \omega n- \chi n^2,\)
so $E_n\rightarrow-\infty$ as $n\rightarrow\infty$. The operator can
nevertheless describe the first few levels of a weakly anharmonic circuit
or oscillator with high accuracy. For the transmon, for example, a local
Duffing or Kerr description follows from expanding a bounded-below
microscopic circuit Hamiltonian near one potential minimum; the truncated
polynomial is a normal form for the low-energy spectrum rather than a
global replacement for the full cosine model
\cite{Koch2007Transmon}.

The distinction becomes unavoidable in equilibrium thermodynamics. On the
full Fock space, the canonical partition sum associated with
Eq.~\eqref{eq:advanced_negative_kerr} is
\begin{equation*}
    Z_{-}(\beta)
    =
    \sum_{n=0}^{\infty}
    \exp
    \left[
        -\beta
        \left(
            \omega n-
            \chi n^2
        \right)
    \right],
\end{equation*}
and it diverges for every $\beta>0$. A finite Fock cutoff makes the sum
finite, but a cutoff introduced only for numerical convenience produces a
cutoff-dependent thermal state and does not define a thermodynamic limit
of Eq.~\eqref{eq:advanced_negative_kerr}. A finite retained space can be
a legitimate model when it follows from a physical constraint or an
independently controlled projection; what fails is treating an arbitrary
numerical boundary as a thermodynamic regularization. The same
Hamiltonian may therefore be appropriate for spectroscopy or finite-time
control in a low-occupation manifold and inappropriate as a global Gibbs
Hamiltonian.

Higher perturbative order does not generically repair this global problem.
The next term can have either sign, the expansion can be asymptotic, and a
polynomial normal form can remain nonuniform at large occupation even when
several low-energy coefficients are accurate. When thermodynamic use is
intended, a controlled calculation normally returns to the bounded
microscopic Hamiltonian, adopts an independently justified finite retained
space, or employs a resummation whose global spectral properties can be
established. The domain of use should be stated through an energy or
occupation window,
\begin{equation*}
    \mathcal M_{E_*}
    =
    \operatorname{span}
    \left\{
        \ket{n}:E_n^{\mathrm{micro}}\leq E_*
    \right\},
\end{equation*}
together with a bound on the population or leakage outside that window.

This separation between local accuracy and global admissibility also
appears in attractive bosonic models, truncated normal forms near
classical equilibria, and effective field theories whose operators are
ordered by a low-energy power counting. In each case, extrapolation to
arbitrarily high excitation is a new assumption. Spectral agreement in a
retained window establishes neither global boundedness nor existence of a
Gibbs state.

\paragraph{Further reading.}
The distinction between the bounded microscopic transmon Hamiltonian and
its weakly anharmonic normal form is developed in
Ref.~\cite{Koch2007Transmon}. General operator-theoretic background on
boundedness, domains, and trace-class functions of self-adjoint operators
can be found in Refs.~\cite{Hall2013QuantumTheory,ReedSimonI}.

\subsection{Operator Domains and Essential Self-Adjointness}
\label{subsec:advanced_self_adjointness}

For unbounded operators, the action of the formal expression and the
choice of domain are inseparable. A densely defined operator $\hat H$ is
symmetric on $\mathcal D(\hat H)$ when
\begin{equation*}
    \langle\psi,\hat H\phi\rangle
    =
    \langle\hat H\psi,\phi\rangle,
    \qquad
    \psi,\phi\in\mathcal D(\hat H),
\end{equation*}
but symmetry alone does not imply that $\hat H$ is self-adjoint. A
self-adjoint operator satisfies both equality of the operator action and
equality of domains,
\begin{equation*}
    \hat H=\hat H^{\dagger},
    \qquad
    \mathcal D(\hat H)=\mathcal D(\hat H^{\dagger}).
\end{equation*}
An operator is essentially self-adjoint on a chosen core when its closure
is self-adjoint. In that case, the core determines a unique unitary group
$e^{-\mathrm{i}t\overline H}$. For a densely defined symmetric operator,
the deficiency indices
\begin{equation*}
    n_{\pm}
    =
    \dim\ker
    \left(
        \hat H^{\dagger}
        \mp
        \mathrm{i}\hat I
    \right)
\end{equation*}
provide a standard diagnostic. Essential self-adjointness corresponds to
$n_{+}=n_{-}=0$, while self-adjoint extensions exist when
$n_{+}=n_{-}$. Nonzero equal indices mean that additional domain or
boundary information is required to select the physical extension
\cite{SA1,SA2,Hall2013QuantumTheory,Nelson1959AnalyticVectors}.

Higher-order bosonic processes provide a direct effective-dynamics
example. Consider
\begin{equation*}
    \hat A_k
    =
    \mathrm{i}
    \left[
        \left(
            \hat a^{\dagger}
        \right)^k
        -
        \hat a^k
    \right]
\end{equation*}
on the finite-Fock core
$\mathcal D_{\mathrm{fin}}=\operatorname{span}\{\ket n:n\geq0\}$.
The displacement case $k=1$ and the ordinary squeezing case $k=2$ are
essentially self-adjoint on the finite-Fock core. By contrast, pure
$k$-photon squeezing with $k\geq3$ is not essentially self-adjoint on
$\mathcal D_{\mathrm{fin}}$: the formal polynomial is symmetric there,
but that core does not select a unique unitary evolution
\cite{fischer2025selfadjoint}.

The broader family
\begin{equation*}
\begin{aligned}
    \hat A_{k,l}[f]
    ={}&
    \xi
    (\hat a^{\dagger})^k
    \hat a^l
    +
    \xi^{*}
    (\hat a^{\dagger})^l
    \hat a^k
    +
    f(\hat n)
\end{aligned}
\end{equation*}
makes the ultraviolet competition explicit. Under the hypotheses of
Ref.~\cite{fischer2025selfadjoint}, if
$f(n)=a_h n^h+\cdots$ with $a_h>0$, then a diagonal term growing faster
than the off-diagonal scale, $2h>k+l$, yields an essentially self-adjoint
operator bounded from below; for $2h<k+l$, essential self-adjointness
fails and the self-adjoint extensions remain unbounded below. At the
borderline $2h=k+l$, the result depends on the ratio
$a_h/|\xi|$: the cited classification proves essential self-adjointness
for $a_h>2|\xi|$ and failure for $a_h<2|\xi|$, while the equality case
requires separate analysis. Thus, the instruction ``add a Kerr term'' is
incomplete: the order and coefficient of the diagonal completion must be
compared with the multiphoton process it is meant to regularize.

The ambiguity is visible in finite-dimensional simulations. Truncating the
Fock basis produces a finite Hermitian matrix and hence a unitary
finite-dimensional evolution for every cutoff. For the generalized
squeezing models analyzed in Ref.~\cite{ashhab2025finite}, even- and
odd-dimensional Fock truncations converge to distinct limiting evolutions
associated with different self-adjoint extensions. More broadly, under
the mild semiboundedness hypotheses of Ref.~\cite{FischerWrongBox2026},
Galerkin truncations converge to the Friedrichs extension of the
Hamiltonian restricted to the chosen basis core, which need not be the
intended physical realization. Numerical stability within one basis and
cutoff sequence is therefore not sufficient when the infinite-dimensional
domain has not been identified.

The preceding asymptotic classification makes precise when a diagonal term
$f(\hat n)$ regularizes an off-diagonal bosonic polynomial. The relevant
condition is not the name or sign of one low-order Kerr coefficient but
its growth relative to the Jacobi coefficients of the multiphoton term
\cite{fischer2025selfadjoint,ashhab2025finite}. The microscopic
Hamiltonian usually supplies the physically meaningful ultraviolet
completion and, with it, the domain or boundary behavior absent from the
isolated polynomial normal form.

The pure-field conclusion should not be transferred indiscriminately to
all multiphoton light--matter Hamiltonians. For the class studied in a
recent mathematical-physics preprint, a nonzero normal matter-coupling
operator leads to multiple self-adjoint extensions for $k\geq3$, whereas
a $k$-photon Jaynes--Cummings interaction with its non-normal lowering
operator remains self-adjoint for every $k$
\cite{Fischer2026KPhoton}. This distinction is relevant when the
higher-order resonant operators of
Sec.~\ref{sec:higher_order_expansions} are promoted from finite retained
manifolds to global bosonic models.

For effective bosonic simulations, a domain-aware convergence study should
therefore include more than increasing one cutoff. Useful tests include:
comparison of different cutoff sequences and bases; monitoring of
population near the numerical boundary; sensitivity to cutoff parity;
convergence after restoring physically motivated high-excitation terms;
and comparison with the bounded microscopic model whenever it is
available. These tests complement, rather than replace, an analytical
statement of the operator domain.

\paragraph{Further reading.}
Pedagogical introductions to operator domains and self-adjoint extensions
are given in Refs.~\cite{SA1,SA2}, while a broader mathematical treatment
is available in Refs.~\cite{Hall2013QuantumTheory,ReedSimonI}. The
classification of higher-order squeezing operators is developed in
Ref.~\cite{fischer2025selfadjoint}; its finite-dimensional consequences
and the wider basis dependence of Galerkin dynamics are studied in
Refs.~\cite{ashhab2025finite,FischerWrongBox2026}. A recent preprint
extends the domain analysis to $k$-photon light--matter Hamiltonians
\cite{Fischer2026KPhoton}.

\subsection{Asymptotic Expansions, Small Denominators, and Optimal Truncation}
\label{subsec:advanced_asymptoticity}

An effective expansion is often written schematically as
\begin{equation*}
    \hat H_{\mathrm{eff}}^{[N]}
    =
    \sum_{r=0}^{N}
    \epsilon^r
    \hat H^{(r)},
    \qquad
    \hat R^{[N]}
    =
    \hat H_{\mathrm{exact}}^{(\mathrm{frame})}
    -
    \hat H_{\mathrm{eff}}^{[N]},
\end{equation*}
but the existence of coefficients to arbitrary order does not imply
convergence of the series. In many applications, the truncated expansion
is an asymptotic normal form: for fixed small $\epsilon$, the remainder
initially decreases with $N$, reaches an optimal order, and can then grow
because of combinatorial proliferation, small denominators, or the
unboundedness of the underlying operators.

For bounded finite-dimensional Hamiltonians, the Magnus expansion admits
standard sufficient convergence criteria, such as an integrated-norm
condition over the interval of interest
\cite{Moan2007,Blanes2009}. In unbounded bosonic systems, these global norm
conditions are unavailable on the full Hilbert space. Control is instead
formulated in an occupied energy sector, through relative bounds, or by
convergence of a numerically truncated calculation together with a
state-dependent tail estimate. The same distinction applies to
Schrieffer--Wolff and projection expansions: a global operator norm can be
irrelevant or divergent even when matrix elements in the physical
manifold remain perturbative.

Small denominators identify a separate mechanism of breakdown. If an
eliminated state $\ket\nu$ approaches a retained state $\ket\mu$, then
\begin{equation*}
    \eta_{\mu\nu}
    =
    \frac{
        \left|
            \langle\nu|\hat V|\mu\rangle
        \right|
    }{
        \left|
            E_\mu-E_\nu
        \right|
    }
\end{equation*}
ceases to be small even when the microscopic coupling is weak. The natural
response is to reorganize the retained manifold so that the resonant or
quasiresonant states are treated together. Increasing the nondegenerate
order without changing the partition generally obscures rather than
resolves the resonance.

Time introduces another nonuniformity. Suppose the first omitted diagonal
term scales as
$\delta H\sim\Omega\epsilon^{N+1}$. Its accumulated phase is
\(\delta\varphi(t) \sim \Omega t\epsilon^{N+1},\)
so a small instantaneous generator error can become an order-one
observable error at
$t\sim(\Omega\epsilon^{N+1})^{-1}$. Conversely, an off-diagonal remainder
can mainly control a small transition amplitude rather than a secular
phase. Reporting the norm and block character of the residual is therefore
more informative than quoting perturbative order alone.

Periodically driven local many-body systems provide the clearest example
of a useful asymptotic expansion without a globally convergent infinite
series. At high driving frequency, a Floquet or van Vleck Hamiltonian
truncated near an optimal order can govern local observables over a
prethermal interval whose duration is exponentially long in the ratio of
the drive frequency to a local many-body energy scale. We denote this
scale by $J_{\mathrm{loc}}$, distinguishing it from the
restricted-manifold scale $h_{\mathcal M}$ and from the perturbation norm
$v_{\mathrm{loc}}$ introduced below. Under the locality and boundedness
assumptions of the cited theorems, the optimal truncation order grows
proportionally to $\omega/J_{\mathrm{loc}}$, and the resulting bounds have
the qualitative form
\begin{equation*}
    \tau_{\mathrm{pre}}
    \gtrsim
    \exp
    \left(
        c
        \frac{\omega}{J_{\mathrm{loc}}}
    \right),
\end{equation*}
with $c>0$ model dependent. The effective Hamiltonian is then a
long-lived prethermal generator, not an exact infinite-time description
\cite{abanin2017rigorous,Abanin2017,kuwahara2016floquet}.

An order study should consequently ask whether spectra, reconstructed
observables, and residuals improve over the target manifold and time
window. The most accurate truncation can vary with state, time, and
observable. This is especially relevant near resonances, near a many-body
gap closing, or in bosonic sectors whose matrix elements grow with
occupation.

\paragraph{Further reading.}
Convergence and practical organization of the Magnus expansion are
reviewed in Refs.~\cite{Moan2007,Blanes2009}. Many-body
Schrieffer--Wolff error and linked-cluster structures are developed in
Ref.~\cite{Bravyi2011}. Rigorous and constructive treatments of Floquet
prethermalization are given in
Refs.~\cite{abanin2017rigorous,Abanin2017,kuwahara2016floquet}.

\subsection{Locality, Extensivity, and Many-Body Effective Dynamics}
\label{subsec:advanced_many_body}

The Jaynes--Cummings benchmark expresses perturbative control through the
occupation-dependent ratio
$g\sqrt{n+1}/|\Delta|$. Extended systems require a broader language. For a
short-range lattice Hamiltonian,
\begin{equation*}
    \hat H
    =
    \hat H_0
    +
    \lambda
    \sum_X
    \hat V_X,
\end{equation*}
the global norm of the perturbation typically grows with system size. A
local interaction scale can instead be defined schematically by
\begin{equation}
    v_{\mathrm{loc}}
    =
    \sup_x
    \sum_{X\ni x}
    \left\|
        \hat V_X
    \right\|.
    \label{eq:advanced_local_norm}
\end{equation}
For a spin or fermionic lattice with finite-dimensional local Hilbert
spaces, $\|\cdot\|$ in Eq.~\eqref{eq:advanced_local_norm} is the operator
norm on the finite support $X$. Bosonic lattices with unbounded local
operators generally require an energy-restricted norm, a weighted norm,
or another model-dependent refinement. Geometric refinements can also
include the range or diameter of $X$. Local norms, Lieb--Robinson bounds,
and linked-cluster expansions allow one to control local observables
without requiring the norm of the full extensive Hamiltonian to remain
small
\cite{BravyiHastingsVerstraete2006,HastingsWen2005,Bravyi2011}.
The distinction among local, collective, and global control parameters is
summarized in Supplemental Material, Sec.~S6.4.

A general retained-manifold measure remains useful,
\begin{equation*}
    \eta_{\mathcal M}
    =
    \sup_{\substack{
        \ket\mu\in\mathcal M\\
        \ket\nu\notin\mathcal M
    }}
    \frac{
        \left|
            \langle\nu|\hat V|\mu\rangle
        \right|
    }{
        \left|
            E_\mu-E_\nu
        \right|
    },
\end{equation*}
but it should be interpreted together with locality and density of states.
In a dense many-body spectrum, the smallest global denominator can vanish
rapidly with size even when local processes remain well separated. A
low-energy effective Hamiltonian can nevertheless be controlled when the
retained sector is separated by a many-body gap and the transformation is
quasi-local. Near a quantum critical point, gap closing enlarges the
relevant manifold and reduces the usefulness of a fixed low-order
partition.

Many-body Schrieffer--Wolff transformations generate more than
renormalized one-body parameters. Virtual processes produce longer-range
exchange, multispin terms, correlated hopping, ring exchange, and
occupation-dependent interactions. At perturbative order $r$, the spatial
support of an induced term can grow with the number of elementary local
processes involved. The linked-cluster structure ensures that disconnected
contributions cancel in an extensive effective Hamiltonian, while
quasi-locality controls the decay of the remaining long-range tails in a
gapped short-range setting
\cite{Bravyi2011,HastingsWen2005}.

Collective light--matter models illustrate a different size dependence.
For the Tavis--Cummings interaction
\begin{equation*}
    \hat V_{\mathrm{TC}}
    =
    g
    \left(
        \hat a\hat J_+
        +
        \hat a^{\dagger}\hat J_-
    \right),
\end{equation*}
the transition matrix element in a Dicke sector is
\begin{align*}
    &\bra{n+1,j,m-1}
    \hat V_{\mathrm{TC}}
    \ket{n,j,m} \\
    &=
    g
    \sqrt{n+1}
    \sqrt{
        (j+m)
        (j-m+1)
    }.
\end{align*}
The perturbative parameter therefore depends on the photon sector, total
spin $j$, polarization $m$, system size, and detuning of the connected
states. Near a polarized state in the symmetric $j=N/2$ sector, the
familiar collective scale is of order $g\sqrt N$. Whether this factor is
already absorbed into the coupling depends on convention: thermodynamic
Dicke models commonly employ a Kac-scaled microscopic coupling
$g/\sqrt N$, whereas cavity-QED discussions often quote the single-emitter
coupling $g$
\cite{Dicke1954,TavisCummings1968}.

Open many-body systems add a spectral scale in Liouville space. For a
primitive finite-dimensional semigroup, the Liouvillian gap characterizes
the asymptotic exponential decay rate. It need not, however, provide the
observed transient relaxation time in a non-normal or boundary-driven
many-body system. Large expansion coefficients, metastable manifolds, and
long crossover times can dominate physically accessible dynamics
\cite{Macieszczak2016Metastability,MoriShirai2020Gap,MoriShirai2023SymmetrizedGap}. Near a dissipative phase transition, closing
of a relevant Liouvillian or metastable spectral separation can invalidate
an elimination that assumed a fixed fast gap.

The many-body extension of effective-model validation should therefore
report: the locality and range of the microscopic terms; the scaling of
induced interactions; the symmetry and excitation sectors populated; the
finite-size behavior of relevant Hamiltonian or Liouvillian gaps; and the
normalization convention for collective couplings. These quantities play
the role occupied by $g\sqrt{n+1}/|\Delta|$ in the single-mode benchmark.

\paragraph{Further reading.}
Quasi-locality and linked-cluster structure in many-body effective
Hamiltonians are developed in
Refs.~\cite{Bravyi2011,BravyiHastingsVerstraete2006,HastingsWen2005}.
Canonical collective light--matter models are introduced in
Refs.~\cite{Dicke1954,TavisCummings1968}. Floquet many-body prethermal
control is discussed in
Refs.~\cite{abanin2017rigorous,Abanin2017,kuwahara2016floquet}, while
Liouvillian metastability and transient relaxation are analyzed in
Refs.~\cite{Macieszczak2016Metastability,MoriShirai2020Gap,MoriShirai2023SymmetrizedGap}.

\subsection{Memory Kernels, Dynamical Maps, and Generator Logarithms}
\label{subsec:advanced_maps_memory}

Projection of environmental or fast degrees of freedom naturally produces
a reduced map before it produces a time-independent generator. A detailed
development centered on the reduced dynamical map is given in
Ref.~\cite{Szankowski2023OpenSystems}. Two exact organizations of this
reduced dynamics are the Nakajima--Zwanzig
memory-kernel equation,
\begin{equation*}
    \frac{\mathrm d}{\mathrm dt}
    \hat\rho_S(t)
    =
    \int_0^t
    \mathcal K(t,s)
    \hat\rho_S(s)
    \,\mathrm ds
    +
    \hat\xi(t),
\end{equation*}
and the time-convolutionless form,
\begin{equation*}
    \frac{\mathrm d}{\mathrm dt}
    \hat\rho_S(t)
    =
    \mathcal L_{\mathrm{TCL}}(t)
    \hat\rho_S(t).
\end{equation*}
The inhomogeneous term $\hat\xi(t)$ contains the effect of initial
components outside the selected projection manifold. The two forms are
complementary representations of reduced dynamics rather than successive
levels of approximation
\cite{Nakajima1958,Zwanzig1960,ChaturvediShibata1979,breuer2002theory}.
The distinction between memory kernels, time-local generators, and initial
slip is summarized in Supplemental Material, Sec.~S6.11.

If the reduced map $\Phi(t,0)$ is differentiable and invertible at
time $t$, it defines the formal map-local generator
\begin{equation*}
    \mathcal L_{\Phi}(t)
    =
    \dot\Phi(t,0)
    \Phi^{-1}(t,0).
\end{equation*}
When the same reduced map is represented by a regular exact
time-convolutionless equation, $\mathcal L_{\Phi}(t)$ is its generator.
It need not, however, coincide term by term with a perturbative TCL
generator derived from a particular projection, initialization, and
truncation scheme. Loss of invertibility can make the map-local generator
singular even when the map and the underlying system--environment
evolution remain regular. Such a singularity can reflect convergence of
distinct initial states, a poor coordinate representation of the reduced
flow, or a genuine obstruction to extending the time-local equation
through that instant
\cite{AnderssonCresserHall2007,HallCresserLiAndersson2014}. The full
model may therefore remain regular even when the time-local representation
becomes singular.

The map--generator distinction is also central in periodic open systems.
For the exact one-period CPTP map $\Phi_T(t_0)$, a formal Floquet generator
can be written on a logarithm branch $k$ as
\begin{equation*}
    \mathcal L_F^{(k)}(t_0)
    =
    \frac{1}{T}
    \operatorname{Log}_k
    \Phi_T(t_0).
\end{equation*}
The logarithm is multivalued, and the resulting candidate generators can
differ by branch choices associated with the eigenvalues of the map.
Invertibility permits a complex matrix logarithm but does not guarantee a
Hermiticity-preserving, trace-preserving, and conditionally completely
positive branch. Even when $\Phi_T(t_0)$ is CPTP, no branch is guaranteed
to be of GKLS form. The physical stroboscopic object is the map; a
time-independent Floquet Lindbladian exists only when the map is
embeddable in a CPTP semigroup
\cite{WolfEisertCubittCirac2008,WolfCirac2008,Schnell2020}.

Searching all logarithm branches is generally difficult. When the full
periodic dynamical map is available rather than only one stroboscopic
snapshot, spectral unwinding and minimization of effective micromotion can
identify useful candidate branches. This is a diagnostic and construction
strategy, not a guarantee that the map is Lindblad embeddable
\cite{Dinc2025SpectralUnwinding}.

A time-dependent GKLS generator can represent a larger class of evolutions
than a time-independent semigroup generator. CP divisibility provides one
widely used notion of time-local Markovianity, whereas memory kernels,
information backflow, and semigroup embeddability characterize different
questions. These notions coincide only under additional assumptions
\cite{RivasHuelgaPlenio2014,HallCresserLiAndersson2014}. The word
\emph{Markovian} should therefore be accompanied by the definition or
construction being used.

Initial conditions form part of the same structure. A projection equation
built around
$\mathcal P\rho=\operatorname{Tr}_B(\rho)\otimes\rho_B^{\mathrm{ref}}$
is exact for arbitrary initial states only when the corresponding
inhomogeneous term is retained. Setting that term to zero assumes that the
initial state already lies in the reference manifold. Initial slip,
slaving maps, and correlated initial states should therefore be treated as
encoding data rather than absorbed silently into the reduced physical
state.

\paragraph{Further reading.}
The foundational projection-operator formulations are
Refs.~\cite{Nakajima1958,Zwanzig1960}, with the time-convolutionless
construction developed in Ref.~\cite{ChaturvediShibata1979}. Modern
reviews of time-local generators and non-Markovianity include
Refs.~\cite{breuer2002theory,RivasHuelgaPlenio2014,HallCresserLiAndersson2014}. Channel divisibility and semigroup
embeddability are treated in Refs.~\cite{WolfEisertCubittCirac2008,WolfCirac2008}, and the Floquet-Lindbladian problem and practical branch
selection are analyzed in
Refs.~\cite{Schnell2020,Dinc2025SpectralUnwinding}.

\subsection{Origins and Interpretations of Non-Hermitian Effective Operators}
\label{subsec:advanced_nonhermitian}

Non-Hermitian effective operators arise in several parts of quantum
physics, but different constructions propagate or diagonalize different
objects. Non-Hermitian Hamiltonian, master-equation,
reservoir-engineering, and feedforward descriptions are compared in
Ref.~\cite{Clerk2022Nonreciprocal}. A complex eigenvalue can describe a
resonance pole, a conditional norm loss, a quasiparticle linewidth, or a
Liouvillian decay mode. These meanings are compared in
Table~\ref{tab:advanced_nonhermitian_origins}.

\subsubsection{Feshbach resonance operators}
\label{subsubsec:advanced_feshbach_nonhermitian}

When the eliminated $Q$ sector contains a continuum, outgoing boundary
conditions give the projected operator
\begin{equation*}
    \hat H_{\mathrm{F}}(E)
    =
    \hat P\hat H\hat P+
    \hat P\hat H\hat Q
    \frac{1}{
        E-
        \hat Q\hat H\hat Q
        +
        \mathrm{i}0^+
    }
    \hat Q\hat H\hat P.
\end{equation*}
Above a continuum threshold, the self-energy acquires a Hermitian shift
and an anti-Hermitian width,
\begin{equation*}
    \hat\Sigma(E)
    =
    \hat\Delta(E)
    -
    \frac{\mathrm{i}}{2}
    \hat\Gamma(E).
\end{equation*}
The complex poles encode resonance positions and lifetimes. Because
$\hat H_{\mathrm{F}}(E)$ depends on the spectral parameter, the resonance
problem is generally a nonlinear eigenvalue problem rather than a
standalone time-local Schr\"odinger generator
\cite{Feshbach1958,Feshbach1962,Moiseyev2011,rotter2009non}.

\subsubsection{Conditional no-jump evolution}
\label{subsubsec:advanced_no_jump}

For a GKLS equation written with the manuscript convention,
\begin{equation*}
    \dot{\rho}
    =
    -\mathrm{i}
    [\hat H,\rho]
    +
    \sum_\ell
    \left(
        \hat J_\ell\rho\hat J_\ell^{\dagger}
        -
        \frac12
        \left\{
            \hat J_\ell^{\dagger}\hat J_\ell,
            \rho
        \right\}
    \right),
\end{equation*}
a chosen quantum-jump unraveling has the unnormalized no-jump
Hamiltonian.\footnote{The rates are absorbed into the jump operators in
this subsection. If the dissipator is instead written using dimensionless
operators $\hat L_\ell$ and explicit rates $\gamma_\ell$, then
$\hat J_\ell=\sqrt{\gamma_\ell}\,\hat L_\ell$.}
\begin{equation*}
    \hat H_{\mathrm{nj}}
    =
    \hat H
    -
    \frac{\mathrm{i}}{2}
    \sum_\ell
    \hat J_\ell^{\dagger}
    \hat J_\ell.
\end{equation*}
The norm of the conditional state is the survival probability for the
specified monitoring record. The unconditional state also contains the
recycling terms
$\hat J_\ell\rho\hat J_\ell^{\dagger}$, so
$\hat H_{\mathrm{nj}}$ alone does not generate the full CPTP dynamics.
It also depends on the chosen unraveling and measurement scheme
\cite{DalibardCastinMolmer1992,daley2014quantum,roccati2022non}.

Effective-model methods enter this construction by generating both the
conditional Hamiltonian and the effective jumps. Adiabatic elimination,
projection, or a dressed-frame transformation can change
$\hat H_{\mathrm{nj}}$ through coherent shifts and through the term
$-\mathrm{i}\sum_\ell\hat J_\ell^{\dagger}\hat J_\ell/2$. Engineering a
target conditional non-Hermitian Hamiltonian therefore requires a
compatible set of recycling channels and is constrained by complete
positivity. The embedding is generally nonunique because different
unravelings can represent the same unconditional Lindblad generator.

\subsubsection{Retarded self-energies and quasiparticle poles}
\label{subsubsec:advanced_green_nonhermitian}

In an interacting many-body system, the retarded Green function
\begin{equation*}
    \hat G^R(\omega)
    =
    \left[
        \omega
        -
        \hat H_0
        -
        \hat\Sigma^R(\omega)
    \right]^{-1}
\end{equation*}
contains a complex, frequency-dependent self-energy. Its Hermitian part
renormalizes quasiparticle energies, while its anti-Hermitian part
produces finite linewidths through scattering into many-body continua.
Writing
$\hat H_{\mathrm{qp}}(\omega)=\hat H_0+\hat\Sigma^R(\omega)$
creates a useful non-Hermitian spectral problem, but it is a response-
function construction rather than a nonunitary evolution law for the full
many-body wavefunction
\cite{michishita2020equivalence,ashida2020non,Bergholtz}.

\subsubsection{Liouvillian spectra and exceptional points}
\label{subsubsec:advanced_liouvillian_nonhermitian}

A Liouvillian acts on operators rather than state vectors. Its complex
eigenvalues describe decay rates and oscillation frequencies of density-
operator modes. Coalescence of Liouvillian eigenvalues and eigenoperators
defines a Liouvillian exceptional point. Quantum jumps can shift, create,
or remove these degeneracies relative to those of a no-jump Hamiltonian;
the two spectra coincide only in restricted limits or under an explicit
postselection construction
\cite{Minganti2019Exceptional,Minganti2020HybridLiouvillian}. Non-normality can also
make transient amplification or relaxation depend strongly on eigenvector
conditioning rather than only on spectral abscissae.

\begin{table*}[!t]
    \centering
    \caption{\justifying
    Distinct origins and meanings of non-Hermitian effective operators.
    Similar complex spectra do not imply that the propagated objects,
    normalization rules, or measured quantities are the same.
    }
    \label{tab:advanced_nonhermitian_origins}
    \footnotesize

    \renewcommand{\arraystretch}{1.24}
    \setlength{\tabcolsep}{4pt}

    \begin{tabularx}{\textwidth}{
        @{}
        >{\RaggedRight\arraybackslash}p{0.25\textwidth}
        >{\RaggedRight\arraybackslash}p{0.24\textwidth}
        >{\RaggedRight\arraybackslash}X
        @{}
    }
        \toprule
        \textbf{Construction and object}
        &
        \textbf{Origin of the imaginary part}
        &
        \textbf{Normalization, omitted information, and physical meaning}
        \\
        \midrule

        \textbf{Feshbach resonance operator.}
        Energy-dependent projected state in a discrete sector
        &
        Outgoing propagation into an eliminated continuum
        &
        Nonlinear spectral problem; continuum amplitudes require
        reconstruction. It predicts resonance positions, widths, and
        scattering poles.
        \\

        \textbf{No-jump Hamiltonian.}
        Unnormalized conditional wavefunction for a chosen unraveling
        &
        Probability loss associated with possible quantum jumps
        &
        Norm equals survival probability; recycling terms and other
        trajectories are omitted. It describes postselected conditional
        dynamics and waiting-time statistics.
        \\

        \textbf{Retarded quasiparticle operator.}
        Pole problem for a single-particle Green function
        &
        Inelastic scattering encoded by the retarded self-energy
        &
        The operator is frequency dependent and does not propagate the full
        many-body wavefunction. It gives renormalized quasiparticle
        energies, linewidths, and spectral functions.
        \\

        \textbf{Liouvillian.}
        Density-operator decay mode
        &
        Dissipative spectrum of the complete master-equation generator
        &
        Trace preservation belongs to the full map; right eigenmodes
        need not represent states, while left eigenmodes encode observables
        or linear functionals. The spectrum describes relaxation,
        oscillation, metastability, and Liouvillian exceptional points.
        \\
        \bottomrule
    \end{tabularx}
\end{table*}

For a closed-system effective Hamiltonian obtained through a unitary
similarity transformation, Hermiticity is preserved when the BCH or
Magnus truncation is organized consistently. An uncompensated
anti-Hermitian term in that setting indicates that the transformation,
order counting, or reconstruction has changed the intended unitary
problem. In projected continua, conditional trajectories, retarded
response functions, and Liouvillian dynamics, the same algebraic feature
can instead have a controlled physical meaning. The origin of the
non-Hermiticity should therefore be identified before its spectrum is
interpreted.

\paragraph{Further reading.}
Feshbach projection and resonance theory originate in
Refs.~\cite{Feshbach1958,Feshbach1962} and are reviewed from a
non-Hermitian perspective in Refs.~\cite{Moiseyev2011,rotter2009non}.
Quantum trajectories are introduced in
Ref.~\cite{DalibardCastinMolmer1992} and reviewed for many-body systems in
Ref.~\cite{daley2014quantum}. Connections and distinctions between
non-Hermitian Hamiltonians and master equations are discussed in
Refs.~\cite{roccati2022non,Minganti2019Exceptional,Minganti2020HybridLiouvillian},
while complex self-energies in correlated systems are considered in
Ref.~\cite{michishita2020equivalence}. Broad reviews of non-Hermitian
spectral physics include Refs.~\cite{ashida2020non,Bergholtz}.

\subsection{Structural Diagnostics for Advanced Effective Models}
\label{subsec:advanced_diagnostics}

The structural issues reviewed above do not replace the validation
hierarchy of Sec.~\ref{sec:common_framework}. They refine it when the
reduced model involves representation-sensitive truncations,
infinite-dimensional operators, extended many-body systems, memory, or
non-Hermitian spectral objects. The audit should be selected according to
the claim: a low-energy spectroscopy calculation need not establish a
Gibbs state, and a finite-dimensional gate model need not solve an
operator-domain problem. Conversely, a thermodynamic, infinite-cutoff,
long-time, or map-level claim must pass the corresponding structural test.

\textbf{Structural audit:}
\begin{itemize}
    \item \textbf{Representation and truncation.}
    State the electromagnetic, block, or Floquet gauge; transform states,
    observables, and bath couplings consistently; and test convergence
    with retained matter levels and field modes.

    \item \textbf{Spectral window.}
    Specify the energy, occupation, or symmetry manifold in which the
    effective Hamiltonian is intended to operate. Check boundedness from
    below and the trace-class character of
    $e^{-\beta H_{\mathrm{eff}}}$ before using the model for equilibrium
    thermodynamics.

    \item \textbf{Operator domain.}
    For unbounded bosonic generators, state a dense domain or core and
    identify whether essential self-adjointness is established. Compare
    cutoff sequences, bases, parity, and boundary populations when
    numerical truncation is used.

    \item \textbf{Perturbative order and time.}
    Examine residuals and reconstructed observables across several orders.
    Report small denominators, retained resonances, an optimal order when
    relevant, and the time interval before omitted phase or transition
    errors become appreciable.

    \item \textbf{Locality and system size.}
    Report local rather than only extensive norms, the range and body
    order of induced interactions, collective normalization conventions,
    and the size dependence of Hamiltonian or Liouvillian gaps.

    \item \textbf{Maps and memory.}
    Distinguish a memory kernel, a time-local generator, and a finite-time
    map. Check map invertibility, initial-slip terms, complete positivity,
    logarithm branches, and semigroup embeddability separately.

    \item \textbf{Non-Hermitian interpretation.}
    Identify the propagated or diagonalized object, the source of the
    imaginary part, the normalization rule, any omitted recycling or
    continuum information, and whether postselection or energy dependence
    is involved.
\end{itemize}

The lesson is not that effective models are intrinsically fragile. Their
mathematical status and physical interpretation are part of the
approximation. A low-energy Hamiltonian can be entirely
reliable within its spectral window, a prethermal Floquet generator can be
predictive over exponentially long times, and a non-Hermitian operator can
faithfully encode a resonance or conditional trajectory. These statements
become precise once the retained manifold, representation, domain,
timescale, and reconstruction data are specified together.

\section{Applications of Effective-Dynamics Methods}
\label{sec:applications}

\subsection*{Reading guide for the applications}

The five applications are worked case studies rather than variations on a
single example.  Each begins with a physical task and asks what must be kept
in the reduced description for that task to be predicted reliably.  The
multiqubit cavity reveals operator sectors lost by a microscopic
rotating-wave approximation.  The three-level mixer uses topology and signed
detunings to select exchange or pair processes.  The spin-chain bus shows how
the retained space changes near a bus resonance.  The lossy mediator ties a
coherent interaction to the dissipative channel generated by the same fast
response.  The driven lattice separates an exact periodic frame from an
approximate Floquet generator and its micromotion.

Each case develops the central derivation, reconstruction rule, validity
conditions, physical interpretation, and a quantitative benchmark.  The
applications may be read in sequence as a comparison of methods, or
individually after the corresponding methods subsection.  The companion
notebooks slow the calculations down and add symbolic checks, convergence
tests, controlled failure cases, parameter scans, and further problems.
Supplemental Material, Sec.~S5, gives the notebook map and reproducibility
guide.  Table~\ref{tab:applications_learning_map} provides a compact route
through the five examples.

\begin{table*}[!ht]
\centering
\caption{Guide to the applications.  The final column joins the physical
question, the principal construction, and the diagnostics used below.}
\label{tab:applications_learning_map}
\footnotesize
\renewcommand{\arraystretch}{1.07}
\setlength{\tabcolsep}{3.5pt}
\begin{tabularx}{\textwidth}{
    @{}
    >{\RaggedRight\arraybackslash}p{0.22\textwidth}
    >{\RaggedRight\arraybackslash}X
    @{}
}
\toprule
\textbf{Application}
&
\textbf{What to follow}
\\
\midrule
Beyond-RWA multiqubit cavity
& Counter-rotating paths generate anomalous operator sectors.  Signed
harmonics, SW dressing, and projection are tested through exchange,
squeezing-sensitive covariance, pair correlations, and reduced-state error.
\\
Three-level quantum mixers
& Topology and detuning signs select exchange or pair processes.  Averaging,
small rotations, projection, and effective operators are tested through
conversion, Stark-shifted resonances, reconstructed auxiliary population,
and scattering.
\\
Spin-chain bus
& The bus spectrum and preparation determine the retained space.  Normal
modes, resolvents, and complex self-energies are tested through transfer,
concurrence, channel fidelity, bus occupation, and repair near a bus pole.
\\
Lossy bosonic mediator
& One fast complex response produces coherent and dissipative terms.  Effective
operators and Langevin response are tested against the full model using trace
distance, reconstruction, and Kossakowski/Choi checks.
\\
Floquet-controlled tunnelling
& An exact periodic frame is separated from the approximate Floquet generator.
One-period propagation and harmonic/Sambe methods test quasienergies,
stroboscopic dynamics, residual tunnelling, current micromotion, and frequency
scaling.
\\
\bottomrule
\end{tabularx}
\end{table*}
\subsection{Beyond-RWA Mediated Interactions: Exchange, Pair Processes, and Squeezing}
\label{subsec:application_beyond_rwa_multi_qubit}

\paragraph{Learning goals and conceptual focus.}
The first problem asks whether counter-rotating couplings merely correct a
familiar dispersive interaction or change the operator content of the
effective theory.  We follow the co- and counter-rotating virtual paths that
generate excitation exchange, qubit-pair terms, and conditional oscillator
squeezing.  This makes it possible to distinguish a rotating-wave
approximation made in the microscopic Hamiltonian from a later secular
approximation applied to operators that have already been generated.  It
also shows why state preparation, observable reconstruction, and the chosen
measurement window enter the validity test.  A complementary,
mechanism-first discussion of parametrically engineered exchange,
conversion, squeezing, and dissipative couplings is given in
Ref.~\cite{Metelmann2023Parametric}.

The distinction is already present in the dispersive quantum Rabi model.
Starting from the full transverse coupling produces an Ising-type
qubit--qubit interaction and a qubit-conditioned quadratic oscillator
term, rather than merely renormalizing the Jaynes--Cummings result
\cite{Zueco2009}.  A related but distinct higher-order sector converts one
photon into two qubit excitations.  It was predicted through
counter-rotating virtual paths \cite{Garziano2016OnePhotonTwoAtoms}, and
circuit proposals showed that competing paths can interfere
constructively or destructively \cite{Wang2017CounterRotatingTwoQubits}.
An ultrastrongly coupled two-flux-qubit device subsequently displayed the
associated avoided crossing spectroscopically
\cite{Tomonaga2025TwoArtificialAtoms}.  That experiment establishes the
observability of excitation-nonconserving multiqubit physics, but it does
not measure the weak-coupling second-order coefficient derived below.

\begin{figure*}[!ht]
\centering
\includegraphics[width=\textwidth]
{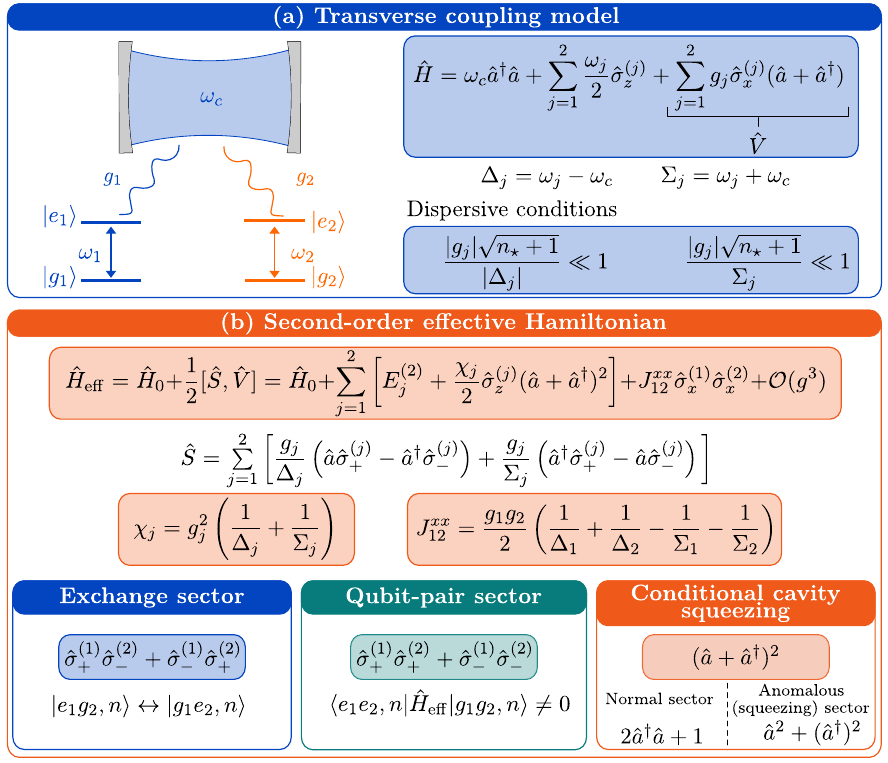}
\caption{\textbf{Beyond-RWA mediated interactions generated by a common cavity
mode.}  (a) Two nonidentical qubits couple transversely to one bosonic
mode through co-rotating coupling terms, controlled by the
difference-frequency denominators
$\Delta_j=\omega_j-\omega_c$, and counter-rotating coupling terms, controlled
by $\Sigma_j=\omega_j+\omega_c$.  Excitation exchange receives both a
$\Delta$-route and a counter-rotating $\Sigma$-route.  The static
qubit-pair sector and the conditional two-boson sector instead arise from
mixed co- and counter-rotating commutators.  With
$\hat U=e^{-\hat S}$, the transformed Hamiltonian is
$\widetilde{\hat H}=\hat U^\dagger\hat H\hat U
=e^{\hat S}\hat H e^{-\hat S}$, so that
$\hat H_{\mathrm{SW}}^{(2)}
=\hat H_0+\tfrac12[\hat S,\hat V]+\mathcal O(g^3)$.
(b) Before any additional secular truncation, the complete second-order
transformed
Hamiltonian contains ordinary exchange, a static qubit-pair operator,
and the conditional oscillator terms
$\hat\sigma_z(\hat a^2+\hat a^{\dagger2})$.  Their presence does not by
itself imply resonant pair conversion or stationary squeezing; those
claims additionally require phase matching and a resolvable observation
window.  The higher-order channel
$\lvert g_1g_2,1\rangle\leftrightarrow
 \lvert e_1e_2,0\rangle$
is shown separately because it is not the second-order static pair
coefficient.  Laboratory predictions require consistent encoding or
observable dressing,
$\widetilde X_\theta=\hat U^\dagger\hat X_\theta\hat U$, and validation
with observables sensitive to the anomalous sector.}
\label{fig:application_multiqubit_schematic}
\end{figure*}

Consider two possibly nonidentical qubits coupled to one cavity mode,
\begin{equation*}
\begin{aligned}
\hat H&=\hat H_0+\hat V,\\
\hat H_0&=
\omega_c\hat a^\dagger\hat a
+\sum_{j=1}^{2}\frac{\omega_j}{2}\hat\sigma_z^{(j)},\\
\hat V&=
\sum_{j=1}^{2}g_j\hat\sigma_x^{(j)}
(\hat a+\hat a^\dagger).
\end{aligned}
\end{equation*}
We define the difference- and sum-frequency denominators
\(\Delta_j=\omega_j-\omega_c\), \(\Sigma_j=\omega_j+\omega_c\).
The local dispersive conditions must be evaluated in the occupied cavity
sector,
\begin{equation}
\begin{aligned}
\epsilon_{j,-}(n_\star)
&=\frac{|g_j|\sqrt{n_\star+1}}{|\Delta_j|}\ll1,
\\
\epsilon_{j,+}(n_\star)
&=\frac{|g_j|\sqrt{n_\star+1}}{\Sigma_j}\ll1.
\end{aligned}
\label{eq:application_multiqubit_local_control}
\end{equation}
They control physically different virtual paths and cannot generally be
replaced by a single bare ratio such as $g_j/\omega_c$.
Figure~\ref{fig:application_multiqubit_schematic} separates operator
generation from the later decision to secularize.  The microscopic
Hamiltonian first supplies four elementary coupling components with distinct
$\Delta_j$ and $\Sigma_j$ denominators.  Their complete second-order
commutator generates exchange, qubit-pair, and conditional two-boson
operators.  Only afterward may a frame- and timescale-dependent filter be
applied.  Making the microscopic RWA first, retaining only
equal-frequency commutator products, or projecting onto a
fixed-excitation space before constructing the effective operator removes
the anomalous sector rather than merely changing its coefficient.
\paragraph{Operational and complementary constructions.}
In the interaction picture, the microscopic coupling separates into
harmonics at $\pm\Delta_j$ and $\pm\Sigma_j$.  James--Jerke bookkeeping is
a rapid operational route from these harmonics to their second-order
commutators \cite{James2007}.  A signed-harmonic construction is
important here: a filter that retains only zero beat frequency removes
mixed products such as
$e^{i\Delta_jt}e^{-i\Sigma_jt}=e^{-2i\omega_ct}$ and therefore deletes the
anomalous oscillator sector by construction.

The SW construction supplies the complementary dressed representation
and its reconstruction map \cite{Bravyi2011}.  We use
the convention
\begin{equation*}
\begin{aligned}
\hat U&=e^{-\hat S},
\qquad
\widetilde H=\hat U^\dagger\hat H\hat U
=e^{\hat S}\hat H e^{-\hat S},
\\
[\hat S,\hat H_0]&=-\hat V.
\end{aligned}
\end{equation*}
The anti-Hermitian first-order generator is
\begin{equation*}
\begin{aligned}
\hat S
={}&\sum_{j=1}^{2}
\frac{g_j}{\Delta_j}
\left(
\hat a\hat\sigma_+^{(j)}
-\hat a^\dagger\hat\sigma_-^{(j)}
\right)
\\
&+\sum_{j=1}^{2}
\frac{g_j}{\Sigma_j}
\left(
\hat a^\dagger\hat\sigma_+^{(j)}
-\hat a\hat\sigma_-^{(j)}
\right).
\end{aligned}
\end{equation*}
The Baker--Campbell--Hausdorff expansion then gives
\begin{equation*}
\hat H_{\mathrm{SW}}^{(2)}
=\hat H_0+\frac12[\hat S,\hat V]+\mathcal O(g^3).
\end{equation*}
Projection or resolvent methods provide a third view: they identify the
individual intermediate states and denominators carrying a selected
matrix element, whereas SW provides a compact global transformation of
the Hamiltonian, states, and observables.

Evaluating the full commutator before any additional secular
approximation yields
\begin{equation}
\begin{aligned}
    \hat H_{\mathrm{SW}}^{(2)}={}&\hat H_0
+\sum_{j=1}^{2}\left[
E_j^{(2)}+
\frac{\chi_j}{2}\hat\sigma_z^{(j)}
(\hat a+\hat a^\dagger)^2
\right]
\\
&+J_{12}^{xx}
\hat\sigma_x^{(1)}\hat\sigma_x^{(2)}
+\mathcal O(g^3),
\label{eq:application_multiqubit_effective_structure}
\end{aligned}
\end{equation}
with
\begin{align*}
&E_j^{(2)}=
\frac{g_j^2}{2}
\left(\frac1{\Delta_j}-\frac1{\Sigma_j}\right),
\\
&\chi_j=g_j^2
\left(\frac1{\Delta_j}+\frac1{\Sigma_j}\right),
\end{align*}
and
\begin{equation*}
J_{12}^{xx}
=\frac{g_1g_2}{2}
\left(
\frac1{\Delta_1}+\frac1{\Delta_2}
-\frac1{\Sigma_1}-\frac1{\Sigma_2}
\right).
\end{equation*}
Equation~\eqref{eq:application_multiqubit_effective_structure} is the
complete Hamiltonian produced by the first-order transformation through
second order, before a further secular or block-diagonalizing step.  It
is not a unique fully block-diagonal normal form: a second generator can
remove the remaining off-diagonal second-order operators and transfer
their effects into higher-order energies and reconstruction maps.

The enlarged operator content is exposed by
\begin{equation*}
(\hat a+\hat a^\dagger)^2
=2\hat a^\dagger\hat a+1
+\hat a^2+\hat a^{\dagger2},
\end{equation*}
and
\begin{equation}
\hat\sigma_x^{(1)}\hat\sigma_x^{(2)}=
\hat\sigma_+^{(1)}\hat\sigma_-^{(2)}
+\hat\sigma_-^{(1)}\hat\sigma_+^{(2)}+\hat\sigma_+^{(1)}\hat\sigma_+^{(2)}
+\hat\sigma_-^{(1)}\hat\sigma_-^{(2)}.
\label{eq:application_multiqubit_spin_decomposition}
\end{equation}
The first line of Eq.~\eqref{eq:application_multiqubit_spin_decomposition}
is excitation exchange; the second is the static qubit-pair sector.  The
operators $\hat a^2+\hat a^{\dagger2}$ form the conditional oscillator
squeezing sector.  A microscopic RWA deletes the coupling terms that contribute
the sum-frequency denominators.  A later secular approximation can retain
the corrected shifts and ordinary exchange while deleting the generated
pair and squeezing dynamics.

\paragraph{Four-model diagnostic hierarchy.}
The distinction is tested by comparing four descriptions: the microscopic
Rabi Hamiltonian, the complete reconstructed SW model, a deliberately
secularized beyond-RWA model, and the ordinary dispersive
Jaynes--Cummings model.  The secularized model is defined explicitly by
\begin{equation*}
\begin{aligned}
(\hat a+\hat a^\dagger)^2
&\longrightarrow 2\hat a^\dagger\hat a+1,\\
\hat\sigma_x^{(1)}\hat\sigma_x^{(2)}
&\longrightarrow
\hat\sigma_+^{(1)}\hat\sigma_-^{(2)}
+\hat\sigma_-^{(1)}\hat\sigma_+^{(2)}.
\end{aligned}
\end{equation*}
The same full SW encoding and reconstruction are used for the complete and
secularized models.  Any remaining anomalous signal in the latter can
therefore be identified as virtual dressing from the reconstruction map,
not dynamics generated by the deleted operators.

\paragraph{Complete physical prediction.}
A laboratory state must first be encoded in the transformed
representation and later reconstructed,
\begin{equation*}
\begin{aligned}
\ket{\widetilde\psi(0)}
&=\hat U^\dagger\ket{\psi(0)},\\
\ket{\psi_{\mathrm{rec}}(t)}
&=\hat U e^{-i\hat H_{\mathrm{SW}}^{(2)}t}
\hat U^\dagger\ket{\psi(0)}.
\end{aligned}
\end{equation*}
Equivalently, a laboratory observable is represented in the SW frame as
\begin{equation*}
\widetilde O
=\hat U^\dagger\hat O\hat U
=e^{\hat S}\hat Oe^{-\hat S}
=\hat O+[\hat S,\hat O]+\cdots.
\end{equation*}
This distinction matters most when the claim concerns a phase-sensitive
or anomalous sector.  For the cavity quadrature
\begin{equation*}
\hat X_\theta=
\frac{\hat a e^{-i\theta}+\hat a^\dagger e^{i\theta}}{\sqrt2},
\end{equation*}
one should reconstruct the covariance matrix and minimize the variance
over $\theta$, rather than infer squeezing from populations alone.

\begin{figure*}[!ht]
\centering
\includegraphics[width=1.0\textwidth]
{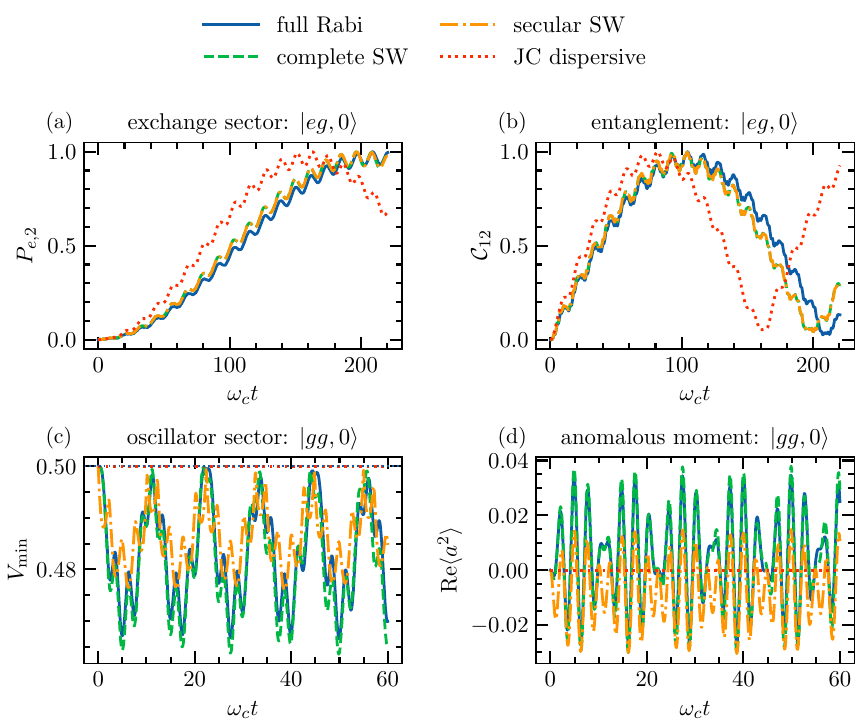}
\caption{\textbf{Claim-matched validation of the beyond-RWA reduction.}  The
microscopic Rabi model is compared with the reconstructed complete SW
model, a reconstructed secularized beyond-RWA model, and the dispersive
Jaynes--Cummings model.  Panels (a) and (b) start from
$\ket{eg,0}$ and show the excitation probability of qubit~2 and the
two-qubit concurrence.  Panels (c) and (d) start from the bare factorized
state $\ket{gg,0}$ and show the minimum cavity-quadrature variance and
$\operatorname{Re}\langle\hat a^2\rangle$.  The parameters are
$\omega_c=1$, $\omega_1=\omega_2=1.5$, $g_1=g_2=0.07$, and
$N_{\mathrm{cav}}=14$.  The lower panels diagnose quench-induced
intracavity dynamics; they are not claims of stationary output squeezing
or ground-state squeezing.}
\label{fig:application_multiqubit_validation}
\end{figure*}

Figure~\ref{fig:application_multiqubit_validation} illustrates why an
exchange curve is not a sufficient validation.  The Jaynes--Cummings
model can produce initially plausible transfer while accumulating a
phase error because its exchange coefficient differs from
$J_{12}^{xx}$.  More decisively, it predicts
$V_{\min}=1/2$ and $\langle\hat a^2\rangle=0$ because the anomalous
oscillator sector is absent.  The complete SW result follows both the
exchange and the squeezing-sensitive observables, whereas the
secularized model preserves improved exchange and static dressing but
cannot reproduce the full anomalous dynamics.  A qubit-pair moment is a
useful check against the JC model, although reconstruction alone can
produce virtual pair coherence; the reduced two-qubit trace distance is
a stricter notebook-level discriminator between the complete and
secularized descriptions.

\paragraph{State-dependent control.}
The number-conserving part of the local quadratic term gives the
Fock-sector-dependent qubit splitting
\begin{equation*}
\widetilde\omega_j(n)
=\omega_j+\chi_j(2n+1).
\end{equation*}
A thermal cavity therefore produces an incoherent average of sectors with
different effective detunings and exchange phases.  The resulting closed-system collapse is inhomogeneous phase averaging, not irreversible
Markovian dephasing.  Both the local parameters in
Eq.~\eqref{eq:application_multiqubit_local_control} and the photon cutoff
must consequently be checked over the occupied distribution rather than
only in the vacuum sector.

\paragraph{Static dressing, activated dynamics, and observability.}
The appearance of an anomalous operator in a static transformed
Hamiltonian does not guarantee a large resonant oscillation.  In the bare
interaction picture, the pair and squeezing operators generally rotate
at sum frequencies.  Off resonance they influence dressed spectra,
virtual correlations, and quench dynamics.  A modulation or a different
retained rotating frame can phase match a selected term and convert it
into a resonant pair-production or parametric process.  Exchange,
quench-induced intracavity squeezing, stationary intracavity squeezing,
propagating output squeezing, and metrologically useful collective
squeezing therefore require different rate-to-linewidth and
observation-window analyses.  No single protocol-independent
``squeezing cooperativity'' certifies all of them.

For $N$ identical qubits, the mediated pair interaction can be reorganized
through
\begin{equation*}
\sum_{j<k}\hat\sigma_x^{(j)}\hat\sigma_x^{(k)}
=2\hat J_x^2-\frac N2\hat I,
\qquad
\hat J_x=\frac12\sum_j\hat\sigma_x^{(j)}.
\end{equation*}
This is a collective quadratic-spin structure, but the large-$N$ control
parameter still depends on the occupied Dicke sector, bright--dark
structure, disorder, bosonic occupation, and the scaling of $g_j$ with
system size.

\paragraph{First failure and repair.}
An incomplete generator that cancels only co-rotating couplings leaves an
order-$g$ residual, whereas the complete first-order transformation has a
cubic remainder.  More importantly, as $g/|\Delta|$ increases, real cavity
occupation and reduced-state error rise together.  Once a nominally
eliminated state becomes dynamically active, increasing perturbative
order is not the primary repair: the state or mode must be promoted to
the retained space.  Rapidly rotating but deliberately targeted pair or
squeezing terms instead require a resonance or observation-window
redesign.  High cavity occupation requires an enlarged Fock cutoff and a
control test over every appreciably occupied sector.

\paragraph{Companion notebook N01.}
Notebook \path{N01_Beyond-RWA-Mediated-Interactions} begins with
James--Jerke harmonic bookkeeping, completes the signed-harmonic and SW
calculations, and uses projection to identify the virtual paths.  It
compares the four-model hierarchy with encoded states and reconstructed
observables; computes transfer, concurrence, quadrature squeezing,
$\langle\hat a^2\rangle$, qubit-pair moments, and reduced-state trace
distance; studies vacuum, Fock, and thermal preparation; and verifies
residual scaling, detuning breakdown, reconstruction-order sensitivity,
and cavity-cutoff convergence.  Its final problems explore
modulation-assisted activation, collective scaling, disorder, and multimode
extensions.


\subsection{Three-Level Quantum Mixers: Topology, Signed Detunings, and Effective Interactions}
\label{subsec:application_three_level_quantum_mixers}
\label{subsec:application_virtual_raman}

\paragraph{Learning goals and conceptual focus.}
Three-level systems add two pieces of structure that a two-level benchmark
cannot show: the topology of the allowed transitions and the signs of the
one-photon detunings.  Together with the retained manifold, they decide
whether the slow process is excitation exchange, correlated pair creation,
or no second-order cross process at all.  We canonicalize
negative-frequency harmonics, derive the selected interactions together
with their photon-number-dependent Stark shifts, and reconstruct the virtual
population and initial layer.  The final part closes the remaining atomic
transition with a classical drive, making clear when an atom-assisted
interaction becomes a field-only converter or squeezer.

Three-level systems underlie Raman transitions, electromagnetically induced
transparency, stimulated Raman adiabatic passage, and cavity-assisted state
control~\cite{Fleischhauer2005,Vitanov2017,Boozer2008}.  When the fields are
quantized, a virtual atomic transition also changes the occupations of the
participating modes.  The resulting dynamics can combine atomic transfer
with photon exchange, frequency conversion, two-photon absorption,
correlated pair emission, photon-number-dependent light shifts, collapse
and revival, and atom--field entanglement
\cite{GerryEberly1990,AlexanianBose1995,Wu1996,ToorZubairy1992}.
Driving the remaining atomic transition can close the interaction loop and
convert these atom-assisted processes into field-only beam-splitter or
squeezing Hamiltonians~\cite{Serra2005,Prado2006}.
Cyclic artificial atoms provide a related route to coherent sum- and
difference-frequency conversion and genuine three-wave mixing
\cite{Liu2014,HoniglDecrinis2018,Wang2015Degenerate,Zhang2019}.

Three questions guide the calculation.  First, the atomic topology decides
which operator products survive the atomic algebra.  Second, the signs of
the one-photon detunings decide which canonical beat can remain slow.  Third,
the partition between the unperturbed and interaction Hamiltonians decides
whether a driven process is counted as third order in three bare couplings or
as second order in the residual couplings of a dressed basis.  James--Jerke
averaging gives the shortest process map; a small rotation adds dressing and
observable reconstruction; and symmetrized projection makes the virtual
paths and unequal endpoint denominators explicit
\cite{James2007,Klimov2000,Klimov2002master,Brion2007}.

\begin{figure*}[t]
\centering
\includegraphics[width=\textwidth]{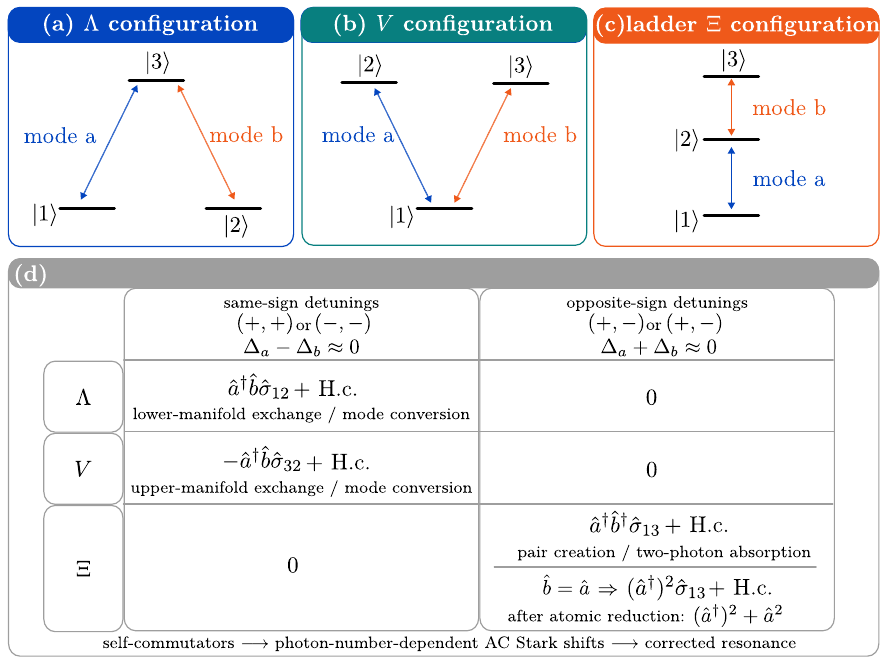}
\caption{\textbf{Atomic topology and detuning-sign selection for two quantized
modes. } Same-sign detunings select exchange-type Raman processes in the
$\Lambda$ and $V$ configurations, while their opposite-sign cross
commutators vanish.  The ladder configuration has the complementary
structure: its same-sign cross terms vanish, whereas opposite signs select
correlated pair emission and two-photon absorption.  Reversing both
detunings preserves the Hermitian interaction class but reverses the signed
denominators, including the AC Stark shifts.  In the degenerate ladder
limit $\hat b=\hat a$, the pair process becomes an atom-assisted
two-photon interaction; a field-only squeezer follows only after the
remaining atomic loop has been closed and a fixed atomic sector has been
selected.}
\label{fig:application_three_level_configuration_map}
\label{fig:application_raman_schematics}
\end{figure*}

\subsubsection{Signed harmonics and process selection}

Let
\begin{equation*}
\hat H_0
=
\sum_{j=1}^{3}E_j\hat\sigma_{jj}
+
\omega_a\hat a^\dagger\hat a
+
\omega_b\hat b^\dagger\hat b,
\qquad
\hat\sigma_{ij}=|i\rangle\langle j|.
\end{equation*}
For an upward transition $|j\rangle\to|i\rangle$, define
\begin{equation}
\Delta_{ij}^{(a)}
=
\omega_a-\omega_{ij},
\qquad
\omega_{ij}=E_i-E_j,
\label{eq:application_three_level_detuning}
\end{equation}
so that $g_a\hat a\hat\sigma_{ij}$ acquires the interaction-picture
phase $e^{-i\Delta_{ij}^{(a)}t}$.  To place every harmonic in the
positive-frequency James--Jerke form, write
\begin{equation*}
\omega_j=|\Delta_j|,
\qquad
\widetilde{\hat h}_j=
\begin{cases}
\hat h_j,&\Delta_j>0,\\
\hat h_j^\dagger,&\Delta_j<0.
\end{cases}
\end{equation*}
The potentially slow second-order cross term is
\begin{equation*}
\begin{aligned}
\hat H_{\mathrm{cross}}^{(2)}(t)
&\propto
\frac{
[\widetilde{\hat h}_a^\dagger,\widetilde{\hat h}_b]
}{
\overline{\omega}_{ab}
}
e^{i(|\Delta_a|-|\Delta_b|)t}
+\mathrm{H.c.},
\\
\frac{1}{\overline{\omega}_{ab}}
&=
\frac12
\left(
\frac1{|\Delta_a|}
+
\frac1{|\Delta_b|}
\right).
\end{aligned}
\end{equation*}
A retained process therefore requires both a nonzero commutator and a slow
canonical beat.  Same-sign detunings satisfy
$\Delta_a-\Delta_b\simeq0$, whereas opposite-sign detunings satisfy
$\Delta_a+\Delta_b\simeq0$.  Figure
\ref{fig:application_three_level_configuration_map} collects the complete
topology-by-sign rule.  In each allowed sector, the displayed monomial
represents the corresponding Hermitian interaction class; reversing both
detunings reverses the signed coefficient and selects the
Hermitian-conjugate orientation, but does not create a new process class.

For the $\Lambda$ configuration,
\begin{equation*}
\hat h_a=g_a\hat a\hat\sigma_{31},
\qquad
\hat h_b=g_b\hat b\hat\sigma_{32},
\end{equation*}
and for positive same-sign detunings
\begin{equation*}
[\hat h_a^\dagger,\hat h_b]
=
g_a^*g_b
\hat a^\dagger\hat b\hat\sigma_{12}.
\end{equation*}
For the $V$ configuration the corresponding commutator is
\begin{equation*}
[\hat h_a^\dagger,\hat h_b]
=
-g_a^*g_b
\hat a^\dagger\hat b\hat\sigma_{32}.
\end{equation*}
The isolated minus sign can be absorbed into a phase convention, but
relative signs become observable when several virtual paths interfere.

For the ladder configuration,
\begin{equation*}
\hat h_a=g_a\hat a\hat\sigma_{21},
\qquad
\hat h_b=g_b\hat b\hat\sigma_{32}.
\end{equation*}
Opposite detuning signs reverse one canonical harmonic and expose
\begin{equation*}
[\hat h_a^\dagger,\hat h_b^\dagger]
=
g_a^*g_b^*
\hat a^\dagger\hat b^\dagger\hat\sigma_{13}.
\end{equation*}

In the slow rotating frame, with the residual two-photon phase absorbed
into the effective detuning, the corresponding cross Hamiltonians are
\begin{align*}
\hat H_{\Lambda,\mathrm{cross}}^{(2)}
&=
\frac{g_a^*g_b}{2}
\left(
\frac1{\Delta_a}
+
\frac1{\Delta_b}
\right)
\hat a^\dagger\hat b\hat\sigma_{12}
+\mathrm{H.c.},
\\
\hat H_{V,\mathrm{cross}}^{(2)}
&=
-\frac{g_a^*g_b}{2}
\left(
\frac1{\Delta_a}
+
\frac1{\Delta_b}
\right)
\hat a^\dagger\hat b\hat\sigma_{32}
+\mathrm{H.c.},
\\
\hat H_{\Xi,\mathrm{cross}}^{(2)}
&=
\frac{g_a^*g_b^*}{2}
\left(
\frac1{\Delta_a}
-
\frac1{\Delta_b}
\right)
\hat a^\dagger\hat b^\dagger\hat\sigma_{13}
+\mathrm{H.c.}.
\end{align*}
For $(-,-)$ in the $\Lambda$ and $V$ systems, and for $(-,+)$ in the
ladder system, canonicalization produces the Hermitian-conjugate
orientation of the same exchange or pair process.

\subsubsection{Stark shifts, dressing, and controlled validity}

Self-commutators occur at the same perturbative order as the cross terms
and cannot be discarded without moving the resonance.  For example, the
$\Lambda$ configuration contains
\begin{align*}
\hat H_{\mathrm{Stark}}^\Lambda
={}&
\frac{|g_a|^2}{\Delta_a}
\left[
\hat n_a\hat\sigma_{11}
-
(\hat n_a+1)\hat\sigma_{33}
\right]
\\
&+
\frac{|g_b|^2}{\Delta_b}
\left[
\hat n_b\hat\sigma_{22}
-
(\hat n_b+1)\hat\sigma_{33}
\right].
\end{align*}
The operator
$\hat a^\dagger\hat b\hat\sigma_{12}$ connects
$|2;n_a,n_b\rangle$ and $|1;n_a+1,n_b-1\rangle$.  In this
sector the corrected Raman condition is
\begin{equation*}
\Delta_a-\Delta_b
+
\frac{|g_a|^2(n_a+1)}{\Delta_a}
-
\frac{|g_b|^2n_b}{\Delta_b}
\simeq0.
\end{equation*}
The Hermitian-conjugate orientation gives the equivalent expression with
the occupations relabeled according to the connected Fock states.  For a
Fock state this is a number-resolved resonance.  Coherent or thermal fields
sample a distribution of Stark shifts and effective couplings, which can
generate conditional phases, dephasing, collapse, and revival.  Reversing
a detuning reverses its Stark contribution even when the Hermitian
interaction class is unchanged.

The perturbative parameter is local to the occupied model space.  An
invariant definition is
\begin{equation*}
\epsilon_{\mathrm{loc}}
=
\max_{\substack{|\alpha\rangle\in P\\|\mu\rangle\in Q}}
\frac{
|\langle\mu|\hat V|\alpha\rangle|
}{
|E_\alpha-E_\mu|
}.
\end{equation*}
For a field state supported below occupations $n_{a,\max}$ and
$n_{b,\max}$, a convenient sufficient bound is
\begin{equation*}
\epsilon_{\mathrm{loc}}
\lesssim
\max\left[
\frac{|g_a|\sqrt{n_{a,\max}+1}}{|\Delta_a|},
\frac{|g_b|\sqrt{n_{b,\max}+1}}{|\Delta_b|}
\right].
\end{equation*}
The exact square-root factors follow from the particular connected Fock
states.  A second-order energy error begins at fourth order in the
couplings, but a small frequency error can accumulate into a visible
long-time phase error.  Perturbative order therefore does not define an
unlimited predictive time window~\cite{ToorZubairy1992,GamelJames2010}.
The first structural failure occurs when a one-photon denominator becomes
small or a Fock-enhanced matrix element is no longer small compared with
it.  The correct repair is to retain the nearly resonant intermediate
state, not to add a higher-order coefficient to an endpoint-only model.

The three closed-system constructions organize this physics differently.
James--Jerke filtering identifies the surviving harmonic most
economically.  A small rotation,
\begin{equation*}
\hat S
=
\sum_j
\frac{\hat h_j^\dagger-\hat h_j}{\Delta_j},
\qquad
\hat H^{(2)}
=
\frac12[\hat S,\hat V],
\end{equation*}
also transforms states and observables.  Symmetrized projection writes
each retained matrix element as
\begin{align*}
\langle\alpha|
\hat H_{\mathrm{proj}}^{(2)}
|\beta\rangle
={}&
\frac12
\sum_{\mu\in Q}
V_{\alpha\mu}V_{\mu\beta}\\
&\times\left(
\frac1{E_\alpha-E_\mu}
+
\frac1{E_\beta-E_\mu}
\right),
\end{align*}
which remains explicitly Hermitian for unequal endpoint denominators and
makes the local Fock-state matrix elements visible.

The same small rotation reconstructs the virtual intermediate state.  In
a classically driven Raman reduction,
\begin{align*}
c_e^{\mathrm{slow}}(t)
\simeq{}&
-
\frac{
\Omega_1c_1(t)+\Omega_2c_2(t)
}{
2\Delta
}\\
&+
\frac{i}{2\Delta^2}
\frac{d}{dt}
\left[
\Omega_1c_1(t)+\Omega_2c_2(t)
\right].
\end{align*}
For a suddenly prepared bare state, this slow component does not by itself
satisfy $c_e(0)=0$.  The leading initial-layer correction is
\begin{equation*}
c_e^{\mathrm{rec}}(t)
=
c_e^{\mathrm{slow}}(t)
-
c_e^{\mathrm{slow}}(0)e^{i\Delta t},
\end{equation*}
showing explicitly that an eliminated level is weakly occupied rather than
identically absent.

\subsubsection{Exact benchmarks}

\begin{figure*}[!ht]
\centering
\includegraphics[width=1.0\textwidth]
{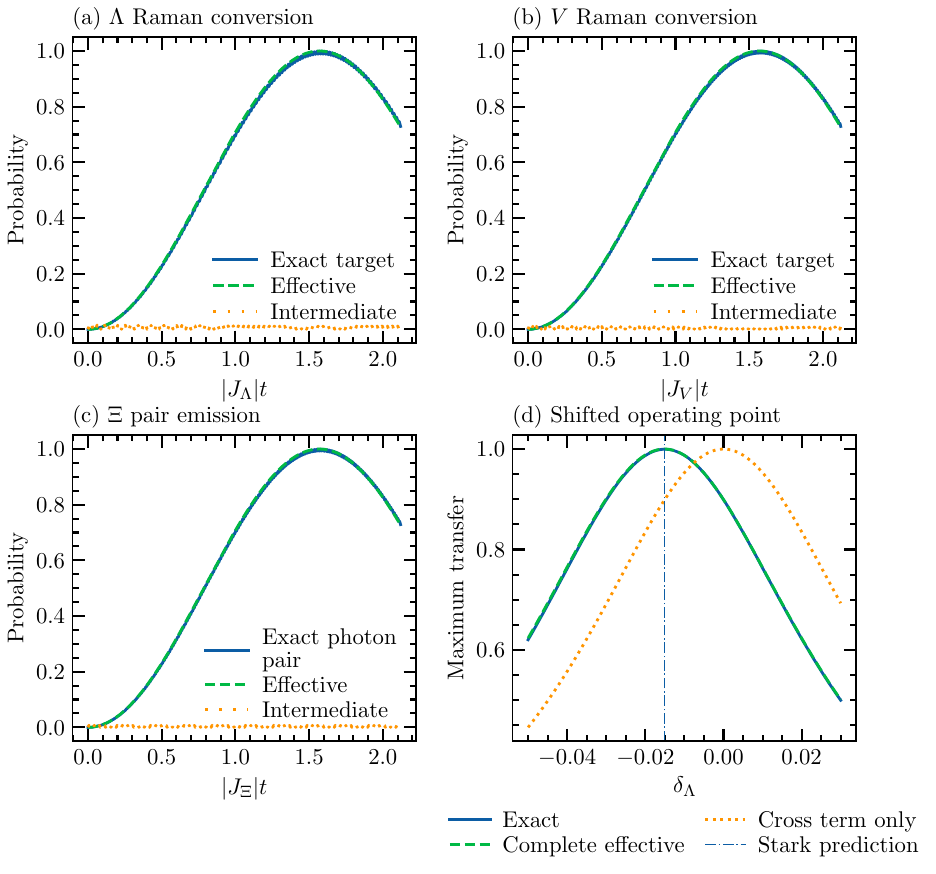}
\caption{\textbf{Physical benchmarks for quantized three-level processes.}
(a) Same-sign detunings select the $\Lambda$ exchange resonance near
$\Delta_b/\Delta_a=+1$, whereas opposite signs select the ladder pair
resonance near $\Delta_b/\Delta_a=-1$.  (b) Exact and complete effective
Raman conversion, together with the small intermediate-state population.
(c) Exact and complete effective correlated pair emission.  (d) The
differential AC Stark shift moves the maximum conversion away from the
bare two-photon resonance; retaining the cross term while deleting the
diagonal shifts misses this displacement.}
\label{fig:application_three_level_quantitative}
\label{fig:application_raman_validation}
\end{figure*}

The exact invariant-sector calculations summarized in
Fig.~\ref{fig:application_three_level_quantitative} test
four complementary consequences of the effective construction:
same-sign versus opposite-sign selection, fully quantized Raman
conversion, spontaneous pair emission, and the Stark displacement of the
two-photon resonance.

The weak but nonzero intermediate population confirms that the eliminated
level is virtual rather than absent.  In the controlled regime the complete
second-order model follows the exact dynamics, whereas a cross-term-only
model predicts the wrong operating point.  The benchmarks also distinguish
a quantitative perturbative error, which can be reduced by increasing the
scale separation, from a structural failure near one-photon resonance,
which requires an enlarged retained space.

\subsubsection{Closing the atomic loop: field-only converters and squeezers}

The second-order atlas produces atom-assisted monomials such as
$\hat a^\dagger\hat b\hat\sigma_{12}$ and
$\hat a^\dagger\hat b^\dagger\hat\sigma_{13}$.  A purely bosonic
Hamiltonian requires one additional physical operation.  A concrete
construction drives the remaining atomic transition and prepares the atom
in a sector left invariant by the effective dynamics
\cite{Serra2005,Prado2006}.

We use the classical-drive convention
\begin{equation*}
\begin{aligned}
\hat H_c^\Lambda(t)
&=
\Omega e^{-i\omega_ct}\hat\sigma_{12}
+\mathrm{H.c.},
\\
\hat H_c^\Xi(t)
&=
\Omega e^{-i\omega_ct}\hat\sigma_{31}
+\mathrm{H.c.},
\end{aligned}
\end{equation*}
with no additional factor of $1/2$ in the definition of $\Omega$.

For a driven $\Lambda$ system, the classical field closes the transition
between the two lower states.  In the equal-detuning weak-drive regime,
and within the prepared upper-state sector $|3\rangle$, the field dynamics
takes the form
\begin{equation}
\begin{aligned}
\hat H_{\mathrm{BS}}^{(3)}(t)
={}&
\chi_a^{(3)}\hat n_a
+
\chi_b^{(3)}\hat n_b\\
&+
\zeta_\Lambda
\hat a\hat b^\dagger e^{-i\delta_ct}
+
\zeta_\Lambda^*
\hat a^\dagger\hat b e^{i\delta_ct},
\label{eq:application_three_level_driven_lambda}
\end{aligned}
\end{equation}
where $\chi_a^{(3)}$ and $\chi_b^{(3)}$ are the signed dispersive shifts
evaluated in the prepared atomic sector and, with the convention above,
\(\zeta_\Lambda = \tfrac{g_ag_b^*\Omega}{\Delta^2}.\)
The residual pump detuning must cancel the differential dispersive shift,
\(\delta_c+\chi_a^{(3)}-\chi_b^{(3)} \simeq0.\)
After this compensation, the remaining interaction is a beam splitter or
coherent frequency converter.

For a driven ladder system, the classical field closes the outer
transition $|1\rangle\leftrightarrow|3\rangle$.  Within the prepared
intermediate-state sector $|2\rangle$, the field dynamics is
\begin{equation}
\begin{aligned}
\hat H_{\mathrm{TMS}}^{(2)}(t)
={}&
\chi_a^{(2)}\hat n_a
+
\chi_b^{(2)}\hat n_b \\
&+
\zeta_\Xi
\hat a\hat b e^{-i\delta_ct}
+
\zeta_\Xi^*
\hat a^\dagger\hat b^\dagger e^{i\delta_ct},
\label{eq:application_three_level_driven_xi}
\end{aligned}
\end{equation}
where the superscript again denotes the prepared atomic sector and
\(\zeta_\Xi = \tfrac{g_ag_b\Omega}{\Delta^2}.\)
Here the pump compensates the sum of the sector-dependent dispersive
shifts,
\(\delta_c+\chi_a^{(2)}+\chi_b^{(2)} \simeq0.\)
The difference-versus-sum rule for the pump correction mirrors the
exchange-versus-pair distinction in the undriven sign atlas.  Equations
\eqref{eq:application_three_level_driven_lambda} and
\eqref{eq:application_three_level_driven_xi} are field-only descriptions
only while the prepared atomic sector remains invariant on the task
timescale.  Residual atomic transfer or atom--field entanglement is then a
leakage error, not a renormalization of the bosonic coupling, and requires
the corresponding atomic states to be restored to the retained model.

These driven processes also illustrate that perturbative order depends on
the partition.  If the classical drive is counted as a weak interaction,
$\zeta\sim g_ag_b\Omega/\Delta^2$ is third order in three bare couplings.
If a strong drive is included in $\hat H_0$ and the atomic subspace is
diagonalized first, the same converter appears at second order in the
remaining quantum couplings, with a characteristic dressed-basis scale
$\sim g_ag_b/\Omega$ and coefficients conditioned on the prepared dressed
state~\cite{Prado2006}.  The effective order is
therefore not a property of the laboratory Hamiltonian alone; it also
depends on the rotating frame, partition, and retained sector.

In the degenerate ladder limit $\hat b\to\hat a$, the undriven atlas gives
the atom-assisted process
\begin{equation*}
\hat H_{\mathrm{deg}}^{(2)}
\propto
\hat a^{\dagger2}\hat\sigma_{13}
+
\hat a^2\hat\sigma_{31},
\qquad
2\omega_a\simeq\omega_{31}.
\end{equation*}
After the driven loop is closed and the atom is fixed in the intermediate
sector, the field-only Hamiltonian becomes
\begin{equation*}
\hat H_{\mathrm{sq}}
=
\zeta\hat a^2+\zeta^*\hat a^{\dagger2}.
\end{equation*}
For
$\hat X_\theta=
(\hat a e^{-i\theta}+\hat a^\dagger e^{i\theta})/2$,
the optimally squeezed quadrature evolves from vacuum or a coherent state
as
\begin{equation*}
\operatorname{Var}(\hat X_{\min})
=
\frac14e^{-2r_{\mathrm{sm}}},
\qquad
r_{\mathrm{sm}}=2|\zeta|\tau.
\end{equation*}

For nondegenerate modes, the driven ladder interaction produces a
two-mode squeezed vacuum,
\begin{equation*}
\begin{aligned}
|\mathrm{TMSV}\rangle
&=
\frac1{\cosh r_{\mathrm{tm}}}
\sum_{n=0}^{\infty}
\left(
e^{i\phi}\tanh r_{\mathrm{tm}}
\right)^n
|n,n\rangle,
\\
r_{\mathrm{tm}}&=|\zeta_\Xi|\tau.
\end{aligned}
\end{equation*}
With
$\hat X_j=(\hat j+\hat j^\dagger)/2$ and
$\hat P_j=(\hat j-\hat j^\dagger)/(2i)$,
\begin{equation*}
\operatorname{Var}(\hat X_a-\hat X_b)
=
\operatorname{Var}(\hat P_a+\hat P_b)
=
\frac12e^{-2r_{\mathrm{tm}}}.
\end{equation*}

The driven $\Lambda$ converter has an equally transparent output.  At a
compensated 50:50 interaction time,
$|\zeta_\Lambda|\tau=\pi/4$, an input $|N,0\rangle$ becomes, up to a
convention-dependent mode phase,
\begin{equation*}
|\psi_N\rangle
=
\frac1{2^{N/2}}
\sum_{n=0}^{N}
(-ie^{i\phi})^n
\sqrt{\binom{N}{n}}
|N-n,n\rangle.
\end{equation*}
Thus the same three-level platform connects microscopic topology not only
to an effective operator but also to directly useful bosonic resources:
coherent conversion, mode entanglement, single-mode squeezing, and EPR
correlations.

\subsubsection{When the eliminated level decays:
coherent Raman coupling and residual scattering}
\label{subsubsec:application_three_level_open_system}

The closed-system derivations above treat the intermediate atomic level as
virtually populated.  If that level can decay, its population may remain
small while still producing real photon-scattering events.  Eliminating a
lossy level must therefore generate two objects simultaneously: an
effective Hamiltonian describing the coherent Stark and Raman processes,
and effective jump operators describing the residual scattering.  Keeping
only the Hamiltonian gives an incomplete reduced dynamics.

Let $P$ project onto the retained atomic manifold and $Q=1-P$ onto the
eliminated manifold.  We separate the interaction into the operators
\begin{equation*}
\hat V_+=Q\hat V P,
\qquad
\hat V_-=P\hat V Q=\hat V_+^\dagger,
\end{equation*}
which excite and de-excite the eliminated manifold, respectively.  Decay
from that manifold is described by jump operators $\hat L_r$.  Its
conditional Hamiltonian is
\begin{equation*}
\hat H_{\mathrm{NH}}
=
Q\hat H Q
-
\frac{i}{2}
\sum_r
\hat L_r^\dagger\hat L_r.
\end{equation*}
The inverse of $\hat H_{\mathrm{NH}}$ is the propagator for a virtual
excursion into the lossy manifold.  The effective-operator construction
then gives~\cite{Reiter2012Effective}
\begin{align*}
\hat H_{\mathrm{eff}}
&=
P\hat H P
-
\frac12
\hat V_-
\left(
\hat H_{\mathrm{NH}}^{-1}
+
\hat H_{\mathrm{NH}}^{-1\dagger}
\right)
\hat V_+,
\\
\hat L_{r,\mathrm{eff}}
&=
\hat L_r
\hat H_{\mathrm{NH}}^{-1}
\hat V_+.
\end{align*}
The reduced density operator consequently obeys
\begin{equation}
\frac{d\hat\rho_P}{dt}
=
-i[\hat H_{\mathrm{eff}},\hat\rho_P]
+
\sum_r
\mathcal D[\hat L_{r,\mathrm{eff}}]\hat\rho_P.
\label{eq:application_three_level_effective_master_equation}
\end{equation}
Here and below,
\begin{equation*}
\mathcal D[\hat L]\hat\rho
=
\hat L\hat\rho\hat L^\dagger
-
\frac12
\{\hat L^\dagger\hat L,\hat\rho\}.
\end{equation*}

The physical content is transparent in a lossy $\Lambda$ system.  Let
$|1\rangle$ and $|2\rangle$ form the retained manifold and let the excited
state $|3\rangle$ decay through
\begin{equation*}
\hat L_r
=
\sqrt{\gamma_r}\,
\hat\sigma_{r3},
\qquad
\Gamma=\sum_r\gamma_r.
\end{equation*}
Writing
\begin{equation*}
\begin{aligned}
\hat V_+
&=
\hat A_1\hat\sigma_{31}
+
\hat A_2\hat\sigma_{32},\\
\hat A_1&=g_a\hat a,
\qquad
\hat A_2=g_b\hat b.
\end{aligned}
\end{equation*}
includes both quantized cavity couplings.  In the simplest equal-detuning
case, and with the convention
\begin{equation*}
Q\hat H Q
=
\Delta\hat\sigma_{33}.
\end{equation*}
Here \(\Delta\) is the signed energy of the eliminated excited manifold in
this rotating frame.  It is not the laboratory detuning defined in
Eq.~\eqref{eq:application_three_level_detuning}: for the absorption term
\(\hat a\hat\sigma_{31}\), the same frame gives
\(\Delta=\omega_{31}-\omega_a=-\Delta_{31}^{(a)}\).  Stating the block
energy explicitly removes the otherwise common sign ambiguity when the
closed- and open-system formulas are compared.  With this convention, the
non-Hermitian propagator is
\begin{equation*}
\hat H_{\mathrm{NH}}^{-1}
=
\frac{\hat\sigma_{33}}
{\Delta-i\Gamma/2}.
\end{equation*}
Its Hermitian part produces
\begin{equation*}
\hat H_{\mathrm{eff}}
=
P\hat H P
-
\frac{\Delta}
{\Delta^2+(\Gamma/2)^2}
\hat V_-\hat V_+,
\end{equation*}
where
\begin{equation*}
\begin{aligned}
\hat V_-\hat V_+
={}&
\hat A_1^\dagger\hat A_1\hat\sigma_{11}
+
\hat A_2^\dagger\hat A_2\hat\sigma_{22}
\\
&+
\hat A_1^\dagger\hat A_2\hat\sigma_{12}
+
\hat A_2^\dagger\hat A_1\hat\sigma_{21}.
\end{aligned}
\end{equation*}
The first line gives the photon-number-dependent Stark shifts.  The second
line gives the coherent Raman exchange.  Both are reduced when the excited
state linewidth is no longer negligible compared with the detuning.

The same virtual paths also produce
\begin{equation}
\hat L_{r,\mathrm{eff}}
=
\frac{\sqrt{\gamma_r}}
{\Delta-i\Gamma/2}
\left(
\hat A_1\hat\sigma_{r1}
+
\hat A_2\hat\sigma_{r2}
\right).
\label{eq:application_three_level_lossy_lambda_Leff}
\end{equation}
The diagonal atomic terms describe Rayleigh-like scattering that can reveal
which retained state was occupied and therefore dephase a superposition.
The off-diagonal terms describe spontaneous Raman transitions between the
retained states.  Thus the same virtual intermediate population that
produces a useful coherent coupling also creates an unavoidable residual
error channel.

In the far-detuned regime,
\begin{equation*}
|\Delta|\gg\Gamma,
\end{equation*}
the characteristic coherent and dissipative scales behave as
\begin{align*}
&\Omega_{\mathrm{Raman}}
\sim
\frac{A^2}{|\Delta|},
\qquad
\Gamma_{\mathrm{sc}}
\sim
\Gamma\frac{A^2}{\Delta^2},
\\
&\frac{\Gamma_{\mathrm{sc}}}
{\Omega_{\mathrm{Raman}}}
\sim
\frac{\Gamma}{|\Delta|},
\end{align*}
where $A$ denotes a representative coupling matrix element.  Increasing
the detuning therefore suppresses scattering relative to the coherent
Raman process, but it also slows the desired interaction.  The operating
point is a compromise between coherent speed, virtual population, and
spontaneous-scattering error.

Replacing
\(\Delta \longrightarrow \Delta-i\Gamma/2\)
only inside an effective Hamiltonian does not reproduce
Eq.~\eqref{eq:application_three_level_effective_master_equation}.  Such a
non-Hermitian Hamiltonian describes conditional evolution in which no
decay event has been observed.  The unconditional state, averaged over
both no-jump evolution and scattering events, requires the effective jump
operators in
Eq.~\eqref{eq:application_three_level_lossy_lambda_Leff}.

\paragraph{When does the elimination fail?}
The relevant separation is now the complex gap of the eliminated
manifold.  A sufficient local condition is
\begin{equation*}
\epsilon_{\mathrm{NH}}
=
\max_{\substack{|\alpha\rangle\in P\\|\mu\rangle\in Q}}
\frac{
|\langle\mu|\hat V|\alpha\rangle|
}{
|\Delta_\mu-i\Gamma_\mu/2|
}
\ll1.
\end{equation*}
When this condition fails, the intermediate state develops appreciable
population or evolves on the same time scale as the retained variables.
Near a one-photon resonance, the corresponding atomic transition must be
promoted into the retained model.  The correct description is then an
explicit three-level master equation, not an endpoint-only Hamiltonian
with a higher-order correction.

A residual two-photon or pump mismatch is a different issue.  It does not
by itself invalidate the elimination, but it must remain in the slow
Hamiltonian together with the differential or summed Stark shifts.  The
effective converter or squeezer is resonant only after these shifts are
compensated.

An analogous warning applies to time averaging.  If the eliminated fast
dynamics is coupled to unresolved fluctuations or dissipative channels,
coarse graining can generate effective decoherence as well as a coherent
Hamiltonian~\cite{GamelJames2010}.  A Hamiltonian-only average is justified
only when the corresponding dissipative contributions are parametrically
negligible.

\paragraph{Companion notebook N06.}
Notebook \path{N06_Three-Level-Quantum-Mixers} derives all four
detuning-sign sectors for the $\Lambda$, $V$, and $\Xi$ configurations;
verifies the complete
commutator map; compares James--Jerke, small-rotation, and projection
constructions; reconstructs the virtual intermediate state and its initial
layer; tests exact versus effective Raman, nondegenerate-pair, and
degenerate-pair dynamics; resolves Stark-shifted resonances and local
validity parameters; and implements the controlled-coherence field-only
extension discussed above.  A final exercise asks the reader to implement
the effective-jump construction; the numerical benchmarks in this
application concern the closed-system processes.

\subsection{Quantum-State Transfer through a Spin-Chain Bus}
\label{subsec:application_spin_chain_bus}

\paragraph{Learning goals and conceptual focus.}
Here the spectrum of a finite quantum bus tells us what the reduced model
must contain.  We derive the normal modes of an open $XX$ chain and compare
two choices: an endpoint-only Hamiltonian when every bus mode is virtual,
and a selected-mode Hamiltonian when one mode approaches resonance.  The
virtual bus profile is then reconstructed and tested through transfer
probability, sender--receiver concurrence, average channel fidelity, and bus
occupation.  Preparation and noise matter independently of this spectral
choice, so polarized, initialization-free, thermal, disordered, and lossy
buses lead naturally to Hamiltonian, master-equation, or finite-time-channel
descriptions.

Spin chains provide a route to long-range connectivity without moving the
qubits or controlling every intermediate site.  The original uniform-chain
proposal established an unmodulated spin chain as a quantum communication
channel~\cite{Bose2003}; engineered spectra can produce perfect transfer
\cite{Christandl2004}, whereas weak endpoint couplings can turn an otherwise
uniform chain into an almost-perfect wire~\cite{Wojcik2005}.  These approaches
represent different uses of the same spectrum: ballistic propagation through
many modes, exact mirror inversion, or perturbative coupling through a small
part of the bus response.

Two effective descriptions arise from the same microscopic
sender--bus--receiver system.  If all normal modes are off resonance, they can
mediate a direct endpoint interaction while remaining only virtually
occupied.  This nonlocal-tunneling regime suppresses exposure to channel loss
at the price of a slower transfer~\cite{MoraesNeto2011,MoraesNeto2012,Almeida2018}.  If one isolated mode is resonant, that mode belongs in the
retained space; the resulting selected-mode protocol is faster but
bus-populating and memory bearing~\cite{Oh2011,Lorenzo2013,Paganelli2013,Gratsea2018}.  Sender--receiver concurrence is therefore as important as
receiver population: the bus may first create a remote Bell pair and only
later complete the state transfer~\cite{Banchi2010}.

\begin{figure*}[!ht]
\centering
\includegraphics[width=\textwidth]
{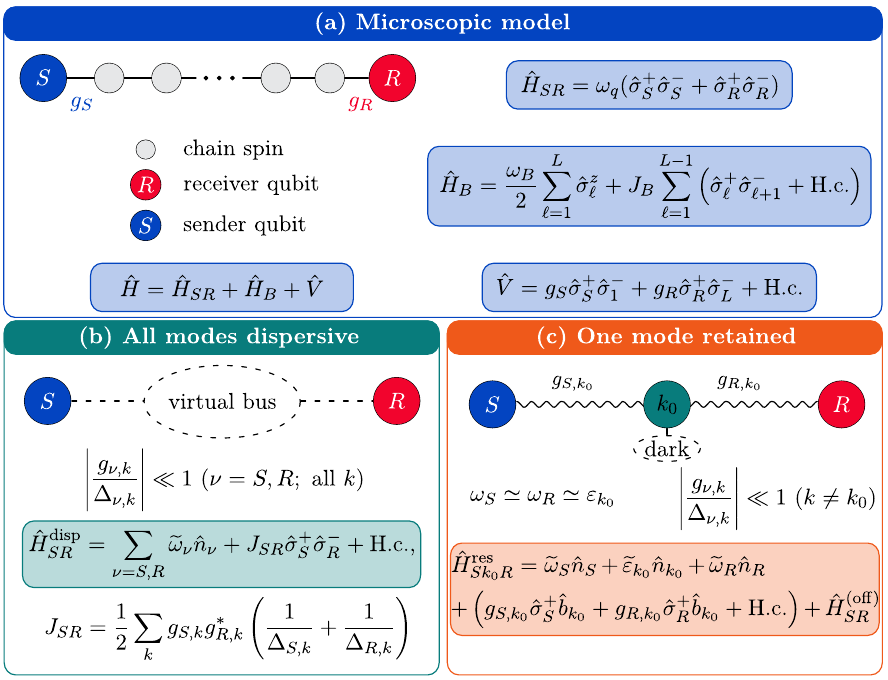}
\caption{\textbf{Retained-space choices for a finite quantum bus.}
(a) Sender and receiver couple to the ends of an open $XX$ chain.
(b) If every relevant mode satisfies
$|g_{\nu,k}/\Delta_{\nu,k}|\ll1$, eliminating the bus gives an endpoint
exchange equal to a signed sum over virtual modes; the bus occupation is
small, of order $|g/\Delta|^2$, but not zero.
(c) A spectrally isolated near-resonant mode must instead be retained.  It
couples to one bright endpoint combination while the orthogonal combination
is dark; its population, loss, and memory remain explicit.}
\label{fig:application_spin_bus_protocol_map}
\end{figure*}

A polarized initial state is sufficient for the elementary
one-excitation derivation, but it is not a universal prerequisite.
Initialization-free protocols may instead use endpoint operations and
feed-forward~\cite{DiFranco2008}, or Jordan--Wigner fermions with a two-spin
encoding in random unpolarized chains~\cite{Yao2011}.  These protocols change
the encoding and retained structure rather than merely correcting the
polarized endpoint Hamiltonian.  Recent semiconductor and superconducting
experiments make these distinctions directly relevant
\cite{Kandel2021,Xiang2024,Roy2025}.
Figure~\ref{fig:application_spin_bus_protocol_map} organizes the retained-space
choices according to the bus spectrum and preparation.  Spectrally, all
appreciably coupled modes may be eliminated, one isolated mode may have to
remain explicit, or a cluster of nearby modes may require a multimode retained
space.  An unknown, excited, or thermal bus defines a separate preparation
and encoding branch rather than a further point along this spectral
hierarchy.

\subsubsection{Microscopic model and boundary response}

We consider sender and receiver qubits locally coupled to the endpoints of an
open $XX$ chain,
\begin{equation*}
\hat H=\hat H_{SR}+\hat H_B+\hat V,
\end{equation*}
with
\begin{equation*}
\hat H_B
=\frac{\omega_B}{2}\sum_{\ell=1}^{L}\hat\sigma_\ell^z
+J_B\sum_{\ell=1}^{L-1}
\left(\hat\sigma_\ell^+\hat\sigma_{\ell+1}^-+\mathrm{H.c.}\right),
\end{equation*}
and
\begin{equation*}
\hat V
=g_S\hat\sigma_S^+\hat\sigma_1^-
+g_R\hat\sigma_R^+\hat\sigma_L^-+\mathrm{H.c.}
\end{equation*}
For a polarized bus and one transferred excitation, the chain maps exactly to
a finite tight-binding problem.  Its normal modes are
\begin{equation*}
\begin{aligned}
\varepsilon_k
&=\omega_B+2J_B\cos\frac{k\pi}{L+1},\\
\phi_k(\ell)
&=\sqrt{\frac2{L+1}}\sin\frac{k\pi\ell}{L+1}.
\end{aligned}
\end{equation*}
with endpoint couplings
$g_{S,k}=g_S\phi_k(1)$ and $g_{R,k}=g_R\phi_k(L)$.  The parity relation
\begin{equation*}
\phi_k(L)=(-1)^{k+1}\phi_k(1)
\end{equation*}
makes the alternating signs of the virtual paths explicit.

Projection onto the endpoints expresses the bus contribution as the boundary
self-energy
\begin{equation*}
\boldsymbol\Sigma(E)
=\hat W^\dagger(E-\hat H_B)^{-1}\hat W,
\end{equation*}
where $\hat W$ couples the sender and receiver to bus sites $1$ and $L$.
The off-diagonal element is the boundary-to-boundary Green function
\(\Sigma_{SR}(E) =g_Sg_R^*\langle L|(E-\hat H_B)^{-1}|1\rangle.\)
For the solvable $XX$ chain this becomes an explicit mode sum, but the
response formulation also extends to interacting buses whose boundary Green
function must be computed numerically.

\subsubsection{All-mode dispersive protocol: nonlocal virtual tunneling}
\label{subsubsec:application_spin_bus_dispersive}

Let $\Delta_{\nu,k}=\omega_\nu-\varepsilon_k$ for
$\nu\in\{S,R\}$.  If every appreciably coupled mode satisfies
$|g_{\nu,k}/\Delta_{\nu,k}|\ll1$, a mode-by-mode small rotation gives
\begin{equation*}
\hat H_{SR}^{\mathrm{disp}}
=\sum_{\nu=S,R}\widetilde\omega_\nu\hat n_\nu
+J_{SR}\hat\sigma_S^+\hat\sigma_R^-+\mathrm{H.c.},
\end{equation*}
where
\begin{equation*}
\delta\omega_\nu
=\sum_k\frac{|g_{\nu,k}|^2}{\Delta_{\nu,k}},
\qquad
\widetilde\omega_\nu=\omega_\nu+\delta\omega_\nu,
\end{equation*}
and
\begin{equation}
J_{SR}
=\frac12\sum_k g_{S,k}g_{R,k}^*
\left(\frac1{\Delta_{S,k}}+\frac1{\Delta_{R,k}}\right).
\label{eq:application_spin_bus_exchange}
\end{equation}
The physical endpoint resonance is
$\widetilde\omega_S=\widetilde\omega_R$.  Equation
\eqref{eq:application_spin_bus_exchange} is a signed coherent sum, not a sum
of rates.  A small exchange may reflect controlled destructive interference,
or cancellation among terms that are individually outside the dispersive
regime.

The same rotation reconstructs the virtual bus amplitudes,
\begin{equation*}
\begin{aligned}
c_k(t)
&\simeq
-\frac{g_{S,k}c_S(t)}{\Delta_{S,k}}
-\frac{g_{R,k}c_R(t)}{\Delta_{R,k}},\\
c_\ell(t)
&=\sum_k\phi_k(\ell)c_k(t).
\end{aligned}
\end{equation*}
Thus eliminating the bus from the slow Hamiltonian does not make its virtual
wavefunction identically zero.  It removes the bus as an independent slow
coordinate.

For a single excitation, the endpoint concurrence is
\(C_{SR}(t)=2|c_S(t)c_R^*(t)|,\)
while $P_R(t)=|c_R(t)|^2$ and
$P_B(t)=\sum_\ell|c_\ell(t)|^2$ measure transfer and bus occupation.  For an
arbitrary input qubit and a vacuum bus, the phase-corrected average transfer
fidelity is
\begin{equation*}
\overline{F}(t)
=\frac12+\frac{|f(t)|}{3}+\frac{|f(t)|^2}{6},
\end{equation*}
where $f(t)$ is the sender-to-receiver amplitude.  Population agreement alone
does not test the transferred phase or the complete quantum channel.

\subsubsection{One isolated resonant mode retained}
\label{subsubsec:application_spin_bus_resonant}

If one mode $k_0$ satisfies
\begin{equation*}
\omega_S\simeq\omega_R\simeq\varepsilon_{k_0},
\qquad
\left|\frac{g_{\nu,k}}{\Delta_{\nu,k}}\right|\ll1
\quad(k\neq k_0),
\end{equation*}
then eliminating $k_0$ is the wrong response to the small denominator.  The
retained space must be enlarged to
$\mathrm{span}\{|S\rangle,|k_0\rangle,|R\rangle\}$.  After eliminating only
the remaining modes,
\begin{equation}
\begin{aligned}
\hat H_{Sk_0R}^{\mathrm{res}}
={}&\widetilde\omega_S\hat n_S
+\widetilde\varepsilon_{k_0}\hat n_{k_0}
+\widetilde\omega_R\hat n_R \\
&+\left(
 g_{S,k_0}\hat\sigma_S^+\hat b_{k_0}
+g_{R,k_0}\hat\sigma_R^+\hat b_{k_0}
+\mathrm{H.c.}\right)\\
&+\hat H_{SR}^{(\mathrm{off})}.
\label{eq:application_spin_bus_three_component} 
\end{aligned}
\end{equation}
For generally complex endpoint couplings, define
\begin{equation*}
\begin{aligned}
G_{k_0}
&=\sqrt{|g_{S,k_0}|^2+|g_{R,k_0}|^2},\\
|B\rangle
&=\frac{g_{S,k_0}|S\rangle+g_{R,k_0}|R\rangle}{G_{k_0}},\\
|D\rangle
&=\frac{g_{R,k_0}^*|S\rangle-g_{S,k_0}^*|R\rangle}{G_{k_0}}.
\end{aligned}
\end{equation*}
With the coupling convention in
Eq.~\eqref{eq:application_spin_bus_three_component}, these states obey
$\langle k_0|\hat H_{Sk_0R}^{\mathrm{res}}|B\rangle=G_{k_0}$ and
$\langle k_0|\hat H_{Sk_0R}^{\mathrm{res}}|D\rangle=0$.
Thus only the bright combination couples directly to the selected mode.
The dark combination can still be mixed by unequal dressed endpoint
energies, the off-resonant endpoint correction
$\hat H_{SR}^{(\mathrm{off})}$, omitted modes, or additional noise.
This is a three-component Hamiltonian with pairwise couplings, not a
genuine three-operator interaction. The protocol is faster than the
all-mode dispersive transfer, but the retained mode is physically populated
and carries memory and loss.

\subsubsection{Lossy buses and protocol-dependent noise exposure}

If the bus has an amplitude-loss matrix $\hat K$, the boundary response
becomes
\begin{equation*}
\boldsymbol\Sigma(E)
=\hat W^\dagger
\left[E-\hat H_B+\frac{i}{2}\hat K\right]^{-1}
\hat W.
\end{equation*}
Its Hermitian part generates endpoint shifts and coherent exchange, while
\begin{equation*}
\boldsymbol\Gamma(E)
=i\left[\boldsymbol\Sigma(E)-\boldsymbol\Sigma^\dagger(E)\right]
\end{equation*}
is the induced endpoint decay matrix.  Diagonalizing
$\boldsymbol\Gamma$ gives the local or collective effective jump operators.
Independent endpoint rates arise only after additional symmetry and secular
approximations.  In the far-detuned limit,
\begin{equation*}
\Gamma_{\mathrm{ind}}
\sim\sum_k\kappa_k\left|\frac{g_k}{\Delta_k}\right|^2,
\end{equation*}
which is the diagonal scaling used in nonlocal dissipative tunneling
\cite{MoraesNeto2012}.  A compact loss-to-exchange ratio is
\begin{equation*}
\mathcal R_{\mathrm{bus}}(E)
=
\frac{\lambda_{\max}[\boldsymbol\Gamma(E)]}
{2|\Sigma_{SR}(E)|}.
\end{equation*}
Smaller values indicate that coherent exchange dominates the induced loss.
This quantity is not a conventional cooperativity: it retains the signed
interference contained in the boundary Green function, and it diverges near
a destructive zero of \(\Sigma_{SR}\), where an endpoint exchange protocol
is ineffective even if every individual mode remains dispersive.

The selected-mode protocol instead retains the mode and its jump operator.
Its bus-loss contribution is controlled by the integrated real occupation,
\begin{align*}
1-F_{\mathrm{res}}
\sim &\kappa_{k_0}\int_0^{t_f}P_{k_0}(t)\,dt \\
&+\text{endpoint contributions}.
\end{align*}
The faster resonant protocol is favored when endpoint decoherence dominates;
the slower virtual protocol can be favored when bus loss dominates.  The
comparison must use a channel metric and place each protocol inside its own
controlled regime.

\paragraph{Method comparison.}
For a quadratic bus, the normal-mode small rotation directly exposes the mode
parities, signed path interference, endpoint shifts, and virtual spatial
profile.  Projection onto the endpoints yields the same second-order
Hamiltonian in terms of the boundary resolvent, while retaining its energy
dependence and pole structure.  The latter identifies whether one mode or a
cluster of modes must remain explicit.  In the presence of loss, the complex
self-energy determines both the coherent endpoint interaction and the local
or correlated induced decay.  Exact finite-chain propagation tests the
accuracy of each reduced model but does not determine the appropriate
retained space.

\subsubsection{Exact benchmarks}

\begin{figure*}[!ht]
\centering
\includegraphics[width=1.0\textwidth]
{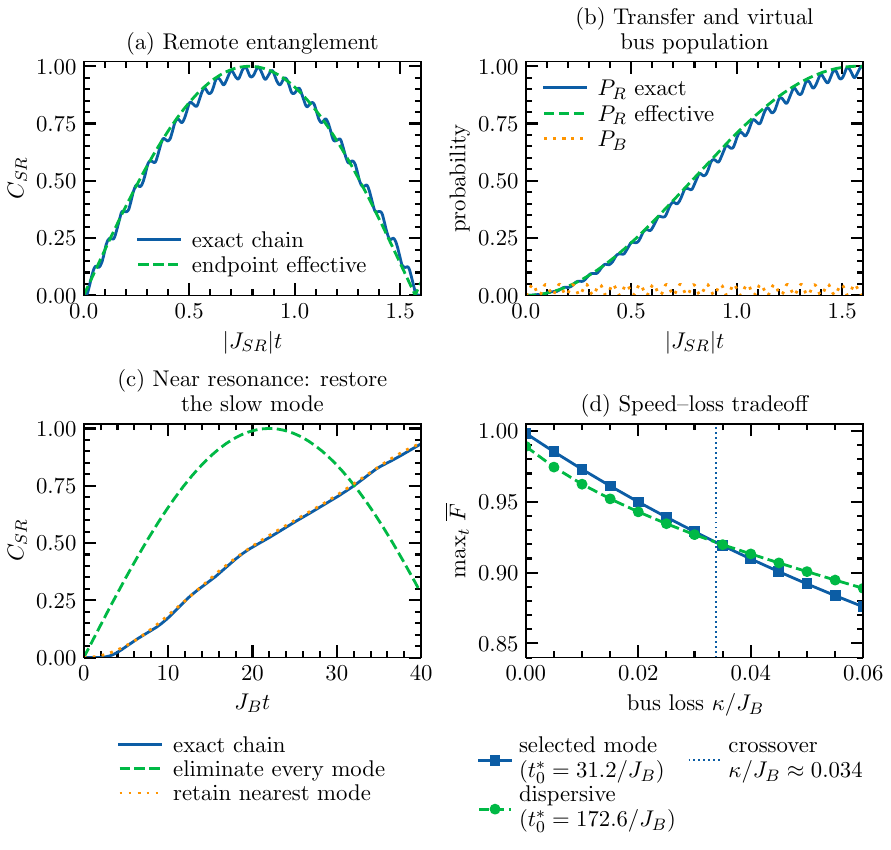}
\caption{\textbf{Physical diagnostics for a spin-chain quantum bus.}  (a) Exact and
endpoint-effective sender--receiver concurrence in the all-mode dispersive
regime.  (b) Receiver population and the small but nonzero bus occupation.
(c) Close to a bus pole, eliminating every mode gives the wrong entanglement
dynamics, while retaining the nearest mode repairs the reduced model.  (d)
Maximum phase-corrected average transfer fidelity versus bus loss for
representative controlled dispersive and selected-mode implementations; a
small endpoint-loss rate is included in both.}
\label{fig:application_spin_bus_validation}
\end{figure*}

Figure~\ref{fig:application_spin_bus_validation} compares the two retained
spaces against exact finite-chain dynamics.  In the dispersive regime the
endpoint model reproduces both remote concurrence and state transfer while
the bus occupation remains small.  Near a bus pole, the endpoint-only model
fails even though it remains a well-defined two-level Hamiltonian; restoring
the nearest mode recovers the entanglement dynamics.  The final panel shows
that speed and bus-loss protection are competing resources rather than a
universal ranking of the two protocols.

\subsubsection{Preparation, statistics, and broader buses}

The elementary derivation assumes a polarized bus,
\begin{equation*}
\rho_B(0)=|\downarrow\cdots\downarrow\rangle
\langle\downarrow\cdots\downarrow|.
\end{equation*}
For this protocol, an occupied spin mode introduces Pauli blocking, reverse
transfer, and occupation-dependent phases.  This does not imply that an
unpolarized chain is unusable.  Initialization-free endpoint protocols
\cite{DiFranco2008} and free-fermion state transfer through random
unpolarized chains~\cite{Yao2011} change the operations, encoding, and parity
handling.  An excited bus therefore invalidates the elementary endpoint map;
it does not rule out a different transfer protocol.

A finite thermal chain is not automatically a Markovian reservoir.  If its
correlations and recurrences are resolved, the appropriate endpoint object is
a finite-time reduced channel.  A time-independent endpoint master equation
requires an additional short-correlation or fast-relaxation assumption.

In the vacuum and one-excitation sectors, an $XX$ spin chain and a bosonic
tight-binding chain have the same single-particle spectrum.  At finite
occupation,
\begin{equation*}
\begin{aligned}
\text{spin/fermionic mode: }&\quad n_k=0,1,\\
\text{bosonic mode: }&\quad n_k=0,1,2,\ldots .
\end{aligned}
\end{equation*}
Thus, a populated spin mode blocks injection whereas a bosonic mode produces the
$\sqrt{n_k+1}$ stimulated matrix element.  The low-density bosonic
approximation is controlled only near a polarized spin state.  The same
boundary-response logic also applies to lossy cavity and polariton networks
\cite{MoraesNeto2011}, and spectrally separated bus modes can support
parallel Raman-like colored channels~\cite{ColoredChannels2013}.

\subsubsection{Validity, spectral crowding, and structural repair}

The reduction should record at least
\begin{equation*}
\begin{aligned}
\eta_{\max}
&=\max_{\nu,k}\left|\frac{g_{\nu,k}}{\Delta_{\nu,k}}\right|,
&\qquad
\Lambda_\nu
&=\sum_k\left|\frac{g_{\nu,k}}{\Delta_{\nu,k}}\right|^2,\\
\Delta_{\min}&=\min_{\nu,k}|\Delta_{\nu,k}|,
&\qquad
\delta_{\mathrm{iso}}
&=\min_{k\neq k_0}|\varepsilon_k-\varepsilon_{k_0}|.
\end{aligned}
\end{equation*}
The first quantity detects one dangerous mode, the second cumulative virtual
dressing, and the last two diagnose spectral crowding and selected-mode
isolation.  As $L$ grows, adjacent mode spacings shrink.  One dangerous pole
must be retained; a cluster of nearby poles requires a multimode retained
space.  Unknown preparation can require a different encoding or finite-time
channel, and an interacting bus requires the corresponding many-body boundary
response.  These are changes of retained structure, not higher-order
corrections to a fixed endpoint-only Hamiltonian.

\paragraph{Companion notebook N02.}
Notebook \path{N02_Spin-Chain-Quantum-Bus} derives the normal modes and
mode-by-mode small rotation; reconstructs the spatial profile of the virtual
bus excitation;
compares all-mode elimination with selected-mode retention; derives the
complex boundary self-energy and induced decay matrix; benchmarks
concurrence, transfer probability, bus occupation, and average channel
fidelity; and distinguishes polarized, initialization-free, spin, bosonic,
and multiplexed buses.  Exercises extend the calculation to disorder,
mode clusters, encoded unpolarized transfer, and interacting mediators.

\subsection{Eliminating a Lossy Bosonic Mediator}
\label{subsec:application_lossy_mediator}

\paragraph{Learning goals and conceptual focus.}
The aim here is to eliminate a rapidly relaxing bosonic mode without losing
the fluctuations that it carries.  A single complex response produces both
the coherent interaction and the effective jump.  We then resolve
nondegenerate slow transitions, reconstruct the eliminated occupation and
its initial layer, and distinguish conditional no-jump evolution from the
unconditional master equation.  This example also keeps two logically
separate tests in view: whether the reduced generator is physical and
whether it is quantitatively accurate.  When the fast spectral gap closes,
the mediator must return to the retained model.  Related engineered-mediator
constructions and a complementary discussion of the validity hierarchy
behind Markovian reductions are given in
Refs.~\cite{Metelmann2023Parametric,Stefanini2026IsLindblad}.

Rapidly damped auxiliaries are basic components of reservoir engineering.  A
controlled coupling can convert a target operator into an effective jump,
while the dispersive part of the same response generates coherent shifts and
mediated interactions
\cite{Poyatos1996Reservoir,Reiter2012Effective,Kessler2012DissipativeSW,Jager2022BosonicModes,Harrington2022EngineeredDissipation}.  The same principle
underlies collective decay, squeezed reservoirs, and nonlinear confinement of
bosonic states \cite{Kronwald2013Squeezing,Leghtas2015TwoPhoton}.  Recent
implementations extend it to multicomponent cat manifolds through nonlinear
gain--loss interference \cite{Rojkov2026NLRE} and to number-selective quartet
dissipation generated by genuine six-wave mixing
\cite{Vanselow2026Quartets}.  These examples also show why residual coherent
terms and occupation-dependent corrections belong to the effective design,
not to a separate error model.

\begin{figure*}[!ht]
\centering
\includegraphics[width=\textwidth]
{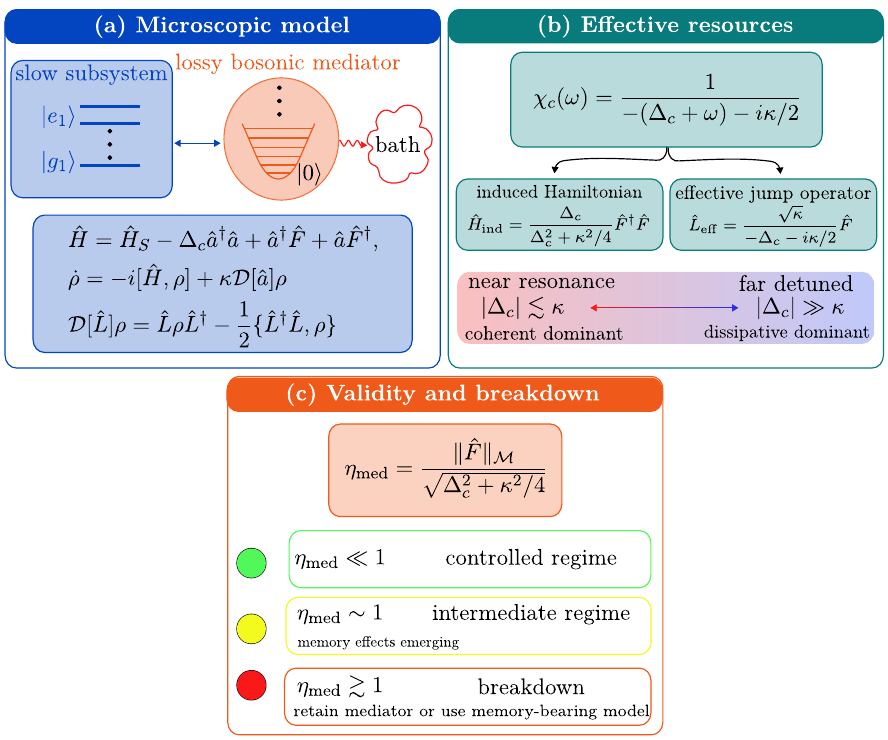}
\caption{\textbf{Schematic summary of lossy-mediator elimination.}
(a) A slow subsystem couples through an operator $\hat F$ to a damped
bosonic mediator $\hat a$, whose microscopic dynamics is governed by
$\hat H=\hat H_S-\Delta_c\hat a^\dagger\hat a+\hat a^\dagger\hat F+\hat a\hat F^\dagger$
and
$\dot\rho=-i[\hat H,\rho]+\kappa\mathcal D[\hat a]\rho$.
The default reference state is the mediator vacuum, while displaced,
thermal, and squeezed reference states require modified encoding and
noise operators.
(b) The same complex mediator response,
$\chi_c(\omega)=\bigl[-(\Delta_c+\omega)-i\kappa/2\bigr]^{-1}$,
generates two effective slow resources together: an induced coherent
Hamiltonian and an effective dissipative jump.  Near resonance the
dissipative response dominates, at large detuning the coherent response
dominates, and for $|\Delta_c|\sim\kappa$ both contribute comparably.
(c) The local control parameter
$\eta_{\mathrm{med}}=\|\hat F\|_{\mathcal M}/\sqrt{\Delta_c^2+\kappa^2/4}$
organizes the validity of the time-local reduction.  In the controlled
regime the mediator can be eliminated and both the induced Hamiltonian
and effective jumps must be retained.  As the complex gap narrows,
memory and spectral structure become important.  In the breakdown
regime, the correct repair is to retain the mediator explicitly or use a
memory-bearing reduced description rather than a time-independent slow
generator.}
\label{fig:application_lossy_schematic}
\end{figure*}

We begin with the harmonic mediator analyzed below.  The
companion notebook later repeats the comparison for a rapidly relaxing qubit,
whose finite excited manifold introduces saturation.  This separates features
fixed by the complex fast response from limitations specific to the auxiliary
statistics.  The elimination problem is summarized in
Fig.~\ref{fig:application_lossy_schematic}.  The key structural point is
that one complex fast response controls two slow resources at once:
coherent mediated interactions and effective dissipation.  The same
schematic also organizes the role of the reference fast state, the
reconstruction of the eliminated mediator, and the distinction between a
controlled time-local reduction and a regime in which the mediator must
be retained explicitly.

Figure~\ref{fig:application_lossy_schematic} also keeps three choices
separate.  First, the microscopic coupling fixes the complex susceptibility
$\chi_c(\omega)$ that governs the fast response.  Second, the choice of
reference mediator state fixes the appropriate encoding and therefore
the form of the effective noise operators.  Third, the validity of the
reduction is controlled not only by the small mixing parameter
$\eta_{\mathrm{med}}$, but also by the separation between the induced
slow rates and the fast Liouvillian gap.  When that complex gap closes,
the mediator is no longer a fast auxiliary and must be promoted into the
retained model.

\subsubsection{Microscopic model, convention, and reference state}

Let a slow subsystem couple linearly to a damped auxiliary mode,
\begin{equation*}
\begin{aligned}
&\hat H
=\hat H_S-\Delta_c\hat a^\dagger\hat a
+\hat a^\dagger\hat F+\hat a\hat F^\dagger,\\
&\dot\rho
=-i[\hat H,\rho]+\kappa\mathcal D[\hat a]\rho,\\
&\mathcal D[\hat L]\rho
=\hat L\rho\hat L^\dagger-
\frac12\{\hat L^\dagger\hat L,\rho\}.
\end{aligned}
\end{equation*}
With this convention, the field amplitude decays at \(\kappa/2\) and the
photon number at \(\kappa\), matching the linewidth convention used in the
complex susceptibility below.

The reference fast state is the mediator vacuum.  A coherent reference should
first be displaced.  Thermal and squeezed references have different
correlation matrices and generally produce both upward and downward, or
phase-sensitive, effective channels.  They therefore require a different
encoding map rather than a change of one denominator in the vacuum result.

\subsubsection{Operational effective-operator derivation}

Define the retained and eliminated projectors
\begin{equation*}
\mathcal P=|0\rangle\langle0|\otimes I_S,
\qquad
\mathcal Q=1-\mathcal P,
\end{equation*}
and split the interaction as
\begin{equation*}
\hat V_+=\hat a^\dagger\hat F,
\qquad
\hat V_-=\hat a\hat F^\dagger.
\end{equation*}
For one relevant fast excitation, the non-Hermitian block is
\begin{equation*}
\hat H_{\rm NH}
=\left(-\Delta_c-i\frac\kappa2\right)|1\rangle\langle1|.
\end{equation*}
The effective-operator formulas \cite{Reiter2012Effective} give
\begin{align*}
&\hat H_{\rm ind}
=-\frac12\hat V_-
\left(\hat H_{\rm NH}^{-1}+\hat H_{\rm NH}^{-1\dagger}\right)
\hat V_+,\\
&\hat L_{\rm eff}
=\sqrt\kappa\,\hat a\hat H_{\rm NH}^{-1}\hat V_+.
\end{align*}
Using $\hat a|1\rangle=|0\rangle$ yields
\begin{align}
&\hat H_{\rm ind}
=\frac{\Delta_c}{\Delta_c^2+\kappa^2/4}
\hat F^\dagger\hat F,
\label{eq:application_lossy_induced_hamiltonian}\\
&\hat L_{\rm eff}
=\frac{\sqrt\kappa}{-\Delta_c-i\kappa/2}\hat F.
\label{eq:application_lossy_effective_jump}
\end{align}
The reduced state obeys
\begin{equation}
\dot\rho_S
=-i[\hat H_S+\hat H_{\rm ind},\rho_S]
+\mathcal D[\hat L_{\rm eff}]\rho_S.
\label{eq:application_lossy_effective_master_equation}
\end{equation}
The induced Hamiltonian and jump are the Hermitian and absorptive components of
one eliminated propagation.  Keeping only Eq.~\eqref{eq:application_lossy_induced_hamiltonian}
is therefore not an order-consistent unconditional theory.

\subsubsection{Langevin response and method comparison}

The Heisenberg--Langevin equation,
\begin{equation*}
\frac{d\hat a}{dt}
=\left(i\Delta_c-\frac\kappa2\right)\hat a
-i\hat F-\sqrt\kappa\,\hat a_{\rm in},
\end{equation*}
separates the homogeneous transient, the driven response, and the input noise.
For a Bohr component $\hat F_\omega e^{-i\omega t}$,
\begin{equation*}
\begin{aligned}
\hat a_\omega
&\simeq-\chi_c(\omega)\hat F_\omega,\\
\chi_c(\omega)
&=
\frac{1}{-(\Delta_c+\omega)-i\kappa/2}.
\end{aligned}
\end{equation*}
At one frequency, Eqs.~\eqref{eq:application_lossy_induced_hamiltonian} and
\eqref{eq:application_lossy_effective_jump} are equivalently
$\hat H_{\rm ind}=-\operatorname{Re}\chi_c\,\hat F^\dagger\hat F$ and
$\hat L_{\rm eff}=\sqrt\kappa\chi_c\hat F$, up to an irrelevant jump phase.

\paragraph{Method comparison.}
A small rotation gives the far-detuned coherent term and transforms the cavity
jump, exposing dressing and virtual occupation.  James--Jerke averaging gives
the same second-order coherent beat and separates it from bounded micromotion,
but its Hamiltonian formula alone does not supply the recycling term.
Langevin theory displays the input noise, initial transient, susceptibility,
and onset of memory.  Effective operators give a direct derivation of the
finite-linewidth Hamiltonian and jump.  Liouvillian projection and dissipative
Schrieffer--Wolff theory reorganize the same response in terms of a reduced
inverse and a fast spectral gap
\cite{Kessler2012DissipativeSW,Saideh2020Projection,GonzalezBallestero2024Projector,Jager2022BosonicModes}.  In the companion
notebook the first four routes are worked explicitly; projection and
dissipative Schrieffer--Wolff theory are used as structural reformulations.

\subsubsection{Several slow transition frequencies}

For a nondegenerate slow Hamiltonian,
\begin{equation*}
\hat F=\sum_\omega\hat F_\omega,
\qquad
[\hat H_S,\hat F_\omega]=-\omega\hat F_\omega.
\end{equation*}
Before secularization, a consistent time-dependent effective jump is
\begin{equation*}
\hat L_{\rm eff}(t)
=\sqrt\kappa\sum_\omega\chi_c(\omega)
\hat F_\omega e^{-i\omega t},
\end{equation*}
with Kossakowski coefficients
\begin{equation*}
\Gamma_{\omega\omega'}(t)
=\kappa\chi_c(\omega)\chi_c^*(\omega')
 e^{-i(\omega-\omega')t}.
\end{equation*}
This is a Gram matrix and is positive semidefinite.  The accompanying
Hamiltonian is
\begin{equation*}
\hat H_{\rm ind}(t)
=-\frac12\sum_{\omega,\omega'}
[\chi_c(\omega')+\chi_c^*(\omega)]
 e^{i(\omega-\omega')t}
\hat F_\omega^\dagger\hat F_{\omega'}.
\end{equation*}
If $|\omega-\omega'|$ is large compared with the induced rates, full
secularization gives
\begin{equation*}
\hat H_{\rm ind}^{\rm sec}
=-\sum_\omega\operatorname{Re}\chi_c(\omega)
\hat F_\omega^\dagger\hat F_\omega,
\end{equation*}
\begin{equation*}
\dot\rho_S
=-i[\hat H_S+\hat H_{\rm ind}^{\rm sec},\rho_S]
+\sum_\omega\kappa|\chi_c(\omega)|^2
\mathcal D[\hat F_\omega]\rho_S.
\end{equation*}
Partial secularization must retain the coherent and dissipative cross terms as
one block.  Mixing diagonal rates from one approximation with an unrelated
cross rate can make the Kossakowski matrix indefinite.

The notebook tests this point with two transitions at
$\Delta_c/\kappa=0.5$, $\omega_{1,2}/\kappa=\mp1.5$, and
$g_1/\kappa=g_2/\kappa=0.12$, starting from $|ee,0\rangle$.  The mixing
parameters are $0.107$ and $0.058$, and the transition separation is much
larger than the induced rates.  The frequency-resolved secular model has a
maximum reduced-state trace distance $2.38\times10^{-2}$, whereas replacing
both responses by $\chi_c(0)$ produces a maximum error $6.95\times10^{-1}$.

\subsubsection{Two qubits: exchange, collective decay, and dark states}

For degenerate qubits,
\begin{equation*}
\hat F=g_1\hat\sigma_-^{(1)}+g_2\hat\sigma_-^{(2)},
\end{equation*}
so $\hat F^\dagger\hat F$ contains individual shifts and coherent exchange,
whereas Eq.~\eqref{eq:application_lossy_effective_jump} gives collective
decay.  For $g_1=g_2=g$,
\begin{equation*}
|B\rangle=\frac{|eg\rangle+|ge\rangle}{\sqrt2},
\qquad
|D\rangle=\frac{|eg\rangle-|ge\rangle}{\sqrt2}
\end{equation*}
are bright and dark with respect to the ideal collective jump.  Starting from
$|eg\rangle=(|B\rangle+|D\rangle)/\sqrt2$, the bright component decays while
the dark component survives.  The ideal long-time state is
$\frac12|D\rangle\langle D|+\frac12|gg\rangle\langle gg|$, whose concurrence
is $1/2$.  Coupling asymmetry, local decay, dephasing, and additional frequency
sectors break this protection.

A Hamiltonian-only model retains the exchange in $\hat F^\dagger\hat F$ but
removes collective decay.  It can therefore predict nearly coherent transfer
and excessive concurrence when the full model relaxes toward the bright--dark
mixture.

\subsubsection{Conditional evolution and the unconditional state}

The no-jump Hamiltonian
\begin{equation*}
\hat H_{\rm nh}
=\hat H-\frac{i}{2}\sum_r\hat L_r^\dagger\hat L_r
\end{equation*}
governs trajectories conditioned on the absence of monitored jumps
\cite{PlenioKnight1998QuantumJumps}.  The unconditional generator also
contains the recycling terms $\sum_r\hat L_r\rho\hat L_r^\dagger$.  These
terms can change the eigenvalues and eigenmodes, so a non-Hermitian exceptional
point is not generally a Liouvillian exceptional point
\cite{Minganti2019Exceptional,Minganti2020HybridLiouvillian,Haack2021LiouvillianEP}.
The notebook evaluates both eigenvector-condition numbers in a minimal
bright--mediator amplitude-damping model.  They peak at the same coupling in
that special model, illustrating a model-dependent coincidence rather than a
general equivalence.

\subsubsection{Mediator reconstruction and the initial layer}

For a slowly varying source,
\begin{equation*}
\hat a_{\rm sl}(t)
\simeq\frac{\hat F(t)}{\Delta_c+i\kappa/2}.
\end{equation*}
A bare vacuum preparation does not generally lie on this instantaneous slow
manifold.  Enforcing $\hat a(0)=0$ gives
\begin{equation*}
\hat a_{\rm rec}(t)
\simeq\hat a_{\rm sl}(t)
-e^{(i\Delta_c-\kappa/2)t}\hat a_{\rm sl}(0).
\end{equation*}
The reconstructed occupation is
$\langle\hat a_{\rm rec}^\dagger(t)\hat a_{\rm rec}(t)\rangle$, including the
cross correlations between $\hat F(t)$ and $\hat F(0)$.  The simpler product
$|1-e^{(i\Delta_c-\kappa/2)t}|^2
\langle\hat F^\dagger(t)\hat F(t)\rangle/
(\Delta_c^2+\kappa^2/4)$ is only a quasistatic approximation.  The numerical
benchmark uses the full reconstructed amplitude and reduces the maximum
mediator-occupation error from $7.81\times10^{-3}$ to approximately
$9.2\times10^{-5}$.

\subsubsection{Exact benchmark}

\begin{figure*}[!ht]
\centering
\includegraphics[width=1.0\textwidth]
{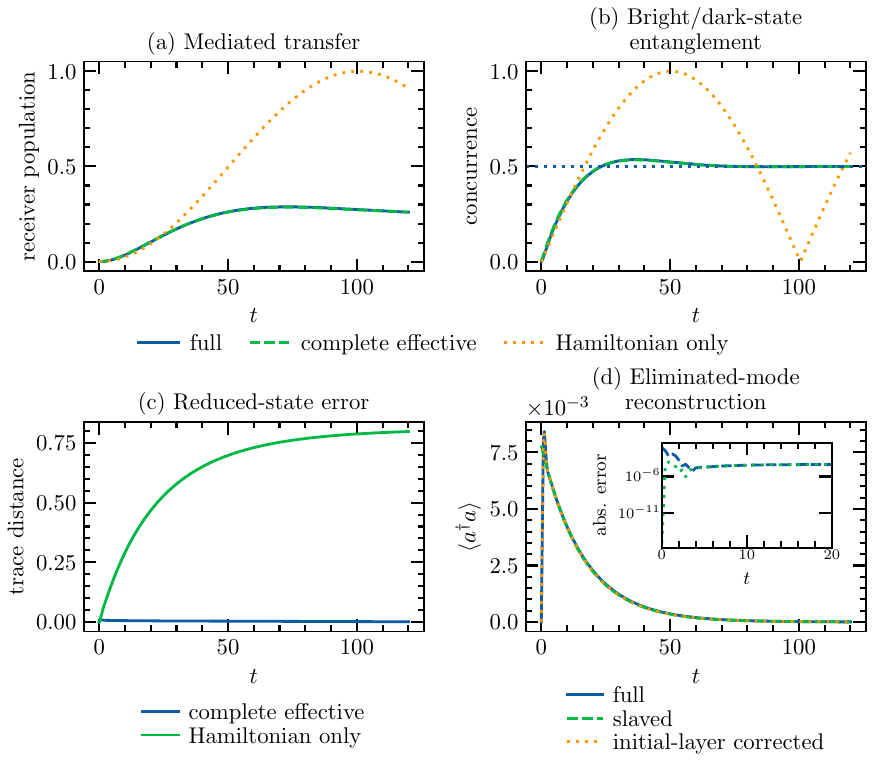}
\caption{\textbf{Validation of the lossy-bosonic-mediator reduction for
$\Delta_c=2$, $\kappa=4$, $g_1=g_2=0.25$, mediator vacuum, and initial qubit
state $|eg\rangle$.}  (a) Receiver population for the full mediator model, the
complete effective master equation, and a Hamiltonian-only truncation.
(b) Sender--receiver concurrence.  The complete reduction captures decay of
the bright component and survival of the dark component, whereas the
Hamiltonian-only model predicts excessive coherent entanglement.  (c) Trace
distance from the exact reduced two-qubit state.  The maximum errors are
$9.4\times10^{-3}$ and approximately $0.80$ for the complete and
Hamiltonian-only models, respectively.  (d) Exact mediator occupation,
instantaneous slaving, and the full initial-layer reconstruction.  The inset
shows the corresponding absolute errors.}
\label{fig:application_lossy_validation}
\end{figure*}

Figure~\ref{fig:application_lossy_validation} uses
$\Delta_c=2$, $\kappa=4$, and $g_1=g_2=0.25$, with the mediator initially in
vacuum and the qubits in $|eg\rangle$.  Time is expressed in the same inverse
frequency units.  The local mixing parameter is
$\eta_{\rm med}=\sqrt{2}g/\sqrt{\Delta_c^2+\kappa^2/4}=0.125$.
The full microscopic model is compared with Eq.~\eqref{eq:application_lossy_effective_master_equation}
and with the Hamiltonian-only truncation using identical parameters.  The
maximum trace-distance error is $9.4\times10^{-3}$ for the complete generator
and approximately $0.80$ when the effective jump is omitted.

The complex susceptibility defines a continuous crossover between operating
regimes.  Near resonance its absorptive part dominates.  Far off resonance,
the coherent shift scales as $1/\Delta_c$, whereas the induced rate is smaller
by $\kappa/|\Delta_c|$; when $|\Delta_c|\sim\kappa$, both components matter.
These are regimes of a single response, not successive approximation orders.

\subsubsection{Physicality, accuracy, and numerical convergence}

An explicitly GKLS generator guarantees complete positivity of its semigroup,
and a finite-time Choi spectrum checks a channel implementation.  Neither
establishes quantitative agreement with the microscopic dynamics.  Accuracy
requires state and observable errors, steady states, and slow spectral poles;
the microscopic reference also requires Hilbert-space and time-discretization
convergence.  At higher order or after inconsistent nonsecular truncation,
complete positivity is not automatic \cite{Tokieda2024CP}.

For two frequency sectors, complete positivity requires a positive
Kossakowski matrix, including
$|\Gamma_{12}|^2\leq\Gamma_{11}\Gamma_{22}$.  The notebook constructs an
inconsistent hybrid using frequency-resolved diagonal rates and a cross rate
from a common zero-frequency response.  Its minimum Kossakowski eigenvalue is
negative, and the corresponding short-time map has a negative Choi
eigenvalue.  This test is more discriminating than exponentiating a generator
already written in GKLS form.

The principal local mixing scale is
\begin{equation*}
\eta_{\rm med}
=\frac{\|\hat F\|_{\mathcal M}}
{\sqrt{\Delta_c^2+\kappa^2/4}},
\end{equation*}
where the norm is evaluated on the occupied slow manifold $\mathcal M$.
Induced rates must also remain small compared with the complex fast gap.  For
the linear two-pole bright--mediator benchmark used in the notebook, the
absolute slow-pole error scales as $G^4$ and the relative error as $G^2$ at
weak coupling.  This is a benchmark-specific expansion, not a universal
fourth-order theorem.

The single-excitation benchmark is exactly contained in a two-level mediator
sector; that observation is a sector-sufficiency check rather than a genuine
bosonic convergence test.  The notebook therefore includes a coherent-state
stress test with a slow bosonic mode, in which the distance to the
$N_c=6$ mediator reference falls from $2.11\times10^{-1}$ at $N_c=2$ to
$6.5\times10^{-4}$ at $N_c=5$, together with a decreasing top-Fock-level
population.

\subsubsection{Structural breakdown and nonlinear extensions}

When the mediator is weakly damped and nearly resonant, it stores population
and phase information on the same time scale as the retained subsystem.  A
single time-independent slow generator then misses transient storage,
oscillatory memory, and revivals.  The correct response is to retain the
mediator, retain the relevant mode cluster, or use a memory kernel or
finite-time map.  Increasing the perturbative order cannot recreate a fast
spectral gap that has closed.

The linear coupling $\hat a^\dagger\hat F+\hat a\hat F^\dagger$ produces a jump
proportional to $\hat F$ at this order.  Nonlinear reservoirs require a
nonlinear target operator, a driven multiwave-mixing interaction, or another
elimination.  For example,
\begin{equation*}
\hat H_{\rm int}=g_2(\hat a^\dagger\hat b^2+\hat a\hat b^{\dagger2})
\end{equation*}
produces $\hat L_{\rm eff}\propto\hat b^2$ after eliminating $\hat a$.
Driving the auxiliary can displace the jump to
$\hat L_{\rm eff}\propto\hat b^2-\alpha^2$, while higher-wave-mixing
processes generate number-selective four-photon channels
\cite{Leghtas2015TwoPhoton,Rojkov2026NLRE,Vanselow2026Quartets}.  The intended
jump, its coherent companion, and residual nonlinear corrections must be
derived together.

\paragraph{Companion notebook N04.}
Notebook \path{N04_Lossy-Bosonic-Mediator} derives the single-frequency bosonic generator by a
small rotation, James--Jerke averaging, Langevin response, and effective
operators, and relates the result to Liouvillian projection and dissipative
Schrieffer--Wolff theory.  It adds a nondegenerate two-frequency benchmark,
full mediator reconstruction, a genuine multi-excitation cutoff test,
Kossakowski and Choi physicality diagnostics, slow-pole scaling, and a
conditional/full-Liouvillian defectiveness comparison.  A lossy-qubit
auxiliary is retained as a finite-dimensional, saturable baseline.  Together,
the two auxiliaries distinguish features fixed by the complex fast response
from effects caused specifically by saturation or an unbounded bosonic
manifold.

\subsection{Floquet Control of Tunnelling in a Driven Lattice}
\label{subsec:application_floquet_tunnelling}

\paragraph{Learning goals and conceptual focus.}
The final application keeps three objects separate from the outset: an exact
periodic change of frame, an approximate Floquet generator, and the complete
laboratory-frame prediction.  From the transformed hopping we derive the
Bessel-renormalized tunnelling amplitude and identify leading-order coherent
destruction and hopping-sign reversal.  Exact one-period propagation then
fixes the quasienergies in a common Floquet zone, while the bond current shows
why observables within a drive period require micromotion reconstruction.
The same example exposes resonant Floquet blocks and the finite-frequency
limits of a high-frequency description.

Coherent destruction of tunnelling provides a transparent setting in which
these distinctions can be tested experimentally and numerically
\cite{Grossmann1991,DellaValle2007,Kierig2008}. More broadly, Floquet
engineering requires the effective Hamiltonian, the periodic micromotion,
and the quasienergy convention to be treated as parts of one physical
construction \cite{Bukov2015,Eckardt2017}. The same separation is important
in driven quantum control: degenerate Floquet perturbation theory can
organize multiphoton couplings, drive-induced shifts, leakage pathways, and
reconstruction operators within a common framework \cite{Huang2026}.
\begin{figure*}[t]
\centering
\includegraphics[width=\textwidth]
{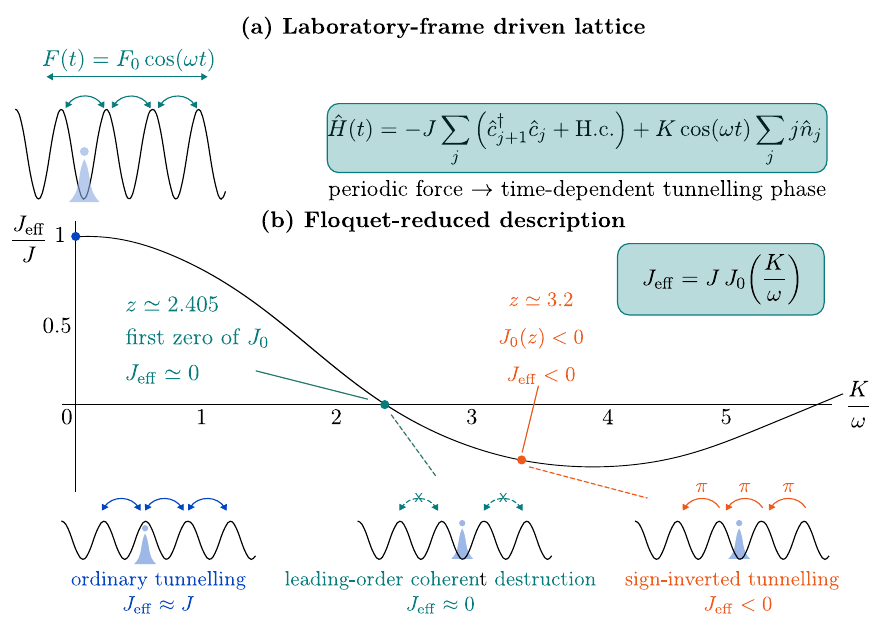}
\caption{\textbf{Floquet control of tunnelling in a periodically driven lattice.}
(a) Laboratory-frame description. A spatially uniform periodic force,
\(F(t)=F_{0}\cos(\omega t)\), produces a time-dependent lattice tilt.
The exact periodic transformation
\(\hat R(t)=\exp[-i(K/\omega)\sin(\omega t)\hat X]\), with
\(\hat X=\sum_j j\hat n_j\), removes the oscillating force and transfers
its effect to a time-dependent phase of the tunnelling matrix element.
(b) Floquet-reduced description. Zeroth-harmonic averaging gives
\(J_{\mathrm{eff}}=J\mathcal J_{0}(K/\omega)\). The marked regimes
correspond to ordinary tunnelling near \(K/\omega=0\), leading-order
coherent destruction at the first zero
\(K/\omega\simeq2.405\), and sign-inverted tunnelling when
\(\mathcal J_{0}(K/\omega)<0\). Vanishing
\(J_{\mathrm{eff}}\) suppresses the leading tunnelling process but does
not generally eliminate finite-frequency corrections. A complete
continuous-time laboratory-frame prediction additionally requires the
periodic frame transformation and Floquet micromotion.}
\label{fig:application_floquet_schematic}
\end{figure*}

For a periodically forced nearest-neighbour lattice, consider
\begin{equation*}
\begin{aligned}
&\hat H(t)
=
-J\sum_j
\left(
\hat c_{j+1}^{\dagger}\hat c_j+\mathrm{H.c.}
\right) \\
&\qquad\qquad+K\cos(\omega t)\hat X
+\hat H_{\mathrm{int}},
\\
&\hat X=\sum_j j\hat n_j .
\end{aligned}
\end{equation*}
Introducing the dimensionless drive strength
\(z=\tfrac{K}{\omega},\)
the periodic transformation
\begin{equation*}
\hat R(t)
=
\exp\left[-iz\sin(\omega t)\hat X\right]
\end{equation*}
removes the oscillating force exactly. With the convention
\begin{equation*}
\widetilde H(t)
=
\hat R^{\dagger}(t)\hat H(t)\hat R(t)
-i\hat R^{\dagger}(t)\dot{\hat R}(t),
\end{equation*}
one obtains
\begin{equation}
\widetilde H(t)
=
-J\sum_j
\left[
e^{iz\sin(\omega t)}
\hat c_{j+1}^{\dagger}\hat c_j
+\mathrm{H.c.}
\right]
+\widetilde H_{\mathrm{int}}(t),
\label{eq:application_floquet_rotating_hamiltonian}
\end{equation}
where
\begin{equation*}
\widetilde H_{\mathrm{int}}(t)
=
\hat R^{\dagger}(t)\hat H_{\mathrm{int}}\hat R(t).
\end{equation*}
On-site and density--density interactions commute with \(\hat X\) and
therefore remain unchanged. More general terms acquire their own periodic
phases. The frame transformation is exact; approximation begins only when
the harmonic structure of
Eq.~\eqref{eq:application_floquet_rotating_hamiltonian} is truncated or
perturbatively block diagonalized.

Using the Jacobi--Anger expansion,
\begin{equation*}
e^{iz\sin(\omega t)}
=
\sum_{m=-\infty}^{\infty}
\mathcal J_m(z)e^{im\omega t},
\end{equation*}
the zeroth-harmonic generator is
\begin{equation*}
\hat H_F^{(0)}
=
-J\mathcal J_0(z)
\sum_j
\left(
\hat c_{j+1}^{\dagger}\hat c_j+\mathrm{H.c.}
\right)
+\hat H_{\mathrm{int}},
\end{equation*}
for interactions invariant under \(\hat R(t)\). The corresponding effective
tunnelling amplitude is
\begin{equation}
J_{\mathrm{eff}}
=J\mathcal J_0(K/\omega).
\label{eq:application_floquet_Jeff}
\end{equation}

This expression exposes three regimes. For \(K/\omega\ll1\), the drive
only weakly renormalizes ordinary tunnelling. Near a zero of
\(\mathcal J_0\), the leading tunnelling term vanishes and coherent
destruction is obtained at zeroth order. When \(\mathcal J_0<0\), the
effective hopping reverses sign. The latter changes band curvature,
interference phases, and frustration relative to other couplings. On an
isolated open bipartite chain containing only nearest-neighbour hopping,
however, the sign can be removed by a staggered gauge transformation. Its
physical significance therefore depends on the complete lattice, boundary
conditions, additional couplings, and measured observables.
Figure~\ref{fig:application_floquet_schematic} marks the point at which the
approximation enters.  The periodic change of frame is exact,
whereas the Bessel-renormalized tunnelling amplitude follows only after
the transformed Hamiltonian is truncated to its zeroth harmonic.  The
same figure also emphasizes that the resulting Floquet generator does
not by itself determine continuous-time laboratory-frame observables:
the periodic frame transformation and the residual kick operator must
also be restored.

\paragraph{Operational and complementary methods.}
The direct operational route is the exact driven-frame transformation
followed by harmonic averaging. It immediately produces
Eq.~\eqref{eq:application_floquet_Jeff} and the basic control intuition.
Floquet--Magnus or van Vleck expansions systematically add higher-order
terms to both the effective generator and the micromotion operator
\cite{GoldmanDalibard2014,Bukov2015,Eckardt2017}. Exact propagation over one
period provides a nonperturbative stroboscopic benchmark, while Sambe-space
truncation makes the couplings between Floquet replicas and the origin of
multiphoton resonances explicit \cite{Sambe1973}.

The exact one-period propagator is
\begin{equation*}
\hat U(T,0)
=
\mathcal T
\exp\left[-i\int_0^T\hat H(t)\,dt\right],
\qquad
T=\frac{2\pi}{\omega}.
\end{equation*}
Its eigenvalues define the quasienergies,
\begin{equation*}
\begin{aligned}
\hat U(T,0)\ket{u_\mu(0)}
&=
e^{-i\varepsilon_\mu T}\ket{u_\mu(0)},
\\
\varepsilon_\mu
&\sim\varepsilon_\mu+\ell\omega,
\qquad \ell\in\mathbb Z.
\end{aligned}
\end{equation*}
Consequently, two raw Floquet spectra cannot be compared before their zones,
branches, and gauge conventions have been aligned. A difference between two
displayed Floquet Hamiltonians need not imply a difference between their
physical predictions.

\paragraph{Micromotion and the complete prediction.}
In a chosen Floquet gauge, the laboratory propagator may be written as
\begin{equation*}
\begin{aligned}
\hat U(t,0)
={}&
\hat R(t)
e^{-i\hat{\mathcal K}(t)}
e^{-i\hat H_Ft}
\times
e^{i\hat{\mathcal K}(0)}
\hat R^{\dagger}(0),
\\
\hat{\mathcal K}(t+T)
={}&\hat{\mathcal K}(t).
\end{aligned}
\end{equation*}
Here \(\hat{\mathcal K}(t)\) is the periodic kick operator. The distinct
symbol \(\hat{\mathcal K}\) avoids confusing the kick with the drive
amplitude \(K\).

At leading order, retaining the exact periodic frame but neglecting residual
kick corrections gives
\begin{equation*}
\ket{\psi_{\mathrm{rec}}(t)}
=
\hat R(t)e^{-i\hat H_F^{(0)}t}\ket{\psi(0)}.
\end{equation*}
This reconstruction restores the large phase motion generated by the
periodic force. It may leave smaller finite-frequency corrections associated
with nonzero harmonics. Thus, a zeroth-order generator can describe
populations accurately at selected or stroboscopic times while failing for
phase-sensitive currents, coherences, or subperiod motion. Every quantitative
claim should state whether it concerns quasienergies, stroboscopic evolution,
or continuous laboratory-frame observables.

For the driven two-site sector used in the companion notebook,
\begin{equation*}
\hat H(t)
=
-J\hat\sigma_x
+\frac{K}{2}\cos(\omega t)\hat\sigma_z,
\end{equation*}
the right-site probability and bond current are
\begin{align*}
P_R(t)
&=
\left|\braket{R|\psi(t)}\right|^2,
\\
I(t)
&=
J\langle\hat\sigma_y\rangle.
\end{align*}
The population measures the slow transfer between the sites, whereas the
current directly probes the relative phase and is therefore much more
sensitive to micromotion.

\begin{figure*}[!ht]
\centering
\includegraphics[width=1.0\textwidth]
{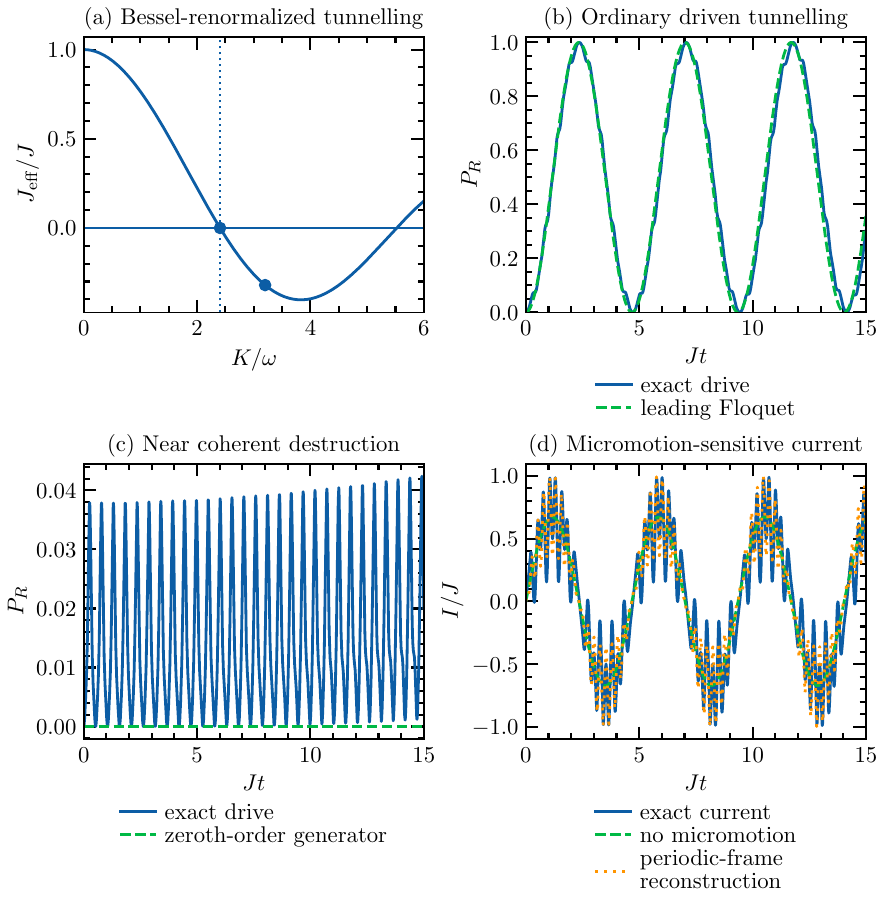}
\caption{\textbf{Floquet-controlled tunnelling in the driven two-site sector.}
(a) Bessel renormalization of the leading tunnelling amplitude,
\(J_{\mathrm{eff}}/J=\mathcal J_0(K/\omega)\), including its first zero and
sign reversal. (b) Right-site population for ordinary driven tunnelling,
comparing exact time-dependent evolution with the leading Floquet generator
at \(\omega/J=12\) and \(K/\omega=1.2\). (c) Residual finite-frequency
tunnelling at the first zero of \(\mathcal J_0\), where the zeroth-order
generator predicts complete suppression. (d) Bond current from the exact
dynamics, from the effective generator without micromotion, and after
restoring the periodic driven-frame transformation. The figure separates
suppression of the leading generator from exact finite-frequency localization
and shows why continuous-time observables require reconstruction.}
\label{fig:application_floquet_validation}
\end{figure*}

Figure~\ref{fig:application_floquet_validation} makes two limitations
visible. First, a zero of \(\mathcal J_0\) eliminates the zeroth-order hopping
but does not generically eliminate every finite-frequency contribution.
Higher-order terms can therefore become the leading residual dynamics near
coherent destruction even when they remain small on the scale of the bare
hopping. Second, the periodic-frame reconstruction recovers the dominant
current micromotion, but the remaining discrepancy records the finite order
of the approximation rather than a failure of the exact frame transformation.

\paragraph{Control scale, first failure, and repair.}
The ratio \(J/\omega\) is sufficient only for the minimal noninteracting
dimer. In a larger or interacting system, the relevant comparison is between
each retained harmonic matrix element and its associated Floquet-replica
denominator. Hopping, interactions, static fields, occupation-dependent
bosonic factors, and the local sector explored by the state can all set the
numerator.

The first structural failure occurs when a Floquet replica approaches
resonance. The resonant sectors must then be retained together rather than
averaged away. Away from isolated resonances, lowering the drive frequency
increases the quantitative error of the leading generator and its
reconstruction. Near a Bessel zero, higher-order terms should be included
because the nominal leading process has been tuned to zero. The corresponding
repairs are a higher-order van Vleck or Floquet--Magnus expansion, an enlarged
Sambe-space block, or direct time-dependent propagation.

In an interacting many-body system, a high-frequency Floquet Hamiltonian may
govern a long prethermal interval without describing the infinite-time state
\cite{Abanin2017}. Heating must therefore be assessed relative to the
experimental or computational observation window, rather than excluded solely
from the existence of a formal high-frequency expansion. Long-lived
prethermal regimes have been observed in driven Bose--Hubbard and spin
systems \cite{RubioAbadal2020,Beatrez2021}. When heating enters the task
window, the claim must be restricted to the prethermal regime or replaced by
a description that resolves the relevant resonant absorption channels.

\paragraph{Companion notebook N03.}
Notebook \path{N03_Floquet-Controlled-Tunnelling} derives the
exact periodic frame and the Bessel-renormalized generator for the driven
two-site model. It constructs the exact one-period propagator, checks
unitarity, extracts quasienergies in a fixed Floquet zone, compares exact and
leading-order tunnelling, resolves residual motion at the first Bessel zero,
and reconstructs the micromotion-sensitive bond current. A drive-frequency
scan demonstrates the recovery of the high-frequency description as
\(\omega/J\) increases. Resonant Sambe blocks, static biases, finite chains,
interacting particles, weak dephasing, and prethermal heating form the final
set of exercises and extensions.


\subsection{Synthesis of the Applications}
\label{subsec:applications_synthesis}

Seen together, the five examples suggest the same working rule without
forcing unlike systems into one mathematical template.  In the multiqubit
cavity, one must first determine which operator sectors are generated before
deciding whether to secularize them.  The three-level mixer adds topology and
signed detunings as process-selection rules.  The finite spin bus changes its
retained space when a bus pole approaches the endpoint manifold.  A lossy
mediator produces coherent and dissipative terms from the same complex
response.  In the driven lattice, the Floquet generator belongs together
with the periodic frame, quasienergy convention, and micromotion
reconstruction.

The relevant control parameter is consequently local and claim dependent.
It may be a populated-sector matrix-element-to-gap ratio, a virtual
occupation, the spectral distance from a bus mode, a complex Liouvillian
gap, a Floquet-replica denominator, or the ratio between an engineered rate
and a decoherence or resolution scale.  A small formal parameter does not
establish uniform validity when a leading process has been tuned through a
destructive zero, an eliminated pole approaches the retained manifold, or
the observation time resolves a nominally weak residual process.

The applications also show why complementary methods must be compared
through complete physical predictions rather than isolated generators.
Harmonic bookkeeping rapidly identifies exchange, pair, Raman, and Floquet
processes; projection and resolvents expose virtual paths, poles, and spatial
response; SW and small rotations supply dressed states and observables;
effective-operator and Liouvillian methods retain the dissipative companion
of a coherent interaction; and exact periodic frames restore laboratory
micromotion.  Agreement becomes meaningful only after the retained spaces,
perturbative orders, encoding maps, gauges, initial preparations, observable
reconstructions, and validity windows have been aligned.

An effective description is therefore useful not because its displayed
generator is short, but because it provides a controlled, interpretable, and
testable prediction for the physical quantity being claimed.  The examples
turn this principle into a practical workflow: identify the process, choose
the retained structure, derive all contributions at the claimed order,
reconstruct the laboratory quantity, and validate the approximation where it
will actually be used.

\section{Discussion and Conclusion}
\label{sec:discussion_conclusion}

Effective-model methods are often introduced as ways to obtain a simpler
Hamiltonian. Their real use is broader. A reduced generator becomes a physical
prediction only when the initial state, observables, eliminated amplitudes,
micromotion, and dissipative channels are treated in the same approximation.
The object to compare across methods is therefore the complete prediction

\begin{equation*}
    \hat\rho_{\mathrm{rec}}(t)
    =
    \left[
        \mathcal R_t
        \circ
        \Phi_{\mathrm{eff}}(t,0)
        \circ
        \mathcal E_0
    \right]
    [\hat\rho(0)],
\end{equation*}

\noindent
not the reduced Hamiltonian in isolation. This common structure allows
projection, block-diagonalization, averaging, adiabatic-elimination, and
Floquet constructions to be compared at the level of measurable quantities.

Several lessons follow from this point of view. An exact change of frame is not
an approximation; the approximation begins when transformed terms, sectors,
frequencies, or memory effects are discarded. Two effective Hamiltonians that
look different may describe the same physics after their frames, gauges, and
endpoint maps are aligned. Conversely, identical reduced Hamiltonians can
produce different laboratory predictions if their states or observables are
reconstructed differently. In an open system, the jump operators must be
transformed or derived together with the coherent Hamiltonian. GKLS form is a
statement about a generator, while complete positivity is a property to be
checked for the corresponding finite-time map.

The dispersive Jaynes--Cummings model made these distinctions especially easy
to see. Its exact excitation blocks reveal both the perturbative orders and the
occupation dependence of the approximation. The familiar parameter
\(g\sqrt{n+1}/|\Delta|\) is one example of a more general rule: the relevant
small parameter is set by the matrix elements and energy gaps of the states
that can actually be occupied. Collective light--matter couplings depend on
the Dicke sector, spin-chain and electronic reductions depend on mode- or
bond-resolved gaps, and many-body Floquet expansions are governed by local
interaction scales rather than the full extensive bandwidth. A ratio such as
\(g/\Delta\) is incomplete until the populated manifold, symmetry sector, and
system size are specified.

Each method keeps the discarded information in a different place. Projection
theory uses a wave or embedding operator. Schrieffer--Wolff and small-rotation
methods place it in dressed states and observables. Magnus and James--Jerke
constructions separate slow evolution from finite-time or filtered
micromotion. Floquet theory adds quasienergy and stroboscopic gauge freedom.
Adiabatic elimination reconstructs the fast variables through a slaving
relation and can produce coherent and dissipative corrections at the same
order. None of these representations is best in every problem. The useful one
is the representation that makes the target process clear and still permits a
controlled return to laboratory observables.

Going to higher order helps only while the expansion remains controlled over
the time window of interest. It cannot rescue a poorly chosen retained space,
a near-resonant state assigned to the eliminated sector, inconsistent gauge
truncation, or extrapolation beyond the spectral window of the model. Even a
small frequency error eventually becomes a large phase error. In driven and
many-body systems the series may be asymptotic, and the effective description
may apply only during a prethermal interval. In that setting, choosing an
appropriate truncation order and stating the useful time window are more
important than simply calculating one more formal order.

Several structural limits also deserve to remain visible. Gauge
transformations and finite truncations need not commute. A low-energy normal
form can reproduce the spectrum accurately without defining a sensible global
Gibbs state. Bosonic polynomial Hamiltonians require attention to operator
domains and to the boundary introduced by a numerical cutoff. A reduced map
may have both memory-kernel and time-local representations without admitting a
time-independent GKLS logarithm. Non-Hermitian effective operators can encode
resonance escape, conditional survival, quasiparticle damping, or Liouvillian
decay; these are related mathematical structures, but they are not the same
physical statement.

Validation is therefore part of the derivation, not a final cosmetic check. A
single fidelity curve is rarely enough. The benchmark must match the claim:
spectrum, transition amplitude, current, correlation function, decay rate, or
stationary state, evaluated in a stated manifold and over a stated time
interval. Exact or numerically converged microscopic dynamics provides the
strongest reference when it is available. Otherwise, residuals, analytical
scaling, small exactly solvable sectors, independent methods, and experiment
provide complementary evidence. Perturbative order, active space, bosonic
cutoff, lattice size, and tests of map physicality are part of the result and
should be reported with it.

The examples showed how these principles change from one regime to another.
Within the quantum Rabi model, near-resonant exchange, dispersive dressing,
displaced-oscillator behavior, selected multiphoton resonances, and the
large-frequency-ratio critical limit require different reorganizations of the
same microscopic Hamiltonian; they are not successive corrections in a single
universal hierarchy. The three-photon resonance provided a sharper
higher-order test: the quadratic resonance shift, cubic avoided-crossing gap,
virtual transition paths, and state reconstruction all had to remain
consistent.

The applications then made the same ideas concrete in other settings. In the
two-qubit cavity, the effective interaction separates exchange, pair, and
squeezing sectors, including their counter-rotating contributions. The
three-level mixer shows how signed detunings and coupling topology select
Raman-conversion and pair processes. The spin-chain bus exposes
mode-resolved denominators and the need to enlarge the retained space when a
bus mode approaches resonance. For a lossy mediator, the coherent shift and
induced decay are the dispersive and absorptive responses encoded by the same
complex susceptibility. In the driven lattice, an exact frame transformation precedes
the Bessel-renormalized tunnelling approximation, while micromotion and
prethermal heating set distinct limits on its use. Quantum-optical models make
many of these mechanisms transparent, but the reasoning carries directly to
spin, lattice, transport, and open-system problems.

In practice, a useful effective-model result should state

\begin{itemize}
    \item the retained and eliminated sectors;
    \item the control parameter, including its dependence on state, symmetry
    sector, occupation, and system size where relevant;
    \item the effective generator or finite-time map and the representation in
    which it is written;
    \item the encoding and reconstruction of physical states and observables;
    \item the leading omitted processes and their expected error scaling;
    \item the analytical and numerical convergence tests; and
    \item the demonstrated region of validity in parameter space and time.
\end{itemize}

The companion notebooks put this checklist into practice. Each central
computational example connects the microscopic model to the derivation, QuTiP
implementation, converged data, physical reconstruction, and at least one
deliberate failure or validity test. They are meant to make the calculation
inspectable and adaptable, while keeping the strength of each conclusion tied
to the evidence that supports it.

Taken together, the methods developed here support three connected abilities.
\emph{Derivation} shows how a controlled effective interaction is obtained.
\emph{Interpretation} identifies the virtual paths, symmetries, response
functions, and approximations that give the interaction physical meaning.
\emph{Discovery} uses that understanding to recognize new resonances,
couplings, and control strategies. Keeping these three abilities together---and
testing them through reconstructed physical predictions---turns effective-model
methods from a collection of formal recipes into practical tools for
engineering quantum dynamics.

\section*{Data and Code Availability}

All code and companion notebooks supporting this work are openly available at
\url{https://github.com/sinara-maker/quantum-engineering-interactions}. The
repository contains detailed, step-by-step derivations of the theoretical
results presented here. The Supplemental Material is also available at the
same link; it contains exact frame transformations and extended algebra in
Secs.~S1--S3, Liouville-space and dynamical-map conventions in Sec.~S4, the
notebook guide in Sec.~S5, and a selective glossary in Sec.~S6.

\section*{Acknowledgments}\label{Acknowledgments}

This work was supported by the São Paulo Research Foundation (FAPESP, Grants Nos. 2022/10218-2, 2022/00209-6, and  2025/15490-0), and by the National Council for Scientific and Technological Development (CNPq), grants No. 311612/2021-0 and No. 302234/2026-8. This work is also part of the MAI/DAI--CNPq Grants No. 141909/2023-4. 
OpenAI ChatGPT Work/Codex assisted with code debugging, deterministic figure-layout scripting, literature organization, and language editing. The authors retain responsibility for the scientific content, calculations, figures, references, and conclusions.
\newpage

\section*{Recurrent Abbreviations}

\begingroup
\centering
    \small
    \renewcommand{\arraystretch}{1.12}
    \begin{tabularx}{\columnwidth}{
        @{}
        >{\RaggedRight\arraybackslash}p{0.30\columnwidth}
        >{\RaggedRight\arraybackslash}X
        @{}
    }
        \toprule
        \textbf{Abbreviation}
        &
        \textbf{Meaning}
        \\
        \midrule
        AE & adiabatic elimination \\
        AMOH & approximate method for oscillating Hamiltonians \\
        CPTP & completely positive and trace preserving \\
        DSC & deep-strong coupling \\
        GKLS & Gorini--Kossakowski--Lindblad--Sudarshan \\
        GRWA & generalized rotating-wave approximation \\
        JC & Jaynes--Cummings \\
        MSR & method of small rotations \\
        NZ & Nakajima--Zwanzig \\
        QRM & quantum Rabi model \\
        RWA & rotating-wave approximation \\
        SW & Schrieffer--Wolff \\
        TCL & time-convolutionless \\
        USC & ultrastrong coupling \\
        \bottomrule
    \end{tabularx}
\par
\endgroup
\appendix
\section{Order-Consistent Encoding, Evolution, and Reconstruction}
\label{app:order_consistent_reconstruction}

This appendix makes precise a rule used throughout these notes: perturbative
order belongs to the complete prediction, not only to its effective generator.
Consider the expansions
\begin{equation*}
\begin{aligned}
    \hat H(\epsilon)
    &=
    \hat H_0
    +\epsilon\hat V_1
    +\epsilon^2\hat V_2
    +\mathcal O(\epsilon^3),
    \\
    \hat S(\epsilon)
    &=
    \epsilon\hat S_1
    +\epsilon^2\hat S_2
    +\mathcal O(\epsilon^3),
    \qquad
    \hat S_j^\dagger=-\hat S_j .
\end{aligned}
\end{equation*}
The retained projector introduced below is assumed to be independent of
\(\epsilon\). For the near-identity unitary
\(\hat W(\epsilon)=e^{\hat S(\epsilon)}\), we use the
physical-to-effective convention
\(
    \widetilde{\hat H}
    =
    \hat W\hat H\hat W^\dagger
\).
The Baker--Campbell--Hausdorff expansion then gives
\begin{align*}
    \widetilde{\hat H}
    ={}&
    \hat H_0
    +\epsilon\bigl(
        \hat V_1+[\hat S_1,\hat H_0]
    \bigr) \\
    &+\epsilon^2\biggl(
        \hat V_2
        +[\hat S_1,\hat V_1]
        +[\hat S_2,\hat H_0]
        \\
        &+\frac12[\hat S_1,[\hat S_1,\hat H_0]]
    \biggr)\\
    &+\mathcal O(\epsilon^3).
\end{align*}

Let \(\hat P\) project onto the retained manifold and
\(\hat Q=\hat I-\hat P\). A construction that is block diagonal through
order \(N\) must satisfy
\begin{equation}
    \hat Q\widetilde{\hat H}^{(N)}\hat P
    =
    \mathcal O(\epsilon^{N+1}),
    \qquad
    \hat H_{\mathrm{eff}}^{(N)}
    =
    \hat P\widetilde{\hat H}^{(N)}\hat P .
    \label{eq:app_block_residual}
\end{equation}
The remainder is understood in a norm appropriate to the stated domain or
converged occupied sector. If the Hamiltonian is Hermitian, the adjoint block
\(\hat P\widetilde{\hat H}^{(N)}\hat Q\) has the same order. The
off-diagonal residual is therefore an independent diagnostic: agreement of a
few retained-space eigenvalues does not by itself show that the eliminated
sector has been consistently decoupled.

The same transformation fixes the preparation and measurement maps. For a
physical initial state and observable, define
\begin{equation}
\begin{aligned}
    \mathcal E_0^{(N)}[\hat\rho]
    &=
    \hat P
    \left(
        \hat W\hat\rho\hat W^\dagger
    \right)_{\!N}
    \hat P,
    \\
    \hat O_{\mathrm{eff}}^{(N)}
    &=
    \hat P
    \left(
        \hat W\hat O\hat W^\dagger
    \right)_{\!N}
    \hat P .
\end{aligned}
\label{eq:app_endpoint_maps}
\end{equation}
Here \((\cdots)_N\) means truncation at the same order used for
\(\hat H_{\mathrm{eff}}^{(N)}\). The encoded state is generally
subnormalized when the physical preparation has weight outside the retained
manifold. A finite-order truncation of a unitary similarity transformation
also need not define an exactly positive or completely positive map by itself;
it is an asymptotic representation to the declared order.

For example, the effective observable through second order is
\begin{align*}
    \hat O_{\mathrm{eff}}^{(2)}
    ={}&\hat P\Bigl\{\hat O
    +\epsilon[\hat S_1,\hat O]+\\
    &\epsilon^2\Big(
        [\hat S_2,\hat O]   
    +\frac12[\hat S_1,[\hat S_1,\hat O]]
    \Big)
    \Bigr\}\hat P .
\end{align*}
Alternatively, one may reconstruct the physical state,
\begin{equation}
    \hat\rho_{\mathrm{rec}}^{(N)}(t)
    =
    \left(
        \hat W^\dagger
        \iota_P[\hat\rho_{\mathrm{eff}}^{(N)}(t)]
        \hat W
    \right)_{\!N},
    \label{eq:app_state_reconstruction}
\end{equation}
where \(\iota_P\) embeds a retained-space operator in the full Hilbert space.
These two measurement routes are equivalent to the controlled order:
\begin{equation}
    \operatorname{Tr}\!\left[
        \hat O\hat\rho_{\mathrm{rec}}^{(N)}(t)
    \right]
    =
    \operatorname{Tr}_{P}\!\left[
        \hat O_{\mathrm{eff}}^{(N)}
        \hat\rho_{\mathrm{eff}}^{(N)}(t)
    \right]
    +\mathcal O(\epsilon^{N+1}).
    \label{eq:app_dual_measurement_routes}
\end{equation}
One should therefore use either the bare observable with the reconstructed
physical state or the transformed observable with the effective state.
Reconstructing the final state and transforming the observable in the same
expectation value would apply the measurement correction twice. The initial
state must still be encoded through \(\mathcal E_0^{(N)}\).

For a time-independent closed effective Hamiltonian, the complete order-\(N\)
prediction can be evaluated in the retained space as
\begin{equation}
\begin{aligned}
    \langle\hat O\rangle_{\mathrm{pred}}(t)
    ={}&
    \operatorname{Tr}_{P}\!\Bigl[
        \hat O_{\mathrm{eff}}^{(N)}
        e^{-\mathrm{i}\hat H_{\mathrm{eff}}^{(N)}t}
    \\
    &\qquad\quad{}\times
        \mathcal E_0^{(N)}[\hat\rho(0)]
        e^{\mathrm{i}\hat H_{\mathrm{eff}}^{(N)}t}
    \Bigr].
    \label{eq:app_complete_prediction}
\end{aligned}
\end{equation}
If \(\mathcal E_0^{(N)}[\hat\rho(0)]\) is left subnormalized, this expression
gives the unconditional contribution carried by the retained manifold.
Dividing by its trace instead defines a conditional prediction and should be
identified as such.

Combining a dressed generator with bare endpoints can introduce errors already
at first order in \(\epsilon\), even when selected retained-space energies are
correct to second or higher order. Conversely, keeping endpoint terms beyond
the controlled order can create apparent accuracy without reducing either the
generator defect or the block-decoupling residual.

The same structure applies to open and periodically driven systems:
\begin{equation*}
    \Phi_{\mathrm{pred}}^{(N)}(t,0)
    =
    \mathcal R_t^{(N)}
    \circ
    \Phi_{\mathrm{eff}}^{(N)}(t,0)
    \circ
    \mathcal E_0^{(N)} .
\end{equation*}
All three factors are Schr\"odinger-picture maps. The reconstruction
\(\mathcal R_t^{(N)}\) may contain static dressing, an initial-layer
correction, eliminated-mode reconstruction, or periodic micromotion. A
time-independent GKLS generator makes the middle propagator a CPTP semigroup,
while a time-local GKLS generator with nonnegative rates gives a CP-divisible
propagator. Neither statement validates the endpoint maps or proves that the
composite prediction agrees with the microscopic channel.

An error budget is meaningful only after all objects have been placed in the
same physical frame and normalization convention. Let
\(\hat\rho_{\mathrm{mic}}(t)\) denote the microscopic state and define
\begin{equation*}
    \delta_O(t)
    =
    \left|
        \operatorname{Tr}\!\left[
            \hat O
            \bigl(
                \hat\rho_{\mathrm{mic}}(t)
                -\hat\rho_{\mathrm{rec}}^{(N)}(t)
            \bigr)
        \right]
    \right| .
\end{equation*}
Relative to an aligned reference factorization
\(\mathcal R_t\circ\Phi(t,0)\circ\mathcal E_0\), introduce
\begin{align*}
    \epsilon_{\mathrm{enc}}
    &=
    \left\|
        \bigl(\mathcal E_0-\mathcal E_0^{(N)}\bigr)[\hat\rho(0)]
    \right\|_1,
    \\
    \epsilon_{\mathrm{dyn}}(t)
    &=
    \left\|
        \bigl(\Phi(t,0)-\Phi_{\mathrm{eff}}^{(N)}(t,0)\bigr)
        \mathcal E_0^{(N)}[\hat\rho(0)]
    \right\|_1,
    \\
    \epsilon_{\mathrm{rec}}(t)
    &=
    \left\|
        \bigl(\mathcal R_t-\mathcal R_t^{(N)}\bigr)
        \Phi_{\mathrm{eff}}^{(N)}(t,0)
        \mathcal E_0^{(N)}[\hat\rho(0)]
    \right\|_1 .
\end{align*}
If the exact intermediate maps are trace-norm contractive on the declared
domain, H\"older's inequality and a telescoping triangle inequality give
\begin{equation}
    \delta_O(t)
    \leq
    \|\hat O\|_{\infty}
    \left[
        \epsilon_{\mathrm{enc}}
        +\epsilon_{\mathrm{dyn}}(t)
        +\epsilon_{\mathrm{rec}}(t)
    \right].
    \label{eq:app_error_budget}
\end{equation}
Leakage already present in the preparation belongs to
\(\epsilon_{\mathrm{enc}}\); leakage generated during the evolution belongs
to \(\epsilon_{\mathrm{dyn}}\). In the dual observable route, the
reconstruction contribution is instead bounded by
\(
    \|\hat O_{\mathrm{eff}}-\hat O_{\mathrm{eff}}^{(N)}\|_\infty
    \|\hat\rho_{\mathrm{eff}}^{(N)}(t)\|_1
\).
Additional stability constants must be stated whenever an intermediate map is
not contractive.

For a bounded finite-dimensional Hamiltonian problem, Duhamel's formula gives
\begin{equation}
\begin{aligned}
    \epsilon_{\mathrm{dyn}}(t)
    &\leq
    2\int_0^t
    \left\|\delta\hat H^{(N)}(s)\right\|_{\infty}
    \mathrm{d}s
    +\epsilon_{\mathrm{leak}}(t)
    \\
    &\leq
    2t\,
    \sup_{0\leq s\leq t}
    \left\|\delta\hat H^{(N)}(s)\right\|_{\infty}
    +\epsilon_{\mathrm{leak}}(t),
    \label{eq:app_duhamel_estimate}
\end{aligned}
\end{equation}
where \(\delta\hat H^{(N)}(s)\) is the retained generator defect in a common
frame. The linear factor in time explains why a spectrally small error can
become a large phase error at long times. For unbounded bosonic models, the
norms in Eqs.~\eqref{eq:app_error_budget}--\eqref{eq:app_duhamel_estimate}
must be restricted to the converged occupied sector rather than interpreted as
global operator-norm bounds.

Equations~\eqref{eq:app_block_residual},
\eqref{eq:app_complete_prediction}, and
\eqref{eq:app_error_budget} give a compact audit rule: state the local control
parameter and retained manifold; test the block or equation-of-motion
residual; encode the preparation and represent the measurement to the declared
order; use one of the two measurement routes in
Eq.~\eqref{eq:app_dual_measurement_routes}; and validate the reconstructed
observable over the stated time window.

\clearpage
\onecolumn
\bibliography{bib}

@article{Abanin2017,
  title = {Effective Hamiltonians,  prethermalization,  and slow energy absorption in periodically driven many-body systems},
  volume = {95},
  ISSN = {2469-9969},
  url = {http://dx.doi.org/10.1103/PhysRevB.95.014112},
  DOI = {10.1103/physrevb.95.014112},
  number = {1},
  journal = {Physical Review B},
  publisher = {American Physical Society (APS)},
  author = {Abanin,  Dmitry A. and De Roeck,  Wojciech and Ho,  Wen Wei and Huveneers,  Fran\c{c}ois},
  year = {2017},
  month = jan,
  pages = {014112}
}

@article{abanin2017rigorous,
  title = {Rigorous Bound on Energy Absorption and Generic Relaxation in Periodically Driven Quantum Systems},
  author = {Mori, Takashi and Kuwahara, Tomotaka and Saito, Keiji},
  journal = {Phys. Rev. Lett.},
  volume = {116},
  issue = {12},
  pages = {120401},
  numpages = {5},
  year = {2016},
  month = {3},
  publisher = {American Physical Society},
  doi = {10.1103/PhysRevLett.116.120401},
  url = {https://link.aps.org/doi/10.1103/PhysRevLett.116.120401}
}

@article{Akbari2025,
  title = {Floquet Engineering the Quantum Rabi Model in the Ultrastrong Coupling Regime},
  volume = {134},
  ISSN = {1079-7114},
  url = {http://dx.doi.org/10.1103/PhysRevLett.134.063602},
  DOI = {10.1103/physrevlett.134.063602},
  number = {6},
  journal = {Physical Review Letters},
  publisher = {American Physical Society (APS)},
  author = {Akbari,  Kamran and Nori,  Franco and Hughes,  Stephen},
  year = {2025},
  month = feb,
  pages = {063602}
}

@article{Albert2011,
  title = {Symmetric rotating-wave approximation for the generalized single-mode spin-boson system},
  author = {Albert, Victor V. and Scholes, Gregory D. and Brumer, Paul},
  journal = {Phys. Rev. A},
  volume = {84},
  issue = {4},
  pages = {042110},
  numpages = {9},
  year = {2011},
  month = {Oct},
  publisher = {American Physical Society},
  doi = {10.1103/PhysRevA.84.042110},
  url = {https://link.aps.org/doi/10.1103/PhysRevA.84.042110}
}

@article{AlexanianBose1995,
  author = {Alexanian, Moorad and Bose, Subir K.},
  title = {Unitary transformation and the dynamics of a three-level atom interacting with two quantized field modes},
  journal = {Physical Review A},
  volume = {52},
  pages = {2218--2224},
  year = {1995},
  doi = {10.1103/PhysRevA.52.2218}
}

@article{Almeida2018,
  author = {Almeida, G. M. A.},
  title = {Interplay between Speed and Fidelity in Off-Resonant Quantum-State Transfer},
  journal = {Physical Review A},
  volume = {98},
  pages = {012334},
  year = {2018},
  doi = {10.1103/PhysRevA.98.012334}
}

@article{AnderssonCresserHall2007,
  author = {Andersson, Erika and Cresser, James D. and Hall, Michael J. W.},
  title = {Finding the Kraus decomposition from a master equation and vice versa},
  journal = {Journal of Modern Optics},
  volume = {54},
  number = {12},
  pages = {1695--1716},
  year = {2007},
  doi = {10.1080/09500340701352581},
  url = {https://doi.org/10.1080/09500340701352581}
}

@article{ArayaDay2025,
  title = {Pymablock: An algorithm and a package for quasi-degenerate perturbation theory},
  ISSN = {2949-804X},
  url = {http://dx.doi.org/10.21468/SciPostPhysCodeb.50},
  DOI = {10.21468/scipostphyscodeb.50},
  journal = {SciPost Physics Codebases},
  publisher = {Stichting SciPost},
  author = {Araya Day,  Isidora and Miles,  Sebastian and Kerstens,  Hugo and Varjas,  Daniel and Akhmerov,  Anton R.},
  year = {2025},
  month = feb,
  pages = {50}
}

@article{arwas2023metrics,
  author = {Arwas, Geva and Manucharyan, Vladimir E. and Ciuti, Cristiano},
  title = {Metrics and properties of optimal gauges in multimode cavity {QED}},
  journal = {Physical Review A},
  volume = {108},
  number = {2},
  pages = {023714},
  year = {2023},
  doi = {10.1103/PhysRevA.108.023714},
  url = {https://doi.org/10.1103/PhysRevA.108.023714}
}

@article{ashhab2025finite,
  title = {Finite-dimensional approximations of generalized squeezing},
  author = {Ashhab, Sahel and Fischer, Felix and Lonigro, Davide and Braak, Daniel and Burgarth, Daniel},
  journal = {Phys. Rev. A},
  volume = {113},
  issue = {1},
  pages = {013703},
  numpages = {17},
  year = {2026},
  month = {1},
  publisher = {American Physical Society},
  doi = {10.1103/9vwp-f35c},
  url = {https://link.aps.org/doi/10.1103/9vwp-f35c}
}

@article{ashida2020non,
  title={Non-Hermitian physics},
  author={Ashida, Yuto and Gong, Zongping and Ueda, Masahito},
  journal={Advances in Physics},
  volume={69},
  number={3},
  pages={249--435},
  year={2020},
  publisher={Taylor \& Francis},
  doi = {10.1080/00018732.2021.1876991}
}

@article{Azouit2017,
  title = {Towards generic adiabatic elimination for bipartite open quantum systems},
  volume = {2},
  ISSN = {2058-9565},
  url = {http://dx.doi.org/10.1088/2058-9565/aa7f3f},
  DOI = {10.1088/2058-9565/aa7f3f},
  number = {4},
  journal = {Quantum Science and Technology},
  publisher = {IOP Publishing},
  author = {Azouit,  R and Chittaro,  F and Sarlette,  A and Rouchon,  P},
  year = {2017},
  month = sep,
  pages = {044011}
}

@article{babiker1983derivation,
  author = {Babiker, M. and Loudon, Rodney},
  title = {Derivation of the {Power--Zienau--Woolley} Hamiltonian in quantum electrodynamics by gauge transformation},
  journal = {Proceedings of the Royal Society of London. A. Mathematical and Physical Sciences},
  volume = {385},
  number = {1789},
  pages = {439--460},
  year = {1983},
  doi = {10.1098/rspa.1983.0022},
  url = {https://doi.org/10.1098/rspa.1983.0022}
}

@article{Banchi2010,
  author = {Banchi, Leonardo and Apollaro, T. J. G. and Cuccoli, Alessandro and Vaia, Ruggero and Verrucchi, Paola},
  title = {Long Quantum Channels for High-Quality Entanglement Transfer},
  journal = {Physical Review A},
  volume = {82},
  pages = {052321},
  year = {2010},
  doi = {10.1103/PhysRevA.82.052321}
}

@article{Beatrez2021,
  author  = {Beatrez, William and Janes, Otto and Akkiraju, Amala and Pillai, Arjun and Oddo, Alexander and Reshetikhin, Paul and Druga, Emanuel and McAllister, Maxwell and Elo, Mark and Gilbert, Benjamin and Suter, Dieter and Ajoy, Ashok},
  title   = {Floquet Prethermalization with Lifetime Exceeding 90 s in a Bulk Hyperpolarized Solid},
  journal = {Physical Review Letters},
  volume  = {127},
  pages   = {170603},
  year    = {2021},
  doi     = {10.1103/PhysRevLett.127.170603}
}

@book{BenderOrszag1999,
  author    = {Bender, Carl M. and Orszag, Steven A.},
  title     = {Advanced Mathematical Methods for Scientists and
               Engineers I: Asymptotic Methods and Perturbation Theory},
  publisher = {Springer},
  address   = {New York},
  year      = {1999},
  doi       = {10.1007/978-1-4757-3069-2}
}

@article{Benito2019,
  author    = {Benito, M. and Petta, J. R. and Burkard, Guido},
  title     = {Three-electron Pauli spin blockade in a double quantum dot},
  journal   = {Physical Review B},
  volume    = {100},
  issue     = {8},
  pages     = {081412(R)},
  year      = {2019},
  doi       = {10.1103/PhysRevB.100.081412}
}

@article{Bergholtz,
  title = {Exceptional topology of non-Hermitian systems},
  author = {Bergholtz, Emil J. and Budich, Jan Carl and Kunst, Flore K.},
  journal = {Rev. Mod. Phys.},
  volume = {93},
  issue = {1},
  pages = {015005},
  numpages = {31},
  year = {2021},
  month = {2},
  publisher = {American Physical Society},
  doi = {10.1103/RevModPhys.93.015005},
  url = {https://link.aps.org/doi/10.1103/RevModPhys.93.015005}
}

@article{Blais2021,
  title = {Circuit quantum electrodynamics},
  volume = {93},
  ISSN = {1539-0756},
  url = {http://dx.doi.org/10.1103/RevModPhys.93.025005},
  DOI = {10.1103/revmodphys.93.025005},
  number = {2},
  journal = {Reviews of Modern Physics},
  publisher = {American Physical Society (APS)},
  author = {Blais,  Alexandre and Grimsmo,  Arne L. and Girvin,  S. M. and Wallraff,  Andreas},
  year = {2021},
  month = may,
  pages = {025005}
}

@article{Blanes2009,
  author = {Blanes, S. and Casas, F. and Oteo, J. A. and Ros, J.},
  title = {The Magnus expansion and some of its applications},
  journal = {Physics Reports},
  volume = {470},
  number = {5-6},
  pages = {151--238},
  year = {2009},
  doi = {10.1016/j.physrep.2008.11.001}
}

@article{Blanes2010,
  author = {Blanes, S. and Casas, F. and Oteo, J. A. and Ros, J.},
  title = {A pedagogical approach to the Magnus expansion},
  journal = {European Journal of Physics},
  volume = {31},
  number = {4},
  pages = {907},
  year = {2010},
  doi = {10.1088/0143-0807/31/4/023}
}

@article{Bloch1958,
  title = {Sur la th{\'e}orie des perturbations des {\'e}tats li{\'e}s},
  volume = {6},
  ISSN = {0029-5582},
  url = {http://dx.doi.org/10.1016/0029-5582(58)90116-0},
  DOI = {10.1016/0029-5582(58)90116-0},
  journal = {Nuclear Physics},
  publisher = {Elsevier BV},
  author = {Bloch,  Claude},
  year = {1958},
  month = mar,
  pages = {329--347}
}

@article{Bloch_Horowitz1958,
      title = {Sur la d{\'e}termination des premiers {\'e}tats d'un syst{\`e}me de fermions dans le cas d{\'e}g{\'e}n{\'e}r{\'e}},
  volume = {8},
  ISSN = {0029-5582},
  url = {http://dx.doi.org/10.1016/0029-5582(58)90136-6},
  DOI = {10.1016/0029-5582(58)90136-6},
  journal = {Nuclear Physics},
  publisher = {Elsevier BV},
  author = {Bloch,  Claude and Horowitz,  Jules},
  year = {1958},
  month = sep,
  pages = {91--105}
}

@article{Boozer2008,
  author = {Boozer, A. D.},
  title = {Theory of Raman transitions in cavity QED},
  journal = {Physical Review A},
  volume = {78},
  pages = {033406},
  year = {2008},
  doi = {10.1103/PhysRevA.78.033406}
}

@article{Bose2003,
  author = {Bose, Sougato},
  title = {Quantum Communication through an Unmodulated Spin Chain},
  journal = {Physical Review Letters},
  volume = {91},
  pages = {207901},
  year = {2003},
  doi = {10.1103/PhysRevLett.91.207901}
}

@article{braak2011integrability,
  title = {Integrability of the Rabi Model},
  author = {Braak, D.},
  journal = {Phys. Rev. Lett.},
  volume = {107},
  issue = {10},
  pages = {100401},
  numpages = {4},
  year = {2011},
  month = {Aug},
  publisher = {American Physical Society},
  doi = {10.1103/PhysRevLett.107.100401},
  url = {https://link.aps.org/doi/10.1103/PhysRevLett.107.100401}
}

@article{Brandow1979,
  title = {Formal theory of effective $\pi$-electron hamiltonians},
  volume = {15},
  ISSN = {1097-461X},
  url = {http://dx.doi.org/10.1002/qua.560150207},
  DOI = {10.1002/qua.560150207},
  number = {2},
  journal = {International Journal of Quantum Chemistry},
  publisher = {Wiley},
  author = {Brandow,  B. H.},
  year = {1979},
  month = feb,
  pages = {207--242}
}

@article{Bravyi2011,
  title = {Schrieffer--Wolff transformation for quantum many-body systems},
  volume = {326},
  ISSN = {0003-4916},
  url = {http://dx.doi.org/10.1016/j.aop.2011.06.004},
  DOI = {10.1016/j.aop.2011.06.004},
  number = {10},
  journal = {Annals of Physics},
  publisher = {Elsevier BV},
  author = {Bravyi,  Sergey and DiVincenzo,  David P. and Loss,  Daniel},
  year = {2011},
  month = oct,
  pages = {2793--2826}
}

@article{BravyiHastingsVerstraete2006,
  author = {Bravyi, Sergey and Hastings, Matthew B. and Verstraete, Frank},
  title = {Lieb--Robinson Bounds and the Generation of Correlations and Topological Quantum Order},
  journal = {Physical Review Letters},
  volume = {97},
  number = {5},
  pages = {050401},
  year = {2006},
  doi = {10.1103/PhysRevLett.97.050401},
  url = {https://doi.org/10.1103/PhysRevLett.97.050401}
}

@book{breuer2002theory,
  title={The theory of open quantum systems},
  author={Breuer, Heinz-Peter and Petruccione, Francesco},
  year={2002},
  publisher={Oxford University Press, USA}
}

@article{Brion2007,
  doi = {10.1088/1751-8113/40/5/011},
  url = {https://doi.org/10.1088/1751-8113/40/5/011},
  year = {2007},
  month = {jan},
  publisher = {{IOP} Publishing},
  volume = {40},
  number = {5},
  pages = {1033-1043},
  author = {E Brion and L H Pedersen and K M{\o}lmer},
  title = {Adiabatic elimination in a lambda system},
  journal = {Journal of Physics A: Mathematical and Theoretical}
}

@article{Brune1990QND,
  author = {Brune, M. and Haroche, S. and Lefevre, V. and Raimond, J. M. and Zagury, N.},
  title = {Quantum Nondemolition Measurement of Small Photon Numbers by Rydberg-Atom Phase-Sensitive Detection},
  journal = {Physical Review Letters},
  volume = {65},
  number = {8},
  pages = {976--979},
  year = {1990},
  doi = {10.1103/PhysRevLett.65.976}
}

@article{Brune1994LightShifts,
  author = {Brune, M. and Nussenzveig, P. and Schmidt-Kaler, F. and Bernardot, F. and Maali, A. and Raimond, J. M. and Haroche, S.},
  title = {From Lamb Shift to Light Shifts: Vacuum and Subphoton Cavity Fields Measured by Atomic Phase-Sensitive Detection},
  journal = {Physical Review Letters},
  volume = {72},
  number = {21},
  pages = {3339--3342},
  year = {1994},
  doi = {10.1103/PhysRevLett.72.3339}
}

@article{Bukov2015,
  author  = {Bukov, Marin and D'Alessio, Luca and Polkovnikov, Anatoli},
  title   = {Universal High-Frequency Behavior of Periodically Driven Systems: from Dynamical Stabilization to Floquet Engineering},
  journal = {Advances in Physics},
  volume  = {64},
  number  = {2},
  pages   = {139--226},
  year    = {2015},
  doi     = {10.1080/00018732.2015.1055918}
}

@misc{Bukov2017Floquet,
  author    = {Marin Georgiev Bukov},
  title     = {Floquet Engineering in Periodically Driven Closed Quantum Systems: From Dynamical Localisation to Ultracold Topological Matter},
  school    = {Boston University, Graduate School of Arts and Sciences},
  year      = {2017},
  address   = {Boston, MA},
  type      = {Dissertation}
}

@article{Cai2021QRMQPT,
  author = {Cai, M.-L. and Liu, Z.-D. and Zhao, W.-D. and Wu, Y.-K. and Mei, Q.-X. and Jiang, Y. and He, L. and Zhang, X. and Zhou, Z.-C. and Duan, L.-M.},
  title = {Observation of a Quantum Phase Transition in the Quantum Rabi Model with a Single Trapped Ion},
  journal = {Nature Communications},
  volume = {12},
  pages = {1126},
  year = {2021},
  doi = {10.1038/s41467-021-21425-8}
}

@article{Chang2020ThreePhotonSPDC,
  author  = {Chang, C. W. Sandbo and Sab{\'i}n, Carlos and Forn-D{\'i}az, P. and Quijandr{\'i}a, Fernando and Vadiraj, A. M. and Nsanzineza, I. and Johansson, G. and Wilson, C. M.},
  title   = {Observation of Three-Photon Spontaneous Parametric Down-Conversion in a Superconducting Parametric Cavity},
  journal = {Physical Review X},
  volume  = {10},
  number  = {1},
  pages   = {011011},
  year    = {2020},
  doi     = {10.1103/PhysRevX.10.011011}
}

@article{ChaturvediShibata1979,
  author = {Chaturvedi, S. and Shibata, F.},
  title = {Time-convolutionless projection operator formalism for elimination of fast variables. Applications to Brownian motion},
  journal = {Zeitschrift f{\"u}r Physik B Condensed Matter},
  volume = {35},
  number = {3},
  pages = {297--308},
  year = {1979},
  doi = {10.1007/BF01319852},
  url = {https://doi.org/10.1007/BF01319852}
}

@article{Chen2017HighOrderSidebands,
  author  = {Chen, Zhen and Wang, Yimin and Li, Tiefu and Tian, Lin and Qiu, Yueyin and Inomata, Kunihiro and Yoshihara, Fumiki and Han, Siyuan and Nori, Franco and Tsai, J. S. and You, J. Q.},
  title   = {Single-Photon-Driven High-Order Sideband Transitions in an Ultrastrongly Coupled Circuit-Quantum-Electrodynamics System},
  journal = {Physical Review A},
  volume  = {96},
  number  = {1},
  pages   = {012325},
  year    = {2017},
  doi     = {10.1103/PhysRevA.96.012325}
}

@article{Chessari2025,
  title = {Unifying Floquet Theory of Longitudinal and Dispersive Readout},
  volume = {134},
  ISSN = {1079-7114},
  url = {http://dx.doi.org/10.1103/PhysRevLett.134.037003},
  DOI = {10.1103/physrevlett.134.037003},
  number = {3},
  journal = {Physical Review Letters},
  publisher = {American Physical Society (APS)},
  author = {Chessari,  A. and Rodr{\'\i}guez-Mena,  E. A. and Abadillo-Uriel,  J. C. and Champain,  V. and Zihlmann,  S. and Maurand,  R. and Niquet,  Y.-M. and Filippone,  M.},
  year = {2025},
  month = Jan,
  pages = {037003}
}

@article{Christandl2004,
  author = {Christandl, Matthias and Datta, Nilanjana and Ekert, Artur and Landahl, Andrew J.},
  title = {Perfect State Transfer in Quantum Spin Networks},
  journal = {Physical Review Letters},
  volume = {92},
  pages = {187902},
  year = {2004},
  doi = {10.1103/PhysRevLett.92.187902}
}

@misc{codigo_python_huang,
    author = {Huang,  Longxiang and Luneau,  Jacquelin and Schirk,  Johannes and Wallner,  Florian and Schneider,  Christian M. F. and Filipp,  Stefan and Liegener,  Klaus and Rabl, Peter},
    title = {Theory of multiphoton processes for applications in quantum control},
    doi = {10.5281/zenodo.15574484},
    url = {https://doi.org/10.5281/zenodo.15574484},
    note = {Code repository},
    year = {2025}
}

@book{cohen1998atom,
  title={Atom-photon interactions: basic processes and applications},
  author={Cohen-Tannoudji, Claude and Dupont-Roc, Jacques and Grynberg, Gilbert},
  year={1998},
  publisher={John Wiley \& Sons}
}

@article{ColoredChannels2013,
  author = {de Moraes Neto, G. D. and de Ponte, M. A. and Moussa, M. H. Y.},
  title = {Colored Channels for High-Fidelity Information Transfer and Processing between Remote Multi-Branch Quantum Circuits},
  journal = {EPL},
  volume = {103},
  pages = {43001},
  year = {2013},
  doi = {10.1209/0295-5075/103/43001}
}

@article{daley2014quantum,
  title={Quantum trajectories and open many-body quantum systems},
  author={Daley, Andrew J},
  journal={Advances in Physics},
  volume={63},
  number={2},
  pages={77--149},
  year={2014},
  publisher={Taylor \& Francis},
  doi = {10.1080/00018732.2014.933502}
}

@article{DalibardCastinMolmer1992,
  author = {Dalibard, Jean and Castin, Yvan and M{\o}lmer, Klaus},
  title = {Wave-function approach to dissipative processes in quantum optics},
  journal = {Physical Review Letters},
  volume = {68},
  number = {5},
  pages = {580--583},
  year = {1992},
  doi = {10.1103/PhysRevLett.68.580},
  url = {https://doi.org/10.1103/PhysRevLett.68.580}
}

@article{DeBernardis24,
author = {Daniele De Bernardis and Alberto Mercurio and Simone De Liberato},
journal = {J. Opt. Soc. Am. B},
number = {8},
pages = {C206--C221},
publisher = {Optica Publishing Group},
title = {Tutorial on nonperturbative cavity quantum electrodynamics: is the Jaynes--Cummings model still relevant?},
volume = {41},
  month = {8},
year = {2024},
url = {https://opg.optica.org/josab/abstract.cfm?URI=josab-41-8-C206},
doi = {10.1364/JOSAB.522786},
}

@article{Meystre1992,
  title = {Theoretical developments in cavity quantum optics: a brief review},
  volume = {219},
  ISSN = {0370-1573},
  url = {http://dx.doi.org/10.1016/0370-1573(92)90141-L},
  DOI = {10.1016/0370-1573(92)90141-l},
  number = {3-6},
  journal = {Physics Reports},
  publisher = {Elsevier BV},
  author = {Meystre,  P},
  year = {1992},
  month = Oct,
  pages = {243–262}
}

@article{Delgado2000,
  doi = {10.1088/1464-4266/2/1/306},
  url = {https://doi.org/10.1088/1464-4266/2/1/306},
  year = {2000},
  month = jan,
  publisher = {{IOP} Publishing},
  volume = {2},
  number = {1},
  pages = {33--40},
  author = {J Delgado and A Luis and L L S{\'{a}}nchez-Soto and A B Klimov},
  title = {Quantum dynamics of the relative phase in second-harmonic generation},
  journal = {Journal of Optics B: Quantum and Semiclassical Optics}
}

@article{DellaValle2007,
  author  = {Della Valle, Giuseppe and Ornigotti, Marco and Cianci, E. and Foglietti, V. and Laporta, P. and Longhi, Stefano},
  title   = {Visualization of Coherent Destruction of Tunneling in an Optical Double Well System},
  journal = {Physical Review Letters},
  volume  = {98},
  pages   = {263601},
  year    = {2007},
  doi     = {10.1103/PhysRevLett.98.263601}
}

@article{Deng2016,
  title = {Dynamics of a two-level system under strong driving: Quantum-gate optimization based on Floquet theory},
  volume = {94},
  ISSN = {2469-9934},
  url = {http://dx.doi.org/10.1103/PhysRevA.94.032323},
  DOI = {10.1103/physreva.94.032323},
  number = {3},
  journal = {Physical Review A},
  publisher = {American Physical Society (APS)},
  author = {Deng,  Chunqing and Shen,  Feiruo and Ashhab,  Sahel and Lupascu,  Adrian},
  year = {2016},
  month = sep,
  pages = {032323}
}

@article{Dicke1954,
  author = {Dicke, R. H.},
  title = {Coherence in Spontaneous Radiation Processes},
  journal = {Physical Review},
  volume = {93},
  number = {1},
  pages = {99--110},
  year = {1954},
  doi = {10.1103/PhysRev.93.99},
  url = {https://doi.org/10.1103/PhysRev.93.99}
}

@article{DiFranco2008,
  author = {Di Franco, C. and Paternostro, M. and Kim, M. S.},
  title = {Perfect State Transfer on a Spin Chain without State Initialization},
  journal = {Physical Review Letters},
  volume = {101},
  pages = {230502},
  year = {2008},
  doi = {10.1103/PhysRevLett.101.230502}
}

@article{Dinc2025SpectralUnwinding,
  author  = {Din{\c{c}}, G{\"o}rkem D. and Eckardt, Andr{\'e} and Schnell, Alexander},
  title   = {Effective Floquet Lindblad generators from spectral unwinding},
  journal = {Physical Review A},
  volume  = {111},
  pages   = {062216},
  year    = {2025},
  doi     = {10.1103/x2qh-12gg}
}

@article{distefano2019resolution,
  author = {Di Stefano, Omar and Settineri, Alessio and Macr{\`i}, Vincenzo and Garziano, Luigi and Stassi, Roberto and Savasta, Salvatore and Nori, Franco},
  title = {Resolution of gauge ambiguities in ultrastrong-coupling cavity quantum electrodynamics},
  journal = {Nature Physics},
  volume = {15},
  number = {8},
  pages = {803--808},
  year = {2019},
  doi = {10.1038/s41567-019-0534-4},
  url = {https://doi.org/10.1038/s41567-019-0534-4}
}

@article{Eckardt2015,
  title = {High-frequency approximation for periodically driven quantum systems from a Floquet-space perspective},
  volume = {17},
  ISSN = {1367-2630},
  url = {http://dx.doi.org/10.1088/1367-2630/17/9/093039},
  DOI = {10.1088/1367-2630/17/9/093039},
  number = {9},
  journal = {New Journal of Physics},
  publisher = {IOP Publishing},
  author = {Eckardt,  Andr{\'e} and Anisimovas,  Egidijus},
  year = {2015},
  month = sep,
  pages = {093039}
}

@article{Eckardt2017,
  author  = {Eckardt, Andr{\'e}},
  title   = {Colloquium: Atomic Quantum Gases in Periodically Driven Optical Lattices},
  journal = {Reviews of Modern Physics},
  volume  = {89},
  pages   = {011004},
  year    = {2017},
  doi     = {10.1103/RevModPhys.89.011004}
}

@book{Economou2006,
  title = {Green's Functions in Quantum Physics},
  DOI = {10.1007/3-540-28841-4},
  ISBN = {978-3-540-28841-1},
  ISSN = {0171-1873},
  publisher = {Springer Berlin Heidelberg},
  author = {Economou, Eleftherios N.},
  year = {2006}
}

@article{Feshbach1958,
  title = {Unified theory of nuclear reactions},
  volume = {5},
  ISSN = {0003-4916},
  url = {http://dx.doi.org/10.1016/0003-4916(58)90007-1},
  DOI = {10.1016/0003-4916(58)90007-1},
  number = {4},
  journal = {Annals of Physics},
  publisher = {Elsevier BV},
  author = {Feshbach,  Herman},
  year = {1958},
  month = dec,
  pages = {357--390}
}

@article{Feshbach1962,
  title = {A unified theory of nuclear reactions. II},
  volume = {19},
  ISSN = {0003-4916},
  url = {http://dx.doi.org/10.1016/0003-4916(62)90221-X},
  DOI = {10.1016/0003-4916(62)90221-x},
  number = {2},
  journal = {Annals of Physics},
  publisher = {Elsevier BV},
  author = {Feshbach,  Herman},
  year = {1962},
  month = aug,
  pages = {287--313}
}

@article{FESHBACH1964,
  title = {Unified Theory of Nuclear Reactions},
  volume = {36},
  ISSN = {0034-6861},
  url = {http://dx.doi.org/10.1103/RevModPhys.36.1076},
  DOI = {10.1103/revmodphys.36.1076},
  number = {4},
  journal = {Reviews of Modern Physics},
  publisher = {American Physical Society (APS)},
  author = {FESHBACH,  HERMAN},
  year = {1964},
  month = oct,
  pages = {1076--1078}
}

@article{FinkelsteinShapiro2020,
  doi = {10.1103/physreva.101.042102},
  url = {https://doi.org/10.1103/physreva.101.042102},
  year = {2020},
  month = apr,
  publisher = {American Physical Society ({APS})},
  volume = {101},
  number = {4},
  author = {Daniel Finkelstein-Shapiro and David Viennot and Ibrahim Saideh and Thorsten Hansen and T{\~{o}}nu Pullerits and Arne Keller},
  title = {Adiabatic elimination and subspace evolution of open quantum systems},
  journal = {Physical Review A},
  pages = {042102}
}

@article{fischer2025selfadjoint,
  author = {Fischer, Felix and Burgarth, Daniel and Lonigro, Davide},
  title = {Self-adjoint realizations of higher-order squeezing operators},
  journal = {Journal of Physics A: Mathematical and Theoretical},
  volume = {59},
  number = {25},
  pages = {255203},
  year = {2026},
  doi = {10.1088/1751-8121/ae7acb},
  eprint = {2508.09044},
  archiveprefix = {arXiv},
  primaryclass = {math-ph},
  url = {https://doi.org/10.1088/1751-8121/ae7acb}
}

@misc{Fischer2026KPhoton,
  author        = {Fischer, Felix and Knapp, Felix and Burgarth, Daniel and Lonigro, Davide},
  title         = {Self-adjoint extensions of {$k$}-photon light--matter Hamiltonians},
  year          = {2026},
  eprint        = {2607.22378},
  archivePrefix = {arXiv},
  primaryClass  = {math-ph},
  note          = {Preprint}
}

@article{FischerWrongBox2026,
  author = {Fischer, Felix and Burgarth, Daniel and Lonigro, Davide},
  title = {Quantum particle in the wrong box (or: the perils of finite-dimensional approximations)},
  journal = {Quantum},
  volume = {10},
  pages = {1985},
  year = {2026},
  month = jan,
  doi = {10.22331/q-2026-01-27-1985},
  eprint = {2412.15889},
  archiveprefix = {arXiv},
  primaryclass = {quant-ph},
  url = {https://doi.org/10.22331/q-2026-01-27-1985}
}

@article{Fleischhauer2005,
  author = {Fleischhauer, Michael and Imamoglu, Atac and Marangos, Jonathan P.},
  title = {Electromagnetically induced transparency: Optics in coherent media},
  journal = {Reviews of Modern Physics},
  volume = {77},
  pages = {633},
  year = {2005},
  doi = {10.1103/RevModPhys.77.633}
}

@article{Floquet1883,
  title = {Sur les {\'e}quations diff{\'e}rentielles lin{\'e}aires {\`a} coefficients p{\'e}riodiques},
  volume = {12},
  ISSN = {1873-2151},
  url = {http://dx.doi.org/10.24033/asens.220},
  DOI = {10.24033/asens.220},
  journal = {Annales scientifiques de l'{\'E}cole normale sup{\'e}rieure},
  publisher = {Societe Mathematique de France},
  author = {Floquet,  G.},
  year = {1883},
  pages = {47--88}
}

@misc{Floquet_handbook,
  doi = {10.48550/ARXIV.2003.08252},
  url = {https://arxiv.org/abs/2003.08252},
  author = {Rudner,  Mark S. and Lindner,  Netanel H.},
  title = {The Floquet Engineer's Handbook},
  publisher = {arXiv},
  year = {2020},
  copyright = {arXiv.org perpetual,  non-exclusive license}
}

@article{FornDiaz2010,
  title={Observation of the Bloch-Siegert shift in a qubit-oscillator system in the ultrastrong coupling regime},
  author={Forn-D{\'\i}az, P. and Lisenfeld, J. and Marcos, D. and Garc{\'\i}a-Ripoll, J. J. and Solano, E. and Harmans, C. J. P. M. and Mooij, J. E.},
  journal={Phys. Rev. Lett.},
  volume={105},
  pages={237001},
  year={2010},
  publisher={APS},
  doi = {10.1103/PhysRevLett.105.237001}
}

@article{FornDiaz2019,
  author  = {Forn-D\'{i}az, P. and Lamata, L. and Rico, E. and Kono, J. and Solano, E.},
  title   = {Ultrastrong coupling regimes of light--matter interaction},
  journal = {Reviews of Modern Physics},
  volume  = {91},
  number  = {2},
  pages   = {025005},
  year    = {2019},
  doi     = {10.1103/RevModPhys.91.025005}
}

@article{GamelJames2010,
  author = {Gamel, Omar and James, Daniel F. V.},
  title = {Time-averaged quantum dynamics and the validity of the effective Hamiltonian model},
  journal = {Physical Review A},
  volume = {82},
  pages = {052106},
  year = {2010},
  doi = {10.1103/PhysRevA.82.052106}
}

@article{Gandon2022,
  title = {Engineering,  Control,  and Longitudinal Readout of Floquet Qubits},
  volume = {17},
  ISSN = {2331-7019},
  url = {http://dx.doi.org/10.1103/PhysRevApplied.17.064006},
  DOI = {10.1103/physrevapplied.17.064006},
  number = {6},
  journal = {Physical Review Applied},
  publisher = {American Physical Society (APS)},
  author = {Gandon,  Anthony and Le Calonnec,  Camille and Shillito,  Ross and Petrescu,  Alexandru and Blais,  Alexandre},
  year = {2022},
  month = jun,
  pages = {064006}
}

@article{GarcaMata2024,
  title = {Effective versus Floquet theory for the Kerr parametric oscillator},
  volume = {8},
  ISSN = {2521-327X},
  url = {http://dx.doi.org/10.22331/q-2024-03-25-1298},
  DOI = {10.22331/q-2024-03-25-1298},
  journal = {Quantum},
  publisher = {Verein zur Forderung des Open Access Publizierens in den Quantenwissenschaften},
  author = {Garc{\'\i}a-Mata,  Ignacio and Corti{\~n}as,  Rodrigo G. and Xiao,  Xu and Ch{\'a}vez-Carlos,  Jorge and Batista,  Victor S. and Santos,  Lea F. and Wisniacki,  Diego A.},
  year = {2024},
  month = mar,
  pages = {1298}
}

@article{Garziano2016OnePhotonTwoAtoms,
  author  = {Garziano, Luigi and Macr{\`i}, Vincenzo and Stassi, Roberto and Di Stefano, Omar and Nori, Franco and Savasta, Salvatore},
  title   = {One Photon Can Simultaneously Excite Two or More Atoms},
  journal = {Physical Review Letters},
  volume  = {117},
  number  = {4},
  pages   = {043601},
  year    = {2016},
  doi     = {10.1103/PhysRevLett.117.043601}
}

@article{GerryEberly1990,
  author = {Gerry, Christopher C. and Eberly, J. H.},
  title = {Dynamics of a Raman coupled model interacting with two quantized cavity fields},
  journal = {Physical Review A},
  volume = {42},
  pages = {6805},
  year = {1990},
  doi = {10.1103/PhysRevA.42.6805}
}

@article{Gleyzes2007QuantumJumps,
  author = {Gleyzes, S. and Kuhr, S. and Guerlin, C. and Bernu, J. and Del{\'e}glise, S. and Busk Hoff, U. and Brune, M. and Raimond, J.-M. and Haroche, S.},
  title = {Quantum Jumps of Light Recording the Birth and Death of a Photon in a Cavity},
  journal = {Nature},
  volume = {446},
  pages = {297--300},
  year = {2007},
  doi = {10.1038/nature05589}
}

@article{GoldmanDalibard2014,
  author  = {Goldman, Nathan and Dalibard, Jean},
  title   = {Periodically Driven Quantum Systems: Effective Hamiltonians and Engineered Gauge Fields},
  journal = {Physical Review X},
  volume  = {4},
  pages   = {031027},
  year    = {2014},
  doi     = {10.1103/PhysRevX.4.031027}
}

@article{GonzalezBallestero2024Projector,
  author = {Gonzalez-Ballestero, C.}, title = {Tutorial: Projector Approach to Master Equations for Open Quantum Systems},
  journal = {Quantum}, volume = {8}, pages = {1454}, year = {2024}, doi = {10.22331/q-2024-08-29-1454}
}

@article{Gorini1976,
  title = {Completely positive dynamical semigroups of N-level systems},
  volume = {17},
  ISSN = {1089-7658},
  url = {http://dx.doi.org/10.1063/1.522979},
  DOI = {10.1063/1.522979},
  number = {5},
  journal = {Journal of Mathematical Physics},
  publisher = {AIP Publishing},
  author = {Gorini,  Vittorio and Kossakowski,  Andrzej and Sudarshan,  E. C. G.},
  year = {1976},
  month = may,
  pages = {821--825}
}

@article{Graham1968,
  title = {The quantum-fluctuations of the optical parametric oscillator. I},
  volume = {210},
  ISSN = {0939-7922},
  url = {http://dx.doi.org/10.1007/BF01379946},
  DOI = {10.1007/bf01379946},
  number = {3},
  journal = {Zeitschrift f\"{u}r Physik A Hadrons and nuclei},
  publisher = {Springer Science and Business Media LLC},
  author = {Graham,  R. and Haken,  H.},
  year = {1968},
  month = jun,
  pages = {276--302}
}

@article{Gratsea2018,
  author = {Gratsea, A. and Nikolopoulos, G. M. and Lambropoulos, P.},
  title = {Photon-Assisted Quantum State Transfer and Entanglement Generation in Spin Chains},
  journal = {Physical Review A},
  volume = {98},
  pages = {012304},
  year = {2018},
  doi = {10.1103/PhysRevA.98.012304}
}

@book{Griffiths_mec_quant,
   title =     {Mec{\^a}nica Qu{\^a}ntica},
   author =    {David J. Griffiths},
   publisher = {Pearson Education},
   isbn =      {978-85-7605-927-1},
   year =      {2011},
   series =    {},
   edition =   {2},
   volume =    {},
   }

@article{Grossmann1991,
  author  = {Grossmann, Frank and Dittrich, Thomas and Jung, Peter and H{\"a}nggi, Peter},
  title   = {Coherent Destruction of Tunneling},
  journal = {Physical Review Letters},
  volume  = {67},
  number  = {4},
  pages   = {516--519},
  year    = {1991},
  doi     = {10.1103/PhysRevLett.67.516}
}

@article{Guerlin2007PhotonCounting,
  author = {Guerlin, C. and Bernu, J. and Del{\'e}glise, S. and Sayrin, C. and Gleyzes, S. and Kuhr, S. and Brune, M. and Raimond, J.-M. and Haroche, S.},
  title = {Progressive Field-State Collapse and Quantum Non-Demolition Photon Counting},
  journal = {Nature},
  volume = {448},
  pages = {889--893},
  year = {2007},
  doi = {10.1038/nature06057}
}

@article{Gustin2023GaugeInvariantTruncation,
  author  = {Gustin, Chris and Franke, Sebastian and Hughes, Stephen},
  title   = {Gauge-invariant theory of truncated quantum light--matter interactions in arbitrary media},
  journal = {Physical Review A},
  volume  = {107},
  pages   = {013722},
  year    = {2023},
  doi     = {10.1103/PhysRevA.107.013722},
  note    = {See also Erratum, Physical Review A 112, 039904(E) (2025)}
}

@article{Haack2021LiouvillianEP,
  author = {Khandelwal, S. and Brunner, N. and Haack, G.}, title = {Signatures of Liouvillian Exceptional Points in a Quantum Thermal Machine},
  journal = {PRX Quantum}, volume = {2}, pages = {040346}, year = {2021}, doi = {10.1103/PRXQuantum.2.040346}
}

@book{Hall2013QuantumTheory,
  author = {Hall, Brian C.},
  title = {Quantum Theory for Mathematicians},
  series = {Graduate Texts in Mathematics},
  volume = {267},
  publisher = {Springer},
  address = {New York},
  year = {2013},
  doi = {10.1007/978-1-4614-7116-5},
  url = {https://doi.org/10.1007/978-1-4614-7116-5}
}

@article{HallCresserLiAndersson2014,
  author = {Hall, Michael J. W. and Cresser, James D. and Li, Li and Andersson, Erika},
  title = {Canonical form of master equations and characterization of non-Markovianity},
  journal = {Physical Review A},
  volume = {89},
  number = {4},
  pages = {042120},
  year = {2014},
  doi = {10.1103/PhysRevA.89.042120},
  url = {https://doi.org/10.1103/PhysRevA.89.042120}
}

@article{Harrington2022EngineeredDissipation,
  author = {Harrington, P. M. and Mueller, E. J. and Murch, K. W.}, title = {Engineered Dissipation for Quantum Information Science},
  journal = {Nature Reviews Physics}, volume = {4}, pages = {660--671}, year = {2022}, doi = {10.1038/s42254-022-00494-8}
}

@article{HastingsWen2005,
  author = {Hastings, M. B. and Wen, Xiao-Gang},
  title = {Quasiadiabatic continuation of quantum states: The stability of topological ground-state degeneracy and emergent gauge invariance},
  journal = {Physical Review B},
  volume = {72},
  number = {4},
  pages = {045141},
  year = {2005},
  doi = {10.1103/PhysRevB.72.045141},
  url = {https://doi.org/10.1103/PhysRevB.72.045141}
}

@article{Heeres2015SNAP,
  author = {Heeres, R. W. and Vlastakis, B. and Holland, E. and Krastanov, S. and Albert, V. V. and Frunzio, L. and Jiang, L. and Schoelkopf, R. J.},
  title = {Cavity State Manipulation Using Photon-Number Selective Phase Gates},
  journal = {Physical Review Letters},
  volume = {115},
  number = {13},
  pages = {137002},
  year = {2015},
  doi = {10.1103/PhysRevLett.115.137002}
}

@article{HoniglDecrinis2018,
  author = {H{\"o}nigl-Decrinis, Teresa and Antonov, Ilya V. and Shaikhaidarov, Rais and Antonov, Vladimir N. and Dmitriev, Aleksei Yu. and Astafiev, Oleg V.},
  title = {Mixing of coherent waves in a single three-level artificial atom},
  journal = {Physical Review A},
  volume = {98},
  pages = {041801},
  year = {2018},
  doi = {10.1103/PhysRevA.98.041801}
}

@article{Houck2008Purcell,
  author = {Houck, A. A. and Schreier, J. A. and Johnson, B. R. and Chow, J. M. and Koch, J. and Gambetta, J. M. and Schuster, D. I. and Frunzio, L. and Devoret, M. H. and Girvin, S. M. and Schoelkopf, R. J.},
  title = {Controlling the Spontaneous Emission of a Superconducting Transmon Qubit},
  journal = {Physical Review Letters},
  volume = {101},
  number = {8},
  pages = {080502},
  year = {2008},
  doi = {10.1103/PhysRevLett.101.080502}
}

@article{Huang2021,
  title = {Engineering Dynamical Sweet Spots to Protect Qubits from $1/f$ Noise},
  volume = {15},
  ISSN = {2331-7019},
  url = {http://dx.doi.org/10.1103/PhysRevApplied.15.034065},
  DOI = {10.1103/physrevapplied.15.034065},
  number = {3},
  journal = {Physical Review Applied},
  publisher = {American Physical Society (APS)},
  author = {Huang,  Ziwen and Mundada,  Pranav S. and Gyenis,  Andr{\'a}s and Schuster,  David I. and Houck,  Andrew A. and Koch,  Jens},
  year = {2021},
  month = mar,
  pages = {034065}
}

@article{Huang2026,
  author = {Huang, Longxiang and Luneau, Jacquelin and Schirk, Johannes and Wallner, Florian and Schneider, Christian M. F. and Filipp, Stefan and Liegener, Klaus and Rabl, Peter},
  title = {Theory of Multiphoton Processes for Applications in Quantum Control},
  journal = {Physical Review A},
  volume = {113},
  number = {3},
  pages = {032620},
  year = {2026},
  doi = {10.1103/tnkf-ckfz}
}

@article{Hwang2015QRMQPT,
  author = {Hwang, Myung-Joong and Puebla, Ricardo and Plenio, Martin B.},
  title = {Quantum Phase Transition and Universal Dynamics in the Rabi Model},
  journal = {Physical Review Letters},
  volume = {115},
  number = {18},
  pages = {180404},
  year = {2015},
  doi = {10.1103/PhysRevLett.115.180404}
}

@article{Irish2007,
  title = {Generalized Rotating-Wave Approximation for Arbitrarily Large Coupling},
  author = {Irish, E. K.},
  journal = {Phys. Rev. Lett.},
  volume = {99},
  issue = {17},
  pages = {173601},
  numpages = {4},
  year = {2007},
  month = {Oct},
  publisher = {American Physical Society},
  doi = {10.1103/PhysRevLett.99.173601},
  url = {https://link.aps.org/doi/10.1103/PhysRevLett.99.173601}
}

@article{Jager2022BosonicModes,
  author = {J{\"a}ger, S. B. and Schmit, T. and Morigi, G. and Holland, M. J. and Betzholz, R.}, title = {Lindblad Master Equations for Quantum Systems Coupled to Dissipative Bosonic Modes},
  journal = {Physical Review Letters}, volume = {129}, pages = {063601}, year = {2022}, doi = {10.1103/PhysRevLett.129.063601}
}

@article{James2000,
  doi = {10.1002/1521-3978(200009)48:9/11<823::aid-prop823>3.0.co;2-m},
  url = {https://doi.org/10.1002/1521-3978(200009)48:9/11<823::aid-prop823>3.0.co;2-m},
  year = {2000},
  month = sep,
  publisher = {Wiley},
  volume = {48},
  number = {9-11},
  pages = {823--837},
  author = {D.F.V. James},
  title = {Quantum Computation with Hot and Cold Ions: An Assessment of Proposed Schemes},
  journal = {Fortschritte der Physik}
}

@article{James2007,
  doi = {10.1139/p07-060},
  url = {https://doi.org/10.1139/p07-060},
  year = {2007},
  month = jun,
  publisher = {Canadian Science Publishing},
  volume = {85},
  number = {6},
  pages = {625-632},
  author = {D F James and J Jerke},
  title = {Effective Hamiltonian theory and its applications in quantum information},
  journal = {Canadian Journal of Physics}
}

@article{JarvisFrain2026Tripartite,
  author  = {Jarvis-Frain, Benjamin and Schang, Andy and Quijandr{\'i}a, Fernando and Nsanzineza, Ibrahim and Dubyna, Dmytro and Chang, C. W. Sandbo and Nori, Franco and Wilson, C. M.},
  title   = {Observation of Genuine Tripartite Non-Gaussian Entanglement from a Superconducting Three-Photon Spontaneous Parametric Down-Conversion Source},
  journal = {Physical Review Letters},
  volume  = {137},
  number  = {4},
  pages   = {040201},
  year    = {2026},
  doi     = {10.1103/4ts4-qj74}
}

@ARTICLE{JaynesCummings1963,
  author={Jaynes, E.T. and Cummings, F.W.},
  journal={Proceedings of the IEEE}, 
  title={Comparison of quantum and semiclassical radiation theories with application to the beam maser}, 
  year={1963},
  volume={51},
  number={1},
  pages={89-109},
  doi={10.1109/PROC.1963.1664}}

@article{Kandel2021,
  author = {Kandel, Y. P. and others},
  title = {Adiabatic Quantum State Transfer in a Semiconductor Quantum-Dot Spin Chain},
  journal = {Nature Communications},
  volume = {12},
  pages = {2156},
  year = {2021},
  doi = {10.1038/s41467-021-22416-5}
}

@article{Karassiov1992,
  doi = {10.1007/bf01121107},
  url = {https://doi.org/10.1007/bf01121107},
  year = {1992},
  publisher = {Springer Science and Business Media {LLC}},
  volume = {13},
  number = {3},
  pages = {188--195},
  author = {V. P. Karassiov},
  title = {New Lie-algebraic structures in nonlinear problems of quantum optics and laser physics},
  journal = {Journal of Soviet Laser Research}
}

@article{Karassiov1994,
  doi = {10.1088/0305-4470/27/1/009},
  url = {https://doi.org/10.1088/0305-4470/27/1/009},
  year = {1994},
  month = jan,
  publisher = {{IOP} Publishing},
  volume = {27},
  number = {1},
  pages = {153--165},
  author = {V P Karassiov},
  title = {G-invariant polynomial extensions of Lie algebras in quantum many-body physics},
  journal = {Journal of Physics A: Mathematical and General}
}

@article{KarassiovandKlimov1994,
  doi = {10.1016/0375-9601(94)90816-8},
  url = {https://doi.org/10.1016/0375-9601(94)90816-8},
  year = {1994},
  month = jun,
  publisher = {Elsevier {BV}},
  volume = {189},
  number = {1-2},
  pages = {43--51},
  author = {Valery P. Karassiov and Andrei B. Klimov},
  title = {An algebraic approach to solving evolution problems in some nonlinear quantum models},
  journal = {Physics Letters A}
}

@book{Kato1980,
  author    = {Tosio Kato},
  title     = {Perturbation Theory for Linear Operators},
  edition   = {2},
  year      = {1980},
  publisher = {Springer-Verlag},
  address   = {New York, NY},
  isbn      = {0-387-07558-5}
}

@article{Keliri2026,
  author       = {Keliri, Andriani and Schir{\`o}, Marco},
  title        = {Sambe Approach to Floquet-Lindblad Open Quantum Systems},
  journal      = {arXiv},
  volume       = {abs/2606.09727},
  year         = {2026},
  eprint       = {2606.09727},
  archivePrefix= {arXiv},
  primaryClass = {quant-ph},
  doi          = {10.48550/arXiv.2606.09727},
  url          = {https://arxiv.org/abs/2606.09727},
}

@article{Kessler2012DissipativeSW,
  author = {Kessler, E. M.}, title = {Generalized Schrieffer-Wolff Formalism for Dissipative Systems},
  journal = {Physical Review A}, volume = {86}, pages = {012126}, year = {2012}, doi = {10.1103/PhysRevA.86.012126}
}

@article{Kierig2008,
  author  = {Kierig, E. and Schnorrberger, U. and Schietinger, A. and Tomkovi{\v c}, J. and Oberthaler, M. K.},
  title   = {Single-Particle Tunneling in Strongly Driven Double-Well Potentials},
  journal = {Physical Review Letters},
  volume  = {100},
  pages   = {190405},
  year    = {2008},
  doi     = {10.1103/PhysRevLett.100.190405}
}

@article{Klein1974,
  title = {Degenerate perturbation theory},
  volume = {61},
  ISSN = {1089-7690},
  url = {http://dx.doi.org/10.1063/1.1682018},
  DOI = {10.1063/1.1682018},
  number = {3},
  journal = {The Journal of Chemical Physics},
  publisher = {AIP Publishing},
  author = {Klein,  D. J.},
  year = {1974},
  month = aug,
  pages = {786--798}
}

@article{Klimov2002master,
  doi = {10.1088/1464-4266/5/1/304},
  url = {https://doi.org/10.1088/1464-4266/5/1/304},
  year = {2002},
  month = dec,
  publisher = {{IOP} Publishing},
  volume = {5},
  number = {1},
  pages = {34--39},
  author = {A B Klimov and J L Romero and J Delgado and L L S~nchez-Soto},
  title = {Master equations for effective Hamiltonians},
  journal = {Journal of Optics B: Quantum and Semiclassical Optics}
}

@article{Koch2007Transmon,
  author = {Koch, Jens and Yu, Terri M. and Gambetta, Jay and Houck, A. A. and Schuster, D. I. and Majer, J. and Blais, Alexandre and Devoret, M. H. and Girvin, S. M. and Schoelkopf, R. J.},
  title = {Charge-insensitive qubit design derived from the Cooper pair box},
  journal = {Physical Review A},
  volume = {76},
  number = {4},
  pages = {042319},
  year = {2007},
  doi = {10.1103/PhysRevA.76.042319},
  url = {https://doi.org/10.1103/PhysRevA.76.042319}
}

@article{Koch2023DSC,
  author = {Koch, Johannes and Hunanyan, Geram R. and Ockenfels, Till and Rico, Enrique and Solano, Enrique and Weitz, Martin},
  title = {Quantum Rabi Dynamics of Trapped Atoms Far in the Deep Strong Coupling Regime},
  journal = {Nature Communications},
  volume = {14},
  pages = {954},
  year = {2023},
  doi = {10.1038/s41467-023-36611-z}
}

@article{Kohn1959,
  title = {Theory of Bloch Electrons in a Magnetic Field: The Effective Hamiltonian},
  volume = {115},
  ISSN = {0031-899X},
  url = {http://dx.doi.org/10.1103/PhysRev.115.1460},
  DOI = {10.1103/physrev.115.1460},
  number = {6},
  journal = {Physical Review},
  publisher = {American Physical Society (APS)},
  author = {Kohn,  Walter},
  year = {1959},
  month = sep,
  pages = {1460--1478}
}

@article{Kronwald2013Squeezing,
  author = {Kronwald, A. and Marquardt, F. and Clerk, A. A.}, title = {Arbitrarily Large Steady-State Bosonic Squeezing via Dissipation},
  journal = {Physical Review A}, volume = {88}, pages = {063833}, year = {2013}, doi = {10.1103/PhysRevA.88.063833}
}

@article{kuwahara2016floquet,
  author = {Kuwahara, Tomotaka and Mori, Takashi and Saito, Keiji},
  title = {Floquet--Magnus theory and generic transient dynamics in periodically driven many-body quantum systems},
  journal = {Annals of Physics},
  volume = {367},
  pages = {96--124},
  year = {2016},
  doi = {10.1016/j.aop.2016.01.012},
  url = {https://doi.org/10.1016/j.aop.2016.01.012}
}

@article{Langford2017DSC,
  author = {Langford, N. K. and Sagastizabal, R. and Kounalakis, M. and Dickel, C. and Bruno, A. and Luthi, F. and Thoen, D. J. and Endo, A. and DiCarlo, L.},
  title = {Experimentally Simulating the Dynamics of Quantum Light and Matter at Deep-Strong Coupling},
  journal = {Nature Communications},
  volume = {8},
  pages = {1715},
  year = {2017},
  doi = {10.1038/s41467-017-01061-x}
}

@book{larson2021jaynes,
  title={The Jaynes--Cummings model and its descendants: modern research directions},
  author={Larson, Jonas and Mavrogordatos, Themistoklis},
  year={2021},
  publisher={IoP Publishing}
}

@article{LeBoit2020,
  doi = {10.1002/qute.201900140},
  url = {https://doi.org/10.1002/qute.201900140},
  year = {2020},
  month = jun,
  publisher = {Wiley},
  volume = {3},
  number = {7},
  pages = {1900140},
  author = {Alexandre Le Boit{\'{e}}},
  title = {Theoretical Methods for Ultrastrong Light{\textendash}Matter Interactions},
  journal = {Advanced Quantum Technologies}
}

@article{Lee2018,
  title = {Effective formalism for open-quantum-system dynamics: Time-coarse-graining approach},
  volume = {97},
  ISSN = {2469-9934},
  url = {http://dx.doi.org/10.1103/PhysRevA.97.012102},
  DOI = {10.1103/physreva.97.012102},
  number = {1},
  journal = {Physical Review A},
  publisher = {American Physical Society (APS)},
  author = {Lee,  Chang-Woo and Noh,  Changsuk and Kim,  Jaewan},
  year = {2018},
  month = jan,
  pages = {012102}
}

@article{Leghtas2015TwoPhoton,
  author = {Leghtas, Z. and others}, title = {Confining the State of Light to a Quantum Manifold by Engineered Two-Photon Loss},
  journal = {Science}, volume = {347}, pages = {853--857}, year = {2015}, doi = {10.1126/science.aaa2085}
}

@article{Li2018,
  title = {Floquet scattering theory based on effective Hamiltonians of driven systems},
  volume = {98},
  ISSN = {2469-9969},
  url = {http://dx.doi.org/10.1103/PhysRevB.98.121101},
  DOI = {10.1103/physrevb.98.121101},
  number = {12},
  journal = {Physical Review B},
  publisher = {American Physical Society (APS)},
  author = {Li,  Huanan and Shapiro,  Boris and Kottos,  Tsampikos},
  year = {2018},
  month = sep,
  pages = {121101}
}

@article{Li2020GaugeTightBinding,
  author = {Li, Jiajun and Gole{\v z}, Denis and Mazza, Giacomo and Millis, Andrew J. and Georges, Antoine and Eckstein, Martin},
  title = {Electromagnetic coupling in tight-binding models for strongly correlated light and matter},
  journal = {Physical Review B},
  volume = {101},
  number = {20},
  pages = {205140},
  year = {2020},
  doi = {10.1103/PhysRevB.101.205140},
  eprint = {2001.09726},
  archiveprefix = {arXiv},
  primaryclass = {cond-mat.str-el},
  url = {https://doi.org/10.1103/PhysRevB.101.205140}
}

@article{Li2023,
  title = {Single-photon switching in a Floquet waveguide-QED system with time-modulated coupling constants},
  volume = {107},
  ISSN = {2469-9934},
  url = {http://dx.doi.org/10.1103/PhysRevA.107.023720},
  DOI = {10.1103/physreva.107.023720},
  number = {2},
  journal = {Physical Review A},
  publisher = {American Physical Society (APS)},
  author = {Li,  Haozhen and Li,  Zerong and Zeng,  Ran and Hu,  Miao and Xu,  Mengmeng and Zhou,  Xuefang and Xia,  Xiuwen and Xu,  Jingping and Yang,  Yaping},
  year = {2023},
  month = feb,
  pages = {023720}
}

@article{Lindblad1976,
  title = {On the generators of quantum dynamical semigroups},
  volume = {48},
  ISSN = {1432-0916},
  url = {http://dx.doi.org/10.1007/BF01608499},
  DOI = {10.1007/bf01608499},
  number = {2},
  journal = {Communications in Mathematical Physics},
  publisher = {Springer Science and Business Media LLC},
  author = {Lindblad,  G.},
  year = {1976},
  month = jun,
  pages = {119--130}
}

@article{Liu2014,
  author = {Liu, Yu-xi and Sun, Hui-Chen and Peng, Z. H. and Miranowicz, Adam and Tsai, J. S. and Nori, Franco},
  title = {Controllable microwave three-wave mixing via a single three-level superconducting quantum circuit},
  journal = {Scientific Reports},
  volume = {4},
  pages = {7289},
  year = {2014},
  doi = {10.1038/srep07289}
}

@article{Lorenzo2013,
  author = {Lorenzo, Salvatore and Apollaro, T. J. G. and Sindona, Antonello and Plastina, Francesco},
  title = {Quantum-State Transfer via Resonant Tunneling through Local-Field-Induced Barriers},
  journal = {Physical Review A},
  volume = {87},
  pages = {042313},
  year = {2013},
  doi = {10.1103/PhysRevA.87.042313}
}

@article{Lowdin1951,
  author    = {L{\"o}wdin, Per-Olov},
  title     = {A Note on the Quantum-Mechanical Perturbation Theory},
  journal   = {The Journal of Chemical Physics},
  volume    = {19},
  number    = {11},
  pages     = {1396--1401},
  year      = {1951},
  doi       = {10.1063/1.1748067}
}

@article{Lugiato1984,
  doi = {10.1103/physreva.29.1438},
  url = {https://doi.org/10.1103/physreva.29.1438},
  year = {1984},
  month = mar,
  publisher = {American Physical Society ({APS})},
  volume = {29},
  number = {3},
  pages = {1438-1452},
  author = {L. A. Lugiato and P. Mandel and L. M. Narducci},
  title = {Adiabatic elimination in nonlinear dynamical systems},
  journal = {Physical Review A}
}

@article{Lupascu2007QND,
  author = {Lupa{\c s}cu, A. and Saito, S. and Picot, T. and de Groot, P. C. and Harmans, C. J. P. M. and Mooij, J. E.},
  title = {Quantum Non-Demolition Measurement of a Superconducting Two-Level System},
  journal = {Nature Physics},
  volume = {3},
  pages = {119--123},
  year = {2007},
  doi = {10.1038/nphys509}
}

@article{Lwdin1962,
  title = {Studies in Perturbation Theory. IV. Solution of Eigenvalue Problem by Projection Operator Formalism},
  volume = {3},
  ISSN = {1089-7658},
  url = {http://dx.doi.org/10.1063/1.1724312},
  DOI = {10.1063/1.1724312},
  number = {5},
  journal = {Journal of Mathematical Physics},
  publisher = {AIP Publishing},
  author = {L\"{o}wdin,  Per-Olov},
  year = {1962},
  month = sep,
  pages = {969--982}
}

@inbook{Lwdin1980,
  title = {Molecular Structure Calculations},
  ISSN = {0065-3276},
  url = {http://dx.doi.org/10.1016/S0065-3276(08)60318-4},
  DOI = {10.1016/s0065-3276(08)60318-4},
  booktitle = {Advances in Quantum Chemistry},
  publisher = {Elsevier},
  author = {L\"{o}wdin,  Per-Olov},
  year = {1980},
  pages = {263--316}
}

@article{Macieszczak2016Metastability,
  author = {Macieszczak, Katarzyna and Gu{\c t}{\u a}, M{\u a}d{\u a}lin and Lesanovsky, Igor and Garrahan, Juan P.},
  title = {Towards a Theory of Metastability in Open Quantum Dynamics},
  journal = {Physical Review Letters},
  volume = {116},
  number = {24},
  pages = {240404},
  year = {2016},
  doi = {10.1103/PhysRevLett.116.240404},
  eprint = {1512.05801},
  archiveprefix = {arXiv},
  primaryclass = {quant-ph},
  url = {https://doi.org/10.1103/PhysRevLett.116.240404}
}

@article{macri2023coarse,
  title={Coarse-grained effective Hamiltonian via the Magnus Expansion for a three-level system},
  author={Macr{\`\i}, Nicola and Giannelli, Luigi and Paladino, Elisabetta and Falci, Giuseppe},
  journal={Entropy},
  volume={25},
  number={2},
  pages={234},
  year={2023},
  publisher={MDPI},
  doi = {10.3390/e25020234}
}

@article{Magnus1954,
  author    = {Wilhelm Magnus},
  title     = {On the exponential solution of differential equations for a linear operator},
  journal   = {Communications on Pure and Applied Mathematics},
  volume    = {7},
  number    = {4},
  pages     = {649--673},
  year      = {1954},
  publisher = {Wiley},
  doi       = {10.1002/cpa.3160070404}
}

@article{Majer2007CavityBus,
  author = {Majer, J. and Chow, J. M. and Gambetta, J. M. and Koch, J. and Johnson, B. R. and Schreier, J. A. and Frunzio, L. and Schuster, D. I. and Houck, A. A. and Wallraff, A. and Blais, A. and Devoret, M. H. and Girvin, S. M. and Schoelkopf, R. J.},
  title = {Coupling Superconducting Qubits via a Cavity Bus},
  journal = {Nature},
  volume = {449},
  pages = {443--447},
  year = {2007},
  doi = {10.1038/nature06184}
}

@article{Manzano2020,
  title = {A short introduction to the Lindblad master equation},
  volume = {10},
  ISSN = {2158-3226},
  url = {http://dx.doi.org/10.1063/1.5115323},
  DOI = {10.1063/1.5115323},
  number = {2},
  journal = {AIP Advances},
  publisher = {AIP Publishing},
  author = {Manzano,  Daniel},
  year = {2020},
  month = feb,
  pages = {025106}
}

@article{MedinaGonzlez2020,
  doi = {10.1103/physreva.101.062108},
  url = {https://doi.org/10.1103/physreva.101.062108},
  year = {2020},
  month = jun,
  publisher = {American Physical Society ({APS})},
  volume = {101},
  number = {6},
  author = {P. U. Medina Gonz{\'{a}}lez and I. Ramos-Prieto and B. M. Rodr{\'{\i}}guez-Lara},
  title = {Heat-flow reversal in a trapped-ion simulator},
  journal = {Physical Review A},
  pages = {062108}
}

@article{michishita2020equivalence,
  title = {Equivalence of Effective Non-Hermitian Hamiltonians in the Context of Open Quantum Systems and Strongly Correlated Electron Systems},
  author = {Michishita, Yoshihiro and Peters, Robert},
  journal = {Phys. Rev. Lett.},
  volume = {124},
  issue = {19},
  pages = {196401},
  numpages = {6},
  year = {2020},
  month = {5},
  publisher = {American Physical Society},
  doi = {10.1103/PhysRevLett.124.196401},
  url = {https://link.aps.org/doi/10.1103/PhysRevLett.124.196401}
}

@article{Mikami2016,
  title = {Brillouin-Wigner theory for high-frequency expansion in periodically driven systems: Application to Floquet topological insulators},
  volume = {93},
  ISSN = {2469-9969},
  url = {http://dx.doi.org/10.1103/PhysRevB.93.144307},
  DOI = {10.1103/physrevb.93.144307},
  number = {14},
  journal = {Physical Review B},
  publisher = {American Physical Society (APS)},
  author = {Mikami,  Takahiro and Kitamura,  Sota and Yasuda,  Kenji and Tsuji,  Naoto and Oka,  Takashi and Aoki,  Hideo},
  year = {2016},
  month = apr,
  pages = {144307}
}

@article{Milonni1995,
  doi = {10.1103/physreva.52.1525},
  url = {https://doi.org/10.1103/physreva.52.1525},
  year = {1995},
  month = aug,
  publisher = {American Physical Society ({APS})},
  volume = {52},
  number = {2},
  pages = {1525--1537},
  author = {P. W. Milonni and D. F. V. James and H. Fearn},
  title = {Photodetection and causality in quantum optics},
  journal = {Physical Review A}
}

@article{Minganti2019Exceptional,
  author = {Minganti, Fabrizio and Miranowicz, Adam and Chhajlany, Ravindra W. and Nori, Franco},
  title = {Quantum exceptional points of non-Hermitian Hamiltonians and Liouvillians: The effects of quantum jumps},
  journal = {Physical Review A},
  volume = {100},
  number = {6},
  pages = {062131},
  year = {2019},
  doi = {10.1103/PhysRevA.100.062131},
  eprint = {1909.11619},
  archiveprefix = {arXiv},
  primaryclass = {quant-ph},
  url = {https://doi.org/10.1103/PhysRevA.100.062131}
}

@article{Minganti2020HybridLiouvillian,
  author = {Minganti, F. and Miranowicz, A. and Chhajlany, R. W. and Arkhipov, I. I. and Nori, F.}, title = {Hybrid-Liouvillian Formalism Connecting Exceptional Points of Non-Hermitian Hamiltonians and Liouvillians via Postselection of Quantum Trajectories},
  journal = {Physical Review A}, volume = {101}, pages = {062112}, year = {2020}, doi = {10.1103/PhysRevA.101.062112}
}

@article{Mirrahimi2009,
  doi = {10.1109/tac.2009.2015542},
  url = {https://doi.org/10.1109/tac.2009.2015542},
  year = {2009},
  month = jun,
  publisher = {Institute of Electrical and Electronics Engineers ({IEEE})},
  volume = {54},
  number = {6},
  pages = {1325--1329},
  author = {M. Mirrahimi and P. Rouchon},
  title = {Singular Perturbations and Lindblad-Kossakowski Differential Equations},
  journal = {{IEEE} Transactions on Automatic Control}
}

@article{Mlmer1999,
  title = {Multiparticle Entanglement of Hot Trapped Ions},
  volume = {82},
  ISSN = {1079-7114},
  DOI = {10.1103/physrevlett.82.1835},
  number = {9},
  journal = {Physical Review Letters},
  publisher = {American Physical Society (APS)},
  author = {M{\o}lmer,  Klaus and S{\o}rensen,  Anders},
  year = {1999},
  month = mar,
  pages = {1835--1838}
}

@article{Moan2007,
  title = {Convergence of the Magnus Series},
  volume = {8},
  ISSN = {1615-3383},
  url = {http://dx.doi.org/10.1007/s10208-007-9010-0},
  DOI = {10.1007/s10208-007-9010-0},
  number = {3},
  journal = {Foundations of Computational Mathematics},
  publisher = {Springer Science and Business Media LLC},
  author = {Moan,  Per Christian and Niesen,  Jitse},
  year = {2007},
  month = nov,
  pages = {291--301}
}

@book{Moiseyev2011,
  author = {Moiseyev, Nimrod},
  title = {Non-Hermitian Quantum Mechanics},
  publisher = {Cambridge University Press},
  address = {Cambridge},
  year = {2011},
  doi = {10.1017/CBO9780511976186},
  url = {https://doi.org/10.1017/CBO9780511976186}
}

@article{MoraesNeto2011,
  author = {de Moraes Neto, G. D. and de Ponte, M. A. and Moussa, M. H. Y.},
  title = {Engineering Interactions for Quasiperfect Transfer of Polariton States through Nonideal Bosonic Networks of Distinct Topologies},
  journal = {Physical Review A},
  volume = {84},
  pages = {032339},
  year = {2011},
  doi = {10.1103/PhysRevA.84.032339}
}

@article{MoraesNeto2012,
  author = {Neto, G. D. M. and de Ponte, M. A. and Moussa, M. H. Y.},
  title = {Nonlocal Dissipative Tunneling for High-Fidelity Quantum-State Transfer between Distant Parties},
  journal = {Physical Review A},
  volume = {85},
  pages = {052303},
  year = {2012},
  doi = {10.1103/PhysRevA.85.052303}
}

@article{MoriShirai2020Gap,
  author = {Mori, Takashi and Shirai, Tatsuhiko},
  title = {Resolving a Discrepancy between Liouvillian Gap and Relaxation Time in Boundary-Dissipated Quantum Many-Body Systems},
  journal = {Physical Review Letters},
  volume = {125},
  number = {23},
  pages = {230604},
  year = {2020},
  doi = {10.1103/PhysRevLett.125.230604},
  eprint = {2006.10953},
  archiveprefix = {arXiv},
  primaryclass = {cond-mat.stat-mech},
  url = {https://doi.org/10.1103/PhysRevLett.125.230604}
}

@article{MoriShirai2023SymmetrizedGap,
  author = {Mori, Takashi and Shirai, Tatsuhiko},
  title = {Symmetrized Liouvillian Gap in Markovian Open Quantum Systems},
  journal = {Physical Review Letters},
  volume = {130},
  number = {23},
  pages = {230404},
  year = {2023},
  doi = {10.1103/PhysRevLett.130.230404},
  eprint = {2212.06317},
  archiveprefix = {arXiv},
  primaryclass = {quant-ph},
  url = {https://doi.org/10.1103/PhysRevLett.130.230404}
}

@article{Nakajima1958,
  author = {Nakajima, Sadao},
  title = {On Quantum Theory of Transport Phenomena: Steady Diffusion},
  journal = {Progress of Theoretical Physics},
  volume = {20},
  number = {6},
  pages = {948--959},
  year = {1958},
  doi = {10.1143/PTP.20.948},
  url = {https://doi.org/10.1143/PTP.20.948}
}

@article{Nelson1959AnalyticVectors,
  author = {Nelson, Edward},
  title = {Analytic Vectors},
  journal = {Annals of Mathematics},
  volume = {70},
  number = {3},
  pages = {572--615},
  year = {1959},
  doi = {10.2307/1970331},
  url = {https://doi.org/10.2307/1970331}
}

@article{Nguyen2024,
  title = {Programmable Heisenberg interactions between Floquet qubits},
  volume = {20},
  ISSN = {1745-2481},
  url = {http://dx.doi.org/10.1038/s41567-023-02326-7},
  DOI = {10.1038/s41567-023-02326-7},
  number = {2},
  journal = {Nature Physics},
  publisher = {Springer Science and Business Media LLC},
  author = {Nguyen,  Long B. and Kim,  Yosep and Hashim,  Akel and Goss,  Noah and Marinelli,  Brian and Bhandari,  Bibek and Das,  Debmalya and Naik,  Ravi K. and Kreikebaum,  John Mark and Jordan,  Andrew N. and Santiago,  David I. and Siddiqi,  Irfan},
  year = {2024},
  month = Jan,
  pages = {240--246}
}

@book{NielsenChuang2010,
  author    = {Michael A. Nielsen and Isaac L. Chuang},
  title     = {Quantum Computation and Quantum Information},
  edition   = {10th Anniversary Edition},
  year      = {2010},
  publisher = {Cambridge University Press},
  address   = {Cambridge},
  isbn      = {978-1-107-00217-3}
}

@article{Nogues1999SinglePhoton,
  author = {Nogues, G. and Rauschenbeutel, A. and Osnaghi, S. and Brune, M. and Raimond, J. M. and Haroche, S.},
  title = {Seeing a Single Photon without Destroying It},
  journal = {Nature},
  volume = {400},
  pages = {239--242},
  year = {1999},
  doi = {10.1038/22275}
}

@article{Ofek2016BosonicQEC,
  author = {Ofek, N. and Petrenko, A. and Heeres, R. and Reinhold, P. and Leghtas, Z. and Vlastakis, B. and Liu, Y. and Frunzio, L. and Girvin, S. M. and Jiang, L. and Mirrahimi, M. and Devoret, M. H. and Schoelkopf, R. J.},
  title = {Extending the Lifetime of a Quantum Bit with Error Correction in Superconducting Circuits},
  journal = {Nature},
  volume = {536},
  pages = {441--445},
  year = {2016},
  doi = {10.1038/nature18949}
}

@article{Oh2011,
  author = {Oh, Sangchul and Wu, Lian-Ao and Shim, Yun-Pil and Fei, Jianjia and Friesen, Mark and Hu, Xuedong},
  title = {Heisenberg Spin Bus as a Robust Transmission Line for Quantum-State Transfer},
  journal = {Physical Review A},
  volume = {84},
  pages = {022330},
  year = {2011},
  doi = {10.1103/PhysRevA.84.022330}
}

@article{Paganelli2013,
  author = {Paganelli, Simone and Lorenzo, Salvatore and Apollaro, T. J. G. and Plastina, Francesco and Giorgi, Gian Luca},
  title = {Routing Quantum Information in Spin Chains},
  journal = {Physical Review A},
  volume = {87},
  pages = {062309},
  year = {2013},
  doi = {10.1103/PhysRevA.87.062309}
}

@article{Paik2016RIPGate,
  author = {Paik, H. and Mezzacapo, A. and Sandberg, M. and McClure, D. T. and Abdo, B. and C{\'o}rcoles, A. D. and Dial, O. and Bogorin, D. F. and Plourde, B. L. T. and Steffen, M. and Cross, A. W. and Gambetta, J. M. and Chow, J. M.},
  title = {Experimental Demonstration of a Resonator-Induced Phase Gate in a Multiqubit Circuit-QED System},
  journal = {Physical Review Letters},
  volume = {117},
  number = {25},
  pages = {250502},
  year = {2016},
  doi = {10.1103/PhysRevLett.117.250502}
}

@article{PhysRevA.91.013814,
  title = {Generalized rotating-wave approximation for the two-qubit quantum Rabi model},
  author = {Zhang, Yu-Yu and Chen, Qing-Hu},
  journal = {Phys. Rev. A},
  volume = {91},
  issue = {1},
  pages = {013814},
  numpages = {7},
  year = {2015},
  month = {Jan},
  publisher = {American Physical Society},
  doi = {10.1103/PhysRevA.91.013814},
  url = {https://link.aps.org/doi/10.1103/PhysRevA.91.013814}
}

@article{PhysRevA.96.063821,
  title = {Analytical solutions by squeezing to the anisotropic Rabi model in the nonperturbative deep-strong-coupling regime},
  author = {Zhang, Yu-Yu and Chen, Xiang-You},
  journal = {Phys. Rev. A},
  volume = {96},
  issue = {6},
  pages = {063821},
  numpages = {7},
  year = {2017},
  month = {Dec},
  publisher = {American Physical Society},
  doi = {10.1103/PhysRevA.96.063821},
  url = {https://link.aps.org/doi/10.1103/PhysRevA.96.063821}
}

@article{PlenioKnight1998QuantumJumps,
  author = {Plenio, M. B. and Knight, P. L.}, title = {The Quantum-Jump Approach to Dissipative Dynamics in Quantum Optics},
  journal = {Reviews of Modern Physics}, volume = {70}, pages = {101--144}, year = {1998}, doi = {10.1103/RevModPhys.70.101}
}

@article{Poyatos1996Reservoir,
  author = {Poyatos, J. F. and Cirac, J. I. and Zoller, P.},
  title = {Quantum Reservoir Engineering with Laser Cooled Trapped Ions},
  journal = {Physical Review Letters},
  volume = {77},
  pages = {4728--4731},
  year = {1996},
  doi = {10.1103/PhysRevLett.77.4728}
}

@article{Prado2006,
  doi = {10.1103/physreva.73.043803},
  url = {https://doi.org/10.1103/physreva.73.043803},
  year = {2006},
  month = apr,
  publisher = {American Physical Society ({APS})},
  volume = {73},
  number = {4},
  author = {F. O. Prado and N. G. de Almeida and M. H. Y. Moussa and C. J. Villas-B{\^{o}}as},
  title = {Bilinear and quadratic Hamiltonians in two-mode cavity quantum electrodynamics},
  journal = {Physical Review A},
  pages = {043803}
}

@article{Qing2025OptimalDynamicalGauge,
  author  = {Qing, Yuqi and Li, Liangsheng and You, Wen-Long and Lan, Yueheng and Liu, Maoxin},
  title   = {Optimal Dynamical Gauge in the Quantum Rabi Model},
  journal = {Chinese Physics Letters},
  volume  = {42},
  number  = {2},
  pages   = {020501},
  year    = {2025},
  doi     = {10.1088/0256-307X/42/2/020501}
}

@article{Rabi1936,
  title = {On the Process of Space Quantization},
  volume = {49},
  ISSN = {0031-899X},
  url = {http://dx.doi.org/10.1103/PhysRev.49.324},
  DOI = {10.1103/physrev.49.324},
  number = {4},
  journal = {Physical Review},
  publisher = {American Physical Society (APS)},
  author = {Rabi,  I. I.},
  year = {1936},
  month = feb,
  pages = {324--328}
}

@article{reascos2024universal,
  author        = {Reascos, Leander and Diotallevi, Giovanni Francesco and Benito, M{\'o}nica},
  title         = {Closed-Form Generators for Perturbative Transformations in Static and Periodically Driven Quantum Systems},
  journal       = {arXiv preprint arXiv:2411.11535},
  year          = {2024},
  eprint        = {2411.11535},
  archivePrefix = {arXiv},
  primaryClass  = {quant-ph},
  doi           = {10.48550/arXiv.2411.11535},
  url           = {https://arxiv.org/abs/2411.11535}
}

@book{ReedSimonI,
  author = {Reed, Michael and Simon, Barry},
  title = {Methods of Modern Mathematical Physics. I: Functional Analysis},
  edition = {Revised and Enlarged},
  publisher = {Academic Press},
  address = {San Diego},
  year = {1980},
  isbn = {9780125850506}
}

@article{Reiter2012Effective,
  author = {Reiter, F. and S{\o}rensen, A. S.}, title = {Effective Operator Formalism for Open Quantum Systems},
  journal = {Physical Review A}, volume = {85}, pages = {032111}, year = {2012}, doi = {10.1103/PhysRevA.85.032111}
}

@article{Ren2019,
  title = {A method to calculate effective Hamiltonians in quantum information*},
  volume = {28},
  ISSN = {1674-1056},
  url = {http://dx.doi.org/10.1088/1674-1056/ab4d43},
  DOI = {10.1088/1674-1056/ab4d43},
  number = {11},
  journal = {Chinese Physics B},
  publisher = {IOP Publishing},
  author = {Ren,  Jun-Hang and Ye,  Ming-Yong and Lin,  Xiu-Min},
  year = {2019},
  month = nov,
  pages = {110305}
}

@article{RivasHuelgaPlenio2014,
  author = {Rivas, {\'A}ngel and Huelga, Susana F. and Plenio, Martin B.},
  title = {Quantum non-Markovianity: characterization, quantification and detection},
  journal = {Reports on Progress in Physics},
  volume = {77},
  number = {9},
  pages = {094001},
  year = {2014},
  doi = {10.1088/0034-4885/77/9/094001},
  url = {https://doi.org/10.1088/0034-4885/77/9/094001}
}

@article{roccati2022non,
  author = {Roccati, Federico and Palma, G. Massimo and Bagarello, Fabio and Ciccarello, Francesco},
  title = {Non-Hermitian Physics and Master Equations},
  journal = {Open Systems \& Information Dynamics},
  volume = {29},
  number = {1},
  pages = {2250004},
  year = {2022},
  doi = {10.1142/S1230161222500044},
  eprint = {2201.05367},
  archiveprefix = {arXiv},
  primaryclass = {quant-ph},
  url = {https://doi.org/10.1142/S1230161222500044}
}

@article{Rodrigues2006,
  doi = {10.1103/physreva.74.063811},
  url = {https://doi.org/10.1103/physreva.74.063811},
  year = {2006},
  month = dec,
  publisher = {American Physical Society ({APS})},
  volume = {74},
  number = {6},
  author = {R. L. Rodrigues and M. H. Y. Moussa and C. J. Villas-B{\^{o}}as},
  title = {Engineering phonon-photon interactions with a driven trapped ion in a cavity},
  journal = {Physical Review A},
  pages = {063811}
}

@article{FornDaz2010,
  title = {Observation of the Bloch-Siegert Shift in a Qubit-Oscillator System in the Ultrastrong Coupling Regime},
  volume = {105},
  ISSN = {1079-7114},
  url = {http://dx.doi.org/10.1103/PhysRevLett.105.237001},
  DOI = {10.1103/physrevlett.105.237001},
  number = {23},
  journal = {Physical Review Letters},
  publisher = {American Physical Society (APS)},
  author = {Forn-Díaz,  P. and Lisenfeld,  J. and Marcos,  D. and García-Ripoll,  J. J. and Solano,  E. and Harmans,  C. J. P. M. and Mooij,  J. E.},
  year = {2010},
  month = Nov 
}

@article{Rojkov2026NLRE,
  author = {Rojkov, I. and Simoni, M. and Zapusek, E. and Reiter, F. and Home, J.}, title = {Stabilization of Cat-State Manifolds Using Nonlinear Reservoir Engineering},
  journal = {Physical Review X}, volume = {16}, pages = {011056}, year = {2026}, doi = {10.1103/d557-9lr4}
}

@article{Rosado2023,
  title = {Comment on ``Generalized James' effective Hamiltonian method''},
  volume = {108},
  ISSN = {2469-9934},
  url = {http://dx.doi.org/10.1103/PhysRevA.108.066201},
  DOI = {10.1103/physreva.108.066201},
  number = {6},
  journal = {Physical Review A},
  publisher = {American Physical Society (APS)},
  author = {Rosado,  W. and Arraut,  Ivan},
  year = {2023},
  month = dec,
  pages = {066201}
}

@article{rotter2009non,
  author = {Rotter, Ingrid},
  title = {A non-Hermitian Hamilton operator and the physics of open quantum systems},
  journal = {Journal of Physics A: Mathematical and Theoretical},
  volume = {42},
  number = {15},
  pages = {153001},
  year = {2009},
  doi = {10.1088/1751-8113/42/15/153001},
  url = {https://doi.org/10.1088/1751-8113/42/15/153001}
}

@article{Roy2025,
  author = {Roy, F. A. and others},
  title = {Parity-Dependent State Transfer for Direct Entanglement Generation},
  journal = {Nature Communications},
  volume = {16},
  pages = {2660},
  year = {2025},
  doi = {10.1038/s41467-025-57818-2}
}

@article{RubioAbadal2020,
  author  = {Rubio-Abadal, Antonio and Ippoliti, Matteo and Hollerith, Simon and Wei, David and Rui, Jun and Sondhi, S. L. and Khemani, Vedika and Gross, Christian and Bloch, Immanuel},
  title   = {Floquet Prethermalization in a Bose--Hubbard System},
  journal = {Physical Review X},
  volume  = {10},
  pages   = {021044},
  year    = {2020},
  doi     = {10.1103/PhysRevX.10.021044}
}

@article{SA1,
    author = {Araujo, Vanilse S. and Coutinho, F. A. B. and Fernando Perez, J.},
    title = {Operator domains and self-adjoint operators},
    journal = {American Journal of Physics},
    volume = {72},
    number = {2},
    pages = {203-213},
    year = {2004},
    month = {02},
    issn = {0002-9505},
    doi = {10.1119/1.1624111},
    url = {https://doi.org/10.1119/1.1624111},
}

@article{Blais2004,
  title = {Cavity quantum electrodynamics for superconducting electrical circuits: An architecture for quantum computation},
  volume = {69},
  ISSN = {1094-1622},
  url = {http://dx.doi.org/10.1103/PhysRevA.69.062320},
  DOI = {10.1103/physreva.69.062320},
  number = {6},
  journal = {Physical Review A},
  publisher = {American Physical Society (APS)},
  author = {Blais,  Alexandre and Huang,  Ren-Shou and Wallraff,  Andreas and Girvin,  S. M. and Schoelkopf,  R. J.},
  year = {2004},
  month = {June} 
}

@article{SA2,
    author = {Bonneau, Guy and Faraut, Jacques and Valent, Galliano},
    title = {Self-adjoint extensions of operators and the teaching of quantum mechanics},
    journal = {American Journal of Physics},
    volume = {69},
    number = {3},
    pages = {322-331},
    year = {2001},
    month = {03},
    issn = {0002-9505},
    doi = {10.1119/1.1328351},
}

@article{Saideh2020Projection,
  author = {Saideh, I. and Finkelstein-Shapiro, D. and No{\^u}s, C. and Pullerits, T. and Keller, A.}, title = {Projection-Based Adiabatic Elimination of Bipartite Open Quantum Systems},
  journal = {Physical Review A}, volume = {102}, pages = {032212}, year = {2020}, doi = {10.1103/PhysRevA.102.032212}
}

@book{Sakurai,
  author    = {Sakurai, J. J. and Napolitano, Jim},
  title     = {Mec{\^a}nica Qu{\^a}ntica Moderna},
  edition   = {2},
  publisher = {Bookman},
  year      = {2013},
  isbn      = {9788565837095}
}

@article{Sambe1973,
  title = {Steady States and Quasienergies of a Quantum-Mechanical System in an Oscillating Field},
  volume = {7},
  ISSN = {0556-2791},
  url = {http://dx.doi.org/10.1103/PhysRevA.7.2203},
  DOI = {10.1103/physreva.7.2203},
  number = {6},
  journal = {Physical Review A},
  publisher = {American Physical Society (APS)},
  author = {Sambe,  Hideo},
  year = {1973},
  month = jun,
  pages = {2203--2213}
}

@article{Sameti2019,
  title = {Floquet engineering in superconducting circuits: From arbitrary spin-spin interactions to the Kitaev honeycomb model},
  volume = {99},
  ISSN = {2469-9934},
  url = {http://dx.doi.org/10.1103/PhysRevA.99.012333},
  DOI = {10.1103/physreva.99.012333},
  number = {1},
  journal = {Physical Review A},
  publisher = {American Physical Society (APS)},
  author = {Sameti,  Mahdi and Hartmann,  Michael J.},
  year = {2019},
  month = Jan,
  pages = {012333}
}

@article{Schafer2020SelfPolarization,
  author = {Sch{\"a}fer, Christian and Ruggenthaler, Michael and Rokaj, Vasil and Rubio, Angel},
  title = {Relevance of the Quadratic Diamagnetic and Self-Polarization Terms in Cavity Quantum Electrodynamics},
  journal = {ACS Photonics},
  volume = {7},
  number = {4},
  pages = {975--990},
  year = {2020},
  doi = {10.1021/acsphotonics.9b01649},
  url = {https://doi.org/10.1021/acsphotonics.9b01649}
}

@article{Schnell2020,
  author = {Schnell, Alexander and Eckardt, Andr{\'e} and Denisov, Sergey},
  title = {Is there a Floquet Lindbladian?},
  journal = {Physical Review B},
  volume = {101},
  number = {10},
  pages = {100301(R)},
  year = {2020},
  doi = {10.1103/PhysRevB.101.100301},
  url = {https://doi.org/10.1103/PhysRevB.101.100301}
}

@article{Schrieffer1966,
  title = {Relation between the Anderson and Kondo Hamiltonians},
  volume = {149},
  ISSN = {0031-899X},
  url = {http://dx.doi.org/10.1103/PhysRev.149.491},
  DOI = {10.1103/physrev.149.491},
  number = {2},
  journal = {Physical Review},
  publisher = {American Physical Society (APS)},
  author = {Schrieffer,  J. R. and Wolff,  P. A.},
  year = {1966},
  month = sep,
  pages = {491--492}
}

@article{Schuster2005,
  title = {ac Stark Shift and Dephasing of a Superconducting Qubit Strongly Coupled to a Cavity Field},
  volume = {94},
  ISSN = {1079-7114},
  url = {http://dx.doi.org/10.1103/PhysRevLett.94.123602},
  DOI = {10.1103/physrevlett.94.123602},
  number = {12},
  journal = {Physical Review Letters},
  publisher = {American Physical Society (APS)},
  author = {Schuster,  D. I. and Wallraff,  A. and Blais,  A. and Frunzio,  L. and Huang,  R.-S. and Majer,  J. and Girvin,  S. M. and Schoelkopf,  R. J.},
  year = {2005},
  month = mar,
  pages = {123602}
}

@article{Schuster2007PhotonNumber,
  author = {Schuster, D. I. and Houck, A. A. and Schreier, J. A. and Wallraff, A. and Gambetta, J. M. and Blais, A. and Frunzio, L. and Majer, J. and Johnson, B. and Devoret, M. H. and Girvin, S. M. and Schoelkopf, R. J.},
  title = {Resolving Photon Number States in a Superconducting Circuit},
  journal = {Nature},
  volume = {445},
  pages = {515--518},
  year = {2007},
  doi = {10.1038/nature05461}
}

@book{scully_zubairy,
   title =     {Quantum Optics},
   author =    {Marlan O. Scully and M. Suhail Zubairy},
   publisher = {Cambridge University Press},
   isbn =      {9780521435956},
   year =      {1997},
   series =    {},
   edition =   {1st},
   volume =    {},
}

@article{Serra2005,
  author = {Serra, R. M. and Villas-B{\^o}as, C. J. and de Almeida, N. G. and Moussa, M. H. Y.},
  title = {Frequency up- and down-conversions in two-mode cavity quantum electrodynamics},
  journal = {Physical Review A},
  volume = {71},
  pages = {045802},
  year = {2005},
  doi = {10.1103/PhysRevA.71.045802}
}

@article{Shao2017,
  doi = {10.1103/physreva.95.032124},
  url = {https://doi.org/10.1103/physreva.95.032124},
  year = {2017},
  month = mar,
  publisher = {American Physical Society ({APS})},
  volume = {95},
  number = {3},
  author = {Wenjun Shao and Chunfeng Wu and Xun-Li Feng},
  title = {Generalized James{\textquotesingle} effective Hamiltonian method},
  journal = {Physical Review A},
  pages = {032124}
}

@article{Shapiro1972,
  title = {Dynamics of Dissociation. I. Computational Investigation of Unimolecular Breakdown Processes},
  volume = {56},
  ISSN = {1089-7690},
  url = {http://dx.doi.org/10.1063/1.1677584},
  DOI = {10.1063/1.1677584},
  number = {6},
  journal = {The Journal of Chemical Physics},
  publisher = {AIP Publishing},
  author = {Shapiro,  Moshe},
  year = {1972},
  month = mar,
  pages = {2582--2591}
}

@article{Shavitt1980,
  title = {Quasidegenerate perturbation theories. A canonical van Vleck formalism and its relationship to other approaches},
  volume = {73},
  ISSN = {1089-7690},
  url = {http://dx.doi.org/10.1063/1.440050},
  DOI = {10.1063/1.440050},
  number = {11},
  journal = {The Journal of Chemical Physics},
  publisher = {AIP Publishing},
  author = {Shavitt,  Isaiah and Redmon,  Lynn T.},
  year = {1980},
  month = Dec,
  pages = {5711--5717}
}

@article{Srensen1999,
  title = {Quantum Computation with Ions in Thermal Motion},
  volume = {82},
  ISSN = {1079-7114},
  DOI = {10.1103/physrevlett.82.1971},
  number = {9},
  journal = {Physical Review Letters},
  publisher = {American Physical Society (APS)},
  author = {S{\o}rensen,  Anders and M{\o}lmer,  Klaus},
  year = {1999},
  month = mar,
  pages = {1971--1974}
}

@article{Srensen2000,
  doi = {10.1103/physreva.62.022311},
  url = {https://doi.org/10.1103/physreva.62.022311},
  year = {2000},
  month = jul,
  publisher = {American Physical Society ({APS})},
  volume = {62},
  number = {2},
  author = {Anders S{\o}rensen and Klaus M{\o}lmer},
  title = {Entanglement and quantum computation with ions in thermal motion},
  journal = {Physical Review A},
  pages = {022311}
}

@article{stokes2019gauge,
  author = {Stokes, Adam and Nazir, Ahsan},
  title = {Gauge ambiguities imply Jaynes--Cummings physics remains valid in ultrastrong coupling {QED}},
  journal = {Nature Communications},
  volume = {10},
  pages = {499},
  year = {2019},
  doi = {10.1038/s41467-018-08101-0},
  url = {https://doi.org/10.1038/s41467-018-08101-0}
}

@article{stokes2022implications,
  author = {Stokes, Adam and Nazir, Ahsan},
  title = {Implications of gauge freedom for nonrelativistic quantum electrodynamics},
  journal = {Reviews of Modern Physics},
  volume = {94},
  number = {4},
  pages = {045003},
  year = {2022},
  doi = {10.1103/RevModPhys.94.045003},
  url = {https://doi.org/10.1103/RevModPhys.94.045003}
}

@article{StokesNazir2024GaugeTruncation,
  author = {Stokes, Adam and Nazir, Ahsan},
  title = {Gauge non-invariance due to material truncation in ultrastrong-coupling quantum electrodynamics},
  journal = {Nature Physics},
  volume = {20},
  number = {3},
  pages = {376--378},
  year = {2024},
  doi = {10.1038/s41567-023-02155-8},
  url = {https://doi.org/10.1038/s41567-023-02155-8}
}

@article{Sun2014ParityTracking,
  author = {Sun, L. and Petrenko, A. and Leghtas, Z. and Vlastakis, B. and Kirchmair, G. and Sliwa, K. M. and Narla, A. and Hatridge, M. and Shankar, S. and Blumoff, J. and Frunzio, L. and Mirrahimi, M. and Devoret, M. H. and Schoelkopf, R. J.},
  title = {Tracking Photon Jumps with Repeated Quantum Non-Demolition Parity Measurements},
  journal = {Nature},
  volume = {511},
  pages = {444--448},
  year = {2014},
  doi = {10.1038/nature13436}
}

@article{Takayanagi2016,
  title = {Effective interaction in unified perturbation theory},
  volume = {364},
  ISSN = {0003-4916},
  url = {http://dx.doi.org/10.1016/j.aop.2015.11.006},
  DOI = {10.1016/j.aop.2015.11.006},
  journal = {Annals of Physics},
  publisher = {Elsevier BV},
  author = {Takayanagi,  Kazuo},
  year = {2016},
  month = jan,
  pages = {200--247}
}

@article{Takayanagi2016_2,
  title = {Unified theory of effective interaction},
  volume = {372},
  ISSN = {0003-4916},
  url = {http://dx.doi.org/10.1016/j.aop.2016.04.011},
  DOI = {10.1016/j.aop.2016.04.011},
  journal = {Annals of Physics},
  publisher = {Elsevier BV},
  author = {Takayanagi,  Kazuo},
  year = {2016},
  month = sep,
  pages = {12--56}
}

@article{TavisCummings1968,
  author = {Tavis, Michael and Cummings, Frederick W.},
  title = {Exact Solution for an {$N$}-Molecule---Radiation-Field Hamiltonian},
  journal = {Physical Review},
  volume = {170},
  number = {2},
  pages = {379--384},
  year = {1968},
  doi = {10.1103/PhysRev.170.379},
  url = {https://doi.org/10.1103/PhysRev.170.379}
}

@article{taylor2020resolution,
  author = {Taylor, Michael A. D. and Mandal, Arkajit and Zhou, Wanghuai and Huo, Pengfei},
  title = {Resolution of Gauge Ambiguities in Molecular Cavity Quantum Electrodynamics},
  journal = {Physical Review Letters},
  volume = {125},
  number = {12},
  pages = {123602},
  year = {2020},
  doi = {10.1103/PhysRevLett.125.123602},
  url = {https://doi.org/10.1103/PhysRevLett.125.123602}
}

@article{TaylorMandalHuo2022,
  author = {Taylor, Michael A. D. and Mandal, Arkajit and Huo, Pengfei},
  title = {Resolving ambiguities of the mode truncation in cavity quantum electrodynamics},
  journal = {Optics Letters},
  volume = {47},
  number = {6},
  pages = {1446--1449},
  year = {2022},
  doi = {10.1364/OL.450228},
  url = {https://doi.org/10.1364/OL.450228}
}

@article{Tokieda2024CP,
  author = {Tokieda, M. and Elouard, C. and Sarlette, A. and Rouchon, P.}, title = {Complete Positivity Violation of the Reduced Dynamics in Higher-Order Quantum Adiabatic Elimination},
  journal = {Physical Review A}, volume = {109}, pages = {062206}, year = {2024}, doi = {10.1103/PhysRevA.109.062206}
}

@article{Tomonaga2025TwoArtificialAtoms,
  author  = {Tomonaga, A. and Stassi, R. and Mukai, H. and Nori, F. and Yoshihara, F. and Tsai, J. S.},
  title   = {Spectral Properties of Two Superconducting Artificial Atoms Coupled to a Resonator in the Ultrastrong Coupling Regime},
  journal = {Nature Communications},
  volume  = {16},
  pages   = {5294},
  year    = {2025},
  doi     = {10.1038/s41467-025-60589-5}
}

@article{ToorZubairy1992,
  author = {Toor, A. H. and Zubairy, M. S.},
  title = {Validity of the effective Hamiltonian in the two-photon atom-field interaction},
  journal = {Physical Review A},
  volume = {45},
  pages = {4951},
  year = {1992},
  doi = {10.1103/PhysRevA.45.4951}
}

@article{ture2024simple,
  title={Simple and General Unitarity Conserving Numerical Real-Time Propagators of the Time-Dependent Schrodinger Equation Based on Magnus Expansion},
  author={Ture, Taner M and Jang, Seogjoo J},
  journal={The Journal of Physical Chemistry A},
  volume={128},
  number={14},
  pages={2871--2882},
  year={2024},
  publisher={ACS Publications},
  doi = {10.1021/acs.jpca.3c07866}
}

@article{Vanselow2026Quartets,
  author  = {Vanselow, A. and Beauseigneur, B. and Lattier, L. and Villiers, M. and Denis, A. and Morfin, P. and Leghtas, Z. and Campagne-Ibarcq, P.},
  title   = {Dissipating Quartets of Excitations in a Superconducting Circuit},
  journal = {Physical Review X},
  volume  = {16},
  number  = {1},
  pages   = {011032},
  year    = {2026},
  doi     = {10.1103/bjpc-8xcf}
}

@article{VanVleck1929,
  title = {On $\sigma$-Type Doubling and Electron Spin in the Spectra of Diatomic Molecules},
  volume = {33},
  ISSN = {0031-899X},
  url = {http://dx.doi.org/10.1103/PhysRev.33.467},
  DOI = {10.1103/physrev.33.467},
  number = {4},
  journal = {Physical Review},
  publisher = {American Physical Society (APS)},
  author = {Van Vleck,  J. H.},
  year = {1929},
  month = apr,
  pages = {467--506}
}

@article{Villanueva2020,
  doi = {10.48550/ARXIV.2005.11298},
  url = {https://arxiv.org/abs/2005.11298},
  author = {Villanueva-Vergara,  L. and Soto-Eguibar,  F. and Moya-Cessa,  H. M.},
  title = {Effect of the nearby levels on the resonance fluorescence spectrum of the atom-field interaction},
  journal = {arXiv},
  year = {2020},
}

@article{Vitanov2017,
  doi = {10.1103/revmodphys.89.015006},
  url = {https://doi.org/10.1103/revmodphys.89.015006},
  year = {2017},
  month = mar,
  publisher = {American Physical Society ({APS})},
  volume = {89},
  number = {1},
  author = {Nikolay V. Vitanov and Andon A. Rangelov and Bruce W. Shore and Klaas Bergmann},
  title = {Stimulated Raman adiabatic passage in physics,  chemistry,  and beyond},
  journal = {Reviews of Modern Physics},
  pages = {015006}
}

@article{Wallraff2005DispersiveReadout,
  author = {Wallraff, A. and Schuster, D. I. and Blais, A. and Frunzio, L. and Majer, J. and Devoret, M. H. and Girvin, S. M. and Schoelkopf, R. J.},
  title = {Approaching Unit Visibility for Control of a Superconducting Qubit with Dispersive Readout},
  journal = {Physical Review Letters},
  volume = {95},
  number = {6},
  pages = {060501},
  year = {2005},
  doi = {10.1103/PhysRevLett.95.060501}
}

@article{Wang2015Degenerate,
  author = {Wang, Z. H. and Sun, C. P. and Li, Yong},
  title = {Microwave degenerate parametric down-conversion with a single cyclic three-level system in a circuit-QED setup},
  journal = {Physical Review A},
  volume = {91},
  pages = {043801},
  year = {2015},
  doi = {10.1103/PhysRevA.91.043801}
}

@article{Wang2017CounterRotatingTwoQubits,
  author  = {Wang, Xin and Miranowicz, Adam and Li, Hong-Rong and Nori, Franco},
  title   = {Observing Pure Effects of Counter-Rotating Terms without Ultrastrong Coupling: A Single Photon Can Simultaneously Excite Two Qubits},
  journal = {Physical Review A},
  volume  = {96},
  number  = {6},
  pages   = {063820},
  year    = {2017},
  doi     = {10.1103/PhysRevA.96.063820}
}

@article{Wang2025OneTwoPhoton,
  author  = {Wang, Shuai-Peng and Mercurio, Alberto and Ridolfo, Alessandro and Wang, Yuqing and Chen, Mo and Wang, Wenyan and Liu, Yulong and Sun, Huanying and Li, Tiefu and Nori, Franco and Savasta, Salvatore and You, J. Q.},
  title   = {Strong Coupling between a Single-Photon and a Two-Photon Fock State},
  journal = {Nature Communications},
  volume  = {16},
  pages   = {8730},
  year    = {2025},
  doi     = {10.1038/s41467-025-63783-7}
}

@book{Winkler2003,
  author    = {Winkler, Roland},
  title     = {Spin-orbit coupling effects in two-dimensional electron and hole systems},
  series    = {Springer Tracts in Modern Physics},
  volume    = {191},
  publisher = {Springer},
  address   = {Berlin},
  year      = {2003}
}

@article{Wojcik2005,
  author = {W{\'o}jcik, Antoni and {\L}uczak, Tomasz and Kurzy{\'n}ski, Pawe{\l} and Grudka, Andrzej and Gdala, Tomasz and Bednarska, Ma{\l}gorzata},
  title = {Unmodulated Spin Chains as Universal Quantum Wires},
  journal = {Physical Review A},
  volume = {72},
  pages = {034303},
  year = {2005},
  doi = {10.1103/PhysRevA.72.034303}
}

@article{WolfCirac2008,
  author = {Wolf, Michael M. and Cirac, J. Ignacio},
  title = {Dividing Quantum Channels},
  journal = {Communications in Mathematical Physics},
  volume = {279},
  number = {1},
  pages = {147--168},
  year = {2008},
  doi = {10.1007/s00220-008-0411-y},
  url = {https://doi.org/10.1007/s00220-008-0411-y}
}

@article{WolfEisertCubittCirac2008,
  author = {Wolf, Michael M. and Eisert, Jens and Cubitt, Toby S. and Cirac, J. Ignacio},
  title = {Assessing Non-Markovian Quantum Dynamics},
  journal = {Physical Review Letters},
  volume = {101},
  number = {15},
  pages = {150402},
  year = {2008},
  doi = {10.1103/PhysRevLett.101.150402},
  url = {https://doi.org/10.1103/PhysRevLett.101.150402}
}

@article{Wu1996,
  author = {Wu, Ying},
  title = {Effective Raman theory for a three-level atom in the Lambda configuration},
  journal = {Physical Review A},
  volume = {54},
  pages = {1586},
  year = {1996},
  doi = {10.1103/PhysRevA.54.1586}
}

@article{Wu2022,
title = {Approximating the two-mode two-photon Rabi model},
journal = {Physics Letters A},
volume = {422},
pages = {127779},
year = {2022},
issn = {0375-9601},
doi = {10.1016/j.physleta.2021.127779},
url = {https://www.sciencedirect.com/science/article/pii/S0375960121006435},
author = {David H. Wu and Victor V. Albert},
}

@article{Xiang2024,
  author = {Xiang, L. and others},
  title = {Enhanced Quantum State Transfer by Circumventing Quantum Chaotic Behavior},
  journal = {Nature Communications},
  volume = {15},
  pages = {4918},
  year = {2024},
  doi = {10.1038/s41467-024-48791-3}
}

@article{Yao2011,
  author = {Yao, N. Y. and Jiang, L. and Gorshkov, A. V. and Gong, Z.-X. and Zhai, A. and Duan, L.-M. and Lukin, M. D.},
  title = {Robust Quantum State Transfer in Random Unpolarized Spin Chains},
  journal = {Physical Review Letters},
  volume = {106},
  pages = {040505},
  year = {2011},
  doi = {10.1103/PhysRevLett.106.040505}
}

@article{Yoshihara2017DSC,
  author = {Yoshihara, Fumiki and Fuse, Tomoko and Ashhab, Sahel and Kakuyanagi, Kosuke and Saito, Shiro and Semba, Kouichi},
  title = {Superconducting Qubit--Oscillator Circuit beyond the Ultrastrong-Coupling Regime},
  journal = {Nature Physics},
  volume = {13},
  pages = {44--47},
  year = {2017},
  doi = {10.1038/nphys3906}
}

@article{Zhang2013,
  title = {Generalized rotating-wave approximation to biased qubit-oscillator systems},
  author = {Zhang, Yu-Yu and Chen, Qing-Hu and Zhao, Yang},
  journal = {Phys. Rev. A},
  volume = {87},
  issue = {3},
  pages = {033827},
  numpages = {6},
  year = {2013},
  month = {Mar},
  publisher = {American Physical Society},
  doi = {10.1103/PhysRevA.87.033827},
  url = {https://link.aps.org/doi/10.1103/PhysRevA.87.033827}
}

@article{Zhang2016,
  title = {Generalized squeezing rotating-wave approximation to the isotropic and anisotropic Rabi model in the ultrastrong-coupling regime},
  author = {Zhang, Yu-Yu},
  journal = {Phys. Rev. A},
  volume = {94},
  issue = {6},
  pages = {063824},
  numpages = {7},
  year = {2016},
  month = {Dec},
  publisher = {American Physical Society},
  doi = {10.1103/PhysRevA.94.063824},
  url = {https://link.aps.org/doi/10.1103/PhysRevA.94.063824}
}

@article{Zhang2019,
  author = {Zhang, Yaxing and Lester, Brian J. and Gao, Yvonne Y. and Jiang, Liang and Schoelkopf, R. J. and Girvin, S. M.},
  title = {Engineering bilinear mode coupling in circuit QED: Theory and experiment},
  journal = {Physical Review A},
  volume = {99},
  pages = {012314},
  year = {2019},
  doi = {10.1103/PhysRevA.99.012314}
}

@article{Zueco2009,
  doi = {10.1103/physreva.80.033846},
  url = {https://doi.org/10.1103/physreva.80.033846},
  year = {2009},
  month = sep,
  publisher = {American Physical Society ({APS})},
  volume = {80},
  number = {3},
  author = {David Zueco and Georg M. Reuther and Sigmund Kohler and Peter H\"{a}nggi},
  title = {Qubit-oscillator dynamics in the dispersive regime: Analytical theory beyond the rotating-wave approximation},
  journal = {Physical Review A},
  pages = {033846}
}

@article{Zwanzig1960,
  author = {Zwanzig, Robert},
  title = {Ensemble Method in the Theory of Irreversibility},
  journal = {The Journal of Chemical Physics},
  volume = {33},
  number = {5},
  pages = {1338--1341},
  year = {1960},
  doi = {10.1063/1.1731409},
  url = {https://doi.org/10.1063/1.1731409}
}

@article{Gambetta2006Measurement,
  author  = {Gambetta, Jay and Blais, Alexandre and Schuster, D. I. and
             Wallraff, A. and Frunzio, L. and Majer, J. and Devoret, M. H. and
             Girvin, S. M. and Schoelkopf, R. J.},
  title   = {Qubit-photon interactions in a cavity: Measurement-induced
             dephasing and number splitting},
  journal = {Physical Review A},
  volume  = {74},
  number  = {4},
  pages   = {042318},
  year    = {2006},
  doi     = {10.1103/PhysRevA.74.042318}
}

@article{Garziano2015Multiphoton,
  author  = {Garziano, Luigi and Stassi, Roberto and Macr{\`{\i}}, Vincenzo and
             Kockum, Anton Frisk and Savasta, Salvatore and Nori, Franco},
  title   = {Multiphoton quantum {Rabi} oscillations in ultrastrong cavity
             {QED}},
  journal = {Physical Review A},
  volume  = {92},
  number  = {6},
  pages   = {063830},
  year    = {2015},
  doi     = {10.1103/PhysRevA.92.063830}
}

@article{Ridolfo2012PhotonBlockade,
  author  = {Ridolfo, Alessandro and Leib, Martin and Savasta, Salvatore and
             Hartmann, Michael J.},
  title   = {Photon Blockade in the Ultrastrong Coupling Regime},
  journal = {Physical Review Letters},
  volume  = {109},
  number  = {19},
  pages   = {193602},
  year    = {2012},
  doi     = {10.1103/PhysRevLett.109.193602}
}

@article{Klimov2000,
  author  = {Klimov, A. B. and S{\'a}nchez-Soto, L. L.},
  title   = {Method of Small Rotations and Effective Hamiltonians in Nonlinear Quantum Optics},
  journal = {Physical Review A},
  volume  = {61},
  number  = {6},
  pages   = {063802},
  year    = {2000},
  doi     = {10.1103/PhysRevA.61.063802}
}

@article{Klimov2002,
  author  = {Klimov, A. B. and S{\'a}nchez-Soto, L. L. and Navarro, A. and Yustas, E. C.},
  title   = {Effective Hamiltonians in Quantum Optics: A Systematic Approach},
  journal = {Journal of Modern Optics},
  volume  = {49},
  number  = {13},
  pages   = {2211--2226},
  year    = {2002},
  doi     = {10.1080/09500340210134675}
}

@article{Klimov2003,
  author  = {Klimov, A. B. and Sainz, I. and Chumakov, S. M.},
  title   = {Resonance Expansion versus the Rotating-Wave Approximation},
  journal = {Physical Review A},
  volume  = {68},
  number  = {6},
  pages   = {063811},
  year    = {2003},
  doi     = {10.1103/PhysRevA.68.063811}
}

@article{ma2015three,
  author  = {Ma, Ken K. W. and Law, C. K.},
  title   = {Three-Photon Resonance and Adiabatic Passage in the Large-Detuning {Rabi} Model},
  journal = {Physical Review A},
  volume  = {92},
  number  = {2},
  pages   = {023842},
  year    = {2015},
  doi     = {10.1103/PhysRevA.92.023842}
}

@article{Stefanini2026IsLindblad,
  author = {Stefanini, Martino and Ziolkowska, Aleksandra A. and Budker, Dmitry and Poschinger, Ulrich and Schmidt-Kaler, Ferdinand and Browaeys, Antoine and Imamoglu, Atac and Chang, Darrick and Marino, Jamir},
  title = {Is Lindblad for me?},
  journal = {SciPost Physics Lecture Notes},
  pages = {129},
  year = {2026},
  doi = {10.21468/SciPostPhysLectNotes.129}
}

@article{Szankowski2023OpenSystems,
  author = {Sza{\'n}kowski, Piotr},
  title = {Introduction to the theory of open quantum systems},
  journal = {SciPost Physics Lecture Notes},
  pages = {68},
  year = {2023},
  doi = {10.21468/SciPostPhysLectNotes.68}
}

@article{Metelmann2023Parametric,
  author = {Metelmann, Anja},
  title = {Parametric couplings in engineered quantum systems},
  journal = {SciPost Physics Lecture Notes},
  pages = {66},
  year = {2023},
  doi = {10.21468/SciPostPhysLectNotes.66}
}

@article{Clerk2022Nonreciprocal,
  author = {Clerk, A. A.},
  title = {Introduction to quantum non-reciprocal interactions: From non-{H}ermitian Hamiltonians to quantum master equations and quantum feedforward schemes},
  journal = {SciPost Physics Lecture Notes},
  pages = {44},
  year = {2022},
  doi = {10.21468/SciPostPhysLectNotes.44}
}

@article{Campaioli2024MasterEquations,
  author = {Campaioli, Francesco and Cole, Jared H. and Hapuarachchi, Harini},
  title = {Quantum Master Equations: Tips and Tricks for Quantum Optics, Quantum Computing, and Beyond},
  journal = {PRX Quantum},
  volume = {5},
  pages = {020202},
  year = {2024},
  doi = {10.1103/PRXQuantum.5.020202}
}

@article{Qvarfort2025LieAlgebra,
  author = {Qvarfort, Sofia and Pikovski, Igor},
  title = {Solving Quantum Dynamics with a {L}ie-Algebra Decoupling Method},
  journal = {PRX Quantum},
  volume = {6},
  pages = {010201},
  year = {2025},
  doi = {10.1103/PRXQuantum.6.010201}
}
\end{document}